\def\Msun{$M_{\sun}$}

\def\Wm{${\rm {W}~{{m^{-2}}}}$}

\def\nh{{n_{\rm H}}}
\def\nh2{{n(\rm H_2)}}
\def\h2{${\rm H_2}$}

\def\3cm{\rm {cm^{-3}}}
\def\2cm{\rm {cm^{-2}}}
\def\s-1{\rm {s^{-1}}}
\def\etal {et al.}
\def\mum {\hbox{$\mu$m}}
\def\kms {\hbox{${\rm km\,s}^{-1}$}}
\def\Kkms {\hbox{${\rm K}\,{\rm km}\,{\rm s}^{-1}$}}
\def\ndv{\hbox{${\rm cm}^{-2}\,{\rm km}^{-1}\,{\rm s} $}}
\def\ergs{{\rm {erg}~{s^{-1}}}}

\def\hcop{\hbox{\rm {HCO$^+$}}}
\def\hthcop{\hbox{{\rm {H}${}^{13}$CO$^+$}}}
\def\hcn{\hbox{\rm {HCN}}}
\def\hthcn{\hbox{{\rm {H}${}^{13}$CN}}}
\def\hnc{\hbox{\rm {HNC}}}

\def\co{\hbox{\rm {CO}}}
\def\twco{\hbox{{\rm $^{12}$CO}}}
\def\thco{\hbox{{\rm $^{13}$CO}}}
\def\co17{\hbox{{\rm {C}${}^{17}$O}}}
\def\ceio{\hbox{{\rm {C}${}^{18}$O}}}

\def\ci{\hbox{{\rm [C {\scriptsize I}]}}}
\def\cii{\hbox{{\rm [C {\scriptsize II}]}}}

\def\h1{\hbox{{\rm H {\scriptsize I}}}}
\def\hii{\hbox{{\rm H {\scriptsize II}}}}
\def\hh{\hbox{H$_2$}}
\def\hho{\hbox{{\rm H$_2$O}}}

\documentclass{aa}
\usepackage[]{natbib}
\bibpunct{(}{)}{;}{a}{}{,} % to follow the A&A style
\usepackage{multirow}
\usepackage{rotating}
\usepackage{graphicx}
\usepackage{epsfig}
\usepackage{txfonts}
\usepackage{psfrag}
\usepackage{units}
\usepackage{mathrsfs}
\usepackage[colorinlistoftodos]{todonotes}
\begin{document}
\title{Disentangling the excitation conditions of the dense gas in M17~SW}
%\subtitle{A SOFIA/GREAT, APEX and IRAM 30m synergy}
%
\author{J.P. P\'{e}rez-Beaupuits\inst{1} \and
        R.~G\"usten\inst{1} \and
        M.~Spaans\inst{3} \and
        V.~Ossenkopf\inst{2} \and
        K.M.~Menten\inst{1} \and        
        M.A.~Requena-Torres\inst{1} \and        
        H.~Wiesemeyer\inst{1} \and                        
        J.~Stutzki\inst{2} \and
        C.~Guevara\inst{2} \and                
        R.~Simon\inst{2}
        %H.-W.~ H\"ubers\inst{3,4} \and
        %O.~Ricken\inst{2,1} \and
        %G.~Sandell\inst{5}        
}
\offprints{J.P. P\'erez-Beaupuits}
\institute{
 Max-Planck-Institut f\"ur Radioastronomie, Auf dem H\"ugel 69, 53121 Bonn, Germany
 \email{jp@mpifr.mpg.de} 
 \and
 I. Physikalisches Institut der Universit\"at zu K\"oln, Z\"ulpicher Stra\ss e 77, 50937 K\"oln, Germany
 \and
% Deutsches Zentrum f\"ur Luft- und Raumfahrt, Institut f\"ur Planetenforschung, Rutherfordstrasse 2, 12489 Berlin, Germany
% \and
% Institut f\"ur Optik und Atomare Physik, Technische Universit\"at Berlin, Hardenbergstraße 36, 10623 Berlin, Germany
% \and
% SOFIA-USRA, NASA Ames Research Center, Mail Stop N211-3, Building N211/Rm. 249, Moffett Field, CA 94035, USA
% \and
 Kapteyn Astronomical Institute, Rijksuniversiteit Groningen, 9747 AV Groningen, The Netherlands
}
\date{Received  / Accepted  24/08/2015 }
\titlerunning{Dense gas in M17~SW}
%\authorrunning{ }

%
\abstract
  {Stars are formed in dense molecular clouds. These dense clouds experience radiative feedback from UV photons, 
  and X-ray from stars, embedded pre-stellar cores, YSOs, and ultra 
  compact \hii\ regions. This radiative feedback affects the chemistry and thermodynamics of the gas.
  }
  {We aim to probe the chemical and energetic conditions created by radiative feedback through observations of 
  multiple CO, \hcn\ and \hcop\ transitions. We measure the spatial distribution and 
  excitation of the dense gas ($n(\rm H_2)>10^4~\3cm$) in the core region of M17~SW and aim to investigate
  the influence of UV radiation fields.
  }
  {We used the dual band receiver GREAT on board the SOFIA airborne telescope to obtain a 
  5$'$.7$\times$3$'$.7 map of the $J=16\rightarrow15$, $J=12\rightarrow11$, and $J=11\rightarrow10$ 
  transitions of \twco\ in M17~SW. We compare these maps with corresponding APEX and IRAM 30m telescope data 
  for low- and mid-$J$ CO, \hcn\ and \hcop\ emission lines, including maps of the 
  \hcn~$J=8\to 7$ and \hcop~$J=9\to 8$ transitions. The excitation conditions of \twco, \hcop\ and \hcn\ are estimated with a two-phase 
  non-LTE radiative transfer model of the line spectral energy distributions (LSEDs) at four selected 
  positions. The energy balance at these positions is also studied.
  }
  {We obtained extensive LSEDs for the CO, \hcn\ and \hcop\ molecules toward M17~SW. 
  %Spatially, the \hcn\ and \hcop\ transitions show similar distributions in the regions with the brightest 
  %emission, but the \hcop~$J_{up}\geq3$ emission is brighter and more extended. 
  %The \hcop/\hcn~$J=1\to0$ line ratio is lower than unity while the higher-$J$ \hcop/\hcn\ line ratios are larger 
  %than unity in the whole region mapped. 
  These LSEDs can be fit simultaneously using the same density and temperature 
  in the two-phase models and to the spectra of all three molecules over a $\sim 12$ square arc 
  minute size region of M17~SW.  
  Temperatures of up to 240~K are found toward the position of the peak emission of the \twco~$J=16\to15$ line. 
  High densities of 10$^6~\3cm$ were found at the position of the peak \hcn~$J=8\to7$ emission.
  }
  {
  We found \hcop/\hcn\ line ratios larger than unity, which can be explained by a lower excitation 
  temperature of the higher-$J$ \hcn\ lines.
  %The isotope ratios of \hcn\ and \hcop\ are about factor two lower than expected for a source at a 
  %Galactocentric distance of $\sim$7~kpc, indicating non-equilibrium chemistry or a high degree of chemical 
  %fractionation due to the high temperature of the gas. 
  The LSED shape, particularly the high-$J$ tail of the CO lines observed with 
  SOFIA/GREAT, is distinctive for the underlying excitation conditions.    
  The cloudlets associated with the cold component of the models are magnetically subcritical and 
  supervirial at most of the selected positions. The warm cloudlets instead are all supercritical 
  and also supervirial. 
  The critical magnetic field criterion implies that the cold cloudlets at two positions are partially 
  controlled by processes that create and dissipate internal motions.  
  Supersonic but sub-Alfv\'enic velocities in the cold component at most selected positions indicates
  that internal motions are likely MHD waves. Magnetic pressure dominates thermal pressure in
  both gas components at all selected positions, assuming random orientation of the magnetic field.
  The magnetic pressure of a constant magnetic field throughout all the gas phases can support 
  the total internal pressure of the cold components, but it cannot support the internal pressure 
  of the warm components. 
  If the magnetic field scales as $B \propto n^{2/3}$, then the evolution of the cold cloudlets at 
  two selected positions, and the warm cloudlets at all selected positions, will be determined by ambipolar 
  diffusion.
  }


\maketitle

\section{Introduction}\label{sec:introduction}

Inhomogeneous and clumpy clouds, or a partial face-on illumination 
in star-forming regions like M17 SW, S106, S140, the Great Nebula 
portion of the Orion Molecular Cloud, and the NGC~7023 Nebula, 
produce extended emission of atomic lines like \ci\ and \cii, and 
suppress the stratification in \cii, \ci\ and CO emission  
expected from standard steady-state models of photon-dominated 
regions (PDRs)
\citep[e.g.][]{keene85, genzel88, stutzki88, gerin98, yamamoto01, schneider02, schneider03, mookerjea03, pb10, pb12}.
The complex line profiles observed in optically thin lines from, e.g., 
\ci, CS, \ceio, \thco\ and their velocity-channel maps, are indicative 
of the clumpy structure of molecular clouds and allow for a robust 
estimation of their clump mass spectra 
\citep[e.g.][]{carr87, loren89, stutzki90, hobson92, kramer98, kramer04, pb10}.
In fact, clumpy PDR models can explain extended \ci\ $^3P_1\to{^3P_0}$ 
emission and a broad LSED for various PDRs and even the whole Milky Way 
\citep{stutzki90,spaans97,cubick08, ossenkopf10}.

Massive star forming regions like the Omega Nebula M17, with a nearly edge-on view, are ideal sources to study 
the clumpy structure of molecular clouds, as well as the chemical and thermodynamic effects of the nearby 
ionizing sources.
At a (trigonometric paralax) distance of $\sim$1.98 kpc \citep{xu11}, 
the south-west region of M17 (M17~SW) concentrates molecular 
material in a clumpy structure.
The structure of its neutral and molecular gas appears to be 
dominated by magnetic rather than by thermal gas pressure, 
in contrast to many other PDRs \citep{pellegrini07}, 
since magnetic fields as strong as $-700~\mu$G have been 
measured from \h1\ and OH Zeeman observations \citep{brogan99, brogan01}.

Models based on far-IR and sub-millimeter observations 
\citep{stutzki88, meixner92} suggest that the distribution 
and intensity of the emissions observed in the M17~SW complex 
can be explained with high density 
($n(\rm H_2)\sim5\times10^5~\3cm$) clumps embedded in a less 
dense interclump medium ($n(\rm H_2)\sim3\times10^3~\3cm$) 
surrounded by a diffuse halo ($n(\rm H_2)\sim300~\3cm$). 

%\todo[inline, color=green!40]{Mention update from latest paper!}

More recent results from the Stratospheric Observatory for Infrared 
Astronomy (SOFIA) and the German REceiver for Astronomy at Terahertz 
frequencies (GREAT)
%\footnote{http://www3.mpifr-bonn.mpg.de/div/submmtech/heterodyne/\newline great/greatmain.html}
showed that the line profiles and distribution of the \cii~158~\mum\ emission 
support the clumpy scenario in M17~SW \citep{pb12}
and that at least 64\% of the \cii~158~\mum\ emission is not 
associated with the star-forming material traced by the \ci\ and \ceio\ 
lines \citep{pb13, pb15a}.

There are more than 100 stars illuminating M17~SW, of which its central 
cluster is NGC~6618 \citep[e.g.][]{lada91, hanson97}. 
The two components of the massive binary CEN1 \citep{kleinmann73, chini80} 
are part of this central cluster. This 
source, originally classified as a double O or early B system by 
\citet{kleinmann73}, is actually composed of two O4 visual binary stars 
with a separation of $\sim 1\farcs8$, 
named CEN 1a (NE component) and CEN 1b 
(SW component), and it appears to be the dominant source of 
photo-ionization in the whole M17 region \citep{hoffmeister08}.

The components of the CEN1 binary are also the brightest X-ray sources 
detected with \textit{Chandra} in the M17 region \citep{broos07}. These 
\textit{Chandra} observations, together with a new long ($\sim300$ ks) 
exposure of the same field, yielded a combined dataset of $\sim2000$ X-ray 
sources, with many of them having a stellar counterpart in IR images. 
\citep{townsley03, getman10}.  
Although the brightest X-ray sources are CEN 1a and CEN 1b from the central 
NGC~6618 cluster, other stellar concentrations of $\sim40$ heavily obscured 
($E_{median}\ge2.5$ keV, $A_V>10$ mag) X-ray sources are distributed 
along the eastern edge of the M17~SW molecular core. The densest 
concentration of X-ray sources coincides with the well known star-forming 
region M17-UC1 and the concentration ends around the Kleinmann-Wright 
Object \citep[][their Fig.~10]{broos07}. The M17-UC1 (G15.04-0.68), 
an ultracompact \hii\ region first studied by \citet{felli80}, harbors an 
ionizing source classified as a B0$-$B0.5 star which, together with its 
southern companion (IRS 5S), correspond to a massive Class I source with an 
observable evolution in the MIR and radio wavebands \citep{chini00}.

In this study we present maps of multiple high-$J$ transitions of \twco, \thco, 
\hcn\ and \hcop. We analyze the underlying excitation conditions from the LSED 
of some of the clumps fitted to their spectral lines.
We are particularly interested in these molecules because their distribution 
and excitation is responsive to energetic radiative environments, which has led 
to their extensive usage for, e.g., disentangling star formation vs. 
black hole accretion and shocks, in the nuclear region of active galaxies 
\citep[e.g.][]{sternberg94, kohno99, kohno01, kohno03, kohno05, usero04, pb07, garcia08, loenen08, krips08, pb09, rosenberg14, garcia14, viti14}

%Besides, the line emission from these (and other species like CN, CS, and HNC) trace the molecular gas in the dense ($\nh2\geq10^5~\3cm$) cores associated with star formation in Galactic molecular clouds, as well as in the nuclear region of active galaxies (e.g. White \etal\ 1982; Sandqyist, Wootten \& Loren 1985; Schilke \etal\ 1992; Solomon, Downes, \& Radford 1992; Hobson 1992; Helfer \& Blitz 1993, 1995; Fuente, Martin-Pintado \& Gaume 1995; Hogerheijde, Jansen \& van Dishoeck 1995; Bergin \etal\ 1997; Usero \etal\ 2004; P\'erez-Beaupuits \etal\ 2007; Garc\'ia-Burillo \etal\ 2008; Loenen \etal\ 2008).

Given the relative youth \citep[$\lesssim 1$ Myr, e.g.][]{lada91, hanson97} of the main ionizing 
cluster (NGC~6618) of M17~SW, and the absence of evolved stars, it is likely that supernovae have not yet 
occurred. This makes M17~SW an ideal place to study the radiative interactions of massive stars with their 
surrounding gas/dust and stellar disks, without the influence of nearby supernovae. This allows us to study the 
effects of UV, IR, and X-ray radiation fields on the emission of CO, HCN, and \hcop\ in a well defined 
environment.

This study of M17~SW can be considered as one more Galactic template for extra-galactic star forming regions. 
%The properties and feedback effects of massive star forming regions, studied at small scale and with high 
%resolution in the Milky Way, are expected to drive the energetics of active galaxies.
Therefore, we expect that our results 
will be important for future high resolution observations where similar regions in extra-galactic sources will be 
studied in great spatial detail with, e.g., ALMA \citep{schleicher10}.

The organization of this article is as follows. In Sect.~2 we describe the observations. The maps of lines 
observed are presented in Sect.~3. The modeling and analysis of the ambient conditions are presented in Sect.~4. 
The conclusions and final remarks are presented in Sect.~5.

\begin{figure*}[!pt]

  \hspace*{\fill}\includegraphics[angle=0,width=0.48\textwidth]{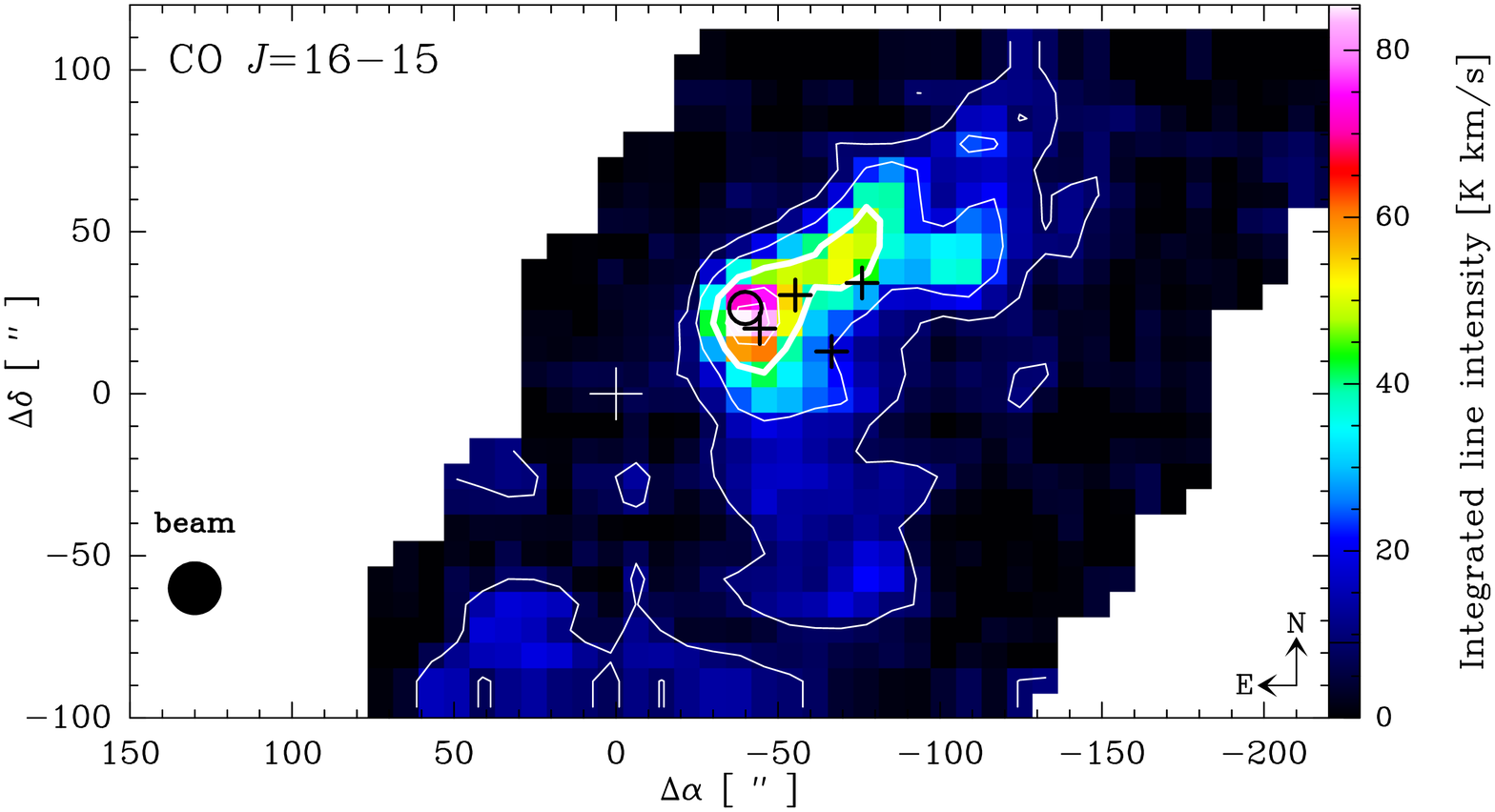}%    
  \hfill\includegraphics[,angle=0,width=0.48\textwidth]{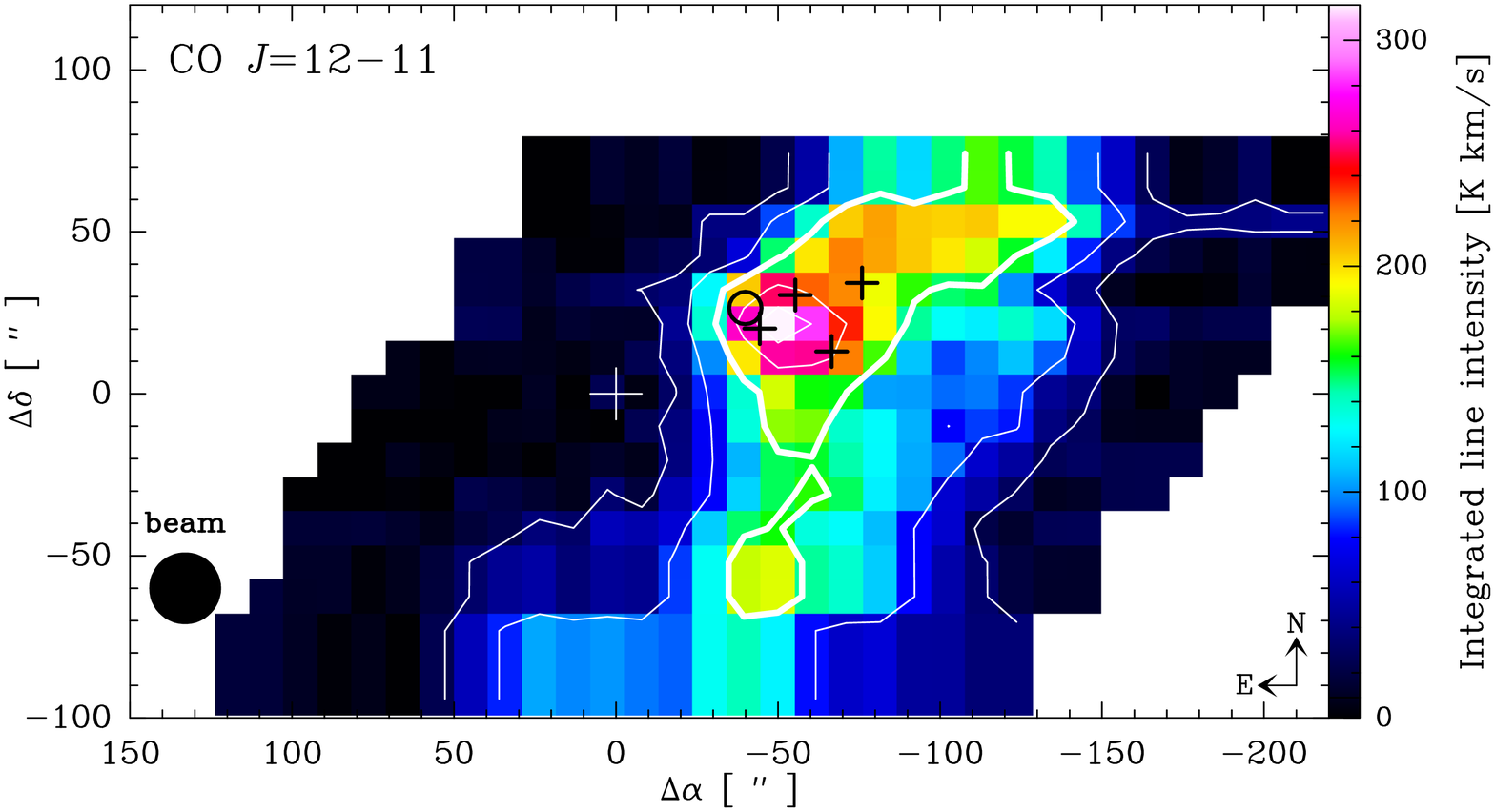}\hspace*{\fill}\\
  
  \hspace*{\fill}\includegraphics[,angle=0,width=0.48\textwidth]{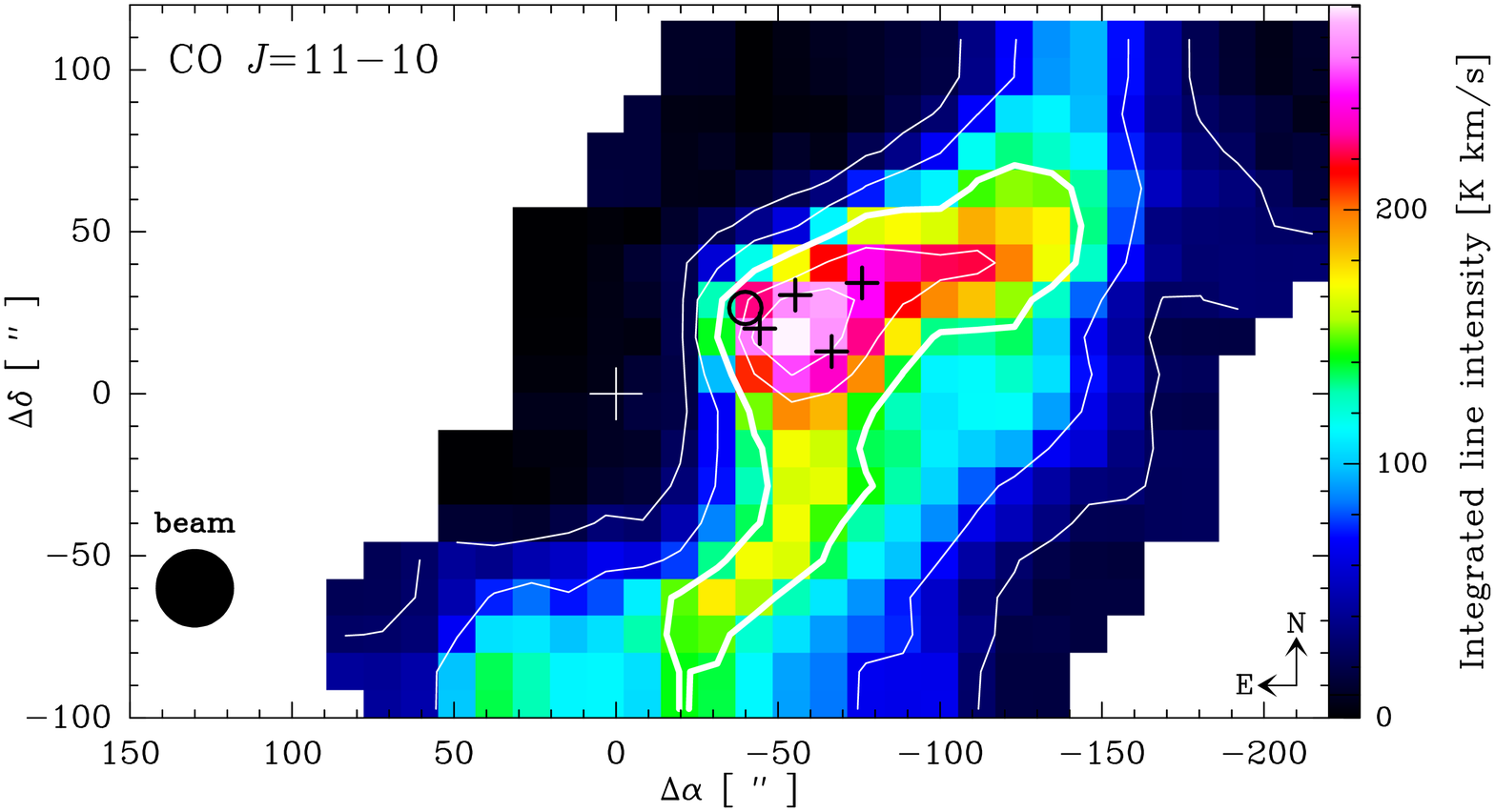}%
  \hfill\includegraphics[,angle=0,width=0.48\textwidth]{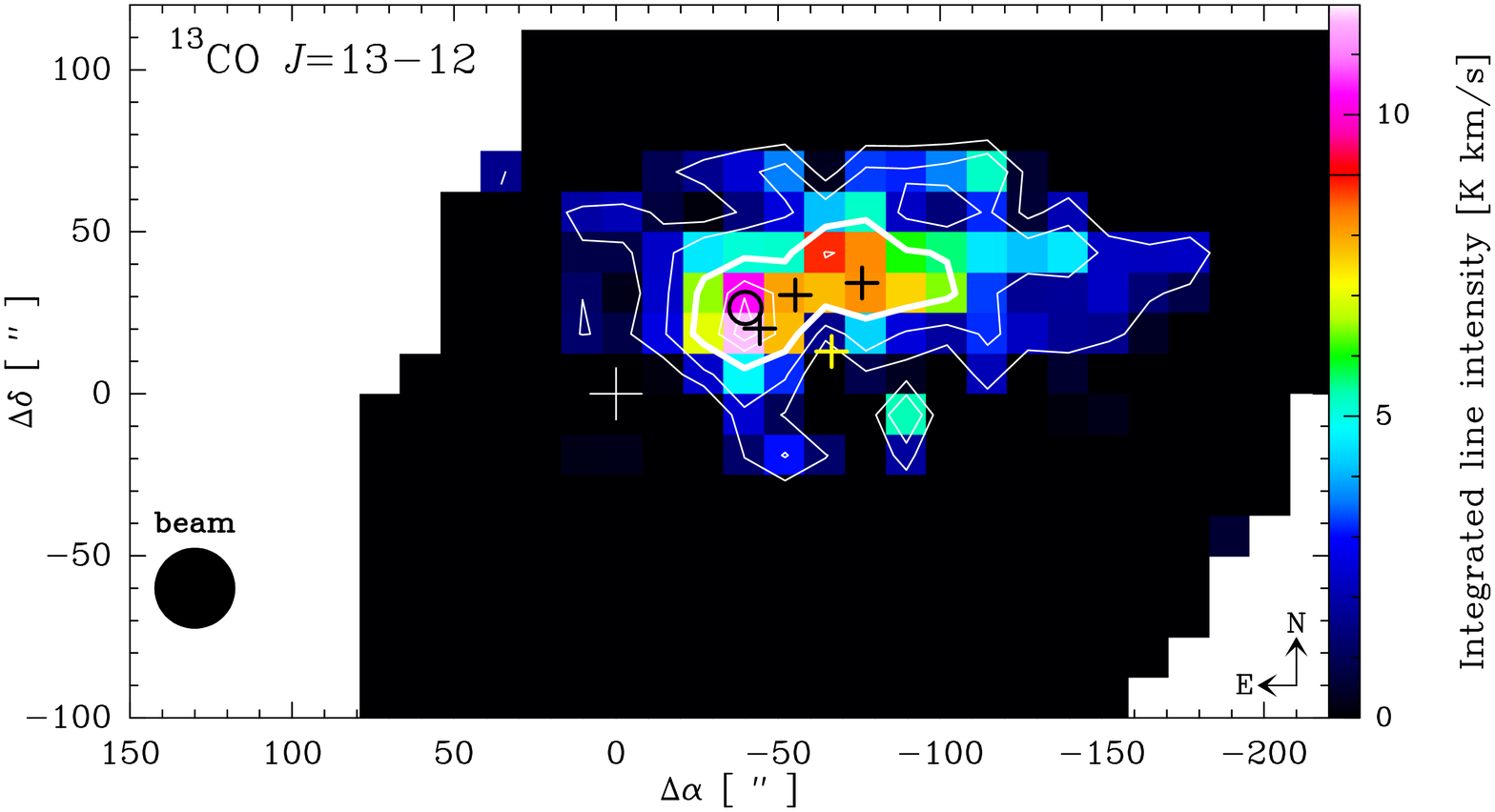}\hspace*{\fill}\\  

  \caption{\footnotesize{Colour maps of the integrated intensity of the $J = 16\to15$, $J = 12\to11$ and $J = 11\to10$ transitions of \twco, and the \thco~$J=13\to12$ line in M17~SW. The contour levels are the 10\%, 25\%, 50\% (thick line), 75\% and 90\% of the peak emissions. The central position ($\Delta \alpha=0$, $\Delta \delta=0$), marked with a cross, corresponds to the star SAO 161357 at R.A(J2000)=$18^h 20^m 27.6^s$ and Dec(J2000)=$-16^{\circ} 12\arcmin 00\farcs9$. The ultracompact \hii\ region M17-UC1 and four H$_2$O masers \citep{johnson98} are marked by the circle and plus symbols, respectively. The \thco~$J=13\to12$ map was convolved with a $25''$ beam to match the resolution of the \twco~$J=12\to11$ and to increase the S/N. All pixels with S/N$<3$ (or an rms$>0.22$~K) were blanked in the \thco\ map.
}}

  \label{fig:GREAT-maps}
\end{figure*}

\begin{figure*}[!pt]

  \hspace*{\fill}\includegraphics[angle=0,width=0.45\textwidth]{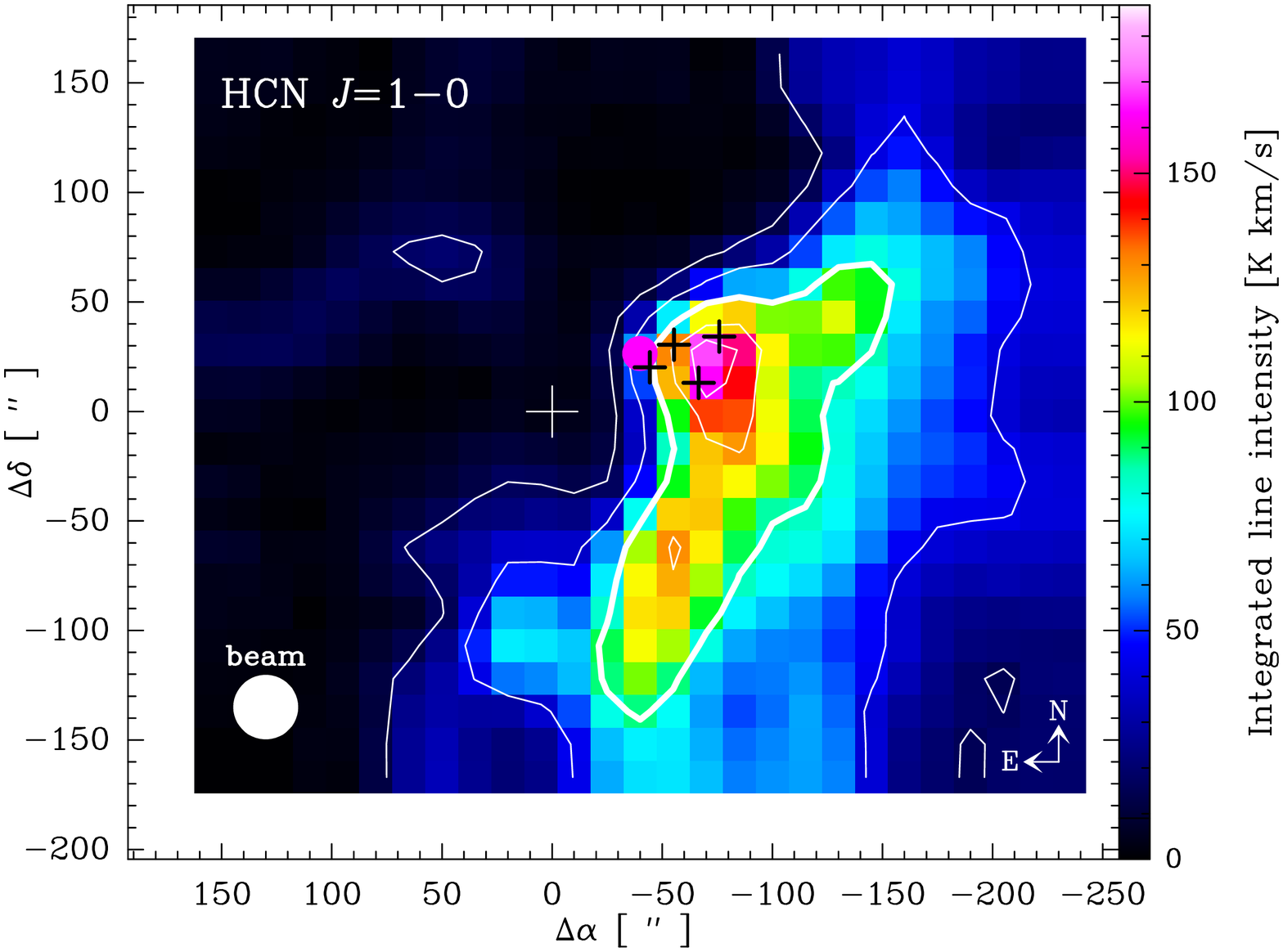}%
  \hfill\includegraphics[angle=0,width=0.45\textwidth]{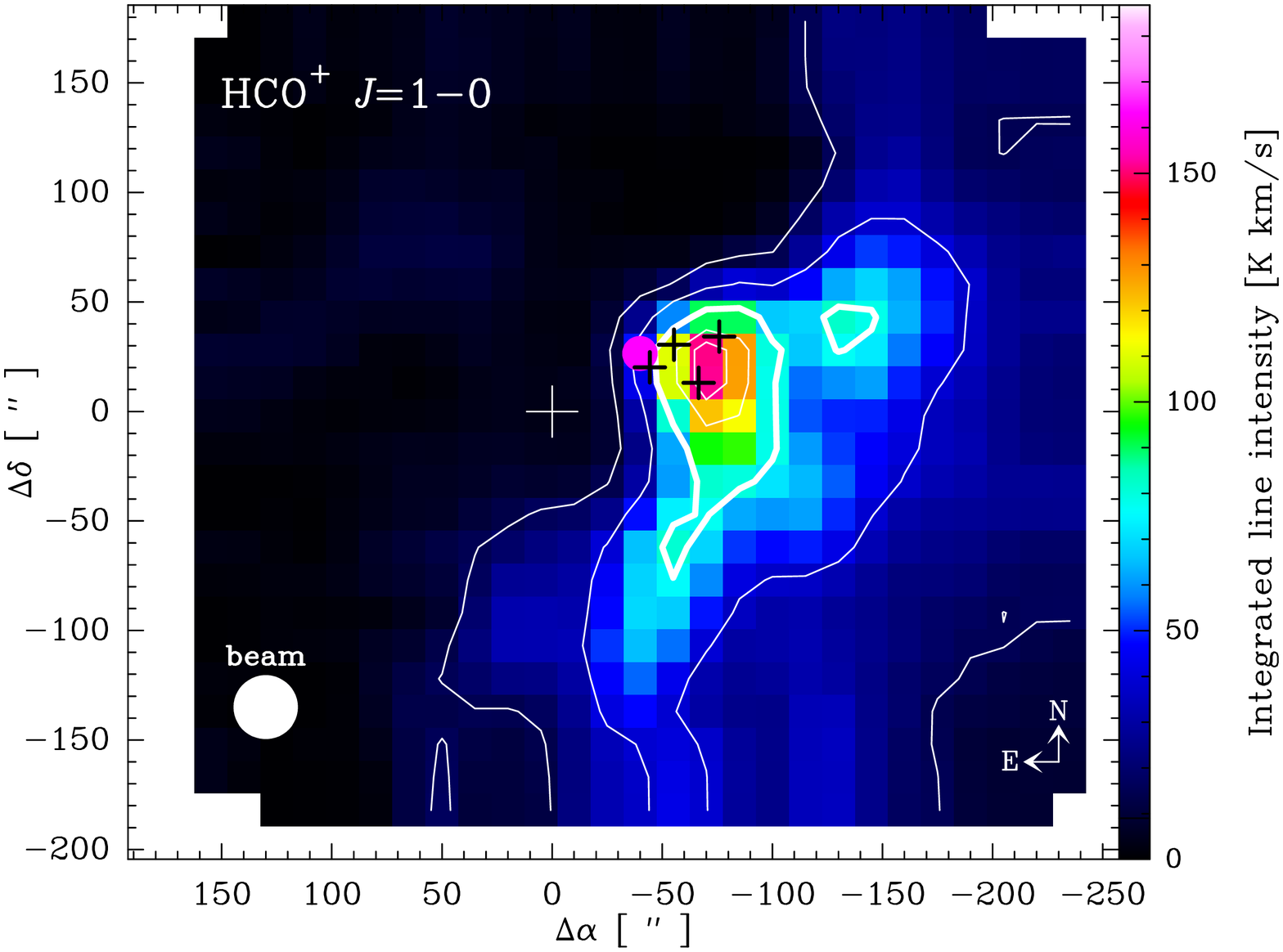}\hspace*{\fill}\\
  
  \vspace*{-0.2cm} 
  
  \hspace*{\fill}\includegraphics[angle=0,width=0.45\textwidth]{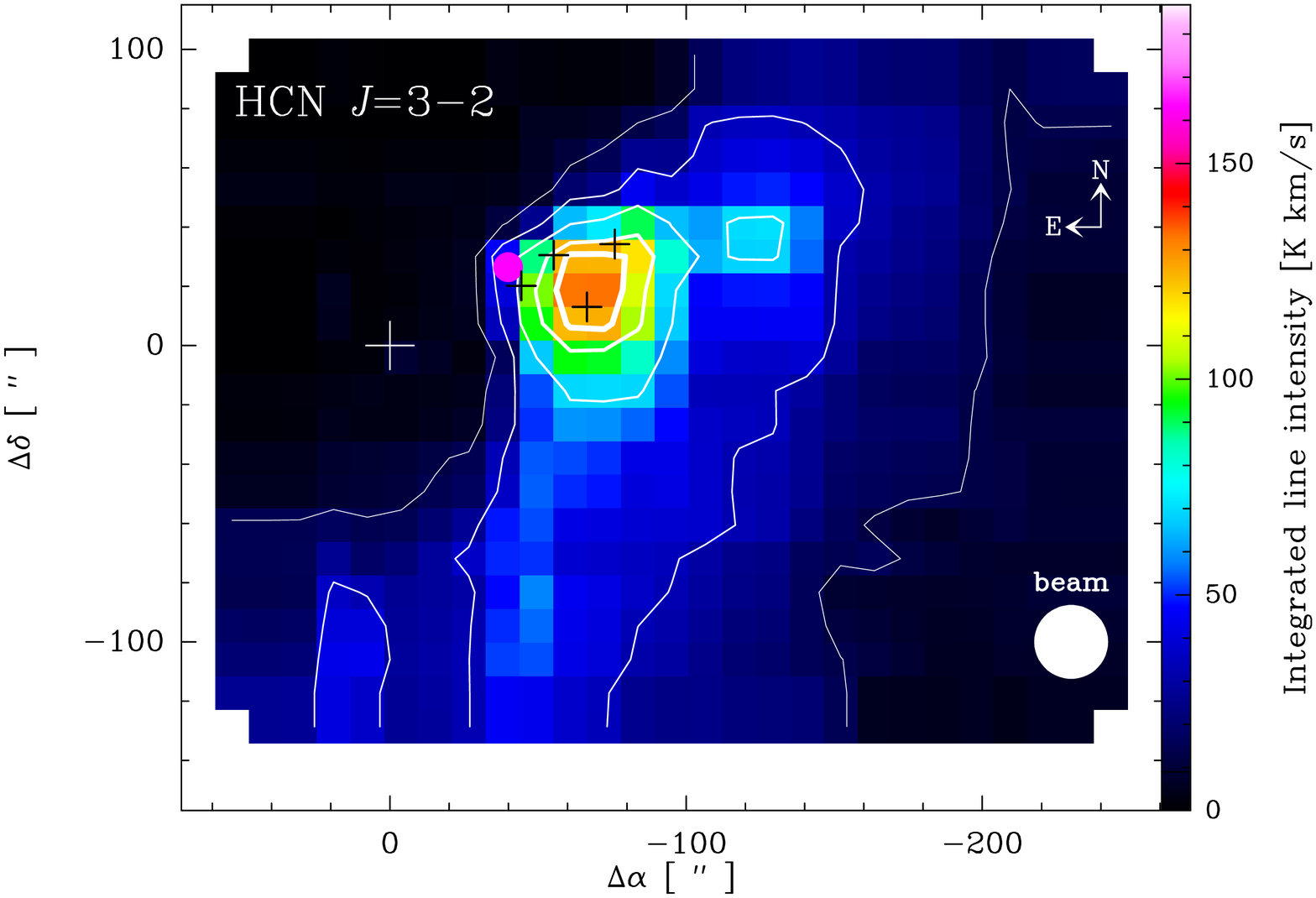}%
  \hfill\includegraphics[angle=0,width=0.45\textwidth]{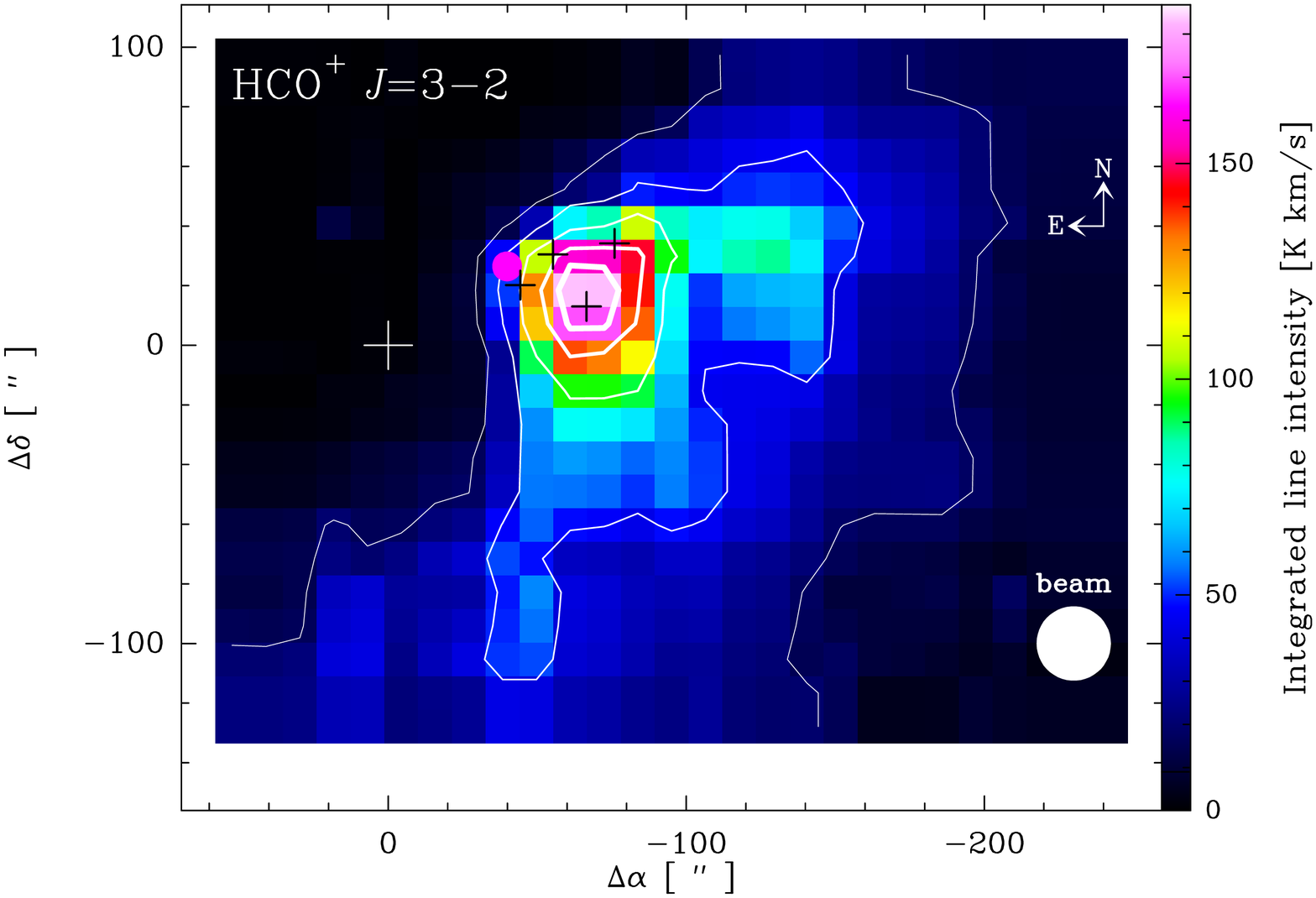}\hspace*{\fill}\\
  
  \vspace*{-0.2cm}     

  \hspace*{\fill}\includegraphics[angle=0,width=0.45\textwidth]{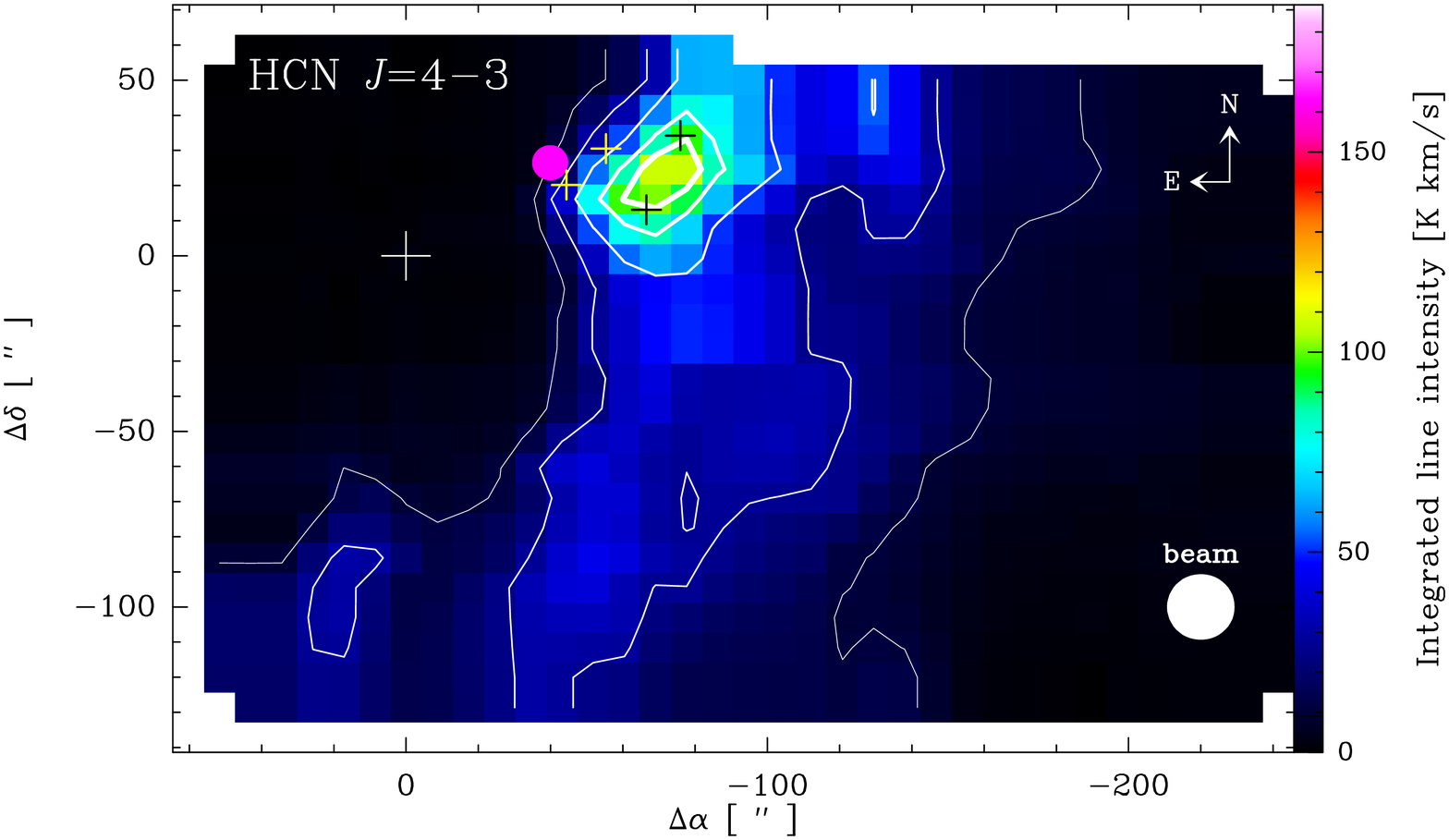}%
  \hfill\includegraphics[angle=0,width=0.45\textwidth]{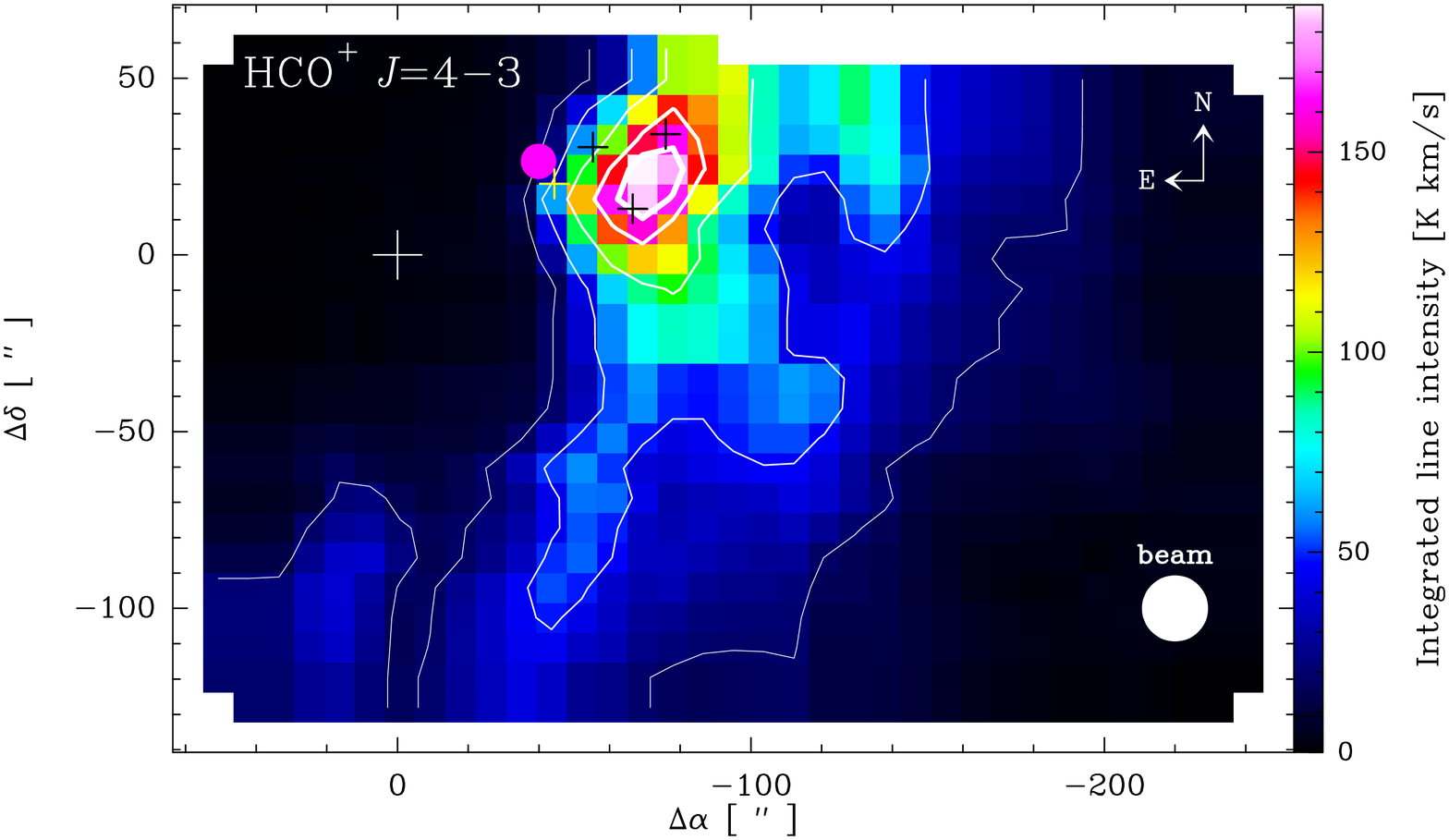}\hspace*{\fill}\\  
  
  \caption{\footnotesize{\textit{Top panels} - Maps of the integrated temperature of the \hcn\ and \hcop~$J = 1\to0$ lines in M17~SW. The peak emissions are $\sim$169~\Kkms\ and $\sim$152~\Kkms, respectively. The contour levels for all panels in this figure are, from thin to thick, the 10\%, 25\%, 50\%, 75\% and 90\% of the peak emission. \textit{Middle panels} - Maps of the integrated temperature of \hcn\ and \hcop~$J=3\to2$ lines, with peak emissions of $\sim$133~\Kkms\ and $\sim$183~\Kkms, respectively. \textit{Bottom panels} - Maps of the integrated temperature of the \hcn\ and \hcop~$J = 4\to3$ lines in M17~SW. The peak emissions are $\sim$107~\Kkms\ and $\sim$187~\Kkms, respectively. 
The reference position ($\Delta\alpha=0$, $\Delta\delta=0$) marked with a cross, and other symbols, are as in Fig.~\ref{fig:GREAT-maps}. The integrated temperature scale is the same in all maps.
}}

  \label{fig:APEX-30m-maps}
\end{figure*}

\begin{figure*}[!pt]
  
  \hspace*{\fill}\includegraphics[angle=0,width=0.45\textwidth]{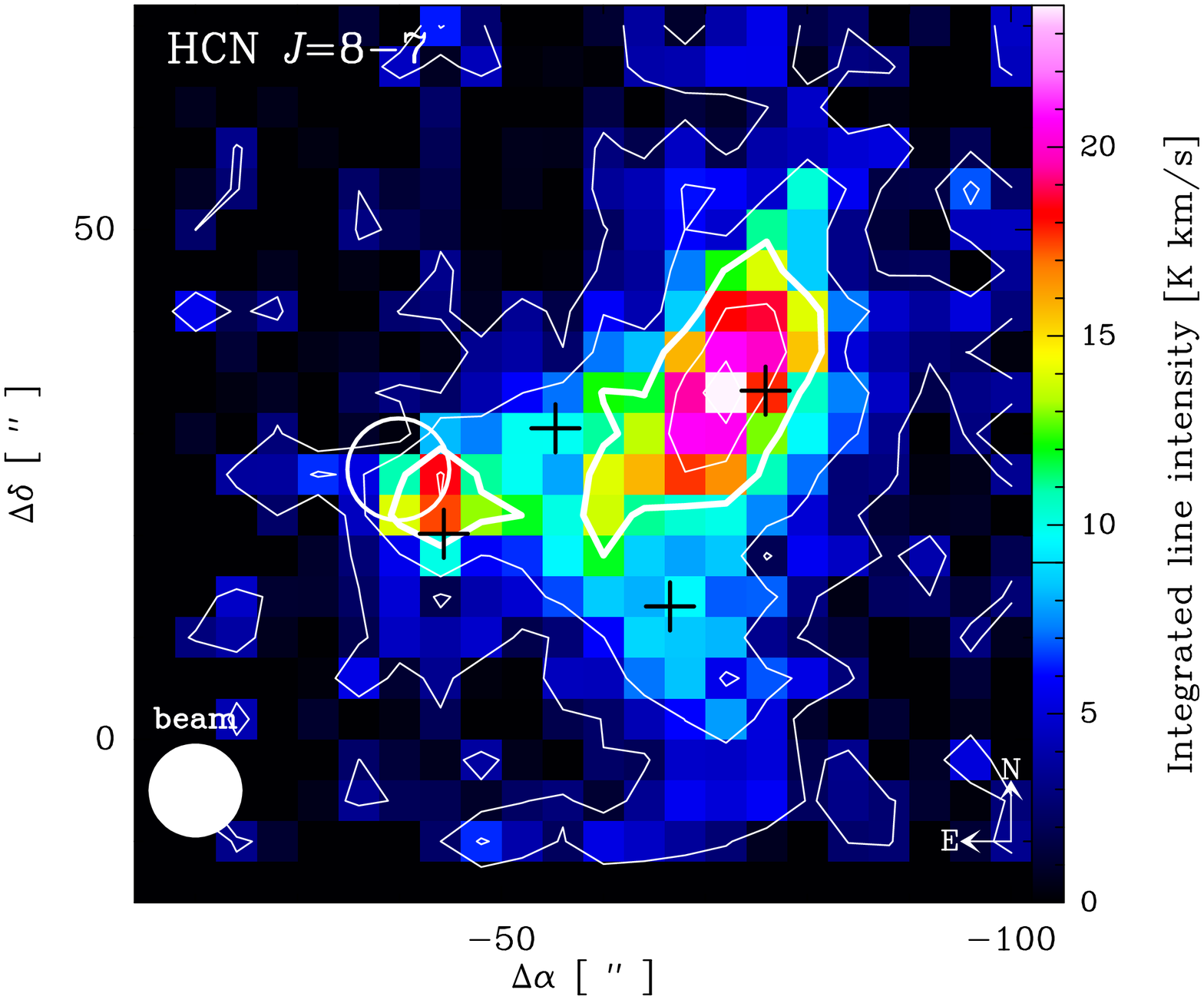}%
  \hfill\includegraphics[angle=0,width=0.45\textwidth]{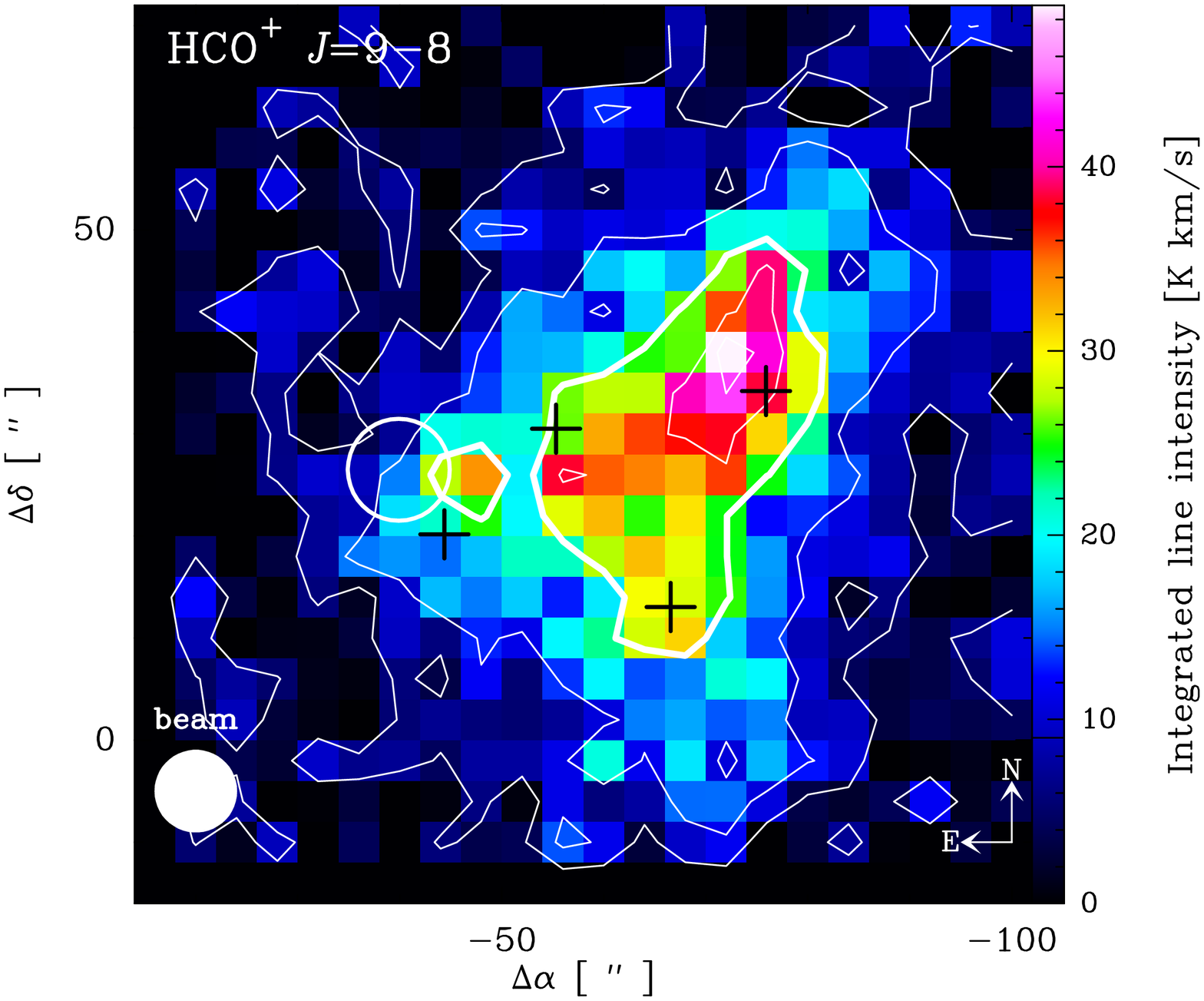}\hspace*{\fill}\\   

  \caption{\footnotesize{Line intensity maps of the \hcn~$J=8\to7$ and \hcop~$J = 9\to8$ transitions in the 
  dense core of M17~SW. The contour levels are the 10\%, 25\%, 50\% (thick contour), 75\% and 90\% of the peak 
  emission. The reference position ($\Delta\alpha=0$, $\Delta\delta=0$) marked with a cross, and other symbols, 
  are as in Fig.~\ref{fig:GREAT-maps} and \ref{fig:APEX-30m-maps}.
}}

  \label{fig:CHAMP-maps}
\end{figure*}

%__________________________________________________ One column table
   \begin{table}[!tp]
      \caption[]{Line parameters of observed transitions.}
         \label{tab:spectral-lines}
         \centering
         \scriptsize
         %\footnotesize
         \setlength{\tabcolsep}{3.5pt} % Default value: 6pt
         \renewcommand{\arraystretch}{1.0} % Default value: 1
         \begin{tabular}{lccccc}
            \hline\hline
	    \noalign{\smallskip}
            Transition & $\nu_0$\tablefootmark{a} & $\theta^*_{mb}$\tablefootmark{b} & $E_{up}$\tablefootmark{a}  &  $n_{crit}$\tablefootmark{c}  & Telescope/Instrument \\
                       &   [GHz] &    [$''$]      &  [K]      &   [$\3cm$]    \\
            \noalign{\smallskip}
            \hline
            \noalign{\smallskip}
            \multicolumn{6}{c} {\twco} \\
            \noalign{\smallskip}
            \hline
            \noalign{\smallskip}

$J=1\to0$ & 115.271 & 22.6 & 5.53 & 2$\times$10$^3$ & IRAM 30m/EMIR\\
$J=2\to1$ & 230.538 & 11.3 & 16.60 & 7$\times$10$^3$ & IRAM 30m/EMIR\\
$J=3\to2$ & 345.796 & 18.9 & 33.19 & 2$\times$10$^4$ & APEX/FLASH$^+$\\
$J=4\to3$ & 461.041 & 14.1 & 55.32 & 4$\times$10$^4$ & APEX/FLASH$^+$\\
$J=6\to5$ & 691.473 & 9.6 & 116.16 & 1$\times$10$^5$ & APEX/CHAMP$^+$\\
$J=7\to6$ & 806.652 & 8.2 & 154.87 & 2$\times$10$^5$ & APEX/CHAMP$^+$\\
$J=11\to10$ & 1267.014 & 24.2 & 364.97 & 8$\times$10$^5$ & SOFIA/GREAT\\
$J=12\to11$ & 1381.995 &  22.2 & 431.29 & 9$\times$10$^5$ & SOFIA/GREAT\\
$J=13\to12$ & 1496.923 &  20.9 & 503.13 & 1$\times$10$^6$ & SOFIA/GREAT\\
$J=16\to15$ & 1841.346 &  16.6 & 751.72 & 2$\times$10$^6$ & SOFIA/GREAT\\

            \noalign{\smallskip}
            \hline
            \noalign{\smallskip}
            \multicolumn{6}{c} {\thco} \\
            \noalign{\smallskip}
            \hline
            \noalign{\smallskip}

$J=1\to0$ & 110.201 & 23.7 &   5.29 & 2$\times$10$^3$ & IRAM 30m/EMIR\\
$J=2\to1$ & 220.399 & 11.8 &  15.87 & 1$\times$10$^4$ & IRAM 30m/EMIR\\
$J=3\to2$ & 330.588 & 19.7 &  31.73 & 3$\times$10$^4$ & APEX/FLASH$^+$\\
$J=6\to5$ & 661.067 & 10.0 & 111.05 & 3$\times$10$^5$ & APEX/CHAMP$^+$\\
$J=13\to12$ & 1431.153 &  21.4 & 481.02 & 2$\times$10$^6$ & SOFIA/GREAT\\

            \noalign{\smallskip}
            \hline
            \noalign{\smallskip}
            \multicolumn{6}{c} {\hcn} \\
            \noalign{\smallskip}
            \hline
            \noalign{\smallskip}

$J=1\to0$ &   88.632 &  29.4 &  4.25 & 2$\times$10$^6$ & IRAM 30m/EMIR\\
$J=3\to2$ &  265.886 &  24.9 & 25.52 & 1$\times$10$^7$ & APEX/HET230\\
$J=4\to3$ &  354.505 &  18.7 & 42.53 & 3$\times$10$^7$ & APEX/FLASH\\
$J=8\to7$ &  708.877 &  9.2 & 153.11 & 2$\times$10$^8$ & APEX/CHAMP$^+$\\

            \noalign{\smallskip}
            \hline
            \noalign{\smallskip}
            \multicolumn{6}{c} {\hthcn} \\
            \noalign{\smallskip}
            \hline
            \noalign{\smallskip}

$J=1\to0$ &  86.339 &  30.2 &  4.14 & 2$\times$10$^6$ & IRAM 30m/EMIR\\
$J=3\to2$ & 259.012 &  25.2 & 24.86 & 5$\times$10$^7$ & APEX/HET230\\
$J=4\to3$ & 345.339 &  19.2 & 41.43 & 1$\times$10$^8$ & APEX/FLASH\\

            \noalign{\smallskip}
            \hline
            \noalign{\smallskip}
            \multicolumn{6}{c} {\hcop} \\
            \noalign{\smallskip}
            \hline
            \noalign{\smallskip}

$J=1\to0$ &   89.189 &  29.2 &  4.28 & 2$\times$10$^5$ & IRAM 30m/EMIR\\
$J=3\to2$ &  267.558 &  24.8 & 25.68 & 3$\times$10$^6$ & APEX/HET230\\
$J=4\to3$ &  356.734 &  18.6 & 42.80 & 6$\times$10$^6$ & APEX/FLASH\\
$J=9\to8$ &  802.458 &  8.1 & 192.58 & 9$\times$10$^7$ & APEX/CHAMP$^+$\\

            \noalign{\smallskip}
            \hline
            \noalign{\smallskip}
            \multicolumn{6}{c} {\hthcop} \\
            \noalign{\smallskip}
            \hline
            \noalign{\smallskip}

$J=1\to0$ &  86.754  &  30.1 & 4.16 & 2$\times$10$^5$ & IRAM 30m/EMIR\\
$J=3\to2$ &  260.255 &  25.1 & 24.98 & 3$\times$10$^6$ & APEX/HET230\\

            \noalign{\smallskip}
            \hline
         \end{tabular}

         \tablefoot{
         \tablefoottext{a}{Rest frequencies and upper level energies are adopted from the 
         Cologne Database for Molecular Spectroscopy, CDMS \citep{mueller05} and the 
         Leiden Atomic and Molecular Database, LAMDA \citep{schoier05}.}
         \tablefoottext{b}{Beam size (resolution) used to create the respective map. This beam is $\sim$6\% 
         larger than the actual HPBW at the corresponding frequencies, as defined by the kernel used in the 
         gridding algorithm of GILDAS/CLASS.}         
         \tablefoottext{c}{Critical densities for temperature ranges 40--300~K (\twco, \thco) and 10--30~K 
         (\hcn, \hthcn, \hcop, \hthcop). As the main collisional transitions for HCN occur with $\Delta J = 2$
the traditional two-level formula for the critical density is not appropriate here. It generally
overestimates the critical density. This also applies to the other molecules where $\Delta J = 2$
transitions are at least comparable. A better estimate for the actual population of the 
molecule is obtained here by adding the coefficients for $\Delta J= 1$ and
$\Delta J = 2$.}
         }

   \end{table}
%__________________________________________________ One column table

\section{Observations}

\subsection{The SOFIA/GREAT data}

The new high-$J$ CO observations were performed with the German Receiver for Astronomy at Terahertz Frequencies \citep[GREAT\footnote{GREAT is a development by the MPI f\"ur Radioastronomie and the KOSMA/ Universit\"at zu K\"oln, in cooperation with the MPI f\"ur Sonnensystemforschung and the DLR Institut f\"ur
Planetenforschung.},][]{heyminck12} on board the Stratospheric Observatory For Infrared Astronomy \citep[SOFIA,][]{young12}.

We used the dual-color spectrometer during the Cycle-1 flight campaign of 2013 July to simultaneously map a 
region of about 310$''\times$220$''$ ($\sim$3.0~pc $\times$ 2.1~pc) in the \twco~$J=11\rightarrow10$ at 
1381.995105~GHz (216.9~\mum) and the \twco~$J=16\rightarrow15$ transition at 1841.345506~GHz (162.8~\mum) 
toward M17~SW.

We also present data for the \twco~$J=12\rightarrow11$ transition at 1381.995105~GHz (200.3~\mum) that was mapped 
during the Early Science flight on 2011 June. Due to lack of time during the observing campaign, the 
northern map could not be extended beyond $\Delta\delta\sim70''$.

The observations were performed in on-the-fly (OTF) total power mode. The area mapped 
consists of six strips, each covering $224''\times32''$ 
($\Delta\alpha \times \Delta\delta$ with a sampling of $8''$, 
half the beamwidth at 1.9 THz.
Hence, each strip consists of four OTF lines containing 28 points each. 
We integrated 1s per dump and 5s for the off-source reference.

All our maps are centered on R.A.(J2000)=$18^h 20^m 27.6^s$ and Dec(J2000)=$-16^{\circ} 12\arcmin 00\farcs9$, 
which corresponds to the star SAO 161357. For better system stability and higher 
observing efficiency we used a nearby reference position at offset (345$''$,$-230''$). 
A pointed observation of this reference position against the 
reference \citep[offset: 1040$''$,$-535''$,][]{matsuhara89} showed that the 
reference is free of \twco\ emission.

Pointing was established with the SOFIA optical guide cameras, 
and was accurate to a few arc seconds during Cycle-1 
observations.
As backends we used the Fast Fourier Transform spectrometers \citep{klein12}, 
which provided 1.5 GHz bandwidth 
with 212 kHz ($\sim$0.03~\kms) spectral resolution.
The calibration of this data in antenna temperature was performed 
with the \textit{kalibrate} task from the 
\textit{kosma\_software} package \citep{guan12}.
Using the beam efficiencies ($\eta_c$) 0.67 for the \twco~$J=16\to15$, $J=11\to10$ and \thco~$J=13\to12$ 
lines,  0.54 for \twco~$J=12\to11$, and the forward efficiency ($\eta_f$) of 0.97\footnote{http://www3.mpifr-
bonn.mpg.de/div/submmtech/heterodyne/great/\newline GREAT\_calibration.html}, we converted all data to main 
beam brightness temperature scale, $T_{mb}=\eta_{f}\times T_{A}^{*}/\eta_{c}$.

\subsection{The APEX data}

%\footnote{This publication is based on data acquired with the Atacama Pathfinder Experiment (APEX)

We have used the lower frequency band of the dual channel DSB receiver FLASH$^+$ 
\citep{heyminck06, klein14} on the Atacama 
Pathfinder EXperiment 12 m submillimeter telescope 
(APEX\footnote{APEX is a collaboration between the Max-Planck-Institut f\"ur 
Radioastronomie, the European Southern Observatory, and the Onsala Space Observatory}; 
\citealt{gusten06}) during 
2012 November, to map the whole M17 region in the $J=3\to2$ and $J=4\to3$ transitions 
of \twco, as well as the \thco~$J=3\to2$ line. The observed region covers about 
$6'.2\times7'.2$ (4.1~pc~$\times$~4.7~pc). In this work, however, we present only a 
smaller fraction matching the region mapped with SOFIA/GREAT.

During 2010 July we performed simultaneous observations of \hcn\ and \hcop\ $J=4\to3$ 
using the 345~GHz band of FLASH$^+$ \citep{klein14}, as well as the $J=3\to2$ 
transitions with the APEX-1 single sideband (SSB) heterodyne SIS receiver 
\citep{risacher06, vassilev08}. 

The regions mapped in \hcn\ and \hcop\ cover about $4.5'\times3.5'$ (3.0~pc $\times$ 2.3~pc) in the $J=3\to2$ 
transition, and about $4.5'\times3.2'$ (3.0~pc $\times$ 2.1~pc) in $J=4\to3$. The maps of \hcn\ and \hcop, of both $J=4\to3$ and $J=3\to2$, were done in raster mode along R.A. with rows of $\sim270''$ long (centered on 
$-95''$), with increments of $15''$ and $10''$ for the lower and higher $J$-lines, respectively. Subsequent scans in declination with the same spacing as in R.A. were done from $-120''$ to $+90''$ for the $J=3\to2$ transition, and from $-120''$ to $+70''$ for $J=4\to3$.

The total power mode was used for the observations, nodding the antenna prior to each 
raster to an off-source position $180''$ east of the star SAO 161357. The latter is 
used as reference position ($\Delta \alpha=0$, $\Delta \delta=0$) in the maps and 
throughout the paper, with R.A.(J2000)=$18^h 20^m 27.6^s$ and Dec(J2000)=$-16^{\circ} 12\arcmin 00\farcs9$.
The telescope pointing was checked with observations of continuum emission from Sgr B2(N) 
and the pointing accuracy was kept below $2''$ 
for all the maps.
Calibration measurements with cold loads were performed every 
$\sim$10 minutes. The data were processed with the APEX real-time calibration software 
\citep{muders06} assuming an image sideband suppression of 10 dB for APEX-1.

\begin{figure*}[!th]

 \begin{tabular}{cc}\setlength{\tabcolsep}{2pt}
  
  \epsfig{file=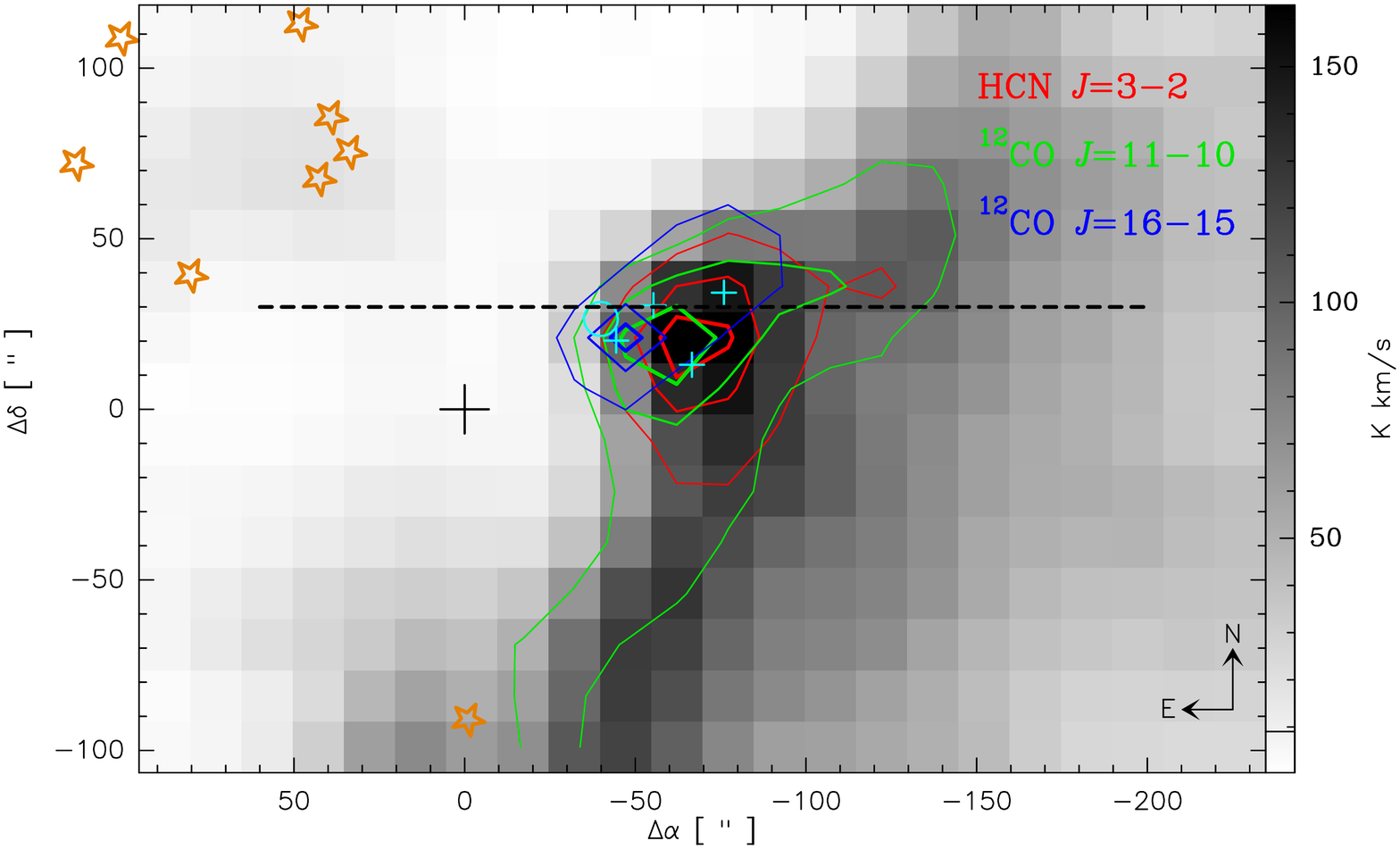,angle=0,width=0.55\linewidth} &
  \hspace{-0.80cm}\epsfig{file=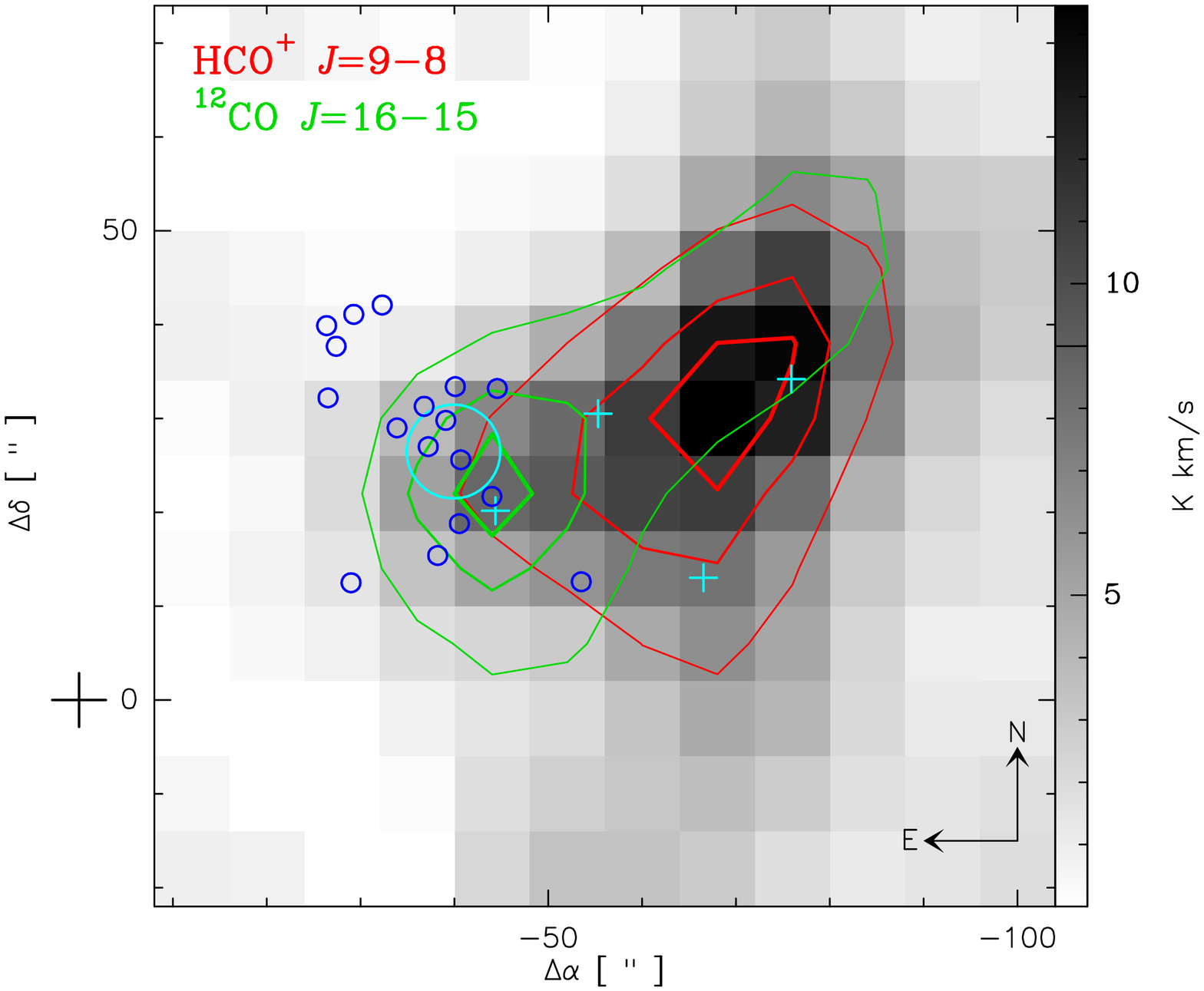,angle=0,width=0.45\linewidth}   
  
 \end{tabular}  

  \caption{\footnotesize{\textit{Left} - Velocity-integrated intensity maps of \hcn~$J=1\to0$ (grey), 
  \hcn~$J=3\to2$ (red contour), \twco~$J=11\to10$ (green contour), and \twco~$J=16\to15$ (blue contour). 
  The contour levels (from thin to thick) are the 50\%, 75\% and 90\% of the respective peak emissions. 
  The \textit{stars} on the top-left show the location of O and B ionizing stars 
  (Beetz \etal\ 1976; Hanson \etal\ 1997). The reference position ($\Delta\alpha=0$, $\Delta\delta=0$) 
  is the same as in Fig~\ref{fig:GREAT-maps}. The ultracompact \hii\ region M17-UC1 and four 
  H$_2$O masers \citep{johnson98}, are marked by the (cyan) circle and plus symbols, respectively. 
  The horizontal dashed line depict the strip lines shown in Fig.~\ref{fig:strip-lines}. 
  All maps are convolved to a $30''$ beam, to match the resolution of the \hcn~$J=1\to0$ map.
  \textit{Right} - Velocity-integrated intensity maps of \hcn~$J=8\to7$ (grey), \hcop~$J=9\to8$ 
  (red contour), and \twco~$J=16\to15$ (green contour). The \textit{blue circles} correspond to the 
  heavily obscured ($E_{median}>2.5$ keV, $A_V\ge10$ mag) population of X-ray sources around the 
  M17-UC1 region (Fig.10 in Broos~\etal\ 2007; coordinates from the VizieR catalogue). In this 
  case, all maps were convolved to the beam size ($16\farcs 6$) of the \twco~$J=16\to15$ map.}}
  \label{fig:HCN-CO-overlay}
\end{figure*}

For the $J=3\to2$ maps of \hcn\ and \hcop\ we used the new FFTS with two sub-sections, each 2.5 GHz width, overlapping 1 GHz in the band center, which gives a total of 4 GHz bandwidth. 
The new FLASH$^+$-345 receiver provides two sidebands separately, each 4 GHz wide. Each IF band is then processed with two 2.5 GHz wide backend, with 1 GHz overlap \citep{klein14}. 
We used 32768 channels that gives a spectral resolution of about 76 kHz in both $J=3\to2$ and $J=4\to3$ transitions (that is, $\sim8.6\times10^{-2}~\kms$ and $\sim6.6\times10^{-2}~\kms$ for the lower and higher $J$ line, respectively, at $v_{\rm LSR}=20~\kms$).
The on-source integration time per dump was 10 seconds for the HCN and \hcop\ $J=4\to3$ maps. For the new FLASH$^+$-345 the $T_{\rm sys}$ was about 180~K, and an average SSB system temperature of 224~K was observed with the APEX-1 receiver for the \hcn\ and \hcop~$J=3\to2$ maps.

We also used the dual color receiver array CHAMP$^+$ on APEX to map a region of about 80$''\times$80$''$ ($\sim$0.8~pc~$\times$~0.8~pc) in the \hcn~$J=8\to7$ (708.9~GHz) and \hcop~$J=9\to8$ (802.5~GHz) towards the dense core of M17~SW. The spatial resolution of these maps varies between $9.2''$ for the low frequency band, and $8.1''$ for the high frequency band.

%Observations toward Jupiter were performed during October 2009 to estimate the beam coupling efficiency ($\eta_c\approx0.59$) of the old FLASH-460 \citep{heyminck06}, assuming a brightness temperature of 158 K for the Jovian planet at 492 GHz, as interpolated from data reported in \citet{griffin86}. 
The coupling efficiency ($\eta_c\approx0.67$) of the new FLASH$^+$-345 was estimated 
from observations at 345~GHz toward Mars, during 2010 
July\footnote{http://www3.mpifr-bonn.mpg.de/div/submmtech/heterodyne/\newline flashplus/flashmain.html}. 
This was extrapolated to 0.659 and 0.656 for the 
\hcn~$J=4\to3$ (354.505 GHz) and \hcop~$J=4\to3$ (356.734 GHz), respectively.
A beam coupling efficiency of 0.72 was assumed for the APEX-1 receiver at the 
frequencies of \hcn\ and \hcop\ $J=3\to2$ \citep[][their Table 2]{vassilev08}.
%This coupling efficiency was chosen because in velocity-space (velocity channels) the size of the M17 clumps is Jupiter-like, which had a size $\sim38.7''$ by the time of the observations.
With these beam coupling efficiencies, and a forward efficiency ($\eta_f$) of 0.95, 
we converted all data to main beam brightness temperature scale, 
$T_{mb}=\eta_{f}\times T_{A}^{*}/\eta_{c}$.

\subsection{The IRAM 30m data}

The IRAM 30m observations toward M17~SW were described in \citet{pb15a}. 
Here we use the $J=1\to0$ and $J=2\to1$ 
maps of \twco\ and \thco\ reported there. In the 32~GHz signal bandwidth 
provided by the IF channels of the broadband EMIR receivers \citep{carter12}, 
we also obtained data for the $J=1\to0$ transitions of the \hcn, \hcop, 
as well as their isotopologues \hthcn\ and \hthcop, in the 3mm band. 
The angular resolutions of the original maps of \hcn, \hcop, \hthcn\ and 
\hthcop\ are $29\farcs4$, $29\farcs2$, $30\farcs2$ and $30\farcs1$, respectively.

The reduction of the calibrated data, as well as the maps shown throughout 
the paper, were done using the GILDAS\footnote{http://www.iram.fr/IRAMFR/GILDAS} 
package CLASS90. All the observed line intensities presented throughout this 
work, as well as the model estimates, are in main beam brightness temperature
obtained as above, but with a main beam coupling efficiencies ($\eta_c$)
and forward efficiencies ($\eta_f$) interpolated for each frequency from 
the latest reported EMIR 
efficiencies\footnote{http://www.iram.es/IRAMES/mainWiki/Iram30mEfficiencies}
We assume Rayleigh-Jeans approximation at all frequencies for consistency.

All the spectral lines used and reported in this work are summarized in 
Table~\ref{tab:spectral-lines}, including the upper-level energy and the 
critical densities associated with each transition.

\begin{figure}[!tp]

  \hfill\includegraphics[width=0.98\linewidth, angle=0]{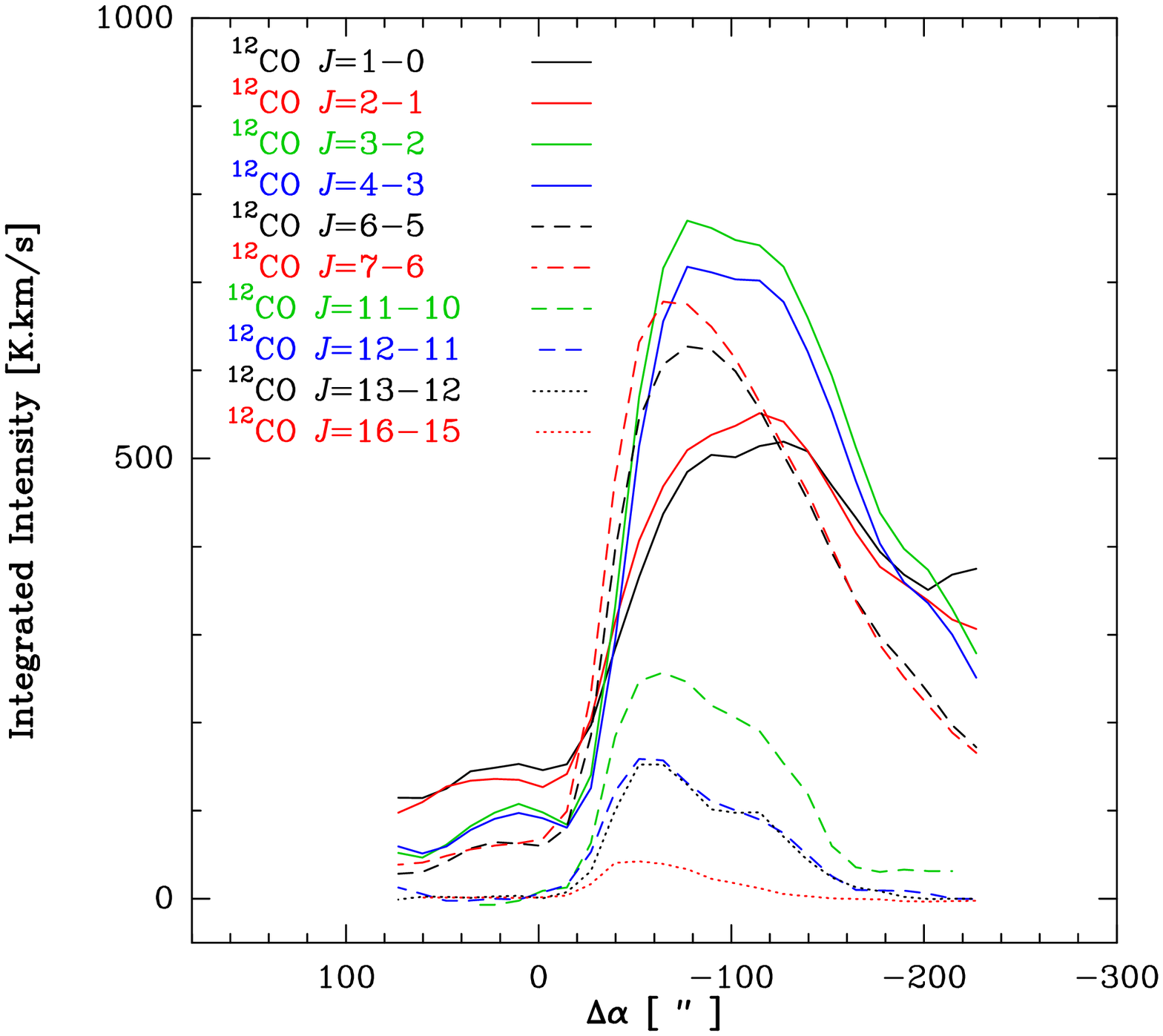}\hspace*{\fill}\\
  
  \hfill\includegraphics[width=0.98\linewidth, angle=0]{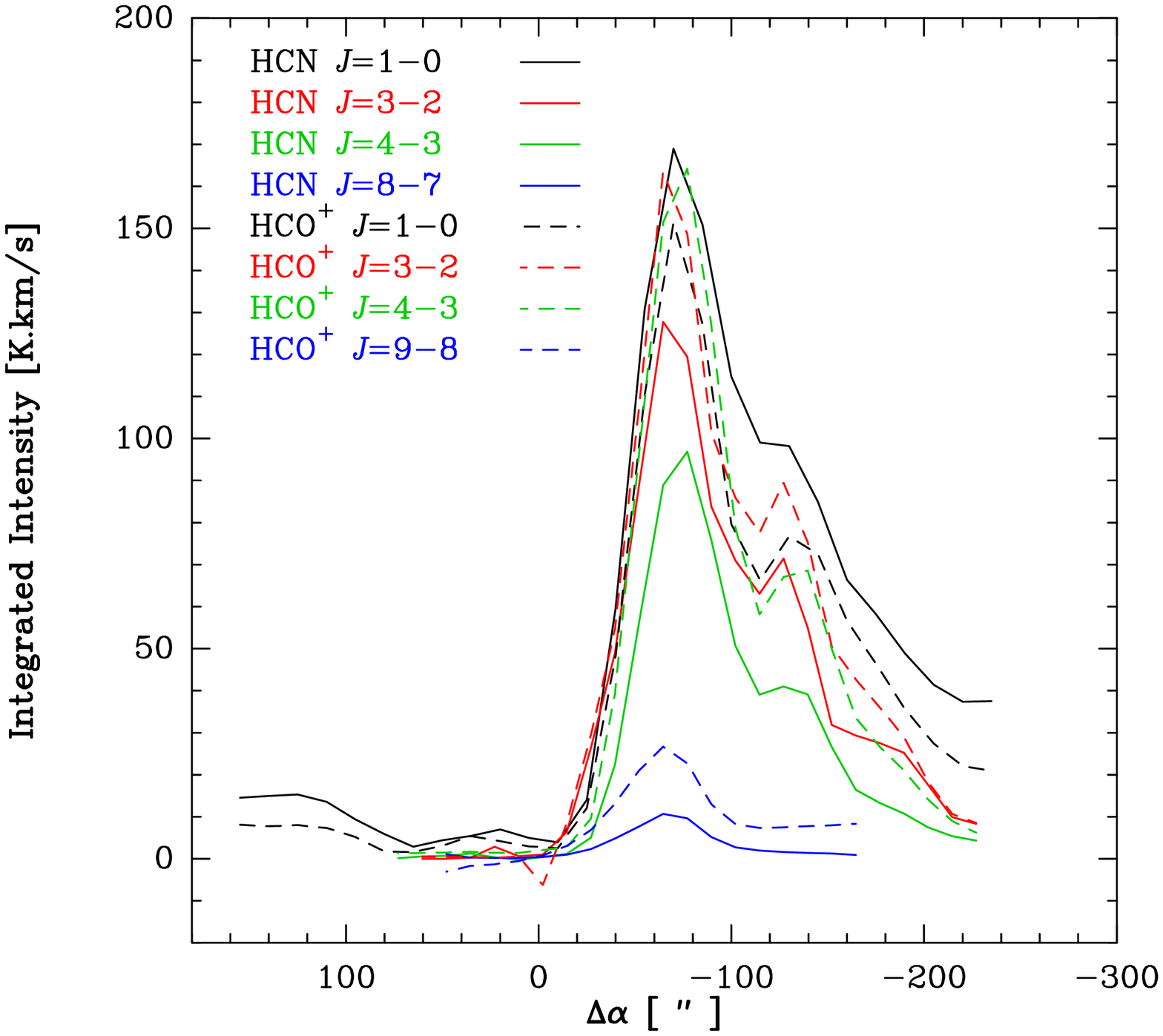}\hspace*{\fill}

  \caption{\footnotesize{Strip lines of the velocity integrated intensities of all the \twco\ (\textit{top}) and the \hcn\ and \hcop\ (\textit{bottom}) lines we have mapped so far, at the same declination $\Delta\delta=30''$ (P.A. $90^{\circ}$) across the ionization front of M17 SW. The \twco~$J=7\to6$ and $J=6\to5$ are from \citet{pb10}, the \twco~$J=13\to12$ is from \citet{pb12}, while \twco~$J=2\to1$ and $J=1\to0$ are from \citet{pb15a}. The rest of the lines are from this work. All maps were convolved to a 25$''$ beam to match the resolution of the \twco~$J=11\to10$ map. The X-axis corresponds to the actual offset in R.A. of the maps shown in Figs.~\ref{fig:GREAT-maps}, \ref{fig:APEX-30m-maps} and \ref{fig:CHAMP-maps}. The offset, $\Delta\alpha=0''$ in R.A. correspond to the reference illuminating star SAO 161357.}}
  \label{fig:strip-lines}
\end{figure}

%_______________________________________________________________
%
\section{Results}

\subsection{The high-$J$ $^{12}$CO line intensity maps}

Figure \ref{fig:GREAT-maps} shows the velocity-integrated (over 10 and 28~\kms) temperature (intensity) maps, 
of the (from \textit{top} to \textit{bottom}) $J = 16\to15$, $J = 12\to11$, and  $J = 11\to10$  
transitions of \twco. Their peak intensities are about 85~\Kkms, 316~\Kkms\ and 281~\Kkms, respectively. The 
\twco\ $J = 12\to11$, and  $J = 11\to10$  follow a similar spatial distribution, and their peaks are found at 
about the same offset position ($\Delta\alpha=-54'', \Delta\delta=17''$). Their emission peaks in between three 
of the four \hho\ masers reported by \citet[][their Table 9]{johnson98}, approximately 0.3~pc ($\sim30\farcs3$ or 
2.5 pixels at P.A. $90^{\circ}$) from the ridge defined by the 25\% contour line. 

On the other hand, the \twco~$J = 16\to15$ emission is much fainter 
and more spatially confined than that observed in the $J = 12\to11$ 
and $J = 11\to10$ transitions, and its spatial distribution is shifted 
towards the north-east, closer to the ionization front, with respect to 
the lower-$J$ lines. In fact, the $J = 16\to15$ emission peaks very close 
($\sim12''$ or 0.1~pc) to the ultra compact \hii\ region M17-UC1, which 
indicates that the \twco~$J = 16\to15$ transition may be excited in warm 
gas facing the ultracompact \hii\ region, not surprising in view of its 
high upper-level energy ($E_u\approx751.72$~K). Note that the peak of the 
\thco~$J=13\to12$ closely follows that of the \twco~$J = 16\to15$ emission.

%At the distance of 1.98 kpc, one arcsec of angular size is about 0.0096 pc.

\subsection{The HCN and HCO$^+$ line intensity maps}

Figure~\ref{fig:APEX-30m-maps} shows (from top to bottom) the velocity-integrated 
(between 0 and 40~\kms to cover the hyperfine structure lines of \hcn) 
temperature maps of the $J=1\to0$, $J = 3\to2$ and $J = 4\to3$ transitions of \hcn\ and \hcop. 
The peak intensities (in $T_{\rm mb}$ scale) of the $J = 1\to0$ lines of \hcn\ and \hcop\ are about
169~\Kkms\ and 152~\Kkms, respectively.
The \hcn\ and \hcop\ $J = 3\to2$ lines have peak intensities of 133~\Kkms\ and 183~\Kkms, and the 
$J = 4\to3$ lines are 107~\Kkms\ and 187~\Kkms, respectively.  
The \hcn~$J=1\to0$ emission is brighter than the \hcop~$J=1\to0$. The opposite 
is true for the higher-$J$ transitions in agreement with the higher critical densities of HCN.
The \hcn\ and \hcop\ $J=4\to3$ lines ($n_{cr}\sim2\times10^8~\3cm$ and $\sim9\times10^6~\3cm$ at 100~K, respectively, and both with $E_u\approx43$ K) are expected to probe much denser and colder regions than the ($J_{up}\geq6$) CO lines which have $E_u\>116$~K (c.f., Table~\ref{tab:spectral-lines}).
We discuss further the \hcop/\hcn\ line intensity ratios in Sec.~\ref{sec:hcn-hcop-ratios}.

The \hcop~$J = 4\to3$ emission seems more extended than the \hcn\ one, particularly towards the northern edge 
of the cloud core, where the ratio is larger. The broader emission of the \hcop\ $J = 4\to3$ line was also 
observed in the $J = 3\to2$ transition by \citet{hobson92}, who reported a more extended emission of \hcop\ 
further into the \hii\ region.
Despite the difference in extension, the overall spatial distribution (morphology) of the \hcn\ and \hcop\ 
emission is similar.

The \hcn~$J=8\to7$ and \hcop~$J=9\to8$ maps obtained with CHAMP$^+$ on APEX are shown in Fig.~\ref{fig:CHAMP-maps}.  
Because these lines are weak and the beam size of the telescope is smaller (with an angular resolution of 
$9\farcs2$ and $8\farcs1$, respectively) we only covered the area around the peak of the \hcn\ and \hcop\ 
$J=3\to2$ emission. The peak intensities of the \hcn~$J=8\to7$ and \hcop~$J=9\to8$ lines are 24~\Kkms\  and
49~\Kkms, respectively, and they are located at offset position ($\Delta\alpha=-73'', \Delta\delta=37''$), 
about $9''$ ($\sim$0.09~pc) from one of the \hho\ masers. Because the upper-level energy of \hcn~$J=8\to7$ and 
\hcop~$J=9\to8$ are $E_u\approx153$~K and $E_u\approx193$~K, respectively, they may be excited by warmer gas 
than the lower-$J$ lines of \hcn\ and \hcop, which can explain the second emission peak (observed in both 
lines) found just next to the M17-UC1 region.

\subsection{Comparison of the morphology}

Comparing with the higher-$J$ \twco\ lines, the left panel of 
Fig.~\ref{fig:HCN-CO-overlay} shows the \hcn~$J=3\to2$, \twco~$J=11\to10$ 
and $J=16\to15$ lines overlaid on the \hcn~$J=1\to0$ map. 
The \hcn~$J=3\to2$ line follows the distribution of the 
\hcn~$J=1\to0$ line, having their peak intensities at about 
the same offset position. Instead, \twco~$J=11\to10$ and 
\twco~$J=16\to15$ show a clear stratification. The peak 
intensity is shifted towards the east relative to the HCN 
lines (by about 48$''$ or $\sim$0.45~pc for \twco~$J=16\to15$). 
The \twco~$J=16\to15$ emission is very compact and peaks close 
to the M17-UC1 region. Since the $J=16\to15$ and $J=11\to10$ 
were observed simultaneously with the dual band receiver GREAT 
onboard SOFIA, the shift between both lines cannot be due to 
pointing errors. A difference in the excitation conditions 
is the most likely reason for the shift between the peak 
intensities of these \twco\ lines. 

In the $30''$ resolution maps, the \twco~$J=16\to15$ line peaks 
right in between the M17-UC1 region and two of the four \hho\ 
masers (Fig.~\ref{fig:HCN-CO-overlay}, left panel). In the higher 
resolution ($16\farcs6$) map shown in the right panel of 
Fig.~\ref{fig:HCN-CO-overlay}, the \twco~$J=16\to15$ emission 
actually peaks closer to the UC1 region and the eastern most 
\hho\ maser, as well as three of the heavily obscured 
($E_{median}>2.5$ keV, $A_V\ge10$ mag) population of X-ray 
sources found around the M17-UC1 region by 
\citet[][their Fig.10, coordinates from the VizieR online 
catalog\footnote{http://vizier.cfa.harvard.edu/viz-bin/VizieR?-source=J/ApJS/ 169/353}]{broos07}.
Because of the proximity to the ultra compact \hii\ region, 
the \hho\ maser, and at least three embedded X-ray sources, 
the \twco~$J=16\to15$ line is very likely excited by warmer 
gas than the $J=11\to10$ transition. This agrees with the fact 
that the upper-level energy ($E_{up}\approx752$~K) of the 
higher-$J$ line is about a factor two higher than that of the 
$J=11\to10$ line ($E_{up}\approx365$~K). 

Figure~\ref{fig:strip-lines} shows the variation of the integrated intensity of the \twco\ 
(\textit{top}) and the \hcn\ (\textit{bottom}) lines across the ionization 
front. This corresponds to the strip line at P.A=$90^{\circ}$ shown by 
\citep[][the \twco, \thco, $J=6\to5$ and $J=7\to6$ are taken from that work]{pb10}.
The cut also quantifies the stratification already discussed for Fig. 4. The first peak appears in the high-$J$ CO lines at $-50''$ while it falls at about $-70''$ for all HCN lines. In contrast to \twco, HCN shows no shift of the peak as a function of the energy level.
The peak of the integrated intensities of the \twco\ lines shows a clear progression from deep into the molecular 
cloud (from about $\Delta\alpha=-120''$) in the $J=1\to0$ and $J=2\to1$ lines towards the molecular ridge and 
the ionization front ($\Delta\alpha$ between $-50''$ and $-60''$) in the $J=12\to11$ and $J=13\to12$ lines. The 
$J=16\to15$ emission shows a smooth increment from the molecular cloud and reaches a plateau about $10''$ 
($\sim$0.096~pc from the peak of the $J=13\to12$ line) closer to the ridge at $\Delta\alpha=-40''$. This 
increment in the \twco~$J=16\to15$ may be indicative of gas getting denser and warmer towards the ionization 
front and close to the UC-\hii\ region.
On the other hand, the \hcn\ and \hcop\ lines show a well defined peak intensity closer to the ionization front, 
between $\Delta\alpha=-60''$ and $-80''$, coinciding with the peak of the \twco~$J=7\to6$ and 
$J=11\to10$ lines. There is a secondary peak 
deeper into the molecular cloud at $\Delta\alpha$ between $-120''$ and $-140''$, which coincides with the peak 
intensity of the lower-$J$ \twco\ lines.

%Every strip line was smoothed spatially (in two pixels with respect to their respective original resolutions) along the strip direction. 

% We do not show here the strip lines 
% at P.A.=$63^{\circ}$ because we do not have \hcn\ and \hcop\ data at lower 
% declinations.

\begin{figure*}[!pt]

  \hspace*{\fill}\includegraphics[angle=0,width=8.5cm]{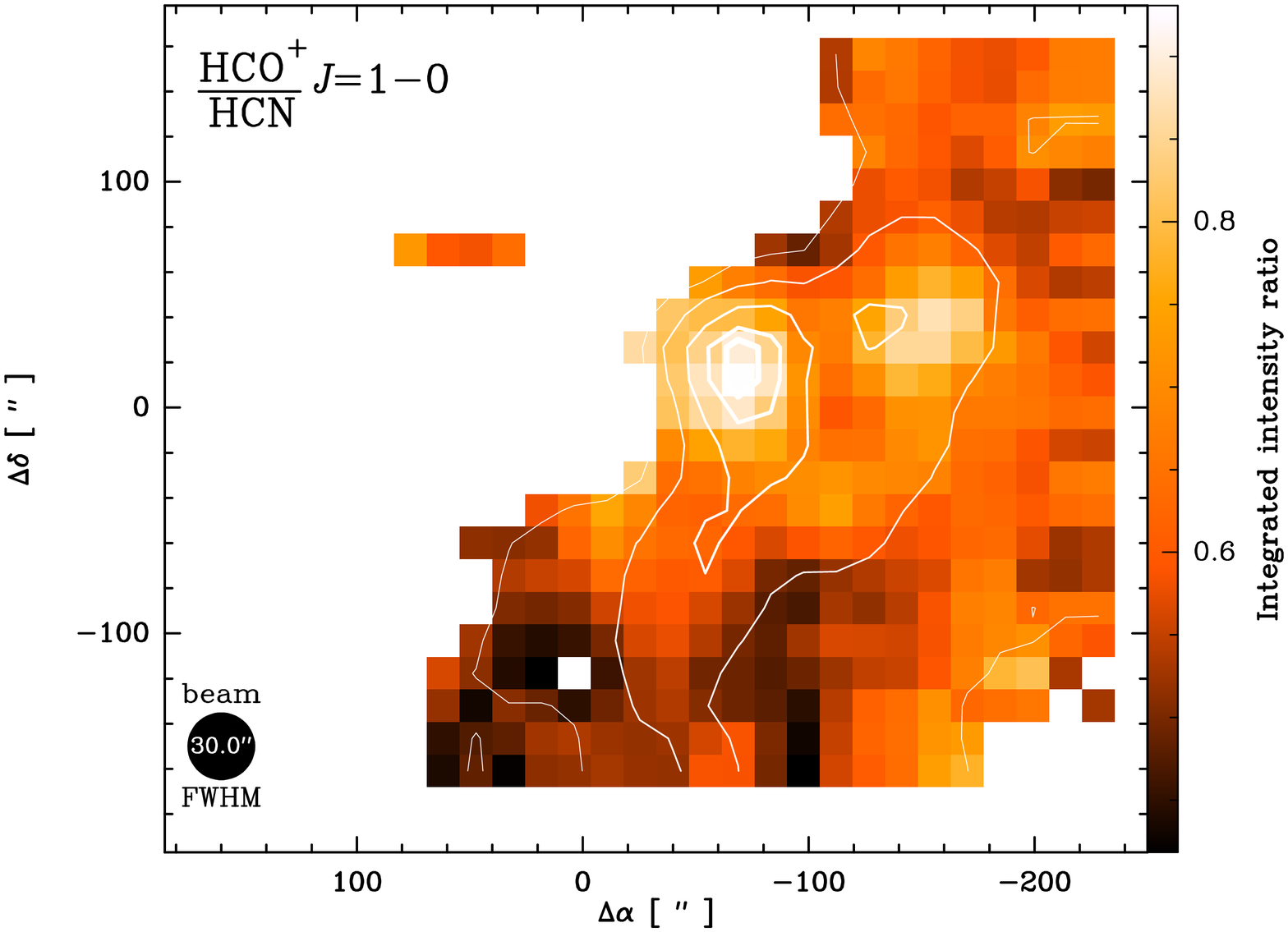}%
  \hfill\includegraphics[angle=0,width=9.0cm]{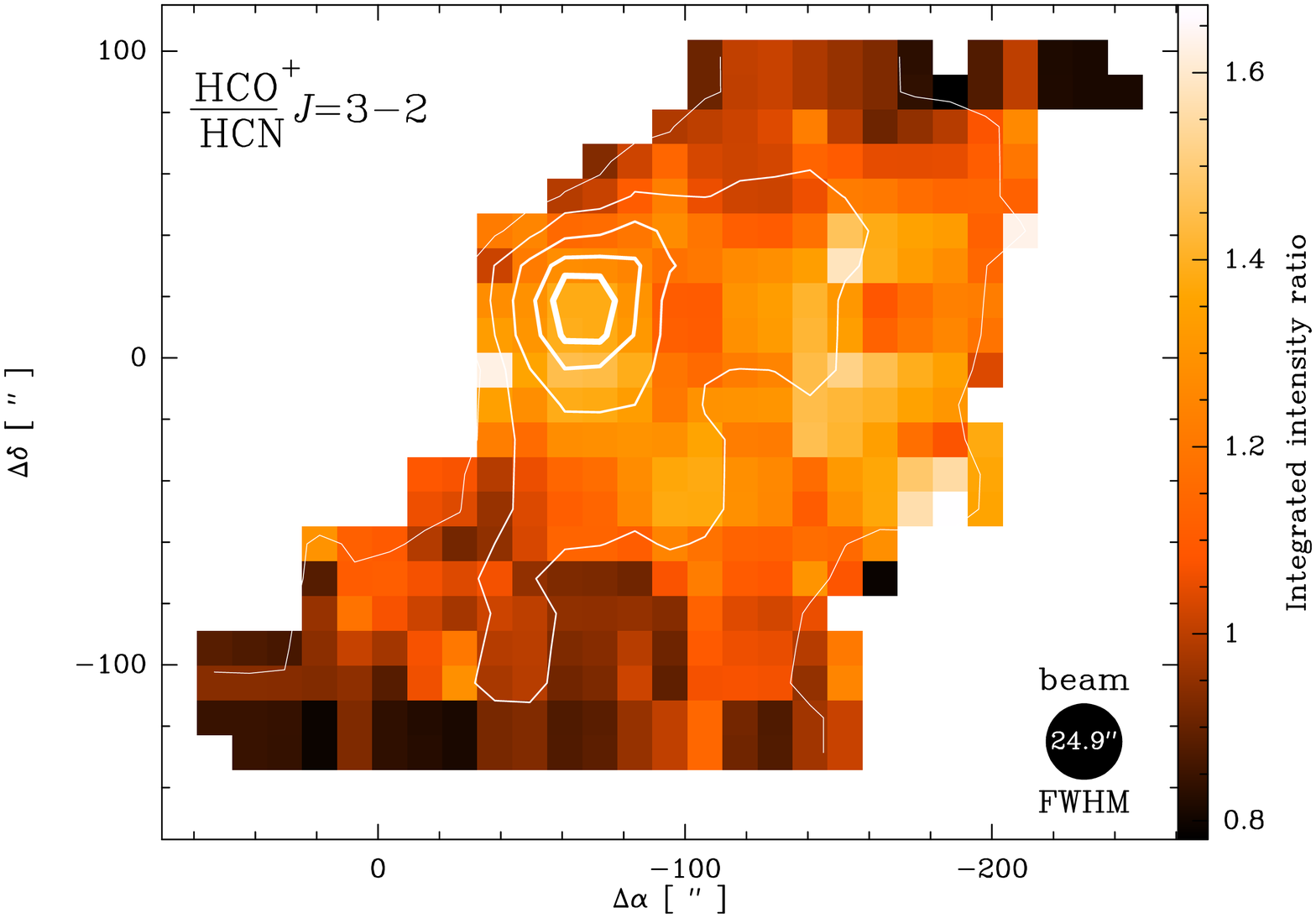}\hspace*{\fill}\\
  
  \hspace*{\fill}\includegraphics[angle=0,width=9.5cm]{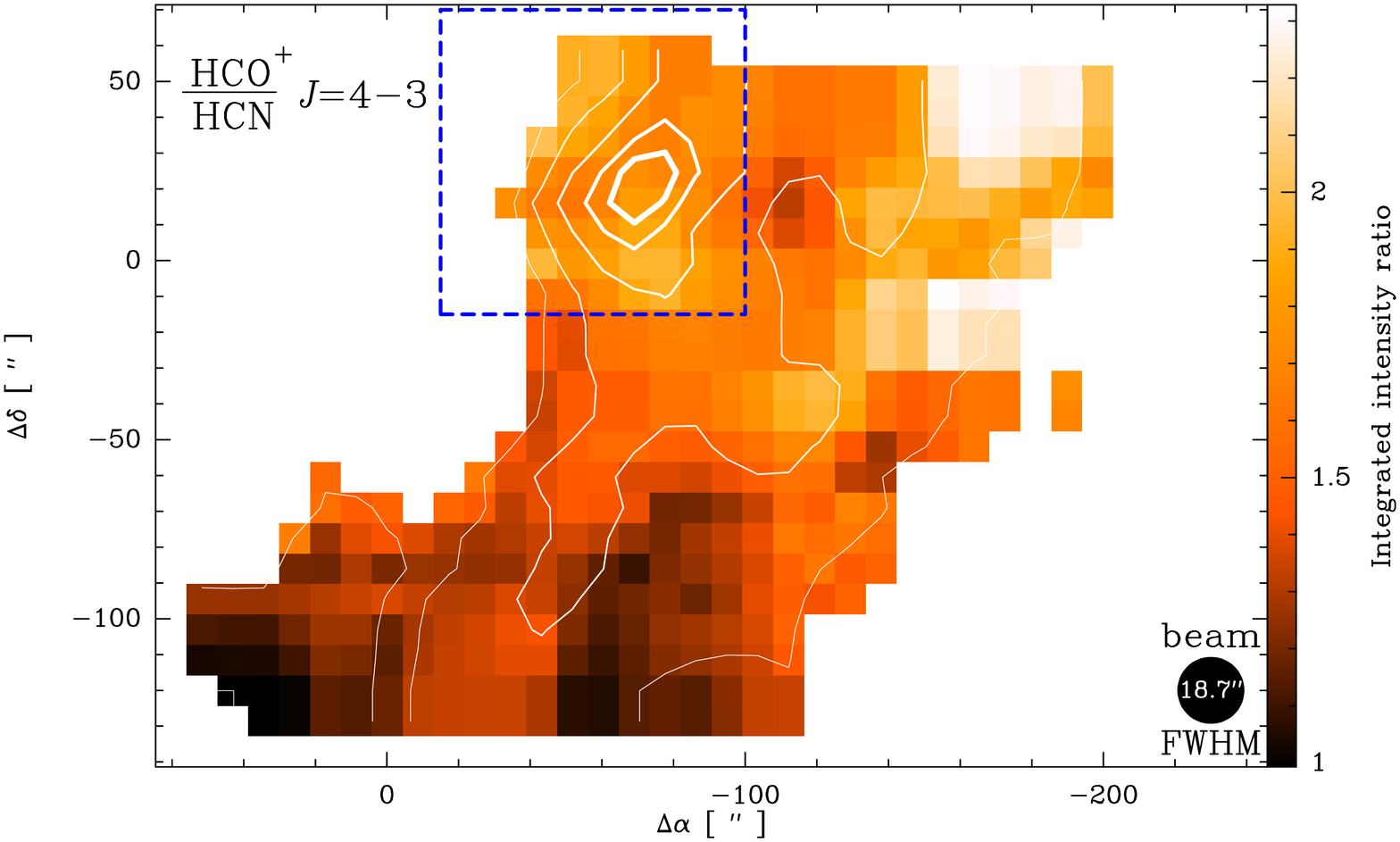}%
  \hfill\includegraphics[angle=0,width=7.5cm]{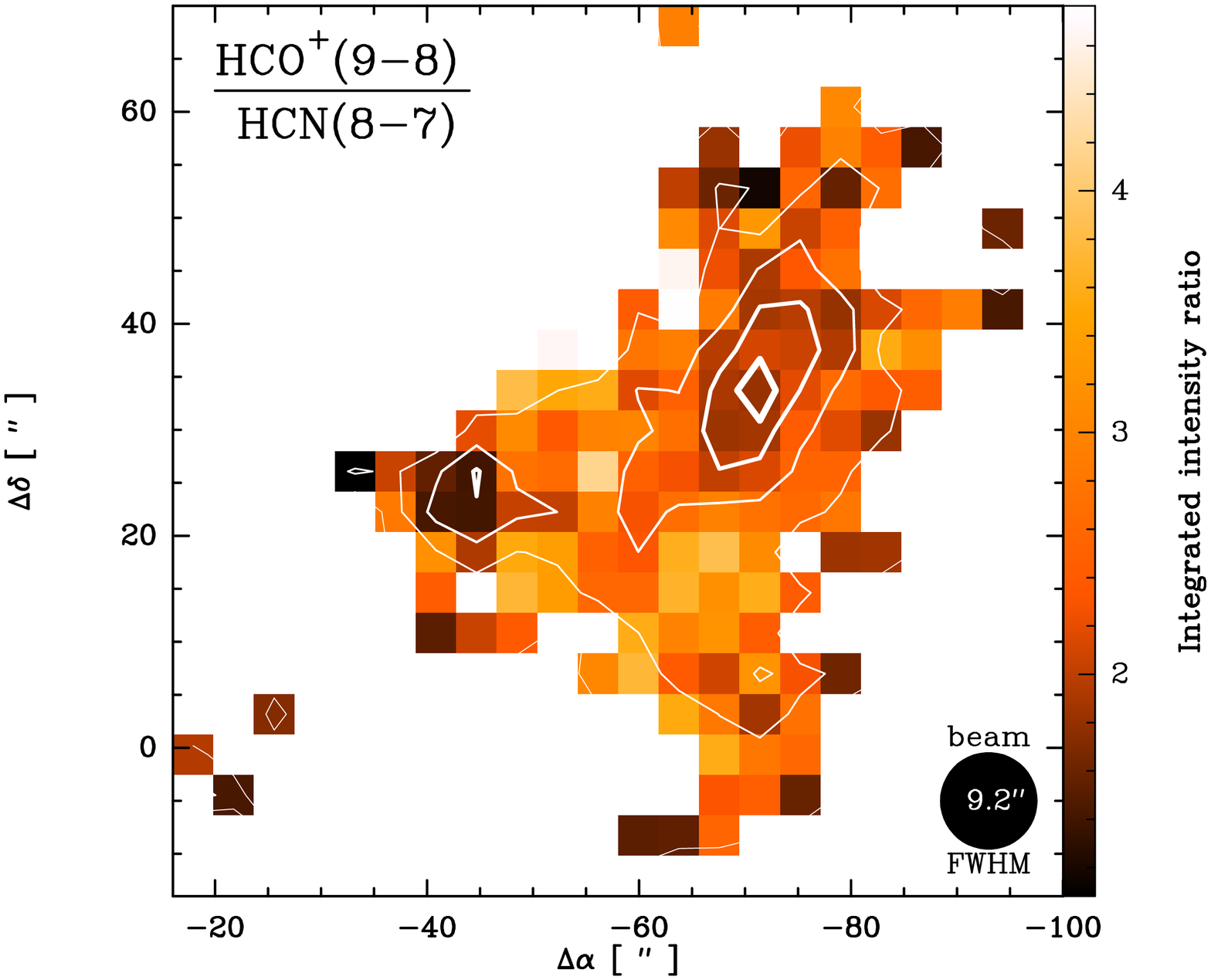}\hspace*{\fill}

  \caption{\footnotesize{Maps of the \hcop/\hcn\ ratio between the $J = 1\to0$ (\textit{top-left}), $J = 3\to2$ (\textit{top-right}) and $J = 4\to3$ (\textit{bottom-left}) line integrated intensities in the region where the lines are brighter than 7\% of their peak values. The $J=1\to0$ ratio varies between about 0.4 and 0.9. The isolated ratios shown at $\Delta\delta\sim70''$ correspond to significant emission of $\sim$30~\Kkms\ in both \hcn\ and \hcop~$J=1\to0$ lines. The $J = 3\to2$ line ratio ranges from $\sim0.8$ to $\sim1.7$ while the $J = 4\to3$ line ratio varies between $\sim1.0$ and $\sim2.3$. The contour lines correspond to the \hcop\ $J = 1\to0$, $J = 3\to2$ and $J = 4\to3$ maps in the respective line ratios, and are defined as in Fig.~\ref{fig:APEX-30m-maps}. The \hcop~$J=9\to8$ to \hcn~$J=8\to7$ line ratio is shown in the \textit{bottom-right} panel, for the region  where the lines are brighter than the 15\% of their peak values. This ratio varies between about unity to $\sim$4.8. The contour lines correspond to the \hcn~$J=8\to7$ maps and are the same as in Fig.~\ref{fig:CHAMP-maps}}. The area of the higher-$J$ \hcn\ and \hcop\ ratio map is depicted with a dashed frame in the bottom-left panel. The maps of the higher frequency lines were convolved to the larger beam of the lower frequency line before determining the respective ratios.}

  \label{fig:ratio-maps}
\end{figure*}

\subsection{The \hcop/\hcn\ line intensity ratios}\label{sec:hcn-hcop-ratios}

The \hcop~$J = 4\to3$ map was convolved with the slightly larger beam size ($17.7''$) of the 
\hcn~$J = 4\to3$ line, in order to have the same number of pixels per map. The pixel size is
$8.9''\times8.9''$ which ensures Nyquist sampling of the spectra in both (R.A. and Dec.) directions. 
Likewise, the \hcop~$J = 3\to2$ map was convolved with the beam size ($23.6''$, and spatial 
resolution of $11.8''\times11.8''$) of the \hcn\ $J = 3\to2$ line before computing the ratio between the 
respective maps. The \hcop~$J=9\to8$ was convolved to the larger beam of the 
\hcn~$J=8\to7$ line to match its spatial resolution ($9.2''$). The same was done for the 
$J=1\to0$ maps, but convolved to a $30''$ beam.

In order to reduce uncertainties and avoid misleading values in zones with relatively low S/N and no clear 
detection, we compute the \hcop/\hcn\ $J = 4\to3$, $J = 3\to2$ and $J = 1\to0$ line intensity ratios only in the 
region where the integrated temperature of both lines (in both species) is brighter than 7\% of their peak 
values. That is, between 5$\sigma$ and 8$\sigma$, where $\sigma$ is the rms of the \hcn\ and \hcop\ $J = 4\to3$ 
and $J = 3\to2$  maps, which ranges between 0.2~K and 0.3~K. In the case of the $J=1\to0$ lines the rms 
is one order of magnitude lower.
Due to the lower S/N of the \hcn~$J=8\to7$ (the fainter line), we used a threshold of 15\% for the 
higher-$J$ line ratio. This corresponds to about 6$\sigma$ in both \hcn~$J=8\to7$ and \hcop~$J=9\to8$, with rms 
values of 0.62~K and 1.24~K, respectively.
Because the \hcn\ and \hcop\ maps were observed simultaneously with practically the same beam size, 
our results are not affected by relative pointing errors.

All the \hcop/\hcn\ line intensity ratios are shown in Fig.~\ref{fig:ratio-maps}.
The \hcop/\hcn\ $J = 1\to0$ line ratio is lower than unity in 
most of the region mapped. The fact that the higher-$J$ \hcn/\hcop\ 
intensity ratio becomes larger than unity may be due to infrared 
pumping of the \hcn~$J=1\to0$ line through a vibrational transition 
at 14~\mum\ wavelength. This mechanism can produce an enhanced 
\hcn~$J=1\to0$ emission in conditions of sub-thermal excitation, 
as discussed in the next section.

The \hcop/\hcn\ $J = 3\to2$ line intensity ratio ranges between 
$\sim0.8$ and $\sim1.7$, while the intensity of the \hcop~$J = 4\to3$ 
line is a factor $\sim1-2$ brighter than that of the \hcn\ line, 
as shown in Fig.~\ref{fig:ratio-maps}. The line profiles of both 
lines match very well in all the positions, which indicates that 
the difference in intensities is not a consequence of different 
line widths.

With a higher spatial resolution ($9\farcs2$) the line intensity 
ratio \hcop($9\to8$)/\hcn($8\to7$) shows the largest values 
(up to $\sim$4.8) of the four ratio maps. The ratio is lowest 
(between $\sim$1 and $\sim$2) at the region of the strongest 
\hcn~$J=8\to7$ emission (as shown by the contours in the bottom-right 
panel of Fig.~\ref{fig:ratio-maps}), which coincides with the M17-UC1 
region (south-east peak) and the peak of the \hcn~$J=8\to7$ line 
(north-west peak). The relatively bright \hcn\ emission in these two 
regions can be explained by dense and warm gas, and by the presence 
of hot cores (or UC-\hii\ regions) 
%which are known to enhance the abundance of fully hydrogenated species due to high temperature chemistry 
\citep[e.g.,][]{prasad87, caselli93, viti99, rodgers01}.

Similar high \hcop/\hcn\ $J = 4\to3$ line ratios have been observed in other Galactic star forming regions 
\citep[e.g., W49A][]{peng07}, while \hcop/\hcn\ $J = 4\to3$ line ratios lower than unity have been found from 
single dish observations of active galaxies like NGC~1068 \citep{pb09} and in recent ALMA observations 
\citep[][note that for the ALMA observations the authors quote the inverse \hcn/\hcop\ line 
ratio instead]{garcia14, viti14}. According to models by \citet[][their Fig.~14]{meijerink07}, however, 
\hcop/\hcn\ $J = 4\to3$ line ratios lower than unity are signatures of a high density ($>10^4~\3cm$) PDR 
environment rather than an X-ray dominated region (XDR), as suggested by \citet{garcia14}. But 
interpretations of the \hcop/\hcn\ line ratios in extragalactic environments may not be as straight forward as 
believed in the past, since these species are now known to be strongly time-dependent in both dense gas 
\citep[e.g.][]{bayet08} and in XDRs \citep[e.g.][]{meijerink13}, as pointed out by \citet[][]{viti14}. Hence, 
the \hcop/\hcn\ line ratios observed in galaxies may reflect departures from chemical equilibrium rather than 
differences in the underlying excitation environment.

\subsubsection{Infrared pumping of the dense molecular gas?}

In environments with high enough UV/X-ray surface brightness to heat the dust to several 
hundred K, like in AGNs, bright starbursts or massive star-forming regions in the Galaxy and the 
circumnuclear disk (CND) around the massive black hole at the center of our Galaxy, 
strong \hcn, \hnc, and \hcop\ emission in the sub-millimeter and millimeter range may 
also be explained by the infrared radiative pumping scenario 
\citep[e.g.][]{aalto95, christopher05, garcia06, guelin07, aalto07a, aalto07b}, 
since hot dust produces strong mid-infrared continuum emission in the 
10--30 \mum\ wavelength range.

If the infrared pumping scenario is at work, absorption features must be 
detected in infrared spectra at 12.1~\mum, 14.0~\mum, and 21.7~\mum, which are the wavelengths of 
the ground vibrational states of \hcop, \hcn, and \hnc, respectively, connected by the transitions 
$\nu_2=1$ of the lowest excited bending states. At the moment we do not have high resolution IR 
spectra to check if this is the case in all the mapped region of M17~SW.
On the other hand, the \hcn~$J=4\to3$, $\nu_2=1$ transition (356.256~GHz) was detected with a peak intensity of $\sim$50~mK towards the CND of the Milky Way \citep{mills13}, indicating radiative pumping of \hcn\ at 14.0~\mum. However, we did not detect this vibrationally excited \hcn\ line, which lies close to our \hcop\ spectra, at an rms level of $\sim$1~mK.
%Besides, temperatures as high as 130~K were estimated for the amorphous carbon grains from ISO/SWS infrared 
%observations (in the $\sim$10--40~\mum\ wave range) toward M17~SW \citep{jones99}. This temperatures 
%imply that the corresponding continuum (black body) radiation peaks at $\lambda_{max}\ga20$ \mum\ 
%(from the Wien's displacement law $\lambda_{max}T=b$, with $b\approx2898$~K~\mum). 
Besides, the fact that the \hcop/\hcn\ line ratio grows 
monotonically with $J$-level can be considered an argument against 14~\mum\ pumping. 
Therefore, we think the \hcn\ and \hcop\ lines in M17~SW are unlikely 
to be affected by infrared pumping.

\subsubsection{Other causes of high \hcop/\hcn\ line ratios}

Due to the smaller physical scales and the different metallicity, 
the time-dependency of the abundance of \hcn\ compared to that of 
\hcop\ should not be an issue in Galactic molecular clouds. However, 
the bulk \hcop\ emission may still arise from gas that does not 
co-exist (in terms of ambient conditions or, equivalently, from a 
different layer within a clump) with the gas hosting the \hcn, as 
previously suggested by \citet[][their Apendix~C.3]{pb09}.

Stronger \hcop\ emission could also be the consequence 
of the higher ionization degree in X-ray dominated regions (XDRs), 
which leads to an enhanced \hcop\ formation rate \citep[e.g.][]{lepp96, meijerink05}.
Using the XDR models by \citet{meijerink05} we estimate that in order to drive an 
XDR, an X-ray source (or a cluster of sources) with a (combined) impinging luminosity 
of at least $10^{32}~\ergs$ would be required to be within a few arcsecs ($<0.03$ pc 
in M17~SW) from the region of the \hcop\ $J=4\to 3$ peak emission, considering that 
the X-ray flux decreases with the square of the distance from the source.

Based on the luminosities estimated from thermal plasma (546 sources) and 
power law fits (52 sources) of the photometrically selected \textit{Chandra}/ACIS 
sources by \citet{broos07}, we estimate that the combined X-ray luminosity 
(integrated between 0.5 keV and 50 keV, and projected on the sky at the position of 
the peak \hcop\ emission) is about three orders of magnitude lower than needed to 
drive an XDR. This result rules out an XDR scenario to explain the relatively 
strong \hcop\ emission. 
However, our estimates are based on an X-ray SED fit, extrapolated up to 50 
keV with no actual observations above 10~keV (the upper band of 
\textit{Chandra}/ACIS). 
Higher energy ($>$10~keV) photons could make a larger contribution to the 
X-ray flux, particularly from sources that have significant power-law tails. 
However, unless such X-ray sources would be located sufficiently close
to (or within) the region with bright \hcop\ emission, they probably could
not increase the required X-ray flux by three orders of magnitude.
In fact, no X-ray source was detected by \textit{Chandra}/ACIS within a 
radius of $\sim$10$''$ ($\sim$0.09~pc) around the peak \hcop\ emission. 
This either rules out the existence of any X-ray source in that region or implies 
that all X-ray photons 
with energy $<$10~keV are heavily absorbed by the large column density of the gas 
($\sim$8$\times$10$^{23}~\2cm$, \citealt{stutzki90}). Therefore, future 
observations sensitive to higher energy ($kT>10$ keV) photons that could escape the 
large column of gas, are required to unambiguously discard 
any heavily obscured X-ray source that may exist in the dense core of M17~SW.

Besides X-rays there are other possible mechanisms that can also lead to a relatively bright \hcop\ emission.
Indeed, if the intensity of emission from \hcop\ is as strong as that from HCN (or 
stronger) it may be due to relatively high kinetic temperatures, strong UV 
radiation fields, and relatively lower densities of the gas from 
where the \hcop\ emission emerges 
\citep[e.g.][]{fuente93, chin97, brouillet05, christopher05, zhang07, meijerink07}.

%Two possible reasons: different gas (probably not) - subthermal HCN excitation due to higher critical density -> sensitive estimate for the density in the considered region.

Since \hcn\ and \hcop\ are expected to co-exist at 
similar depths in a cloud, the kinetic temperature of 
their surrounding gas must also be similar. But the 
critical density of the \hcop\ lines is one order of 
magnitude lower than that of \hcn\ 
(cf., Table~\ref{tab:spectral-lines}), which makes 
\hcop\ more easily excited (collisionally) in single-phase 
molecular gas. 
Because the \hcop/\hcn\ line ratio reflects 
the effects of the combination of abundance and excitation 
temperature, the increasing ratio with $J$-line can be 
indicative of a lower excitation temperature in the 
higher-$J$ transitions of \hcn.
%as the explanation for the unusually high \hcop/\hcn\ 
%line ratios observed in the $J_{up}\geq3$ transitions.
We explore this alternative through excitation models 
in Sect.\ref{sec:analysis}.

%---------------------------------------------------------------
\begin{figure}[!tp]
 \centering
 \begin{tabular}{c}
  \hspace{-0.3cm}\epsfig{file=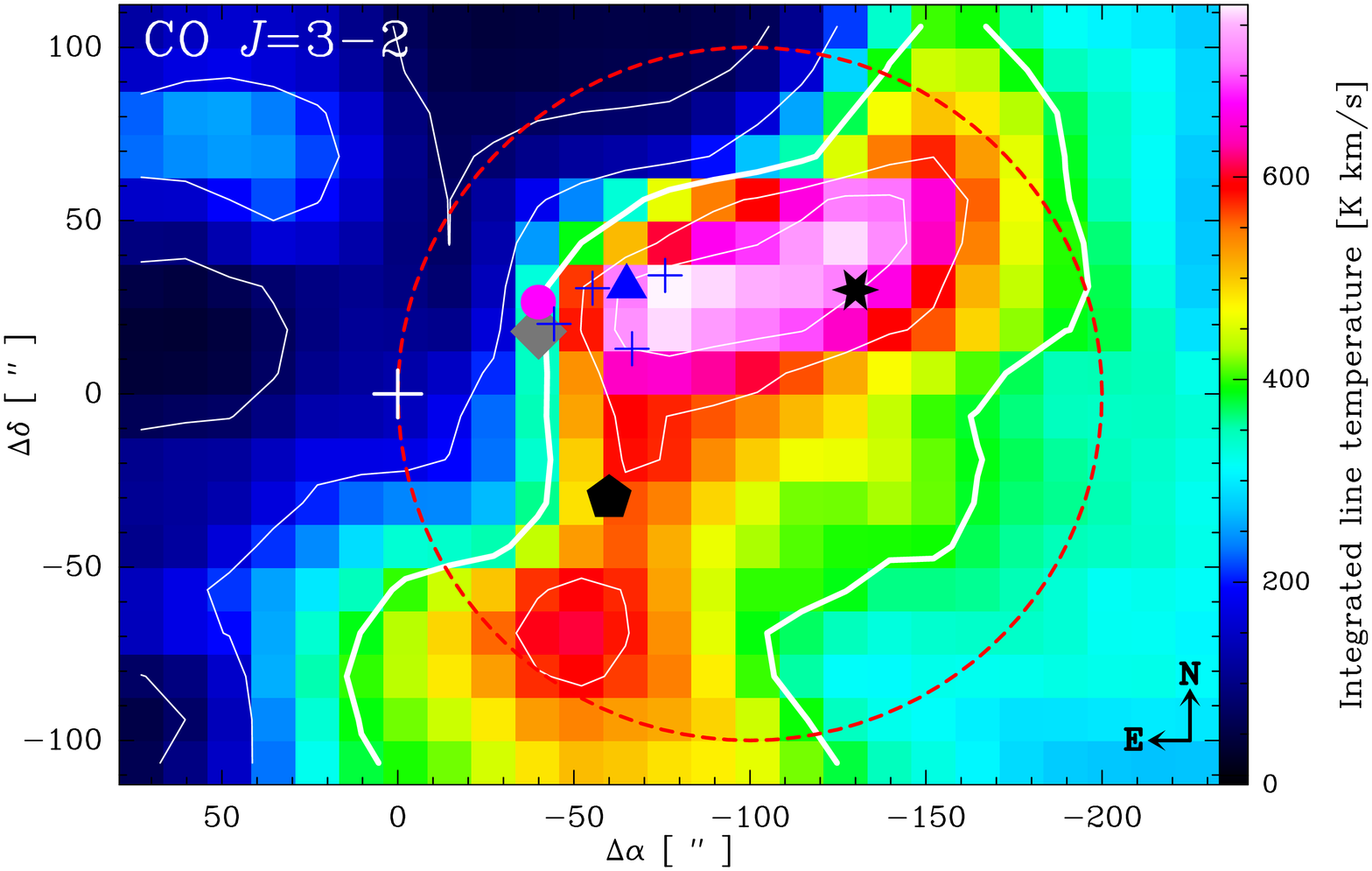,angle=0,width=0.90\linewidth} 
 \end{tabular}

  \caption{\footnotesize{Velocity-integrated intensity map of \twco~$J=3\to2$, 
  convolved a $25''$ beam to match the resolution of the \twco~$J=11\to10$ map. 
  The contour levels are the 10\%, 25\%, 50\% (thick line), 75\% and 90\% of the 
  peak emission (770~\Kkms). The central position ($\Delta \alpha=0$, $\Delta \delta=0$) 
  marked with a white cross, is the same as in Fig.~\ref{fig:GREAT-maps}. The ultracompact 
  \hii\ region M17-UC1 and four H$_2$O masers \citep{johnson98} are marked by the filled 
  circle and plus symbols, respectively. The positions of the peak intensities of 
  \hcn~$J=8\to7$ and \twco~$J=16\to15$, as well as the offset positions at ($-60''$,$-30''$) 
  and ($-130''$,$+30''$) analyzed in Sec.~\ref{sec:key-positions}, are indicated with 
  a triangle, a diamond, a pentagon and a star, respectively.
  }}

  \label{fig:CO3-2_map_B25}
\end{figure}
%---------------------------------------------------------------

%---------------------------------------------------------------
\begin{figure}[!tp]
 \centering
 \begin{tabular}{c}
  \epsfig{file=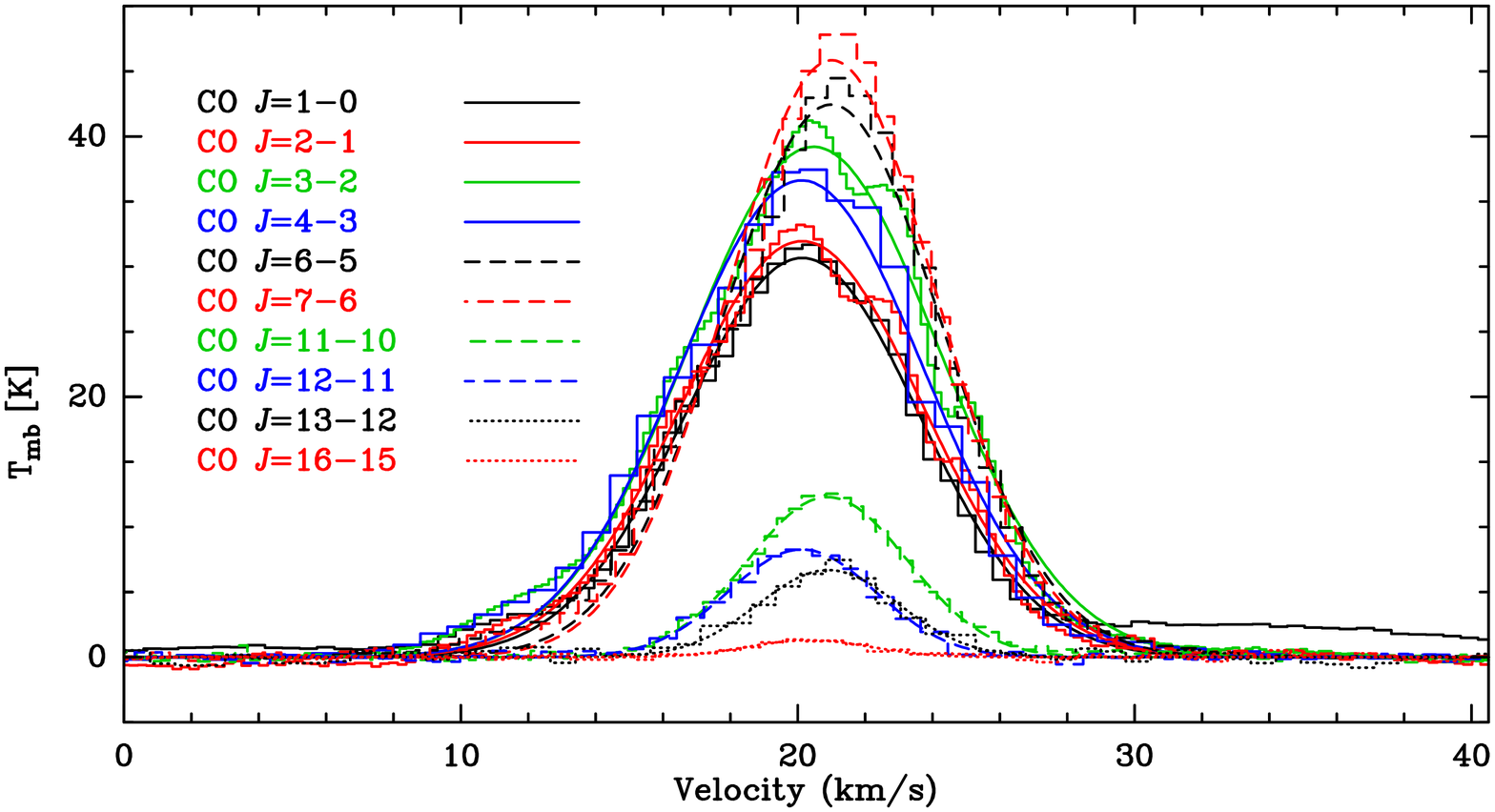,angle=0,width=0.80\linewidth} \\
  \epsfig{file=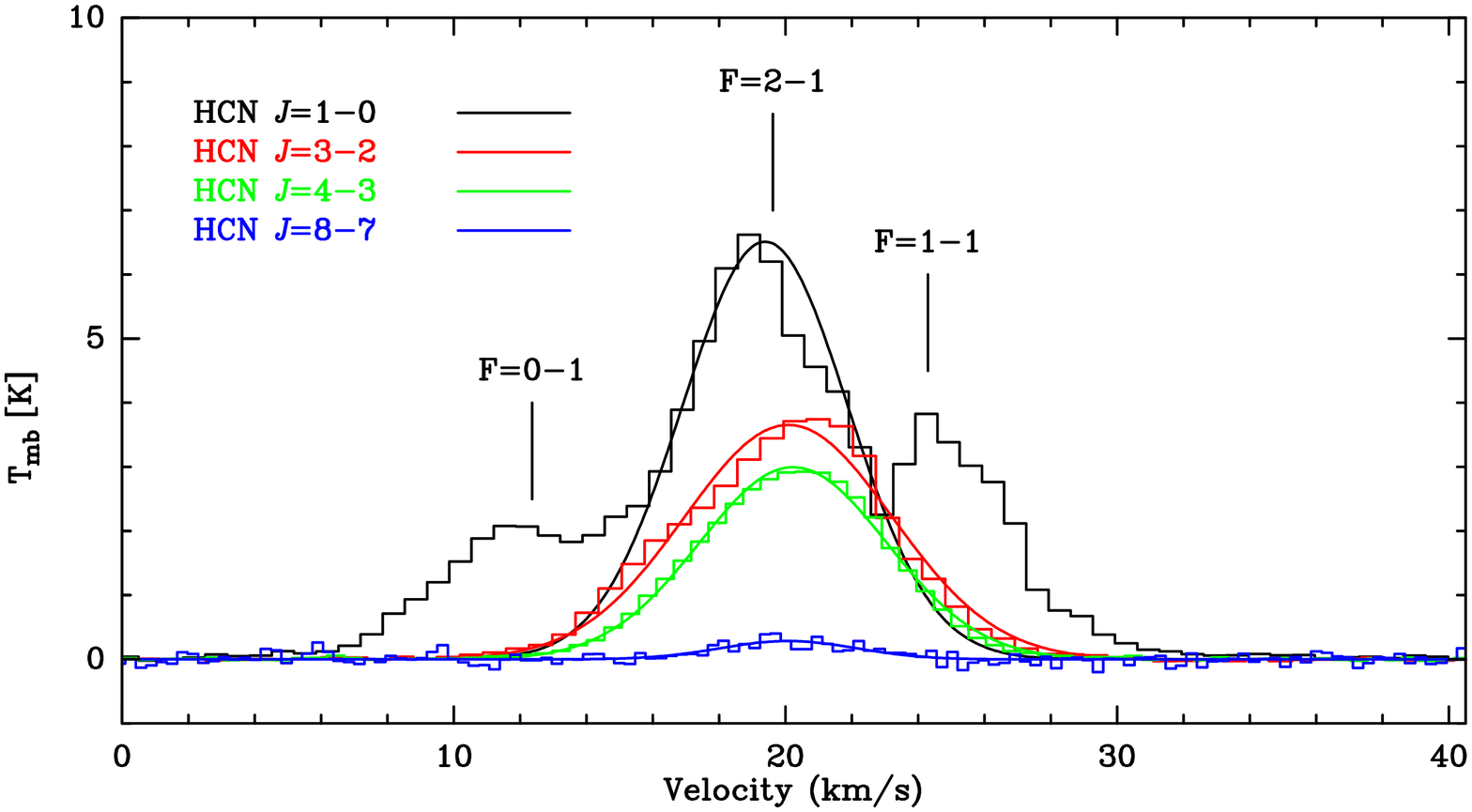,angle=0,width=0.80\linewidth} \\
  \epsfig{file=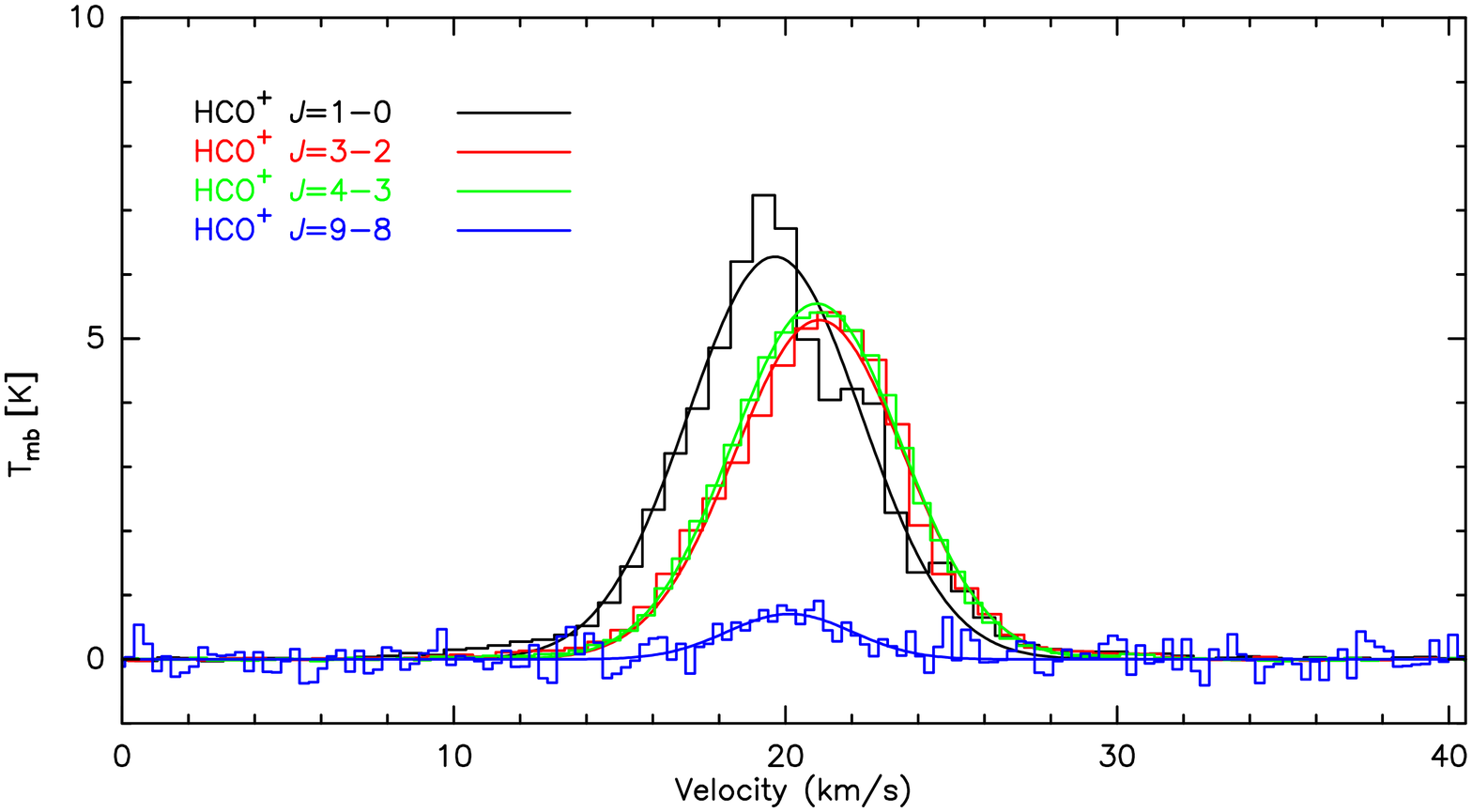,angle=0,width=0.80\linewidth} 
   
 \end{tabular}

  \caption{\footnotesize{Average spectra of \twco\ (\textit{top}), \hcn\ (\textit{middle} - 
  indicating the hyperfine structure lines of \hcn~$J=1\to0$ ) and \hcop\ 
  (\textit{bottom}), obtained from the $200''$-beam area (dashed circle in Fig.~\ref{fig:CO3-2_map_B25}) toward 
  M17SW, centered at offset position ($-100''$,$0''$). This correspond to a region of $\sim$1.92~pc in diameter 
  at the distance of 1.98~kpc \citep{xu11}.}}

  \label{fig:average-spectra}
\end{figure}
%---------------------------------------------------------------

\section{Analysis of the LSEDs}\label{sec:analysis}

We first computed the average spectra of a 200~arcsec$^2$ region to mimic the results 
obtained when observing extragalactic sources with ALMA. These average spectra can be 
fit using a single Gaussian component, as shown in Fig.~\ref{fig:average-spectra}.

Then we analyze the spectra from the 25$''$ resolution maps at four selected positions in 
M17~SW. 
%The analysis of these spectra is described in the sections below.
The profiles of the CO lines show a very rich structure due to the clumpiness of the 
source (several velocity components along the line of sight) and to optical depth effects 
(self-absorption) of the lower-$J$ transitions 
(cf., Fig.~\ref{fig:CO-spectra-3positions}).
While the lower- and mid-$J$ \twco\ lines can 
be fit with up to five Gaussian components, the higher-$J$ ($J_{up}\geq11$) can be fit with 
just one or two Gaussian components. In order to ensure we are comparing the line fluxes 
corresponding to the same velocity component, we chose the Gaussian component of the lower- 
and mid-$J$ lines that is the closest related to the central velocity and line width obtained 
for the single Gaussian used to fit the \twco\ $J=16\to15$ line.

The profiles of the \hcn\ and \hcop\ lines also show structures that could be associated with different 
overlapping cloudlets along the line of sight. 
However, contrarily to the \thco\ lines, the \hthcn\ and \hthcop\ lines 
do not show the same structure as the main isotope lines. 
(cf., Figs.~\ref{fig:HCN-spectra-3positions} and \ref{fig:HCOp-spectra-3positions}).
With the exception of the $J=1\to0$ transitions, the 
higher-$J$ $^{13}$C bearing lines seem to be made of a single 
component that can be fit with one Gaussian. This is 
indicative of self-absorption (or optical depth effects) in the $^{12}$C lines. Therefore, instead of using 
several Gaussian components in the \hcn\ and \hcop\ lines, we fit a single component trying to reproduce the 
missing flux by using just the apparently less self-absorbed emission in the line wings and masking the line 
centers for the Gaussian fit. The fluxes obtained in this way are significantly larger (between 13\% and 95\%, 
depending on the line and the offset position) for the \hcn\ lines, compared to the flux obtained by using two or 
more Gaussian components. The \hcop\ lines show a lower degree of self-absorption.

After obtaining the velocity-integrated intensities from the Gaussian fit, we convert the intensities to fluxes in units of erg~s$^{-1}$~cm$^{-2}$~sr$^{-1}$ as 
$F_{\nu_0}^{\prime} = 10^5 k \left(\frac{\nu_0}{c}\right)^3 \int T_{mb} dv$,
where $k$ is the Boltzmann's constant, $\nu_0$ is the rest frequency of the transition, $c$ is the speed of light and the factor $10^5$ is to convert from \kms\ to cm s$^{-1}$. This flux is then converted to units of W~m$^2$ by
$F_{\nu_0} = 10^{-7} F_{\nu_0}^{\prime} \Omega_{25''} 10^4$
with $\Omega_{25''}$ the 25$''$ beam solid angle (resolution) used in our maps, the factor $10^4$ is to convert from cm$^2$ to m$^2$, and the factor $10^{-7}$ to convert from erg s$^{-1}$ to Watts.

We model the observed fluxes using the modified 
RADEX\footnote{http://www.sron.rug.nl/$\sim$vdtak/radex/index.shtml} 
code \citep{vdtak07} that uses a background radiation field, as done 
by \citet{poelman05} and \citet{pb09}, and considering a 
dust temperature of 50~K found by \citet{meixner92} toward M17~SW.

We chose this modified background radiation field because interaction with far- and 
mid-infrared radiation (mainly from dust emission in circumstellar material or in star-
forming regions) can be important for molecules with widely spaced rotational energy levels 
(e.g., the \textit{lighter hydrides} OH, H$_2$O, H$_3$O$^+$ and NH$_2$), and particularly for 
the higher-$J$ levels of \textit{heavy molecular rotors} such as CO, CS, HCN, HCO$^+$ and 
H$_2$CO, that are observable with Herschel and SOFIA in the far- and mid-infrared regime.
%Since the dust is usually at higher temperatures than 2.73 K (the cosmic microwave 
%background), the diluted black body radiation field of dust will peak at shorter wavelengths 
%($<$1.871 mm), increasing the radiative excitation of the higher-$J$ levels. 
However, the actual effect of the dust IR emission on the redistribution among rotational 
levels of a molecule depends on the local ambient conditions of the emitting gas. 
That is because at high densities (or temperatures) collisions are expected to dominate 
the excitation of the mid- and high-$J$ levels of molecules, such as CO, while at lower 
densities (or temperatures) radiative excitation, as well as spontaneous decay from 
higher-$J$ levels, are expected to be the dominant component driving the redistribution of the 
level populations. Since we do not know a priori what the ambient conditions are at the 
selected positions in M17~SW, we include the modified background radiation field for 
completeness.

The RADEX code computes the local escape probability assuming a uniform 
temperature and density of the collision partners. For the case of M17~SW the escape 
probability can be computed using two independent methods implemented in RADEX: the large 
velocity gradient (LVG) formalism and assuming an homogeneous sphere geometry. We tested both 
methods against each other with a test fit of the LSEDs using low enough densities 
and column densities to ensure convergence.
The results obtained with both methods were not significantly different for the CO 
LSEDs; the rms (or $\chi^2$) values obtained from the observed fluxes and those 
predicted by the two formalisms differ by just $\sim$2\% between the methods, with 
the largest differences pertaining for $J_{up}>10$ CO lines. For the \hcn\ lines, however, the 
LVG method gives about 34\% higher intensities, while for the \hcop\ lines the LVG method 
gives between 15\% and 20\% higher intensities.
In order to obtain similar \hcn\ and \hcop\ line intensities 
as with the LVG method the uniform sphere formalism requires larger 
column densities and slightly larger filling factors.
When using relatively large densities ($>$4$\times$10$^5~\3cm$) and column densities 
($>$2$\times$10$^{15}~\2cm$) for \hcn\ the LVG method did not converge. Since these parameters 
ranges turned out to be relevant for a good fit of the LSEDs, we used the uniform sphere 
geometry method in our models.
The physical conditions were modeled using the collisional data available in the 
LAMDA\footnote{http://www.strw.leidenuniv.nl/$\sim$moldata/} database \citep{schoier05}. 
The collisional rate coefficients for \twco\ are adopted from \citet{yang10}.

\subsection{The Spectral Line Energy Distribution}\label{sec:SEDs}

We had to fit the full LSEDs of the \twco, \hcn, and \hcop\ 
lines by assuming a cold and a warm component. A single 
component does not fit all the fluxes and three components 
require more free parameters than the available number of 
observations (particularly for the \hcn\ and \hcop\ lines). 
The model we use is described by:
\begin{equation}\label{eq:LSED-model}
 F_{tot}(\nu) = \Phi_{cold}F_{cold} + \Phi_{warm}F_{warm}
\end{equation}
\noindent
where $\Phi_{i}$ are the beam area filling factors and $F_{i}$ are the estimated fluxes for each component in 
units of \Wm. The estimated fluxes of the main isotopologues (i.e., ${}^{12}$C-bearing molecules) are a function of 
four parameters per component: the beam area filling factor $\Phi$, the density of the collision partner 
$n(\rm H_2)$ ($\3cm$), the kinetic temperature of the gas $T_k$ (K), and the column density per 
line width $N/\Delta V$ ($\ndv$) of the molecule in study. We used the average FWHM obtained from the Gaussian 
fit of the $J_{up}>2$ lines for the line widths $\Delta V$  of a given species (the $J_{up}\leq2$ transitions 
were not considered because their emission was collected with a beam size larger than 25$''$).

The two component fit of the LSED requires eight free parameters.
We use the simplex method \citep[e.g.,][]{nelder65, kolda03} to minimize 
the error between the observed and estimated fluxes, using sensible initial values and constraints of the 
input parameters as described below. The uncertainty in the parameters is obtained by allowing 
solutions of the LSED fit in a $3\sigma$ range around the estimated fluxes, where $\sigma$ includes 
the standard deviation of the fluxes obtained from the Gaussian fit and 20\% of calibration uncertainties.

We also included all the available \thco\ fluxes to constrain mainly the column density. 
We use the same excitation conditions (density and temperature) as for \twco. 
The same was done for the \hthcn\ and \hthcop\ lines. In order to reduce the number of free 
parameters, we use the same filling factors for the $^{13}$C as for the 
$^{12}$C bearing lines in the cold and warm component. This is a reasonable 
assumption since the $^{13}$C lines, although usually fainter and less extended
when shown in linear scale, appear as extended as the $^{12}$C lines when shown 
in logarithmic scale.

Because for \hcn\ and \hcop\ we have fewer transitions than for CO, we first 
fit the CO LSED and then used the same density and temperature found for the 
cold and warm components from CO for fitting the \hcn\ and \hcop\ LSEDs.

%---------------------------------------------------------------
\begin{figure*}[!pt]
 \centering
 \begin{tabular}{ccc}
  \hspace{-0.3cm}\epsfig{file=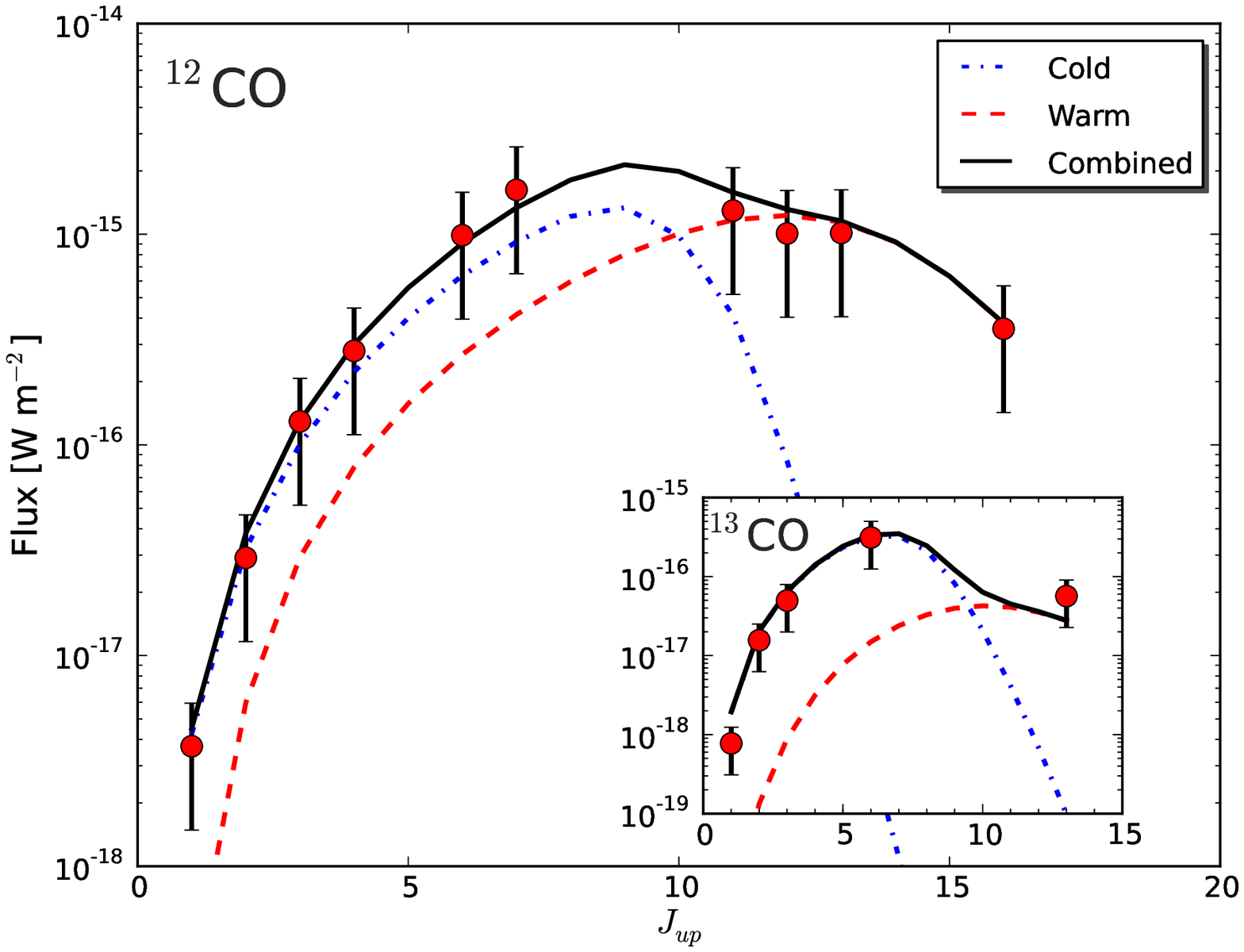,angle=0,width=0.33\linewidth} &
  
  \hspace{-0.3cm}\epsfig{file=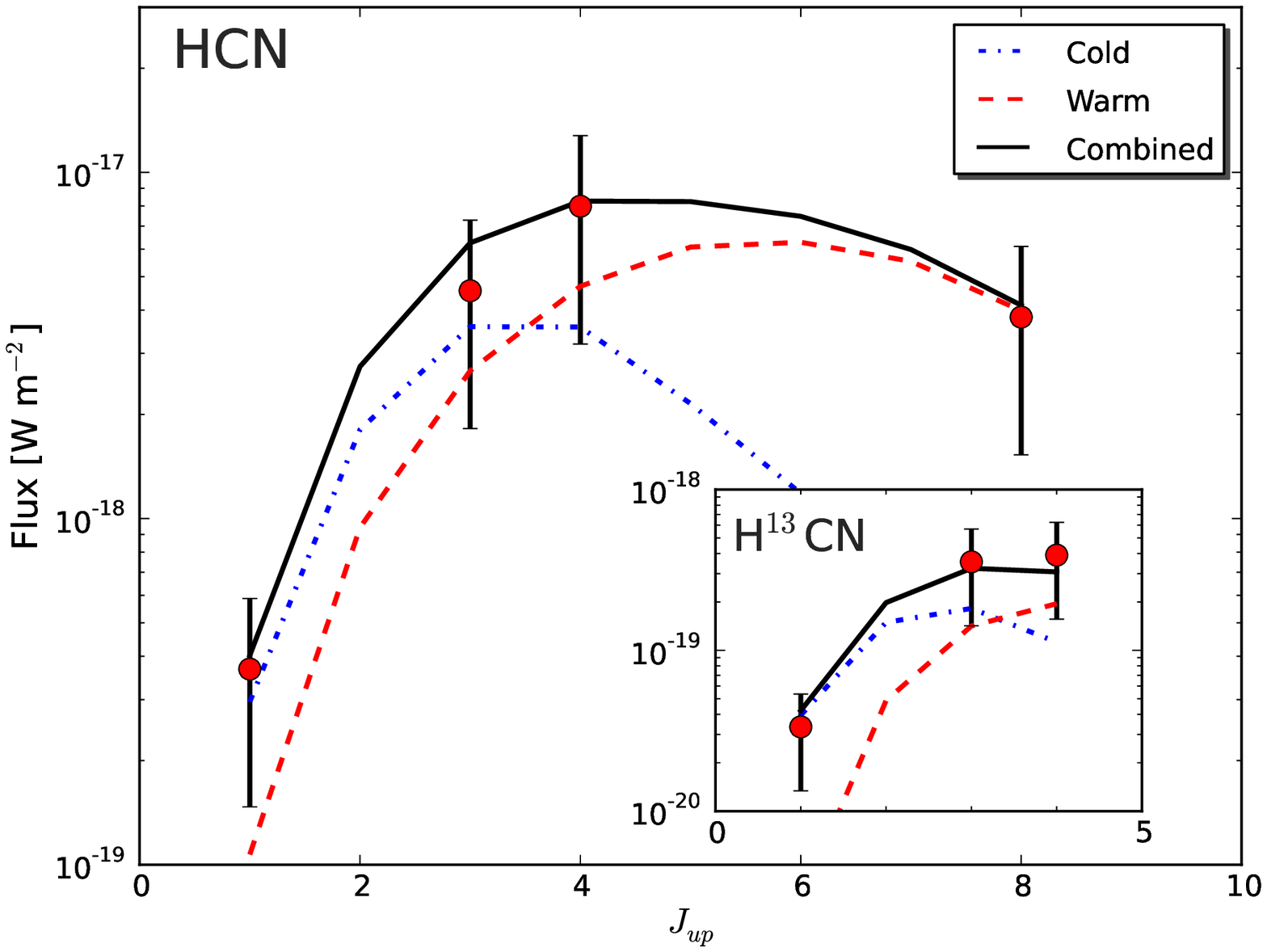,angle=0,width=0.33\linewidth} &
  
  \hspace{-0.3cm}\epsfig{file=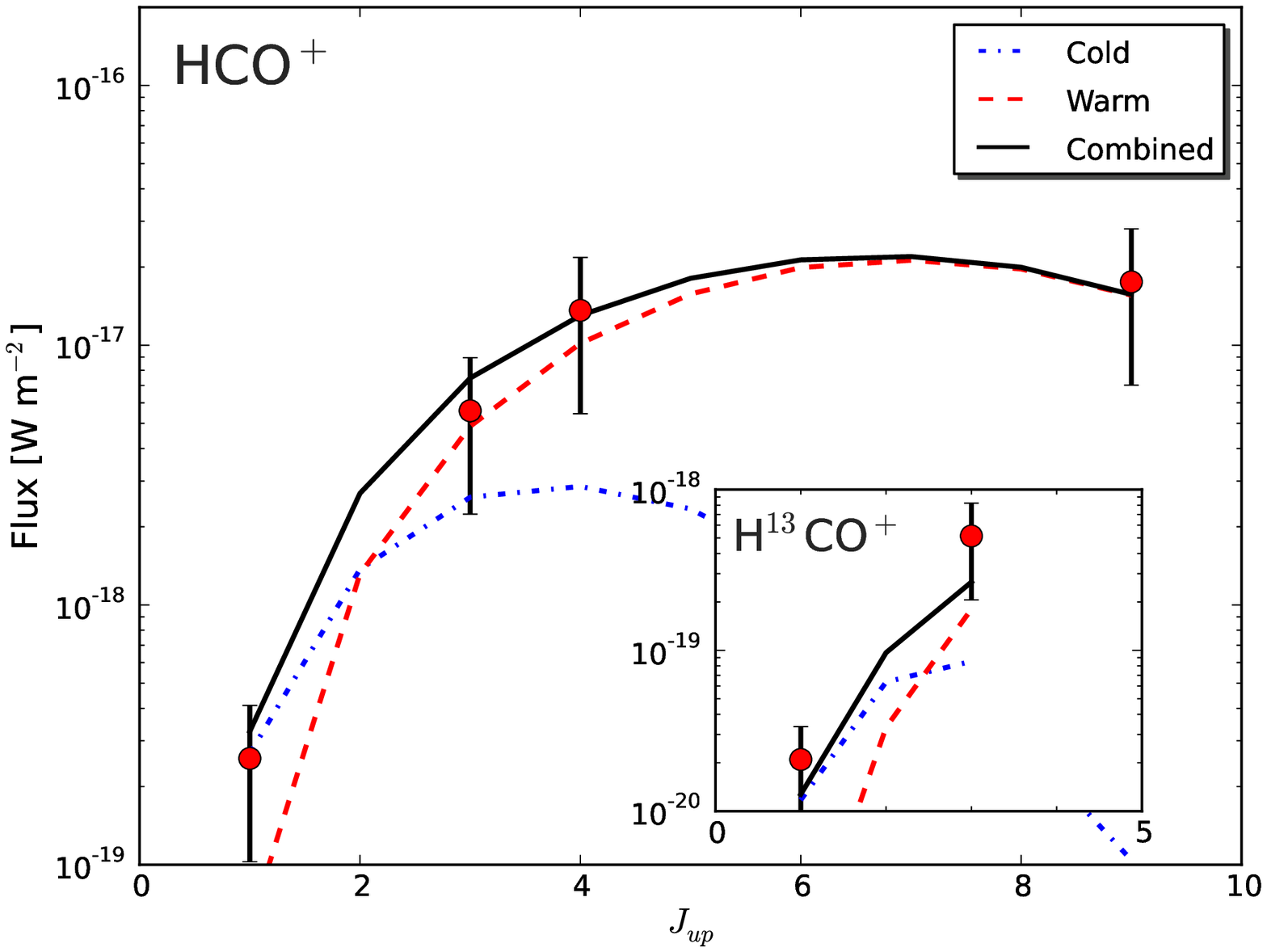,angle=0,width=0.33\linewidth} 
 \end{tabular}

  \caption{\footnotesize{Two component fit of the line spectral energy distribution of the CO, \hcn, and \hcop\ 
  species, for the average spectra toward M17~SW. The fit of the $^{13}$C bearing isotopologues are shown in the 
  insets. The cold and warm components are shown in dotted and dashed lines, respectively. The error bars 
  correspond to the 3$\sigma$ range used to estimate the standard deviation of the excitation 
  parameters, where $\sigma$ is the flux uncertainties obtained from the Gaussian fit.}}

  \label{fig:average-SED-fit}
\end{figure*}
%---------------------------------------------------------------

%__________________________________________________ One column table
   \begin{table}[!tp]
      \caption[]{Gaussian fit parameters for the average spectra.}
         \label{tab:average-gaussian-fit}
         \centering
         \scriptsize
         %\footnotesize
         \setlength{\tabcolsep}{3.5pt} % Default value: 6pt
         \renewcommand{\arraystretch}{1.0} % Default value: 1
         \begin{tabular}{lcccc}
            \hline\hline
	    \noalign{\smallskip}
            Transition & $\int T_{mb} dV$ & $T_{mb}^{peak}$  &  $V_0$  & $\Delta V$ (FWHM) \\
                       &   [\Kkms]        &    [K]           &  [\kms]  &    [\kms]  \\
            \noalign{\smallskip}
            \hline
            \noalign{\smallskip}
            \multicolumn{5}{c} {\twco} \\
            \noalign{\smallskip}
            \hline
            \noalign{\smallskip}

$J=1\to0$ & 264.0$\pm$  5.2 & 30.66$\pm$0.95 & 19.89$\pm$ 0.08 &  8.09$\pm$ 0.19\\ 
$J=2\to1$ & 282.9$\pm$  1.0 & 31.95$\pm$0.18 & 19.88$\pm$ 0.02 &  8.32$\pm$ 0.04\\ 
$J=3\to2$ & 360.4$\pm$  0.0 & 39.20$\pm$0.01 & 20.21$\pm$ 0.00 &  8.64$\pm$ 0.00\\ 
$J=4\to3$ & 328.0$\pm$  0.1 & 36.62$\pm$0.01 & 19.86$\pm$ 0.00 &  8.41$\pm$ 0.00\\ 
$J=6\to5$ & 342.7$\pm$  0.0 & 42.43$\pm$0.00 & 20.73$\pm$ 0.00 &  7.59$\pm$ 0.00\\ 
$J=7\to6$ & 354.3$\pm$  1.0 & 45.84$\pm$0.20 & 20.75$\pm$ 0.01 &  7.26$\pm$ 0.02\\ 
$J=11\to10$ &  72.0$\pm$  0.5 & 12.30$\pm$0.15 & 20.63$\pm$ 0.02 &  5.50$\pm$ 0.05\\ 
$J=12\to11$ &  44.9$\pm$  0.7 &  8.28$\pm$0.22 & 19.87$\pm$ 0.04 &  5.10$\pm$ 0.10\\ 
$J=13\to12$ &  34.7$\pm$  1.2 &  6.69$\pm$0.38 & 20.65$\pm$ 0.08 &  4.87$\pm$ 0.21\\ 
$J=16\to15$ &   5.9$\pm$  0.4 &  1.31$\pm$0.16 & 20.10$\pm$ 0.16 &  4.28$\pm$ 0.41\\

            \noalign{\smallskip}
            \hline
            \noalign{\smallskip}
            \multicolumn{5}{c} {\thco} \\
            \noalign{\smallskip}
            \hline
            \noalign{\smallskip}

$J=1\to0$ &  63.0$\pm$  0.7 & 10.88$\pm$0.19 & 19.74$\pm$ 0.03 &  5.44$\pm$ 0.07\\ 
$J=2\to1$ & 169.0$\pm$  0.0 & 26.91$\pm$0.01 & 19.96$\pm$ 0.00 &  5.90$\pm$ 0.00\\ 
$J=3\to2$ & 157.6$\pm$  0.4 & 24.35$\pm$0.09 & 20.09$\pm$ 0.01 &  6.08$\pm$ 0.02\\ 
$J=6\to5$ & 124.6$\pm$  0.0 & 21.67$\pm$0.00 & 20.57$\pm$ 0.00 &  5.40$\pm$ 0.00\\ 
$J=13\to12$ &   1.9$\pm$  0.5 &  0.28$\pm$0.13 & 20.39$\pm$ 0.91 &  6.62$\pm$ 2.55\\

            \noalign{\smallskip}
            \hline
            \noalign{\smallskip}
            \multicolumn{5}{c} {\hcn} \\
            \noalign{\smallskip}
            \hline
            \noalign{\smallskip}

$J=1\to0$ &  40.1$\pm$  1.8 &  6.51$\pm$0.29 & 19.15$\pm$ 0.20 &  5.80$\pm$ 0.00\\ 
$J=3\to2$ &  28.3$\pm$  0.0 &  3.65$\pm$0.00 & 19.84$\pm$ 0.00 &  7.29$\pm$ 0.00\\ 
$J=4\to3$ &  20.9$\pm$  0.0 &  3.00$\pm$0.01 & 19.98$\pm$ 0.01 &  6.57$\pm$ 0.02\\ 
$J=8\to7$ &   1.4$\pm$  0.1 &  0.29$\pm$0.04 & 19.84$\pm$ 0.25 &  4.77$\pm$ 0.53\\ 

            \noalign{\smallskip}
            \hline
            \noalign{\smallskip}
            \multicolumn{5}{c} {\hthcn} \\
            \noalign{\smallskip}
            \hline
            \noalign{\smallskip}
            
$J=1\to0$ &   5.7$\pm$  0.1 &  0.60$\pm$0.02 & 19.61$\pm$ 0.09 &  9.01$\pm$ 0.23\\ 
$J=3\to2$ &   2.1$\pm$  0.0 &  0.39$\pm$0.03 & 19.44$\pm$ 0.11 &  5.06$\pm$ 0.26\\ 
$J=4\to3$ &   1.0$\pm$  0.0 &  0.23$\pm$0.01 & 19.56$\pm$ 0.06 &  4.47$\pm$ 0.14\\ 

            \noalign{\smallskip}
            \hline
            \noalign{\smallskip}
            \multicolumn{5}{c} {\hcop} \\
            \noalign{\smallskip}
            \hline
            \noalign{\smallskip}
            
$J=1\to0$ &  40.7$\pm$  0.1 &  6.28$\pm$0.02 & 19.43$\pm$ 0.00 &  6.09$\pm$ 0.01\\ 
$J=3\to2$ &  33.2$\pm$  0.0 &  5.29$\pm$0.02 & 20.76$\pm$ 0.00 &  5.91$\pm$ 0.02\\ 
$J=4\to3$ &  34.9$\pm$  0.0 &  5.54$\pm$0.01 & 20.69$\pm$ 0.00 &  5.92$\pm$ 0.01\\ 
$J=9\to8$ &   3.3$\pm$  0.4 &  0.71$\pm$0.16 & 19.88$\pm$ 0.25 &  4.40$\pm$ 0.79\\ 

            \noalign{\smallskip}
            \hline
            \noalign{\smallskip}
            \multicolumn{5}{c} {\hthcop} \\
            \noalign{\smallskip}
            \hline
            \noalign{\smallskip}
            
$J=1\to0$ &   3.7$\pm$  0.0 &  0.80$\pm$0.02 & 19.51$\pm$ 0.03 &  4.37$\pm$ 0.07\\ 
$J=3\to2$ &   3.4$\pm$  0.1 &  0.75$\pm$0.03 & 20.19$\pm$ 0.06 &  4.32$\pm$ 0.15\\             

            \noalign{\smallskip}
            \hline
         \end{tabular}

   \end{table}
%__________________________________________________ One column table

%__________________________________________________ One column table
   \begin{table}[!tp]
      \caption[]{LSED fit parameters\tablefootmark{a} for the average spectra.}
         \label{tab:average-LSED-fit}
         \centering
         \scriptsize
         %\footnotesize
         \setlength{\tabcolsep}{3.5pt} % Default value: 6pt
         \renewcommand{\arraystretch}{1.0} % Default value: 1
         \begin{tabular}{lcccc}
            \hline\hline
	    \noalign{\smallskip}
            Parameter & CO & HCN  &  HCO$^+$  \\
            \noalign{\smallskip}
            \hline
            \noalign{\smallskip}

$\Phi_{cold}(^{12}{\rm C})$  &   1.00 $\pm$  0.13  &   0.60 $\pm$  0.07  &   0.60 $\pm$  0.07 \\ 
$n_{cold}(\rm H_2)$ [cm$^{-3}$]  &   4.80 $\pm$  0.25  &   4.80 $\pm$  0.49  &   4.80 $\pm$  0.54 \\ 
$T_{cold}$ [K]  &  42.00 $\pm$  3.71  &  42.00 $\pm$  3.91  &  42.00 $\pm$  4.38 \\ 
$N_{cold}$ [cm$^{-2}$]  &  19.50 $\pm$  0.48  &  15.40 $\pm$  1.19  &  14.20 $\pm$  0.88 \\ 
 &  &  &  \\ 
$\Phi_{warm}(^{12}{\rm C})$  &   0.10 $\pm$  0.01  &   0.10 $\pm$  0.01  &   0.10 $\pm$  0.01 \\ 
$n_{warm}(\rm H_2)$ [cm$^{-3}$]  &   6.00 $\pm$  0.60  &   6.00 $\pm$  0.77  &   6.00 $\pm$  0.48 \\ 
$T_{warm}$ [K]  &  135.00 $\pm$  8.46  &  135.00 $\pm$ 12.09  &  135.00 $\pm$ 15.51 \\ 
$N_{warm}$ [cm$^{-2}$]  &  18.10 $\pm$  0.36  &  14.70 $\pm$  0.94  &  14.40 $\pm$  0.50 \\ 
 &  &  &  \\ 
$\Phi_{cold}(^{13}{\rm C})$  &   1.00 $\pm$  0.12  &   0.60 $\pm$  0.06  &   0.60 $\pm$  0.05 \\ 
$\Phi_{warm}(^{13}{\rm C})$  &   0.10 $\pm$  0.00  &   0.10 $\pm$  0.01  &   0.10 $\pm$  0.01 \\ 
 &  &  &  \\ 
%$^{12}$C/$^{13}$C  &  50.00 $\pm$  2.51  &  50.00 $\pm$  4.65  &  50.00 $\pm$  5.39 \\ 
 &  &  &  \\ 
$\Delta V(^{12}{\rm C})$ [km s$^-1$]  &   8.00   &   6.00   &   6.00  \\ 
$\Delta V(^{13}{\rm C})$ [km s$^-1$]  &   6.00   &   6.00   &   4.00  \\
		
            \noalign{\smallskip}
            \hline
         \end{tabular}
         
         \tablefoot{
         \tablefoottext{a}{The density and column density values are given in $log_{10}$ scale.}
         }

   \end{table}
%__________________________________________________ One column table

\subsection{Constraints on the model parameters}\label{sec:pars-constraints}

Two of the most critical parameters of the models are the column densities 
and the beam area filling factors. The filling factors were discussed above,
while the column densities depend on the observed fluxes and the excitation
conditions (temperature and density) of the models.

%The lower limits of the column densities are given by the observed fluxes, 
%while their upper limit is set by the optical depth obtained from the 
%radiative transfer solution. Since most radiative transfer models are known to have 
%convergence uncertainties when dealing with large opacities, we have 
%imposed a limit of $\tau_J\lesssim400$ for all the $J$-transitions of the 
%molecules. 

Another important parameter of the models is the carbon-12 to carbon-13 
isotope ratio, which couples the column densities of the $^{12}$C and 
$^{13}$C lines since we assume 
$N({\rm ^{13}CO})=N({\rm ^{12}CO})/\mathcal{R}_{12/13}$, with 
$\mathcal{R}_{12/13}$=[$^{12}$C]/[$^{13}$C]. 
The isotope ratio $\mathcal{R}_{12/13}$ is known to vary from source 
to source in the Milky Way, depending on the distance to the 
Galactic center \citep[e.g.][]{henkel82, henkel85, langer90, milam05}.
At a distance of 1.98~kpc \citep{xu11}, M17~SW is expected to have a 
CO isotope ratio of $\mathcal{R}_{12/13}=51.6\pm7.9$, according to 
Eq.~(3) by \citet{milam05}. In our models we use $\mathcal{R}_{12/13}=50$. 
For \hcn\ and \hcop\ the same isotope 
ratio as for CO can be considered. However, there is theoretical 
and observational evidence that the actual \hcn\ isotope ratio is 
closer to that obtained from observations of H$_2$CO, which provides 
an upper limit to $\mathcal{R}_{12/13}$ while CO provides a lower 
limit, and isotope ratios derived from \hcop\ lead to intermediate 
values \citep[e.g.][]{langer84, henkel85, milam05}. 
Thus, from Eqs.~(5) and (4) by \citet{milam05}, the isotope ratios 
for \hcop\ and \hcn\ in M17~SW should be $\sim$56 and $\sim$63, 
respectively. Using these higher $\mathcal{R}_{12/13}$ values, however, 
we could not fit the \hcn\ and \hcop\ LSEDs simultaneously with the 
\hthcn\ and \hthcop\ lines. We were able to fit the \hcn\ and \hcop\ 
LSEDs using the same isotope ratio of 50 as for the CO lines, although 
smaller $\mathcal{R}_{12/13}$ of 30 and even 20, would lead to a 
better fit of the rare \hcn\ and \hcop\ isotopologues. 
This would be in agreement with the more recent results found by 
\citet{roellig13} for \hcop, but not for the HCN isotope ratio.
The discrepancy we found for \hcn\ is likely due to optically thick 
lines which are also heavily affected by self-absorption, as mentioned
in Sect.~\ref{sec:analysis}.

%\mathscr{E} % uses `mathrsfs`
%\mathcal{E}

\subsection{Excitation of the Average Emission in M17~SW}\label{sec:average-emission}

%---------------------------------------------------------------
\begin{figure*}[!pt]

 \begin{tabular}{cc}
  %\vspace{-0.6cm}\\
  \epsfig{file=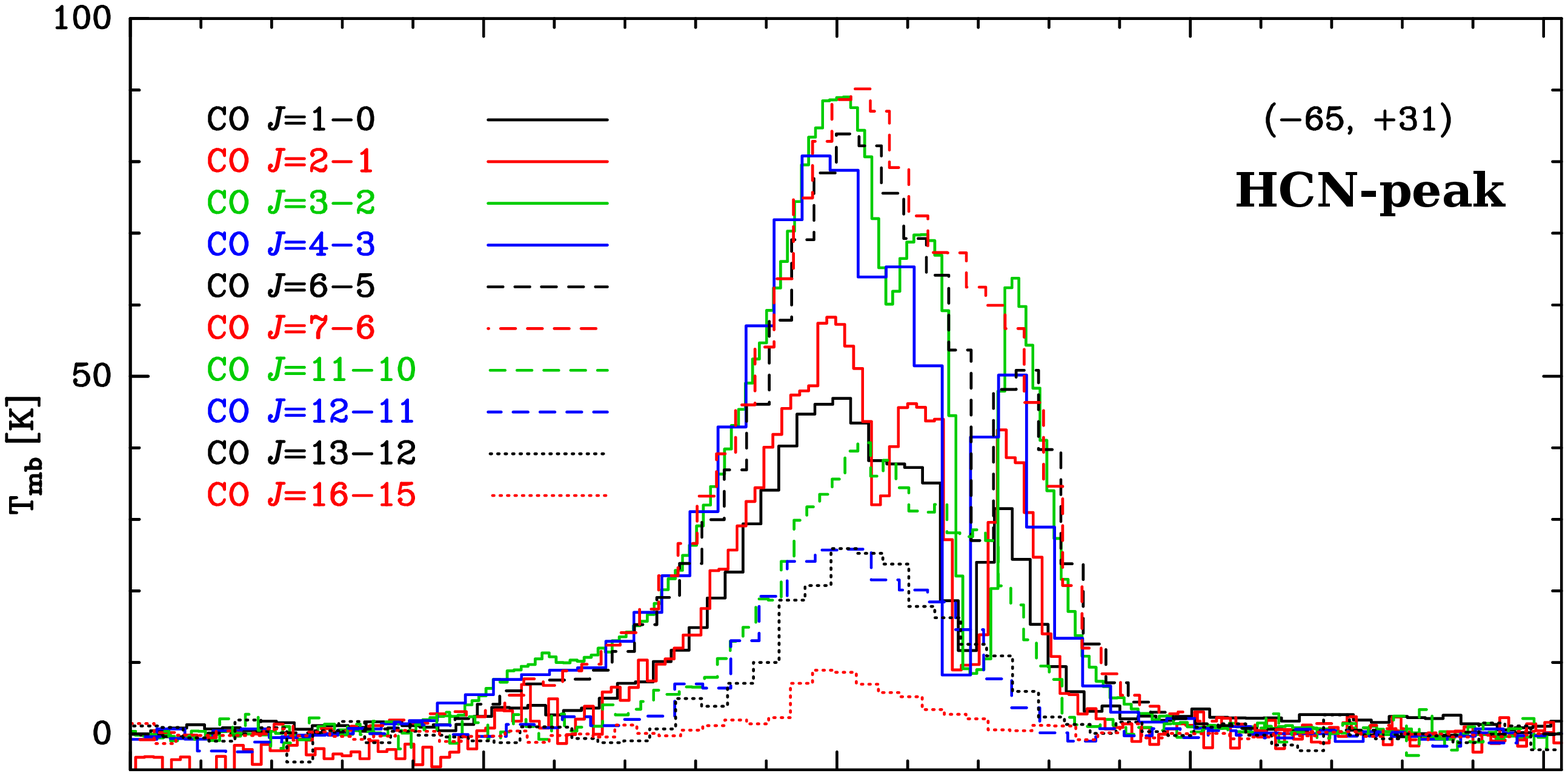,angle=0,width=0.46\linewidth} &
  \epsfig{file=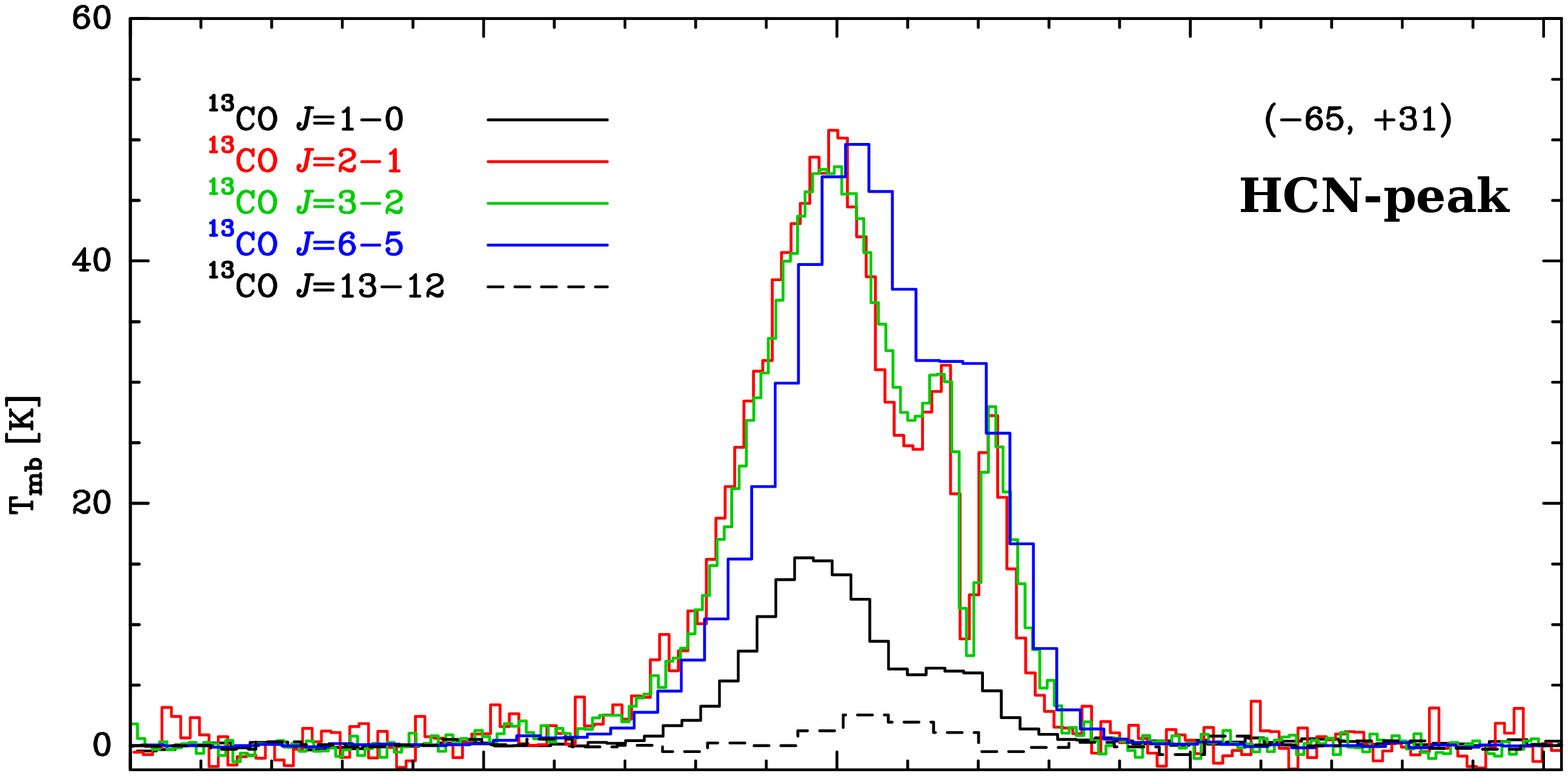,angle=0,width=0.46\linewidth}\\  
  \vspace{-0.54cm}\\

  \epsfig{file=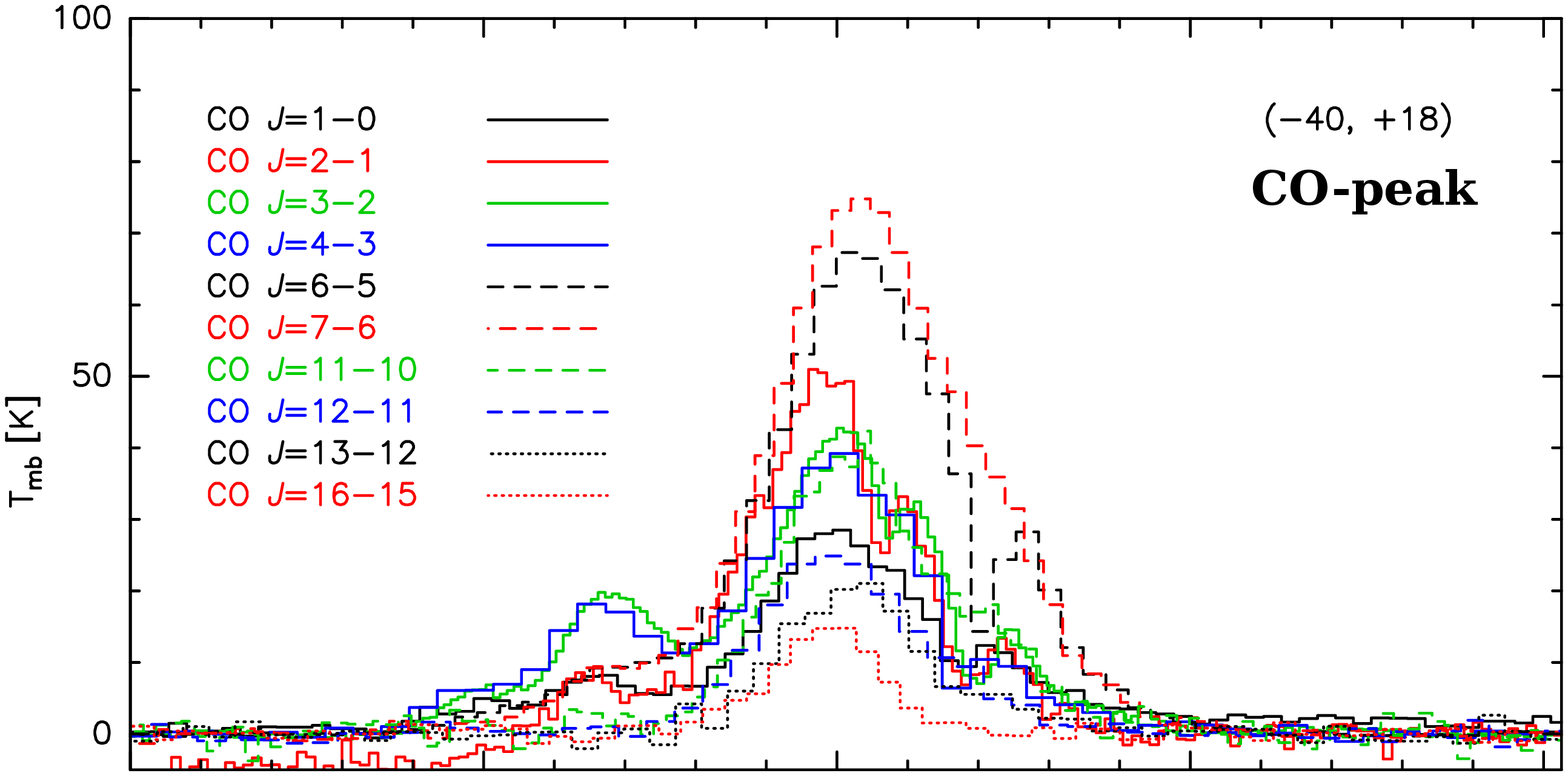,angle=0,width=0.46\linewidth} &
  \epsfig{file=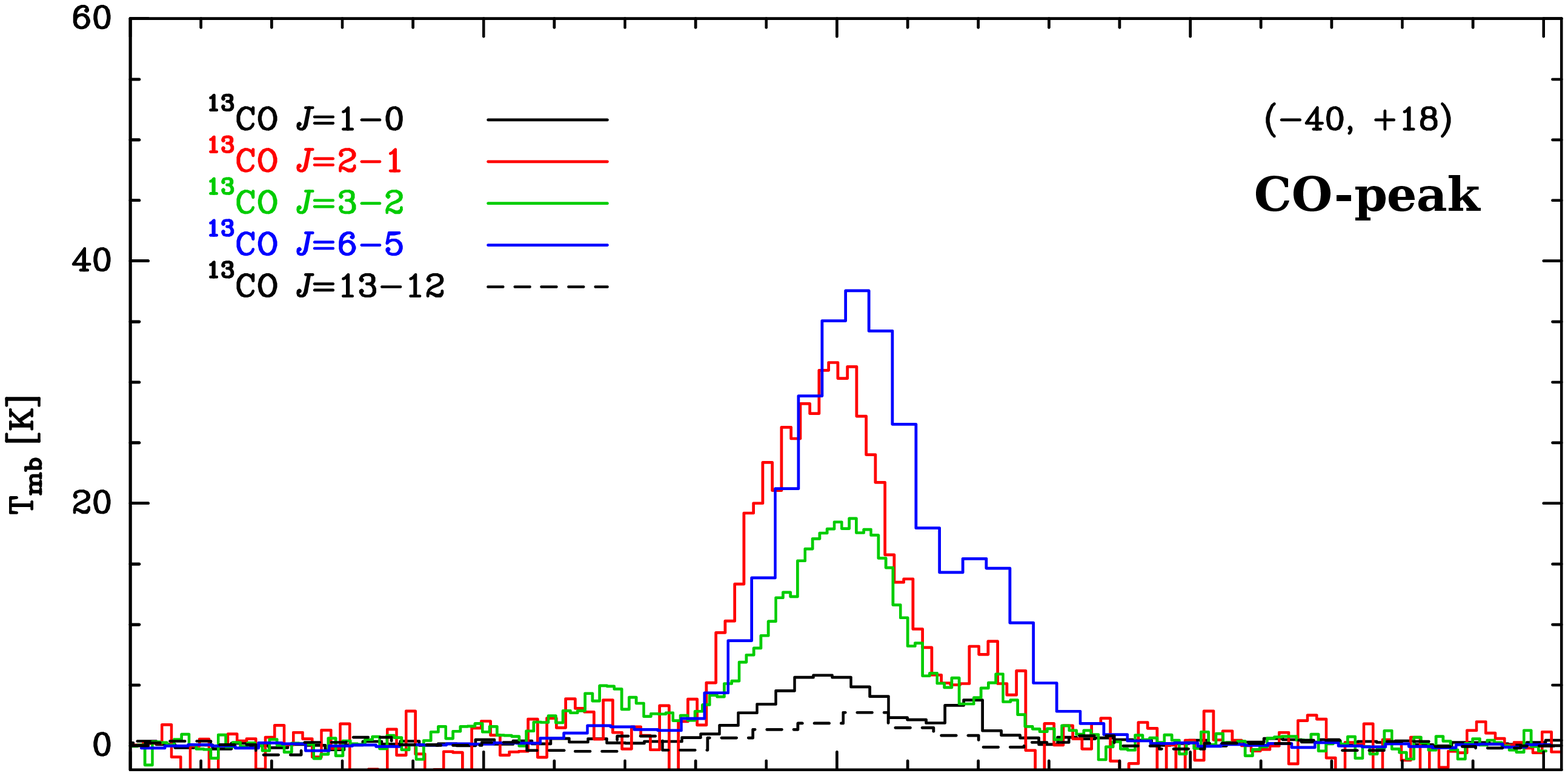,angle=0,width=0.46\linewidth}\\  
  \vspace{-0.54cm}\\

  \epsfig{file=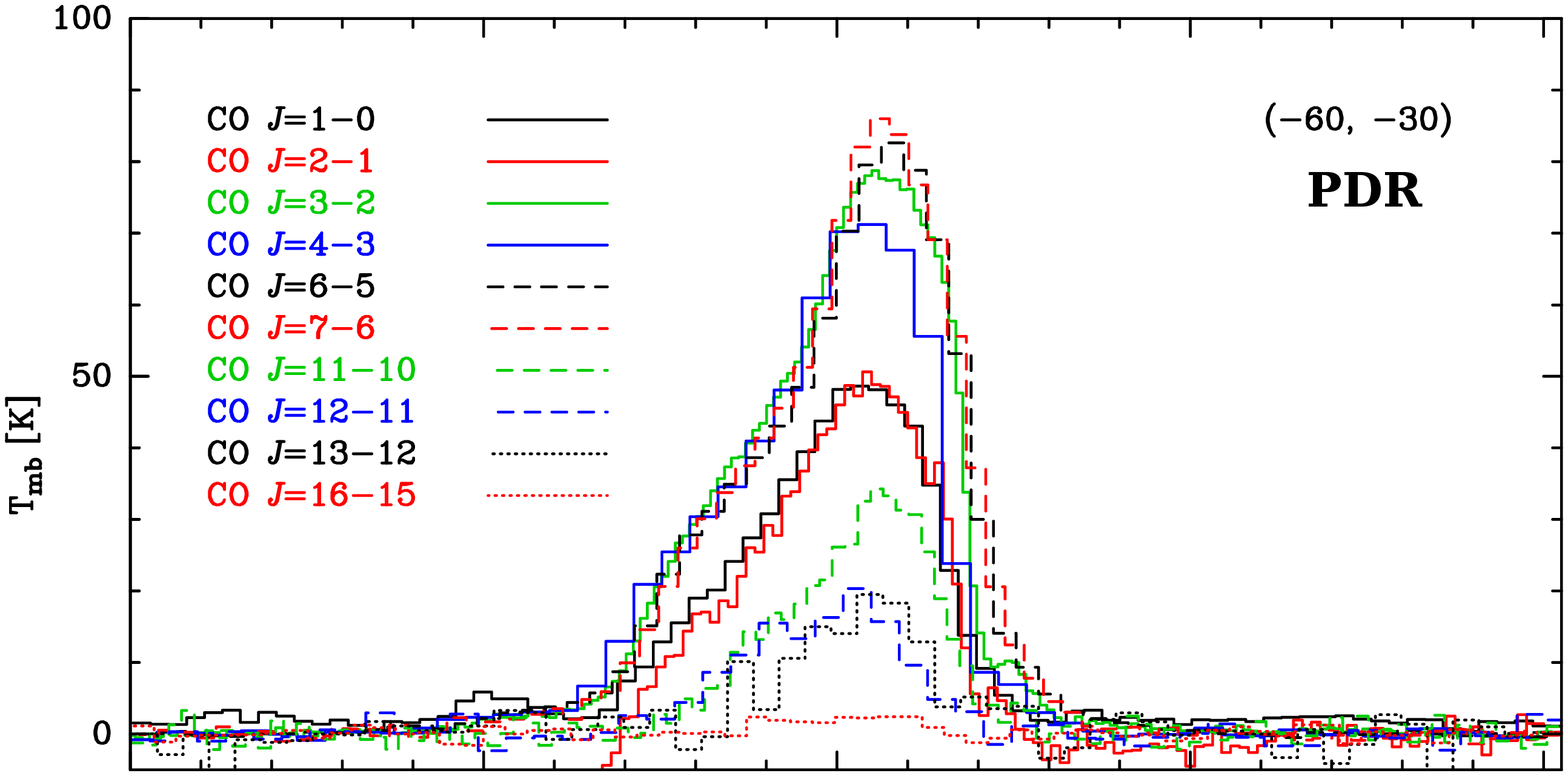,angle=0,width=0.46\linewidth} &
  \epsfig{file=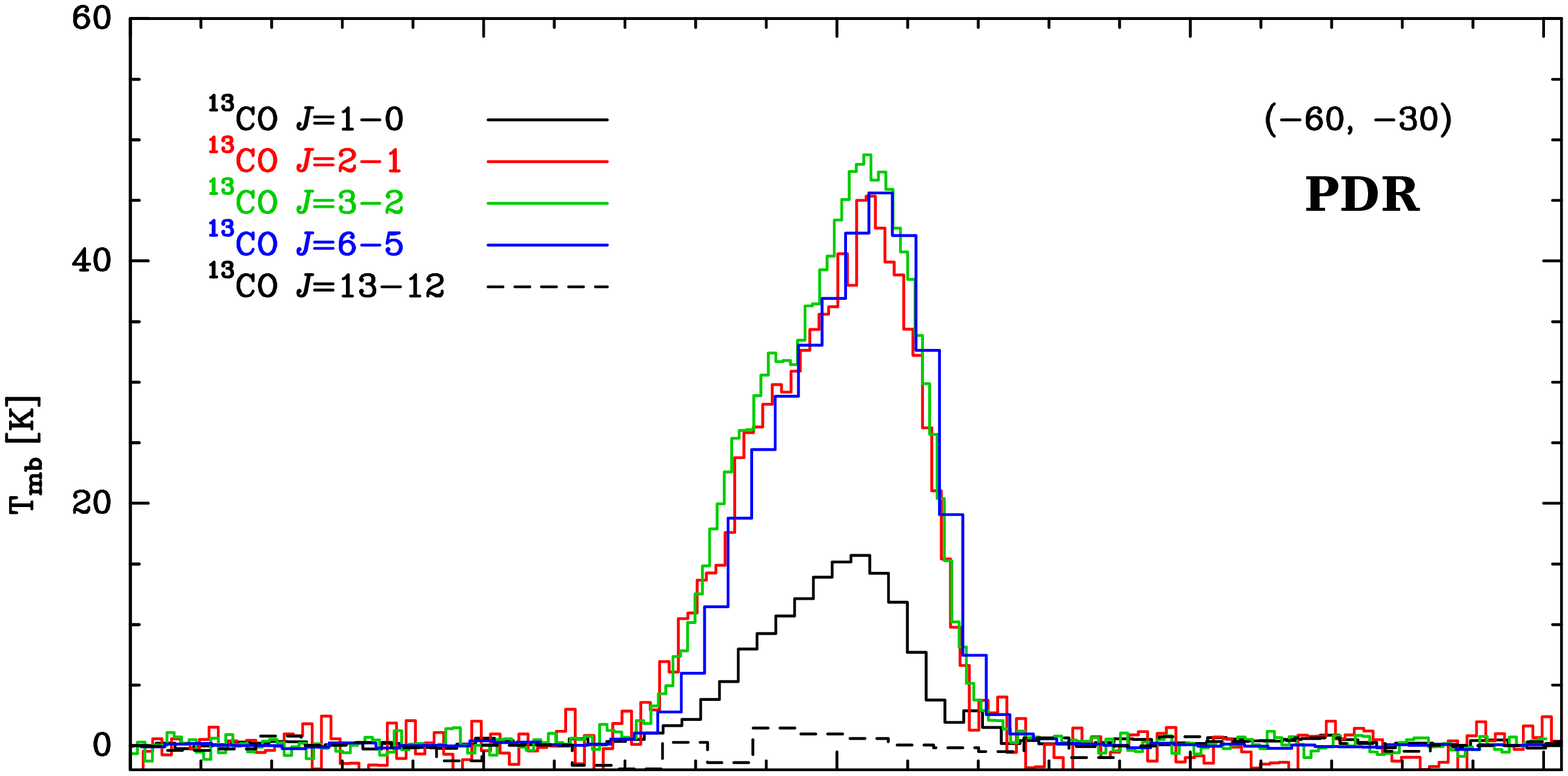,angle=0,width=0.46\linewidth}\\  
  \vspace{-0.54cm}\\
  
  \epsfig{file=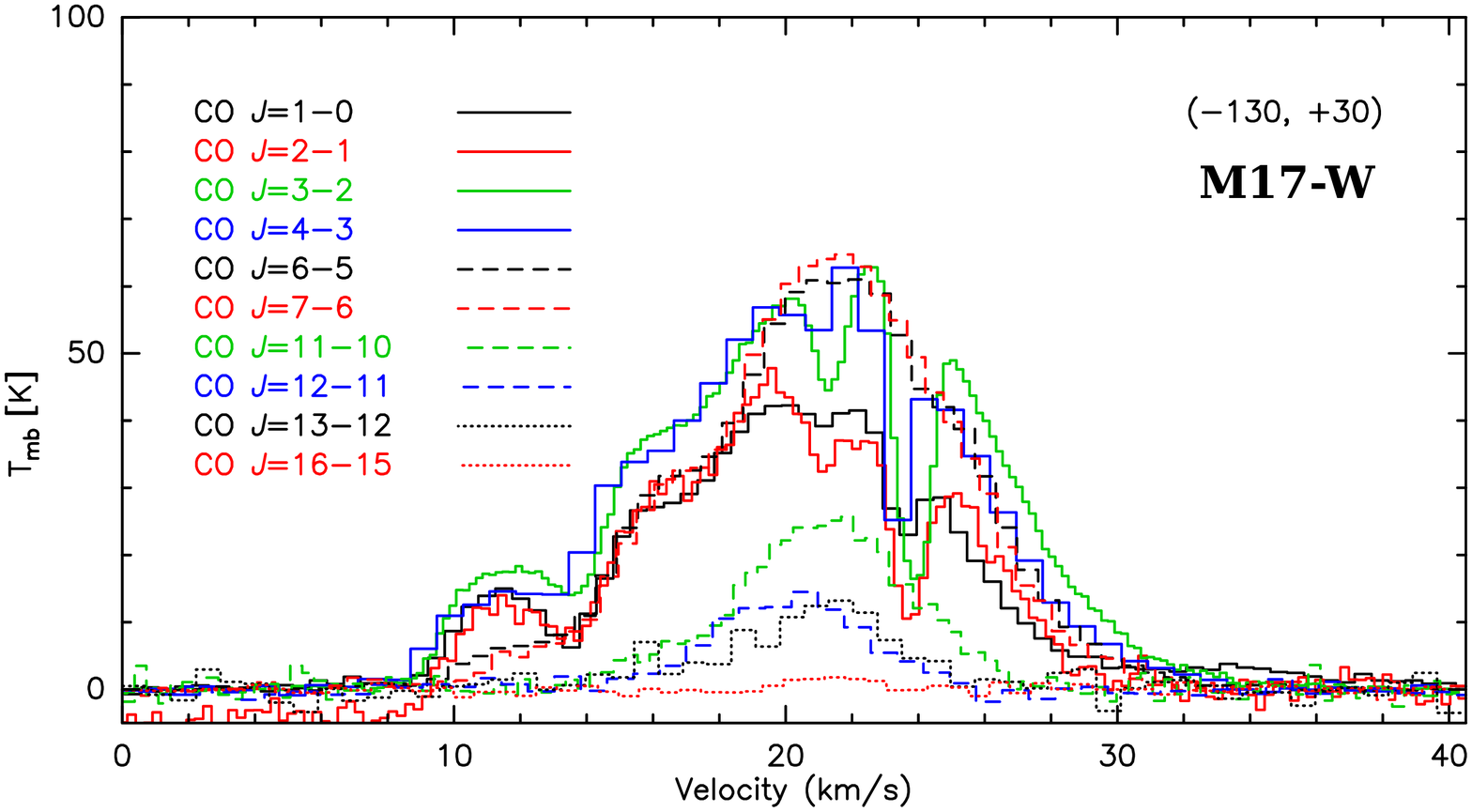,angle=0,width=0.46\linewidth} &
  \epsfig{file=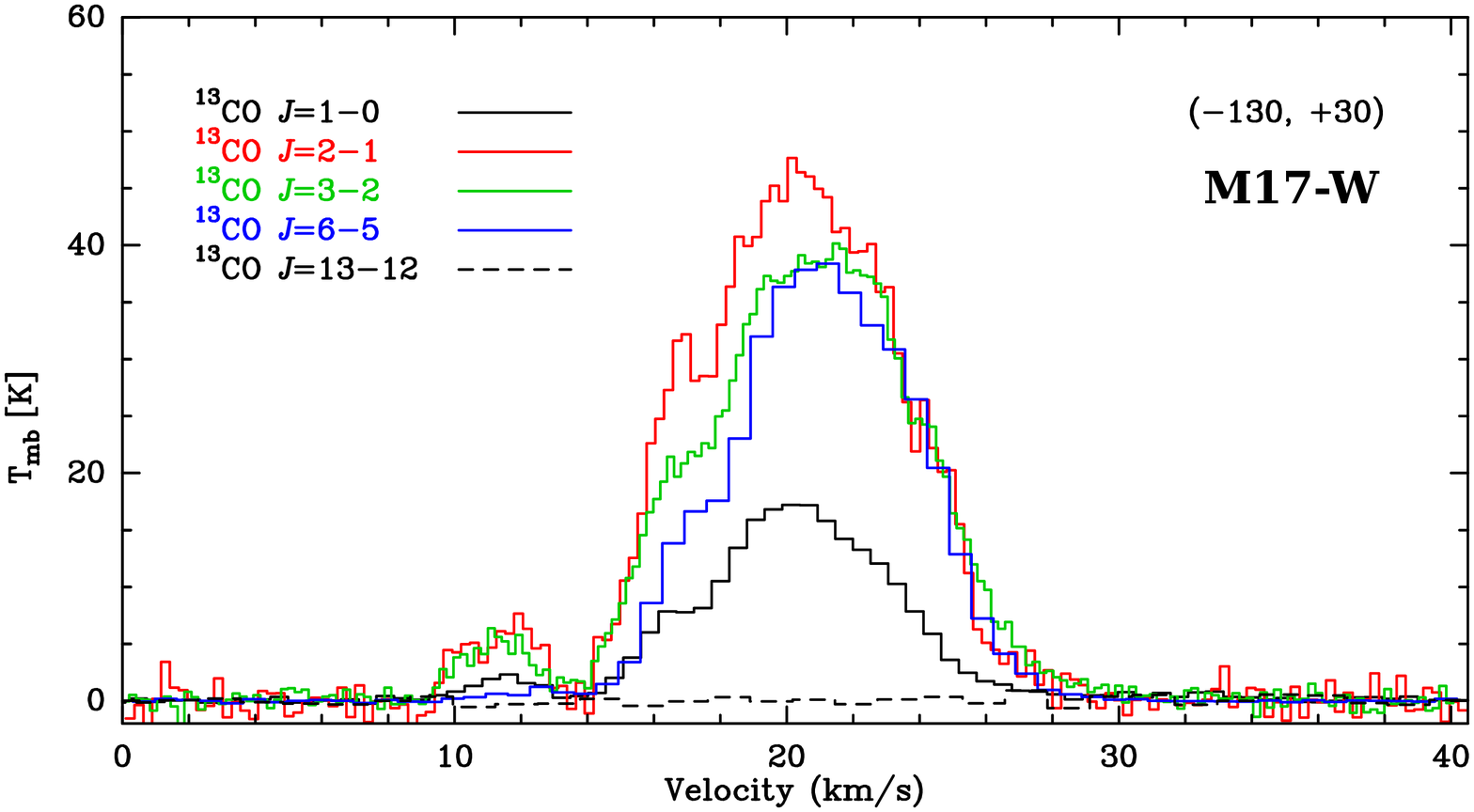,angle=0,width=0.46\linewidth}\\  
  \vspace{-0.54cm}\\

 \end{tabular}

  \caption{\footnotesize{Spectra of all \twco\ (\textit{left panel}) and \thco\ (\textit{right panel}) lines 
  observed at the four selected offset (within $\pm$3$''$) positions shown in Fig.~\ref{fig:CO3-2_map_B25}. 
  All the spectra, except the $J=1\to0$ 
  and $J=2\to1$ have been convolved with a 25$''$ beam, corresponding to the \twco~$J=11\to12$ map.}}

  \label{fig:CO-spectra-3positions}
\end{figure*}
%---------------------------------------------------------------

%---------------------------------------------------------------
\begin{figure*}[!pt]

 \begin{tabular}{cc}
  %\vspace{-0.6cm}\\
  \epsfig{file=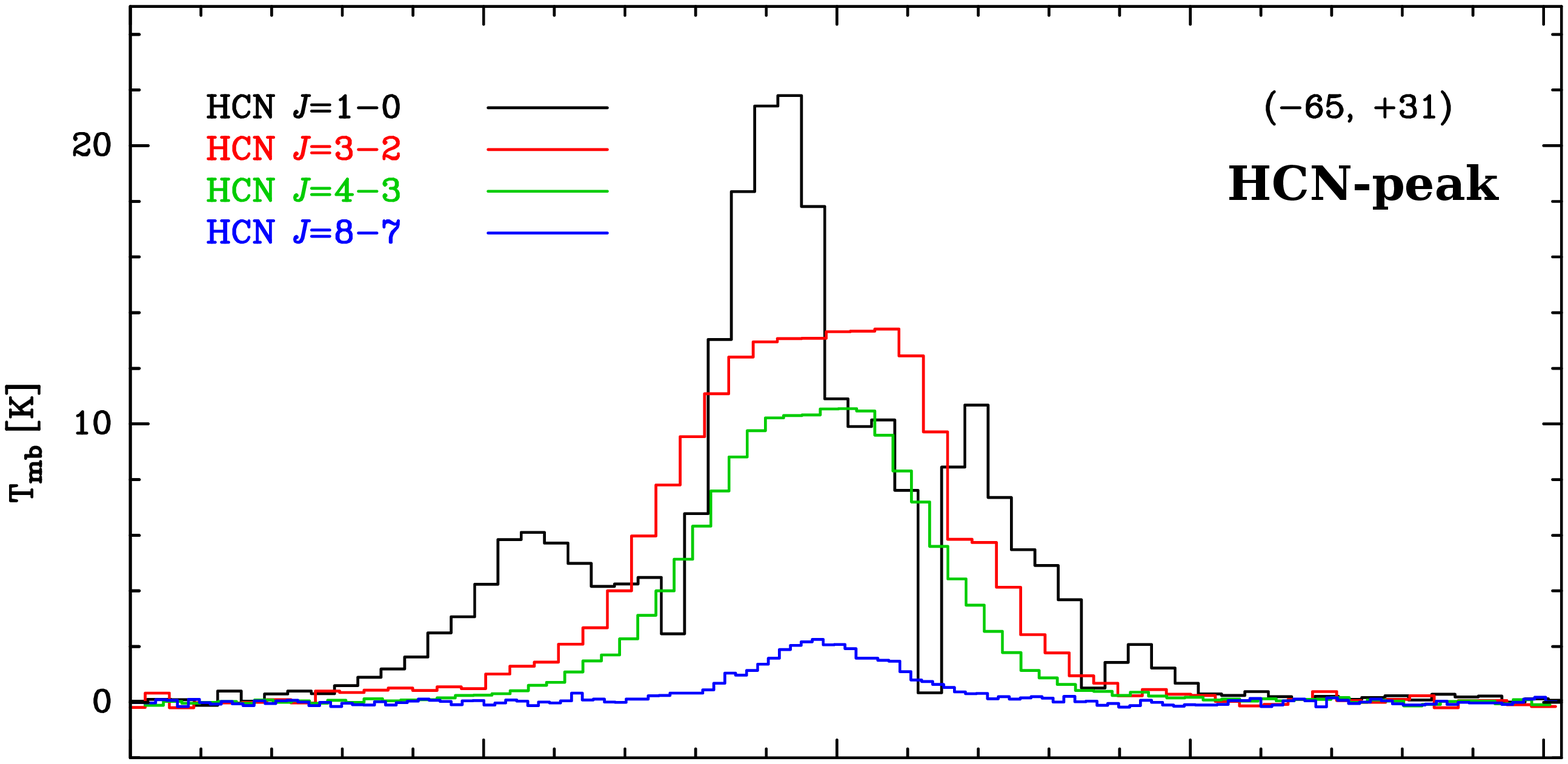,angle=0,width=0.46\linewidth} &
  \epsfig{file=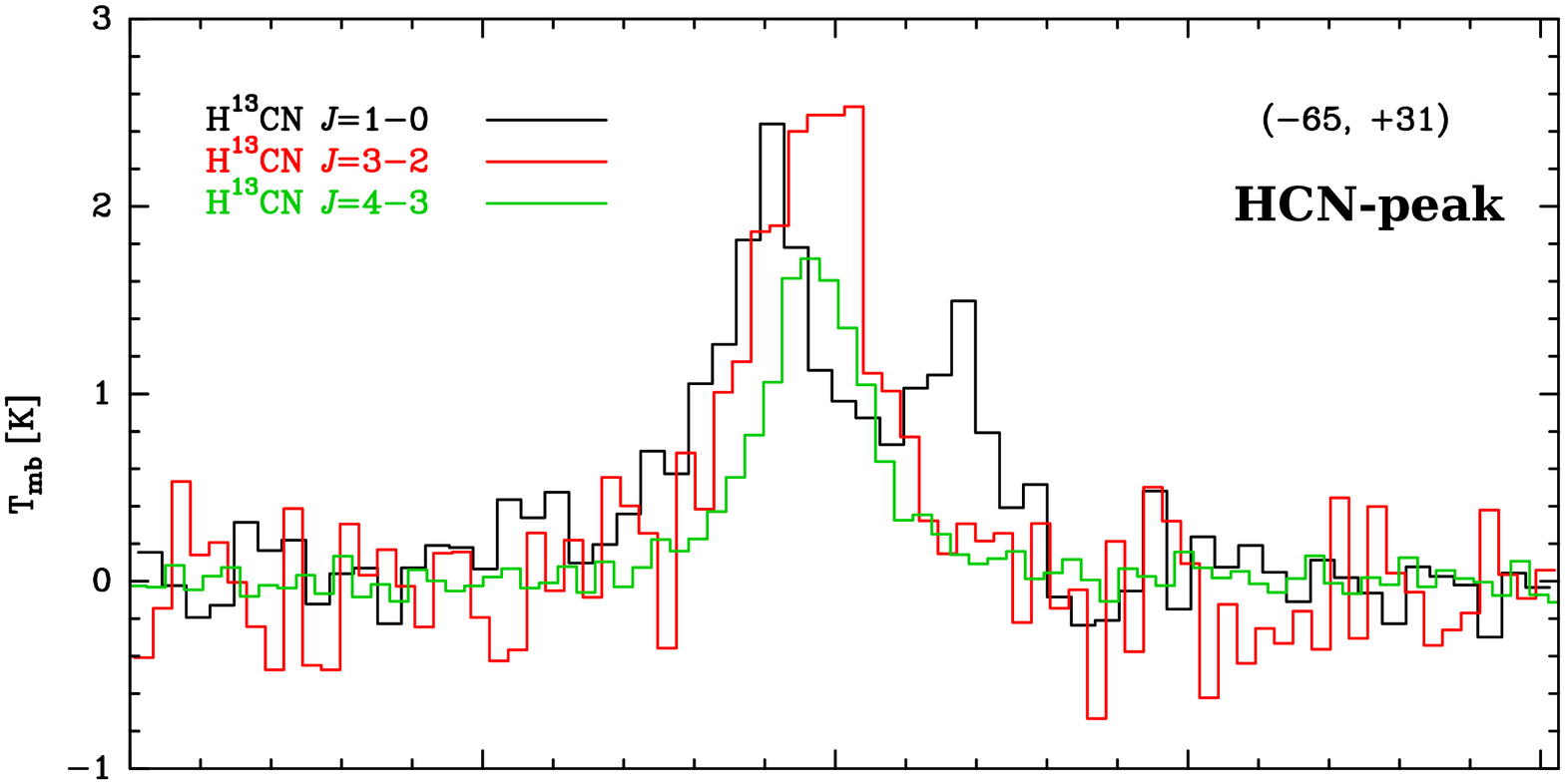,angle=0,width=0.46\linewidth}\\  
  \vspace{-0.54cm}\\

  \epsfig{file=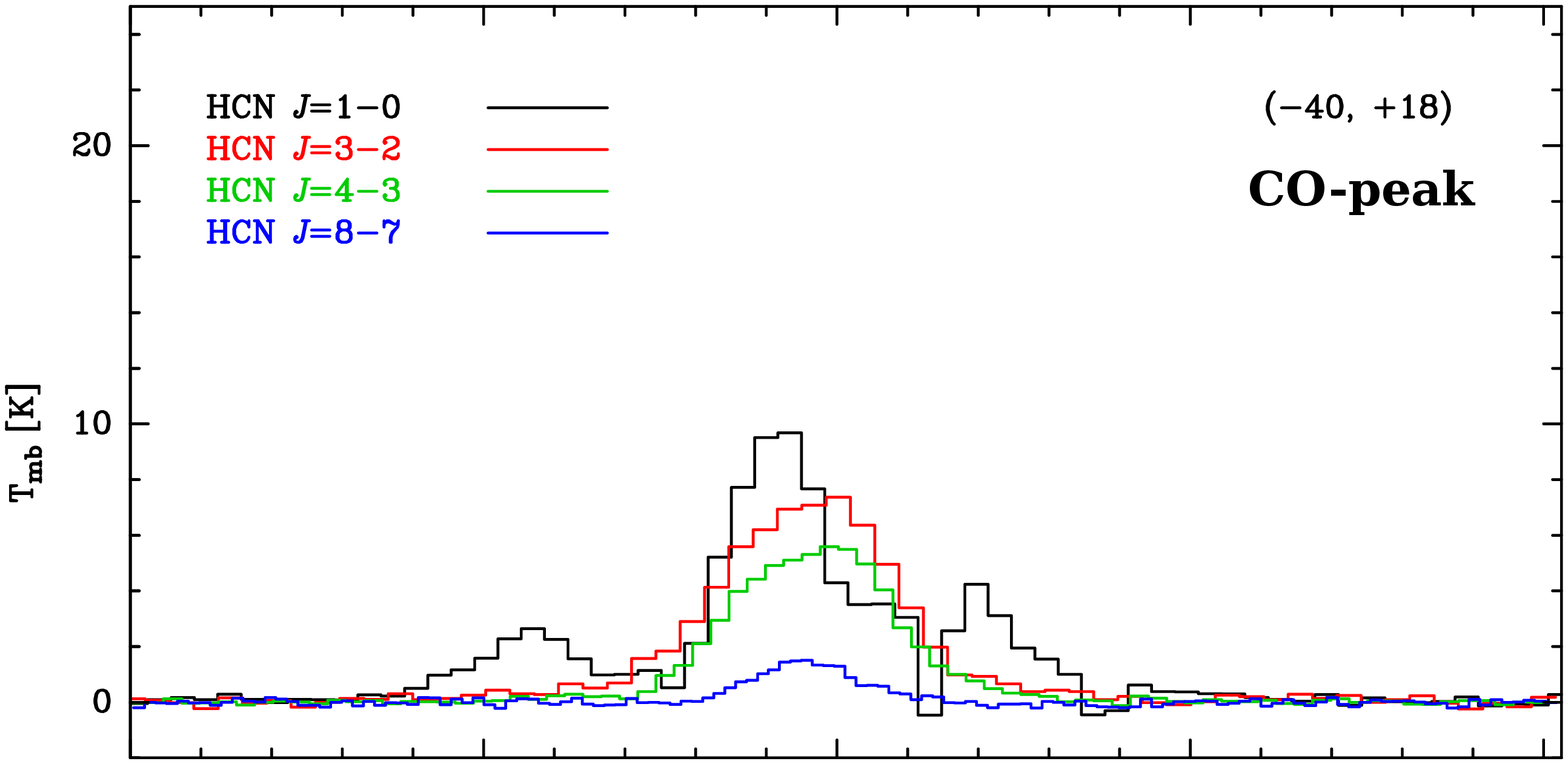,angle=0,width=0.46\linewidth} &
  \epsfig{file=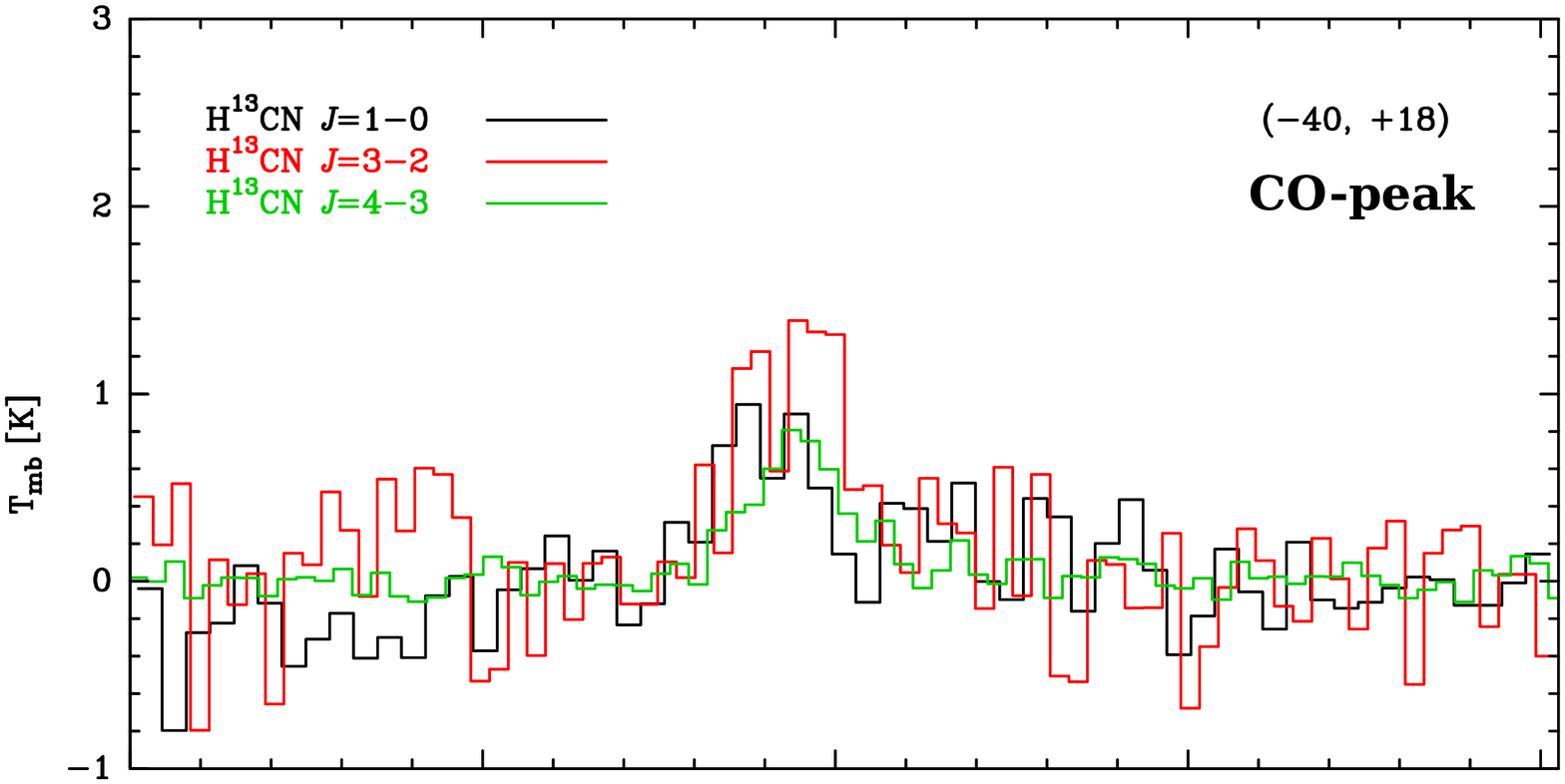,angle=0,width=0.46\linewidth}\\  
  \vspace{-0.54cm}\\

  \epsfig{file=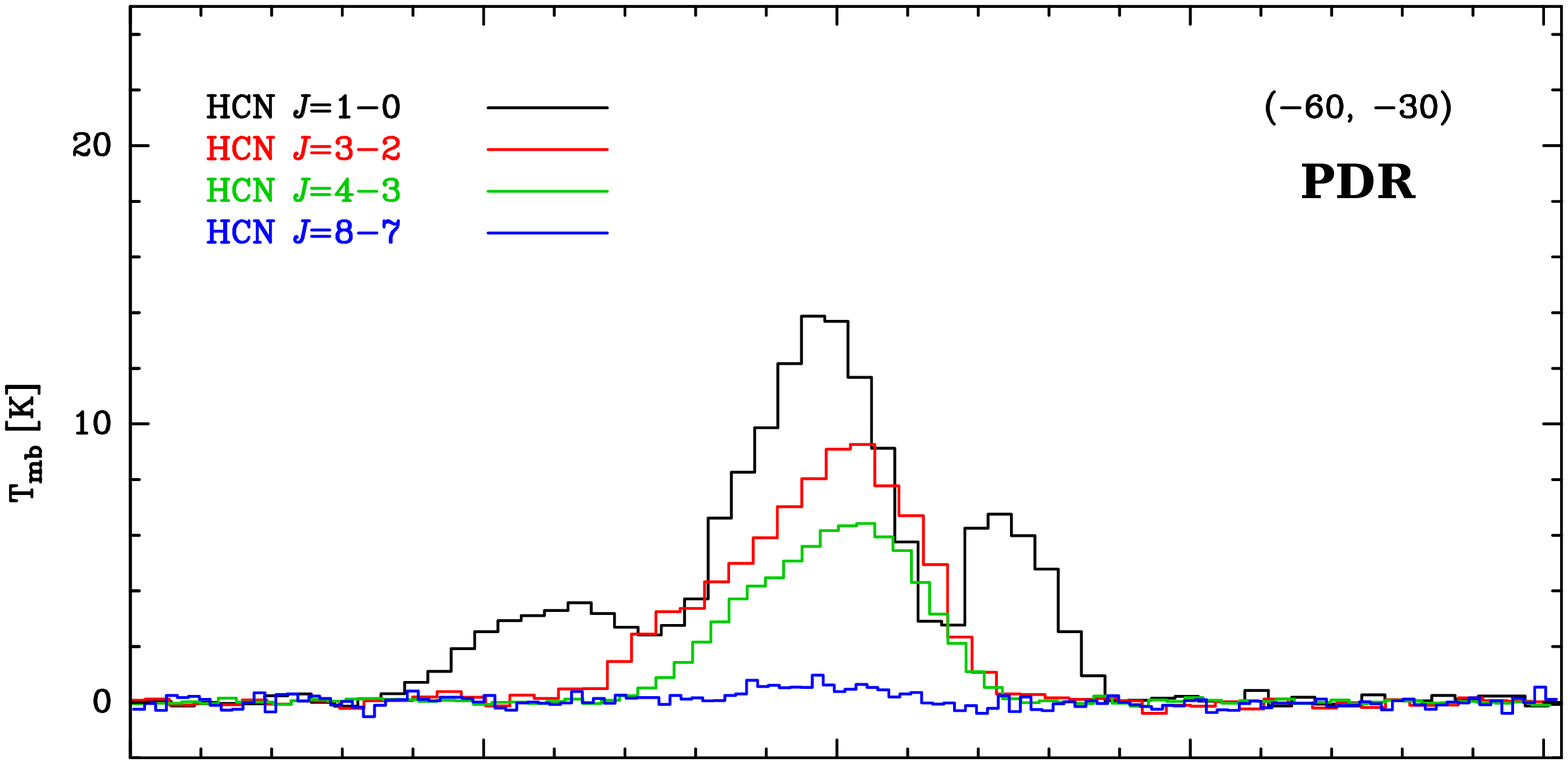,angle=0,width=0.46\linewidth} &
  \epsfig{file=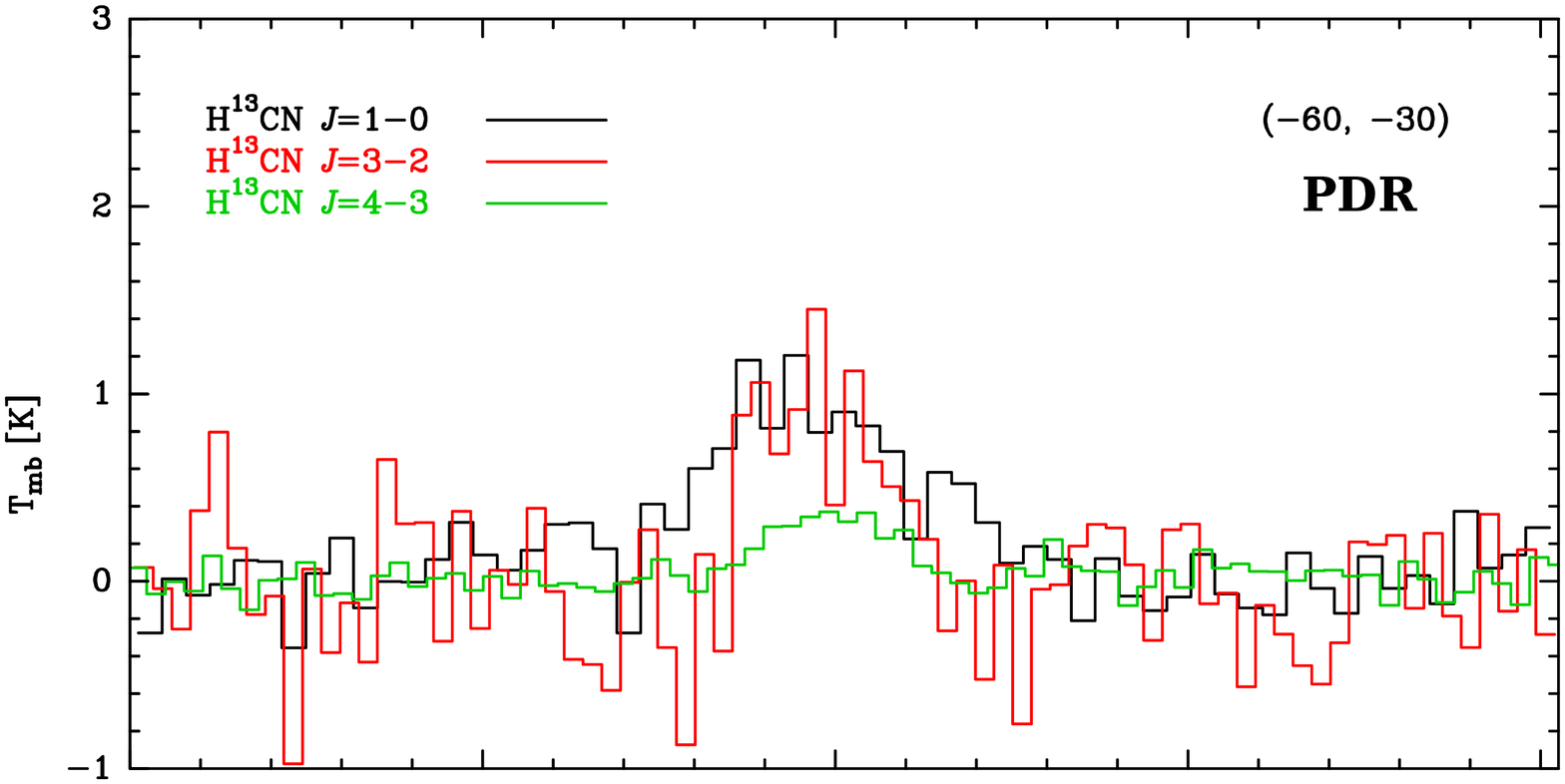,angle=0,width=0.46\linewidth}\\  
  \vspace{-0.54cm}\\
  
  \epsfig{file=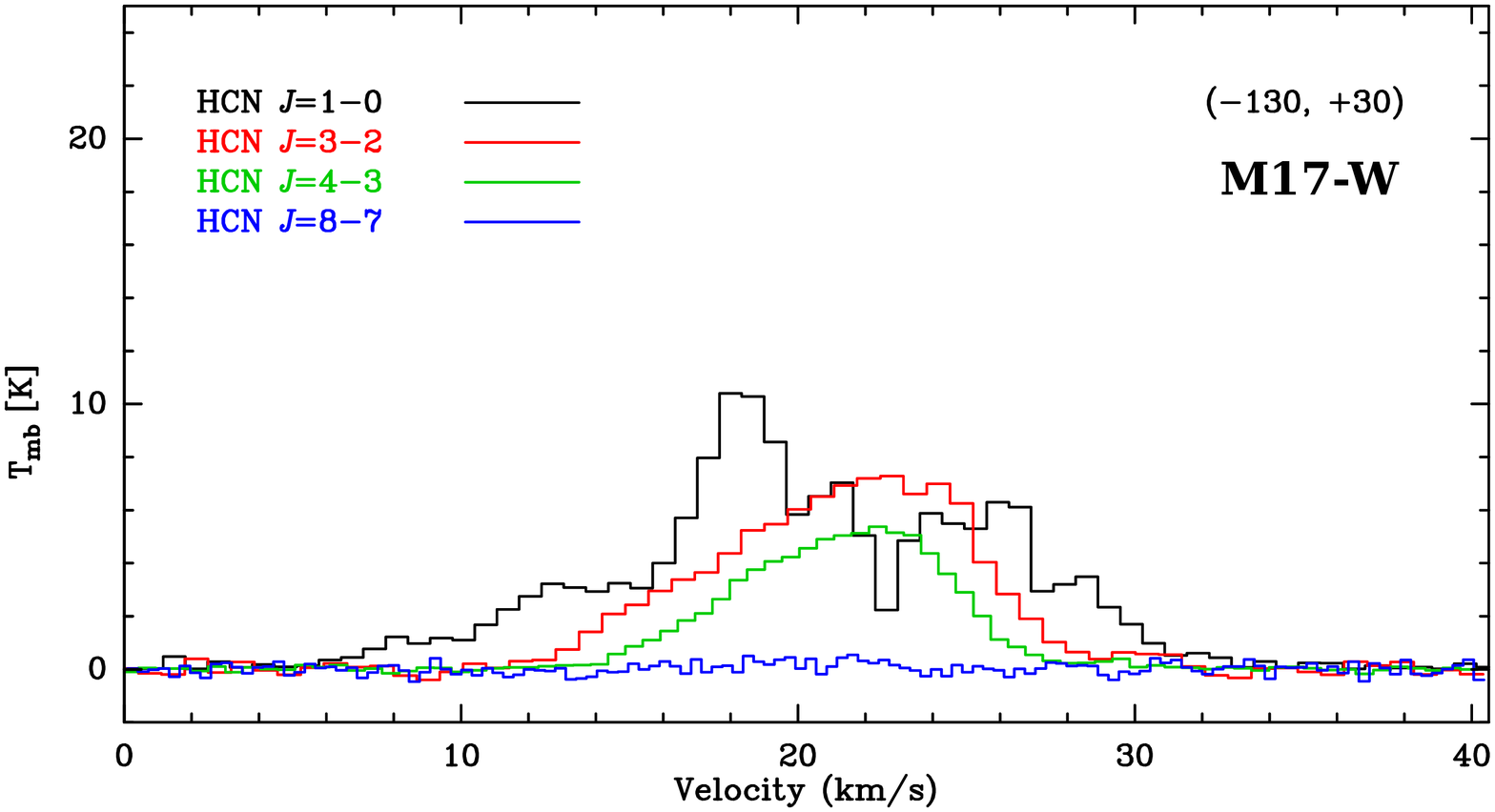,angle=0,width=0.46\linewidth} &
  \epsfig{file=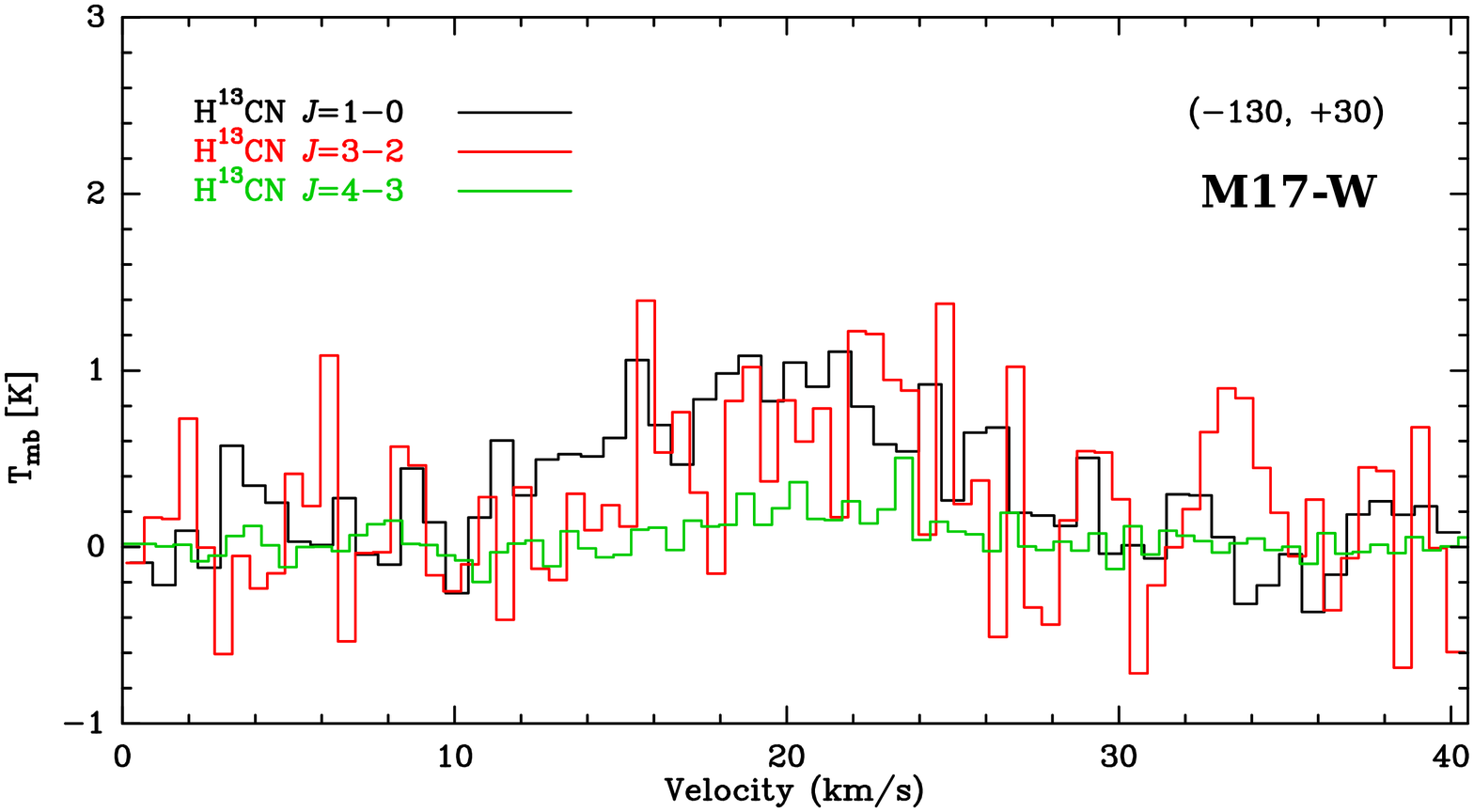,angle=0,width=0.46\linewidth}\\  
  \vspace{-0.54cm}\\

 \end{tabular}

  \caption{\footnotesize{Spectra of all \hcn\ (\textit{left panel}) and \hthcn\ (\textit{right panel}) lines 
  observed at the four selected offset (within $\pm$3$''$) positions shown in Fig.~\ref{fig:CO3-2_map_B25}. 
  All the spectra, except the $J=1\to0$ 
  and $J=2\to1$ have been convolved with a 25$''$ beam, corresponding to the \twco~$J=11\to12$ map.}}

  \label{fig:HCN-spectra-3positions}
\end{figure*}
%---------------------------------------------------------------

%---------------------------------------------------------------
\begin{figure*}[!pt]

 \begin{tabular}{cc}
  %\vspace{-0.6cm}\\
  \epsfig{file=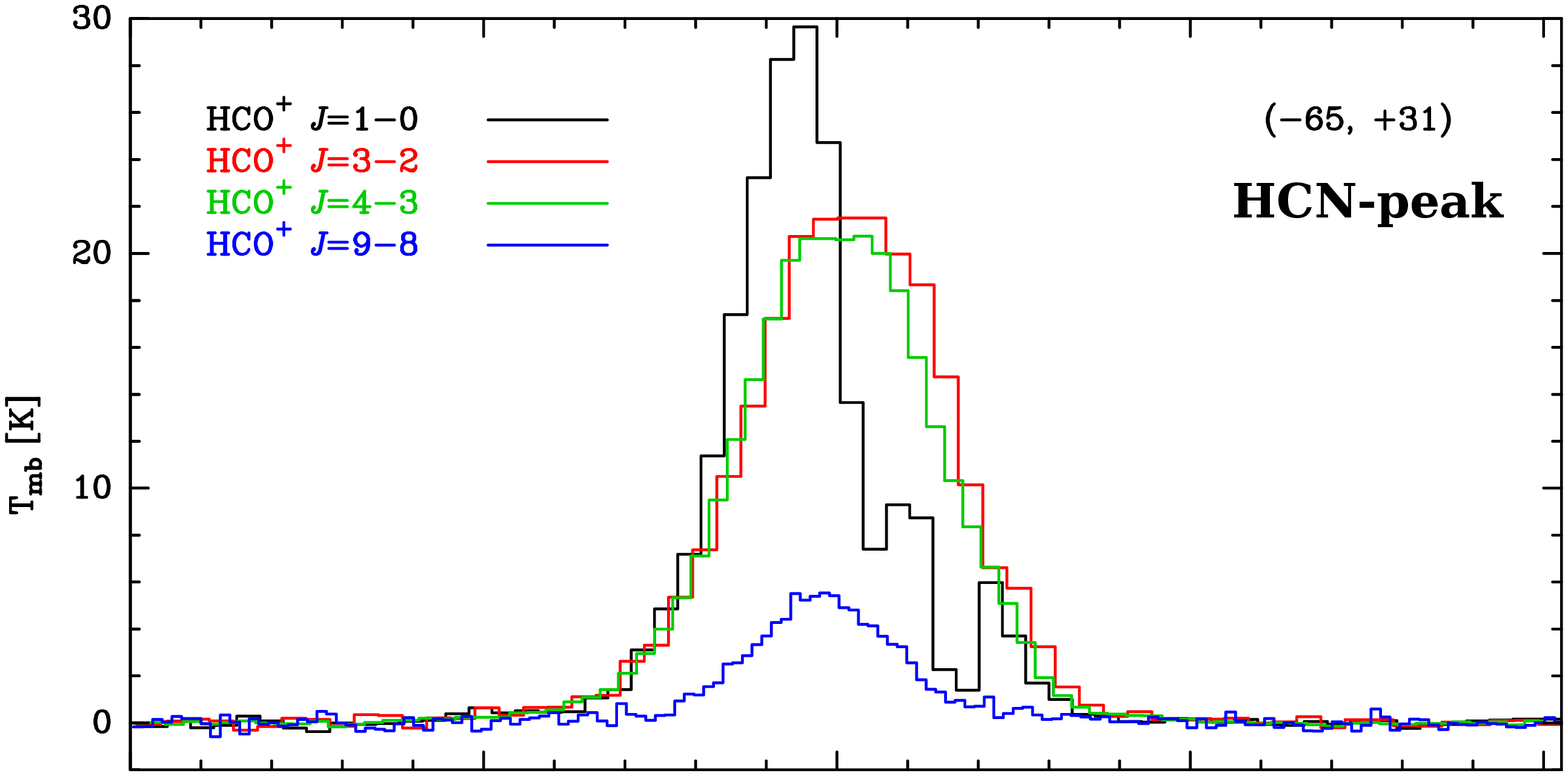,angle=0,width=0.46\linewidth} &
  \epsfig{file=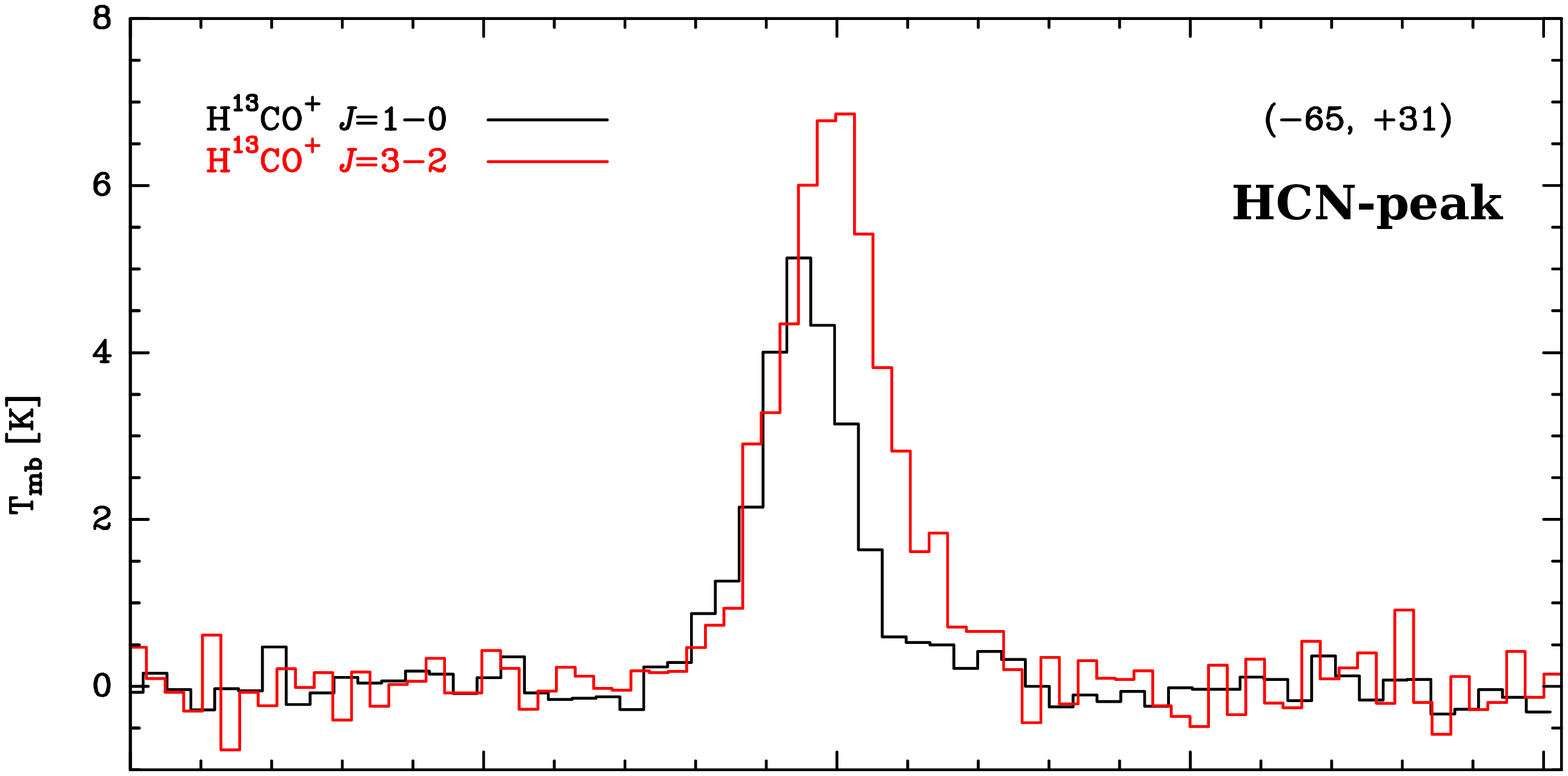,angle=0,width=0.46\linewidth}\\  
  \vspace{-0.54cm}\\

  \epsfig{file=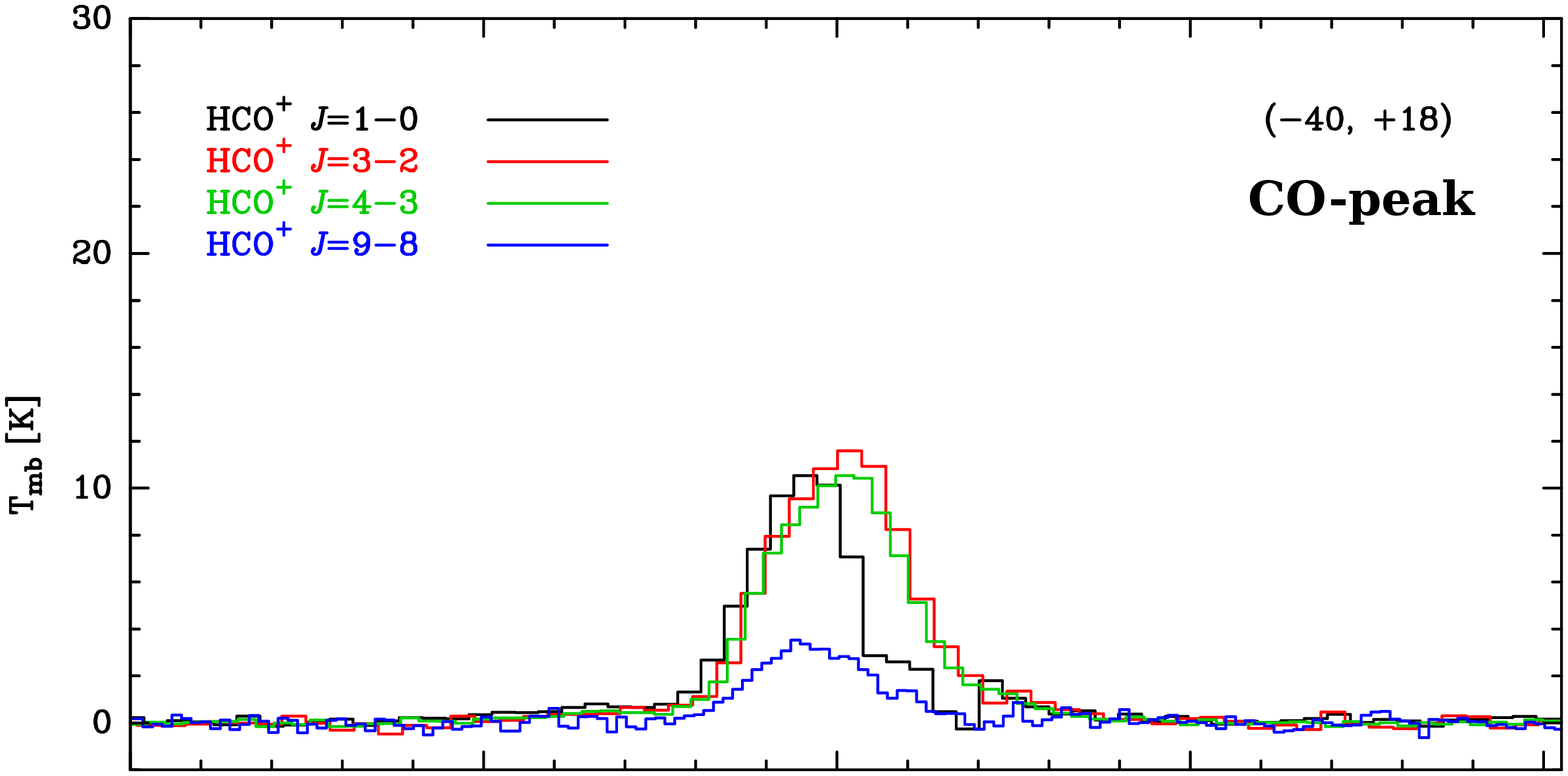,angle=0,width=0.46\linewidth} &
  \epsfig{file=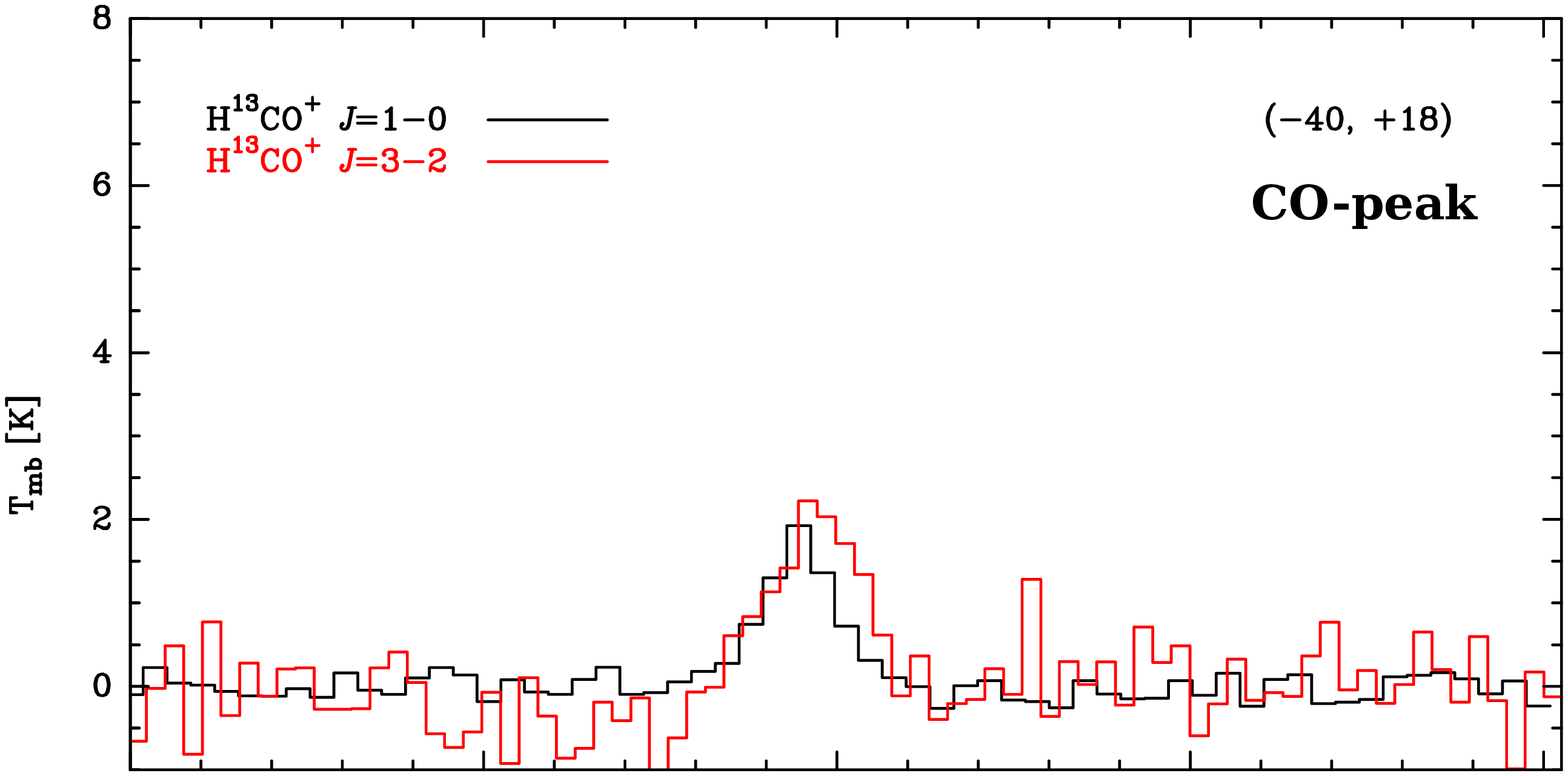,angle=0,width=0.46\linewidth}\\  
  \vspace{-0.54cm}\\

  \epsfig{file=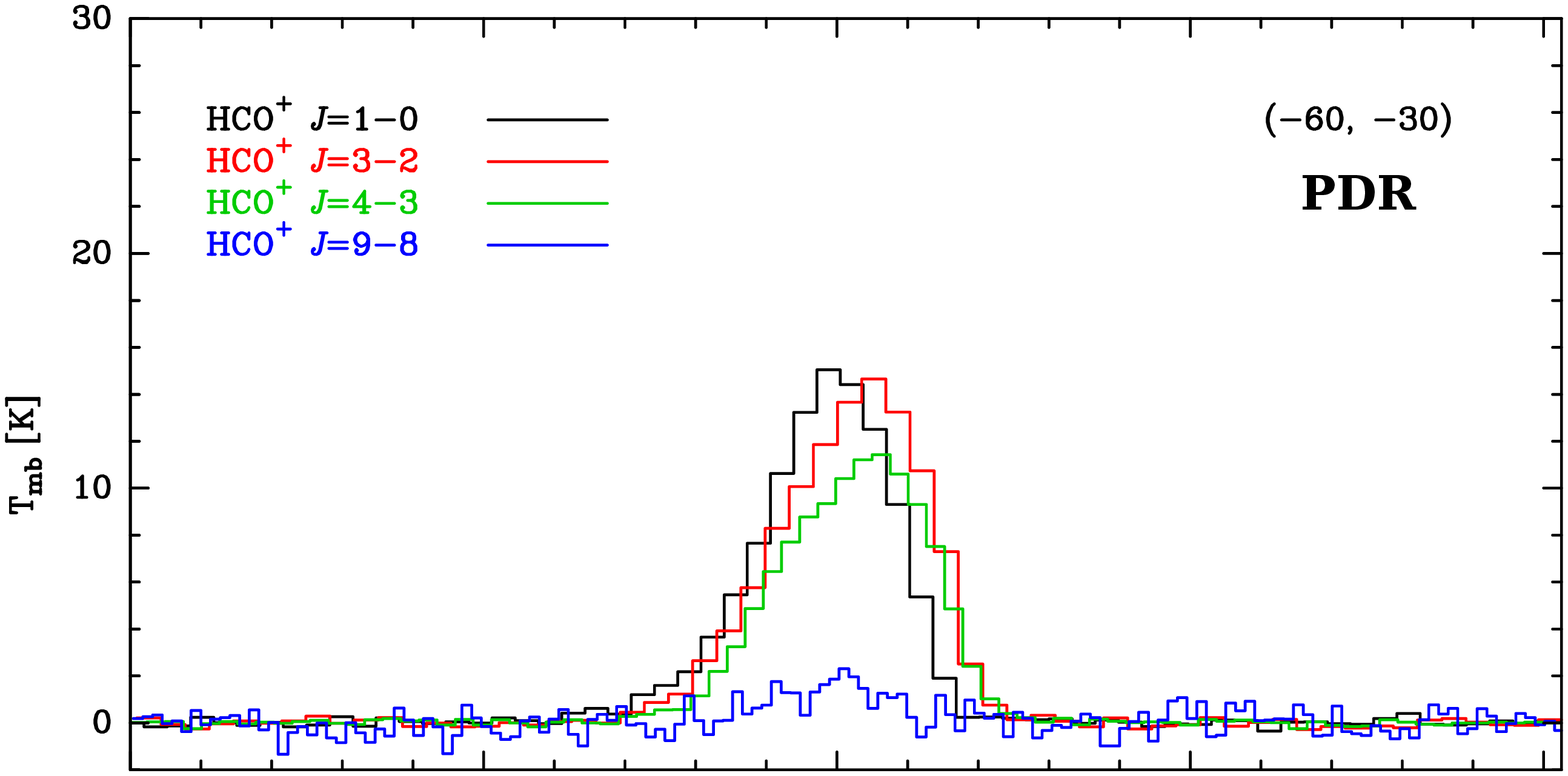,angle=0,width=0.46\linewidth} &
  \epsfig{file=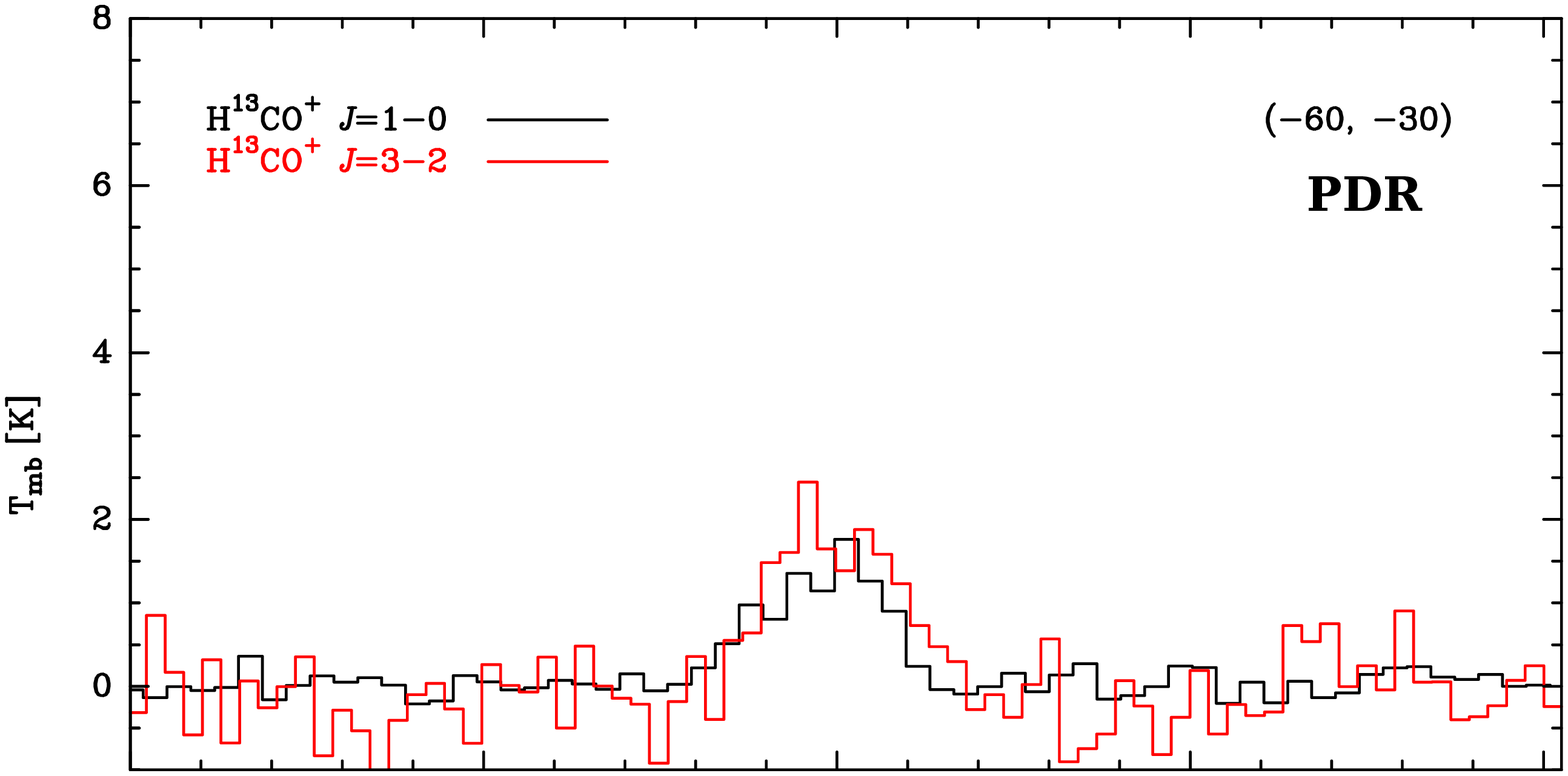,angle=0,width=0.46\linewidth}\\  
  \vspace{-0.54cm}\\
  
  \epsfig{file=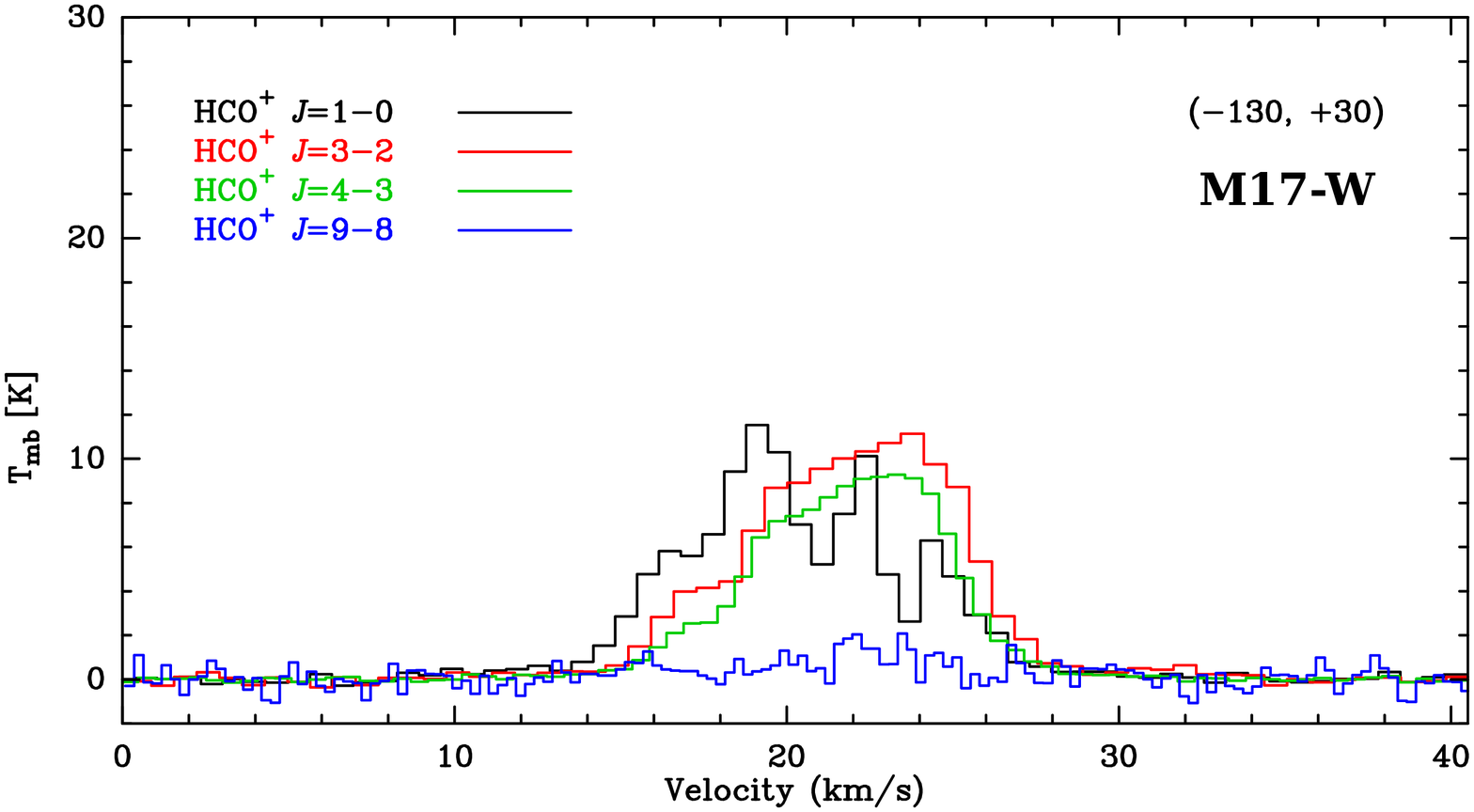,angle=0,width=0.46\linewidth} &
  \epsfig{file=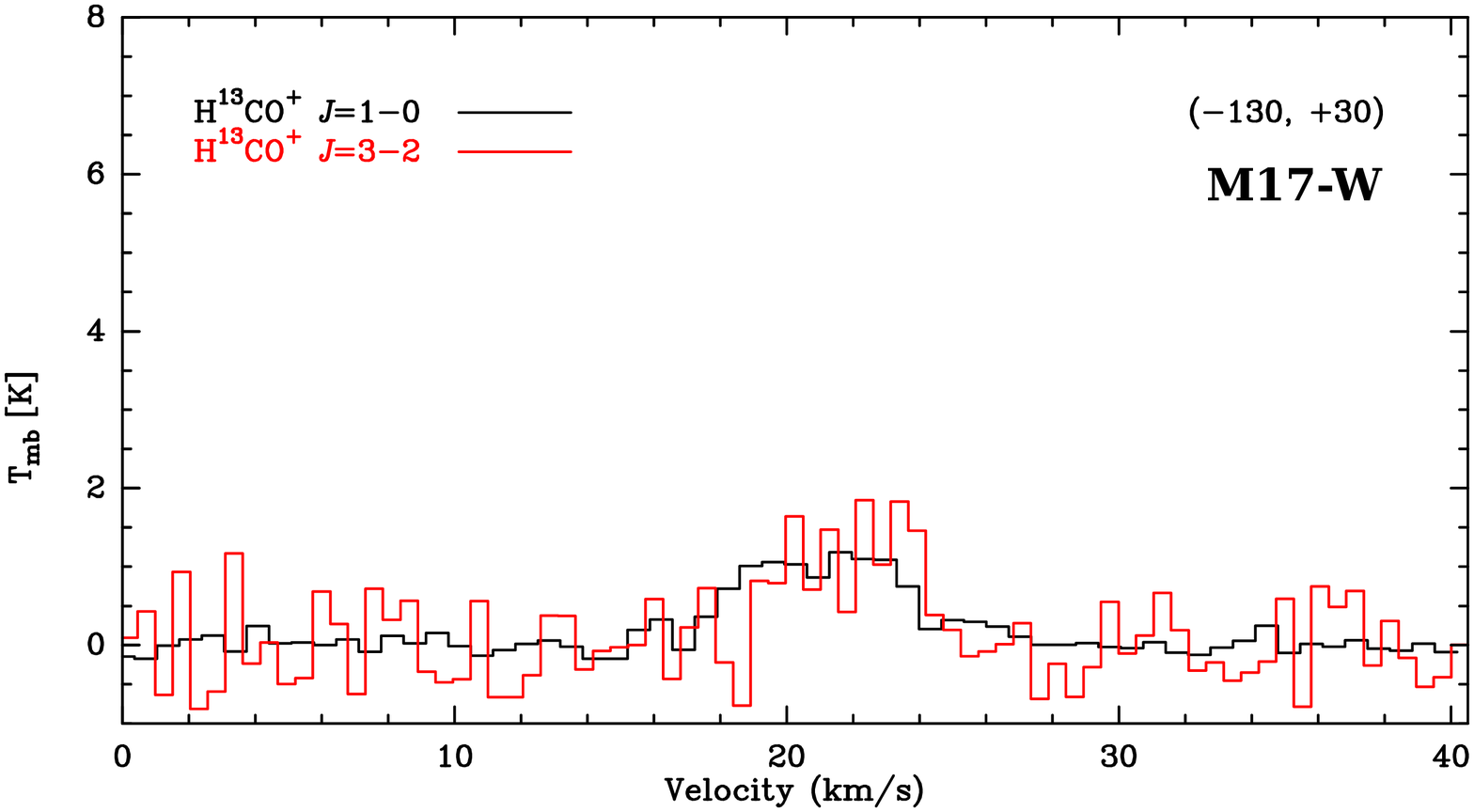,angle=0,width=0.46\linewidth}\\  
  \vspace{-0.54cm}\\

 \end{tabular}

  \caption{\footnotesize{Spectra of all \hcop\ (\textit{left panel}) and \hthcop\ (\textit{right panel}) lines 
  observed at the four selected offset (within $\pm$3$''$) positions shown in Fig.~\ref{fig:CO3-2_map_B25}. 
  All the spectra, except the $J=1\to0$ 
  and $J=2\to1$ have been convolved with a 25$''$ beam, corresponding to the \twco~$J=11\to12$ map.}}

  \label{fig:HCOp-spectra-3positions}
\end{figure*}
%---------------------------------------------------------------

We computed the average spectrum in a region associated with a 200$''$ beam 
size, centered at offset position ($-100''$,$0''$), in order to estimate 
the excitation derived from ALMA observations of nearby Galaxies.
In M17~SW the $200''$ beam covers a spatial scale of $\sim$2~pc. This 
correspond to the spatial scale that would be resolved by ALMA with the finest achievable 
angular resolution ($0.03''$ with the capabilities available in Cycle 3) of the bands 6 
(230~GHz) and 10 (870~GHz) towards a galaxy like NGC~1068 at a distance of $\sim$14.4~Mpc. 
Coarser angular resolution would be sufficient to resolve the same spatial scale in closer 
galaxies like NGC~253 (at $\sim$3.5~Mpc).

%For the $J=1\to0$ transition of \hcn\ and \hcop, we computed the 
%average emission from the same 200$''$ region, but at their respective 
%original resolutions ($>25''$). We did not take into account these low-$J$ 
%transitions in the fitting process, but we show them in the final fit for 
%completeness.
 
Figure~\ref{fig:CO3-2_map_B25} shows the 
\twco~$J=3\to2$ map convolved with a 25$''$ beam, and the dashed circle 
depicts the area where the average emission in all the maps was estimated 
from. 
The average spectra of the \twco, \hcn\ and \hcop\ lines are shown in 
Fig.~\ref{fig:average-spectra}. We computed the intensities integrating the 
spectra over the 5--35~\kms\ velocity range. We also fit one Gaussian 
component to estimate the (FWHM) line width. The velocity-integrated 
intensities and the total intensity obtained from the Gaussian fit are 
similar within a few percent.

The best two components model fit of Eq.~\ref{eq:LSED-model} is shown in 
Fig.~\ref{fig:average-SED-fit} for \twco, \hcn\ and \hcop. The LSED fit of 
the isotopologues, \thco, \hthcn\ and \hthcop, is shown in the insets. 
The error bars correspond to the $3\sigma$ range around the fluxes used to 
estimate the standard deviation of the parameters, where $\sigma$ is the 
uncertainty of the estimated fluxes obtained from the Gaussian fit. Even 
though the fluxes of the \hcn\ fine structure lines are not considered in 
the Gaussian fit, the flux estimated for the \hcn~$J=1\to0$ line is similar 
than predicted by the model (Fig.~\ref{fig:average-SED-fit}, middle panel). 
%The models predict relatively well the fluxes of the $J=1\to0$ lines of 
%\hcop, \hthcn\ and \hthcop, despite that they are observed with larger 
%beams.

The parameters of the LSED models are summarized in 
Table~\ref{tab:average-LSED-fit}. Since the CO emission is more extended 
than the \hcn\ and \hcop, the beam area filling factors of both CO 
components need to be larger as well. 
The same excitation conditions (density and temperature) as found for CO, 
were sufficient to reproduce the \hcn\ and \hcop\ fluxes, considering the 
same filling factors for these two species but with 37\% and 84\% lower 
column densities for the cold and warm component of \hcop, respectively.
At first look, this might be counter-intuitive with the fact that the 
emission from the higher-$J$ ($J_{up}\geqslant3$) transitions of \hcop\ 
are brighter and seem to be more extended than those of \hcn\ 
(cf., Figs.~\ref{fig:APEX-30m-maps} and \ref{fig:CHAMP-maps}). But this 
can be explained because \hcop\ is easier to excite than \hcn, given 
their different critical densities (see Table~\ref{tab:spectral-lines}), 
and the larger optical depths of the \hcn\ lines.

The densities and temperatures found for the cold component of CO are 
comparable with those found in starburst galaxies like, e.g. NGC~253 
\citep[][their Tables~2--4]{rosenberg14}, but the warm component requires 
excitation conditions with at least one order of magnitude higher density 
and column density. Similar discrepancies are found when comparing with the 
excitation conditions estimated for Seyfert galaxies like NGC~1068 
\citep[][their Table~3]{spinoglio12}. These differences may be due to an 
underestimated beam dilution effect in the extragalactic observations, as 
well as the fact that in their larger beams they collect emission emerging 
from a variety of molecular clouds with different sources of heating. 
Another factor affecting the determination of the excitation conditions in 
extragalactic observations is the dust extinction that seems to affect the 
high-$J$ CO lines at the large ($>$10$^{24}~\2cm$) column densities observed 
towards their centers \citep[e.g.][]{pineda10, etxaluze13}, which was not 
taken into account in the extragalactic studies. 

We do not correct the higher-$J$ CO lines for dust extinction either since 
the dust extinction in M17~SW is expected to be less significant than towards
circumnuclear regions, given that the total column densities estimated towards 
M17~SW are between one and two orders of magnitude smaller than towards our 
Galactic Center \citep[e.g.,][]{stutzki90, meixner92} and the centers of other galaxies.
%we do not have detailed estimates of the dust opacities and spectral 
%indices matching our map resolutions. But they will be included in a follow 
%up work using dust continuum observations with the new A-MKID camera for 
%APEX, scheduled for commissioning in March 2015. According to estimates 
%done for the Galactic Center, the \twco~$J=16\to15$ line could be between 3 
%and 30 times brighter, depending on whether the gas and dust are well 
%mixed, or if the CO emission arises from the inner regions of the 
%star-forming cores \citep{etxaluze13}. Given that \twco~$J=16\to15$ is at 
%163~\mum, the shielding column must be larger than 10$^{24}~\2cm$ (with 
%solar metallicity) for such attenuation. Therefore, the excitation 
%conditions we report here for the warm CO component in terms of 
%temperature, density, and column density, should be considered lower limits.

\subsection{Excitation at key positions in M17~SW}\label{sec:key-positions}

Using the 25$''$ resolution maps we estimated the fluxes from the Gaussian 
fit of the line wings, as described in the third paragraph of Sect.~\ref{sec:analysis}.
Then we fit the LSED of the three molecular species at four particular locations toward 
M17~SW to study the variation in excitation conditions. 

We chose the peak of the \hcn~$J=8\to9$ emission at 
about the offset position ($-65''$,$+31''$) which is expected to be 
dominated by dense gas, the peak of the 
\twco~$J=16\to15$ emission at about ($-40''$,$+18''$), close to the UC \hii\ region, 
the emission at about ($-60''$,$-30''$), deeper into the PDR with an
expected fair mixture of excitation conditions, and the emission at about ($-130''$,$+30''$)
west from the northern concentration of \hho\ masers, corresponding to the 
\textit{shoulder} or secondary peak of the \hcn\ and \hcop\ strip lines of 
Fig.~\ref{fig:strip-lines}. We refer to these positions as: \hcn-peak, CO-peak, 
PDR, and M17-W, respectively, and indicated in Fig.~\ref{fig:CO3-2_map_B25}.

\begin{figure*}[!pt]
 \centering
 \begin{tabular}{ccc}
  \hspace{-0.3cm}\epsfig{file=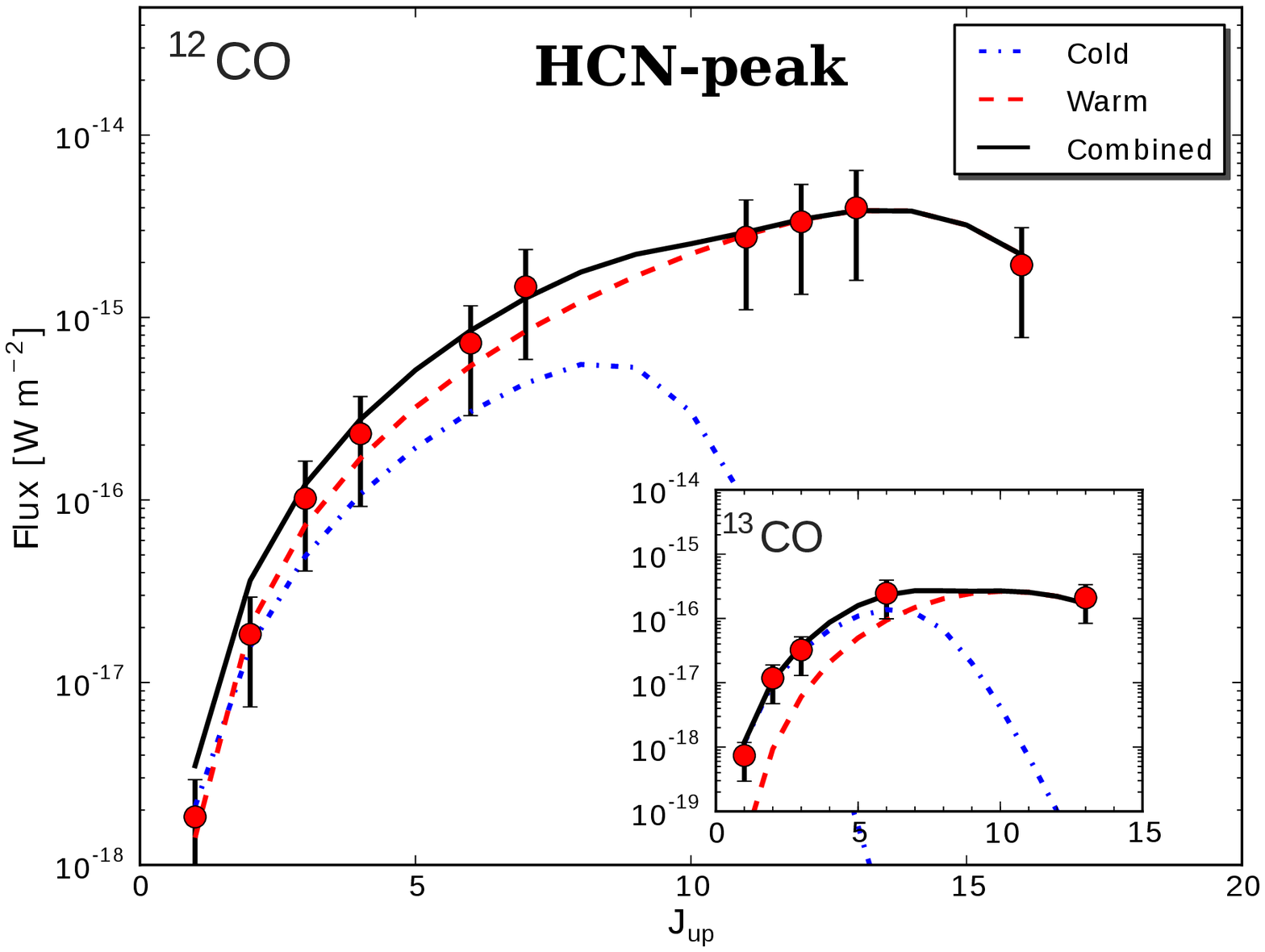,angle=0,width=0.33\linewidth} &
  \hspace{-0.3cm}\epsfig{file=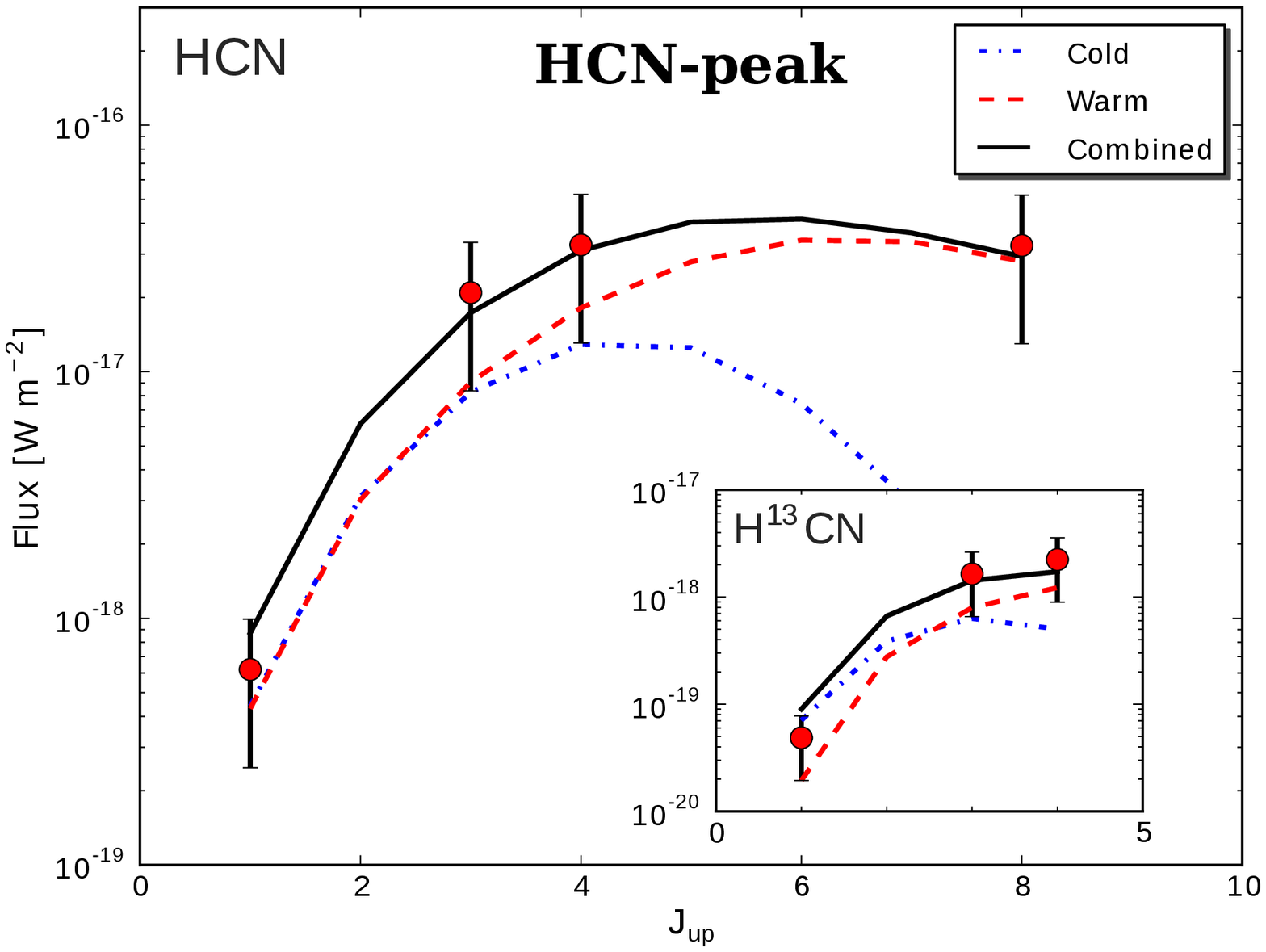,angle=0,width=0.33\linewidth} &
  \hspace{-0.3cm}\epsfig{file=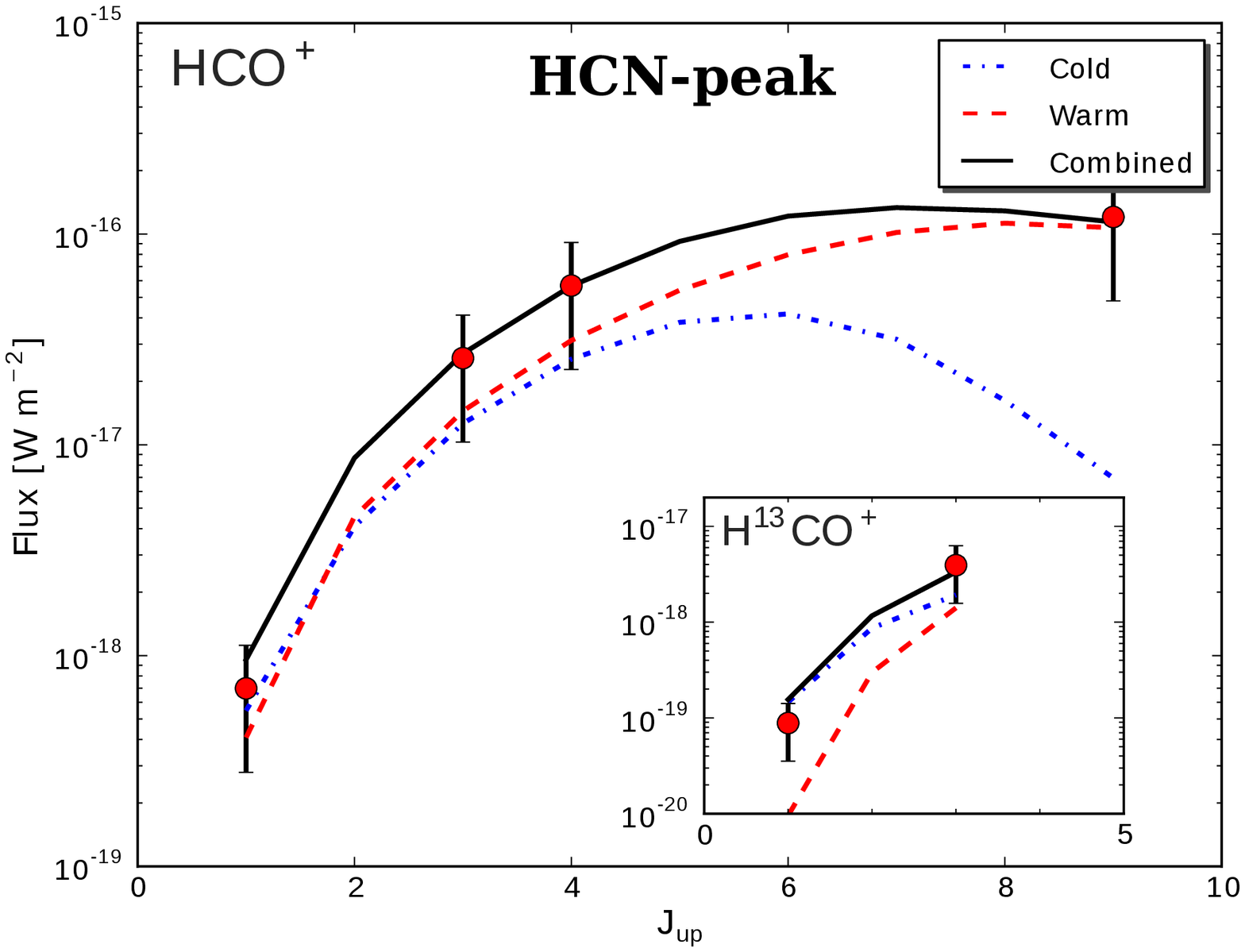,angle=0,width=0.33\linewidth}\\
  
  \hspace{-0.3cm}\epsfig{file=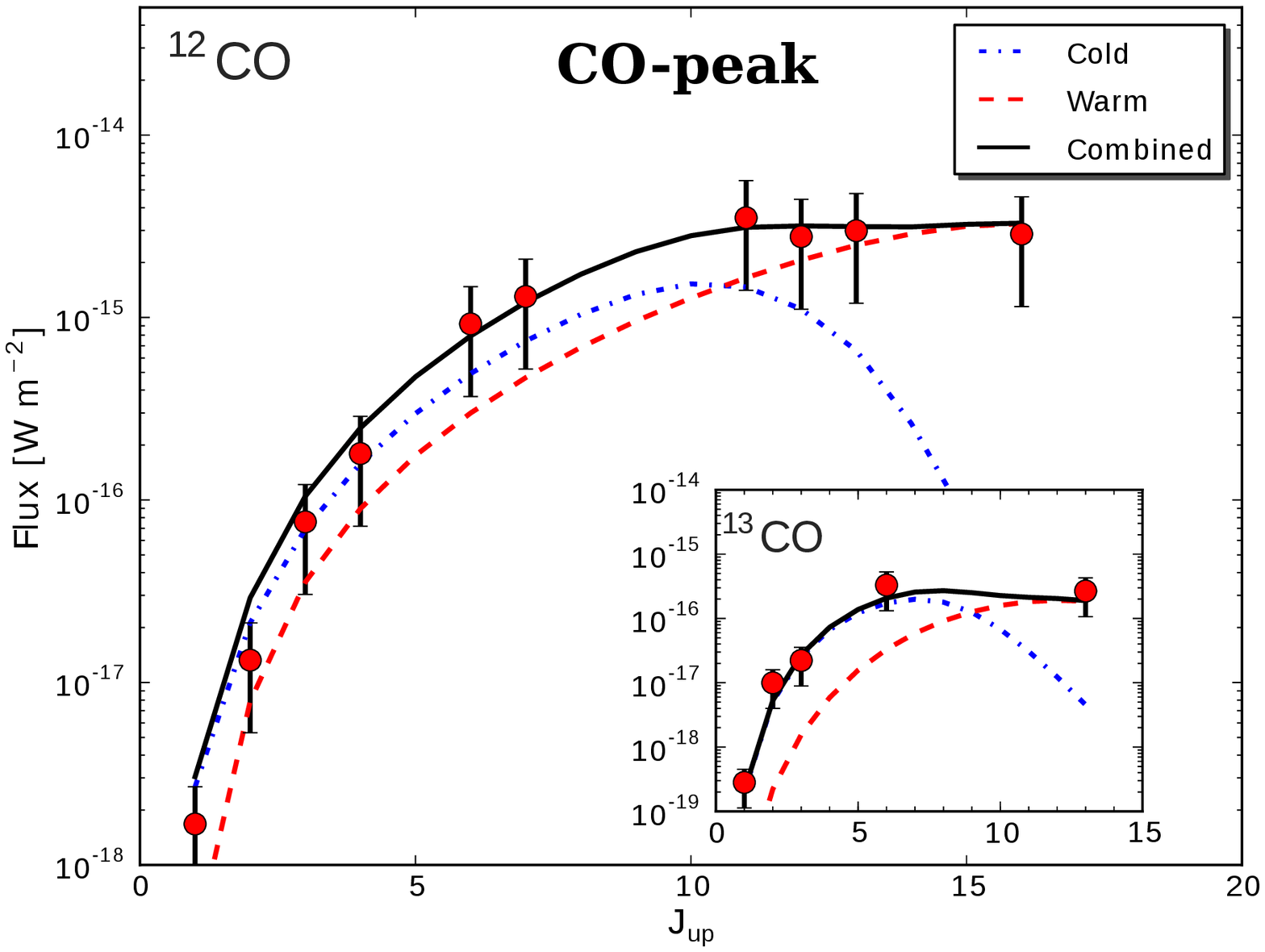,angle=0,width=0.33\linewidth} &
  \hspace{-0.3cm}\epsfig{file=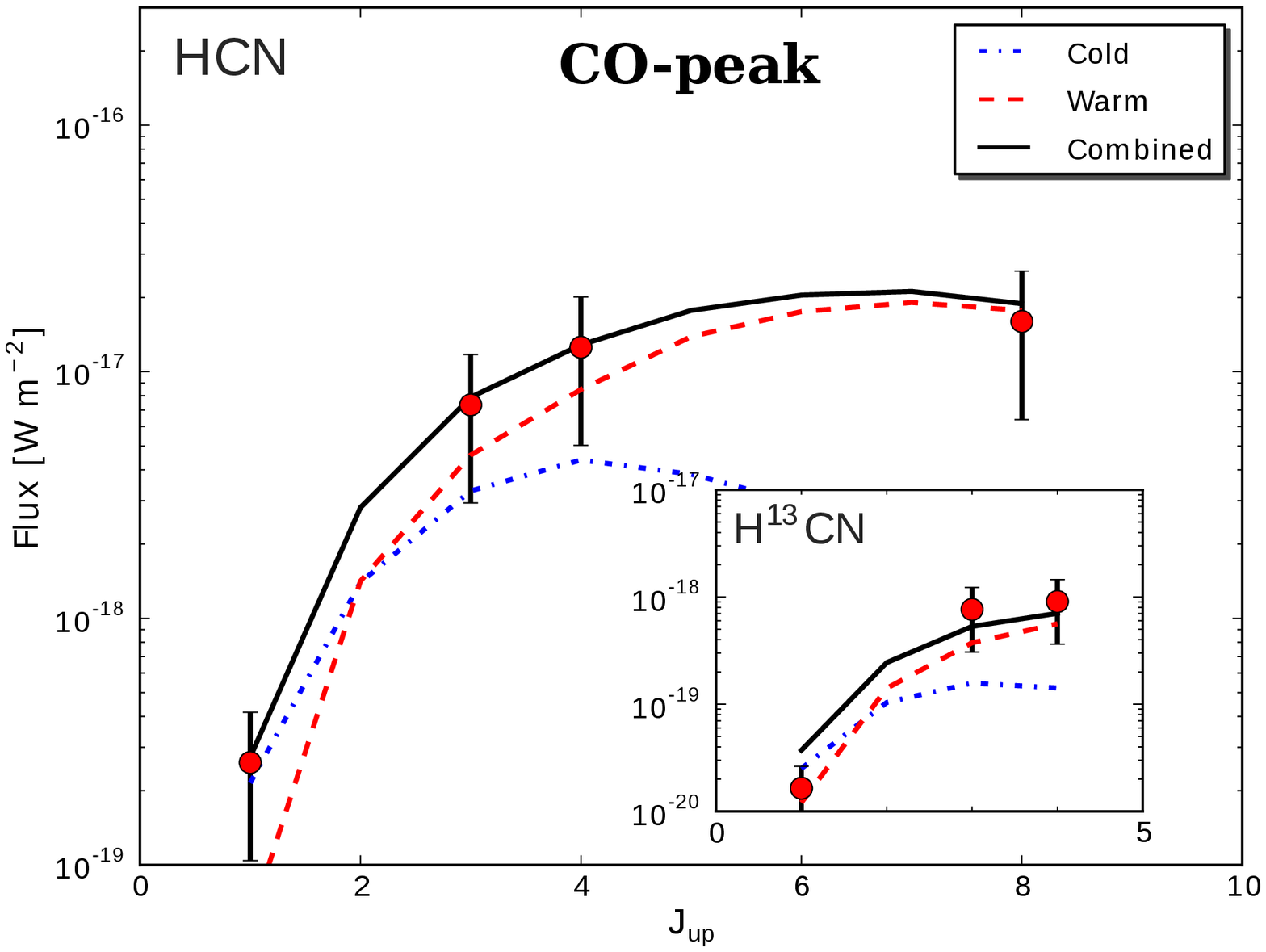,angle=0,width=0.33\linewidth} &
  \hspace{-0.3cm}\epsfig{file=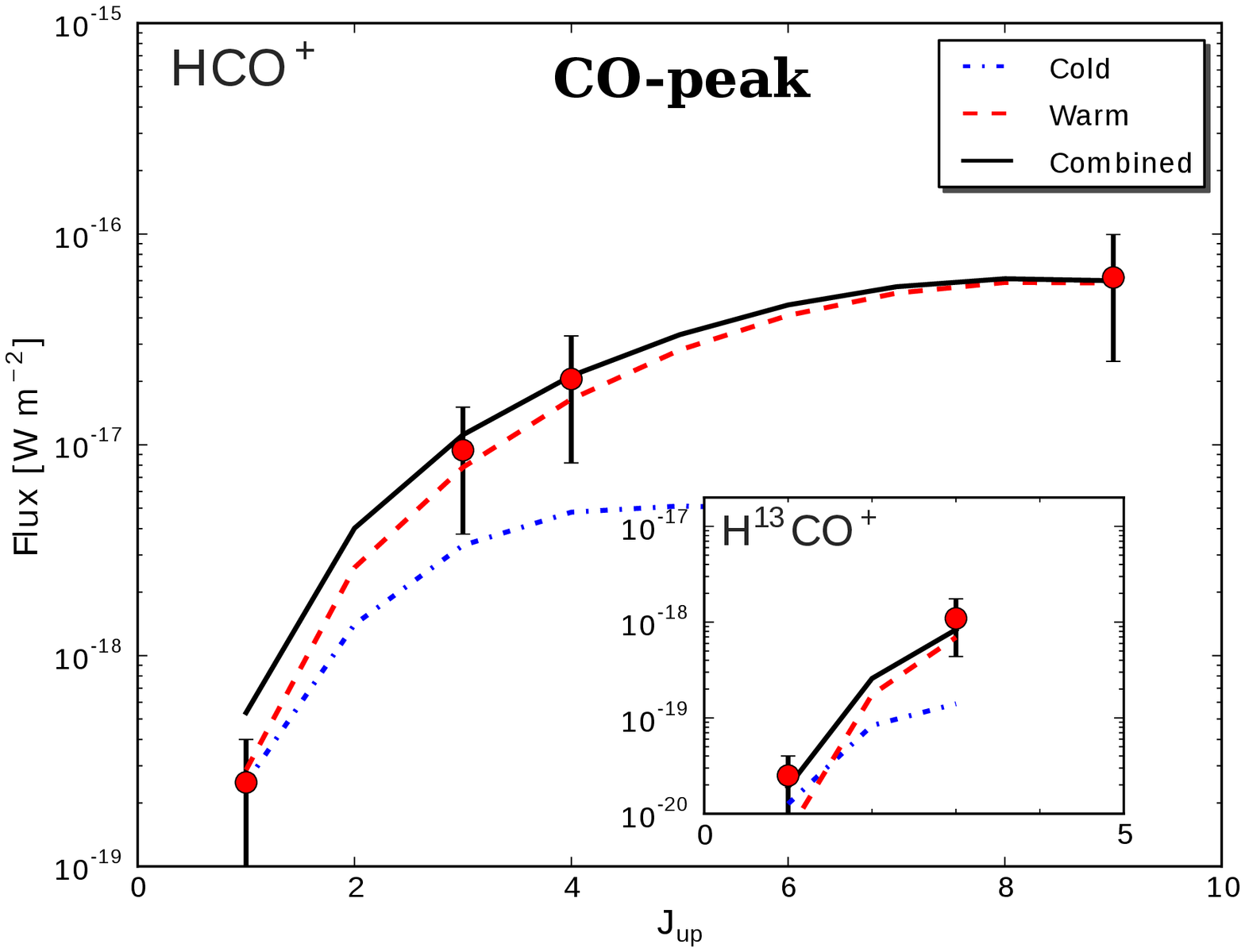,angle=0,width=0.33\linewidth}\\
  
  \hspace{-0.3cm}\epsfig{file=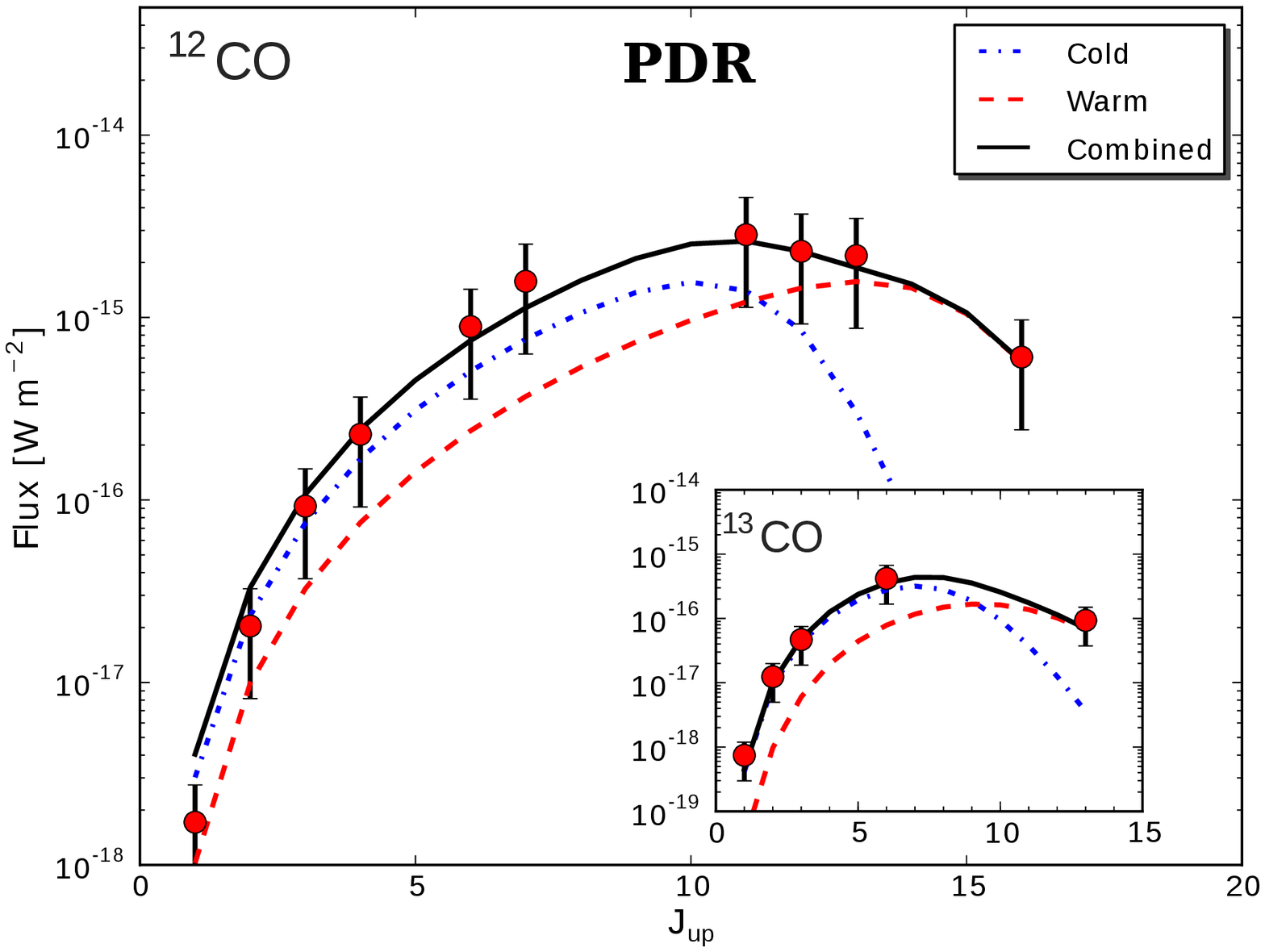,angle=0,width=0.33\linewidth} &
  \hspace{-0.3cm}\epsfig{file=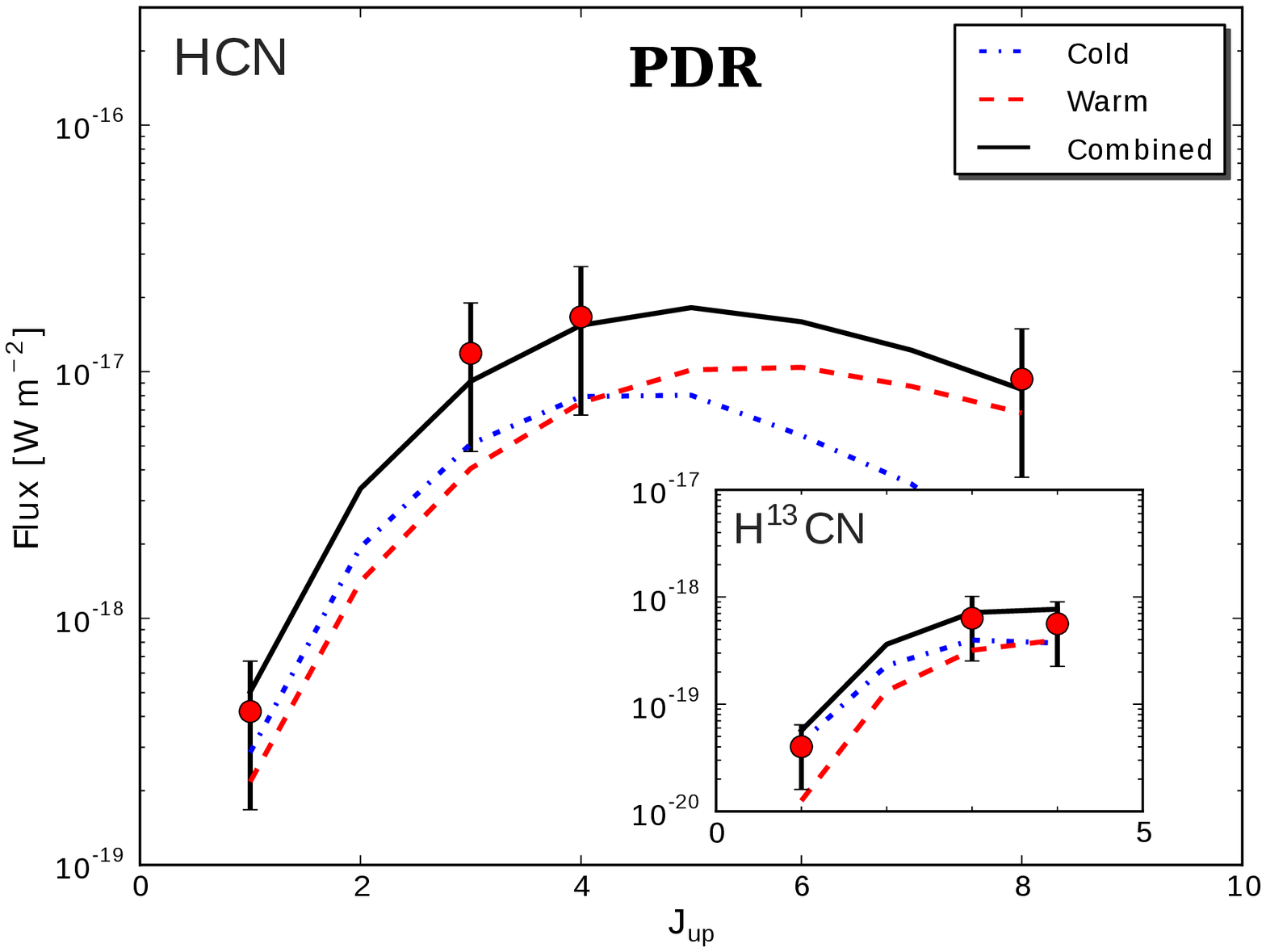,angle=0,width=0.33\linewidth} &
  \hspace{-0.3cm}\epsfig{file=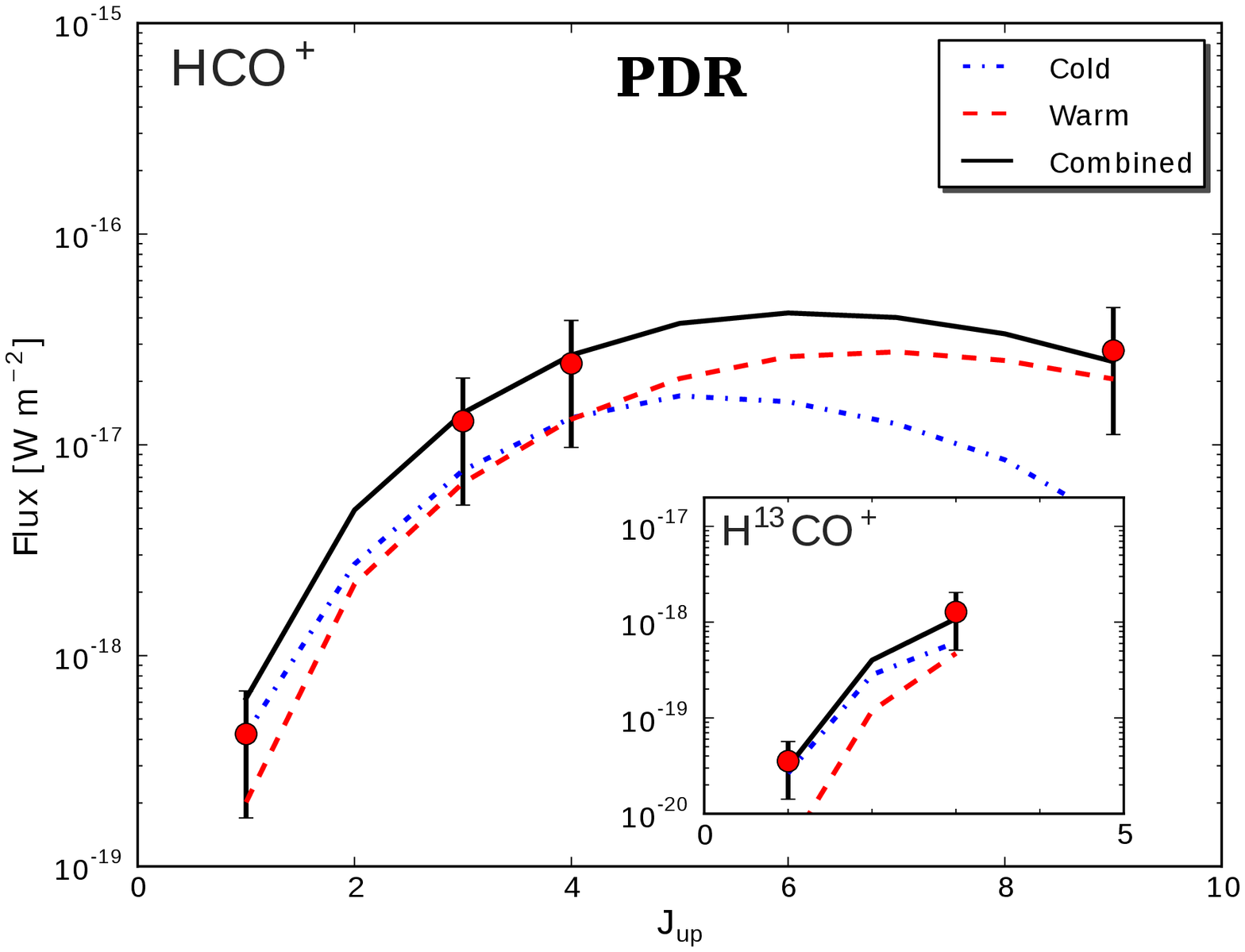,angle=0,width=0.33\linewidth}\\
  
  \hspace{-0.3cm}\epsfig{file=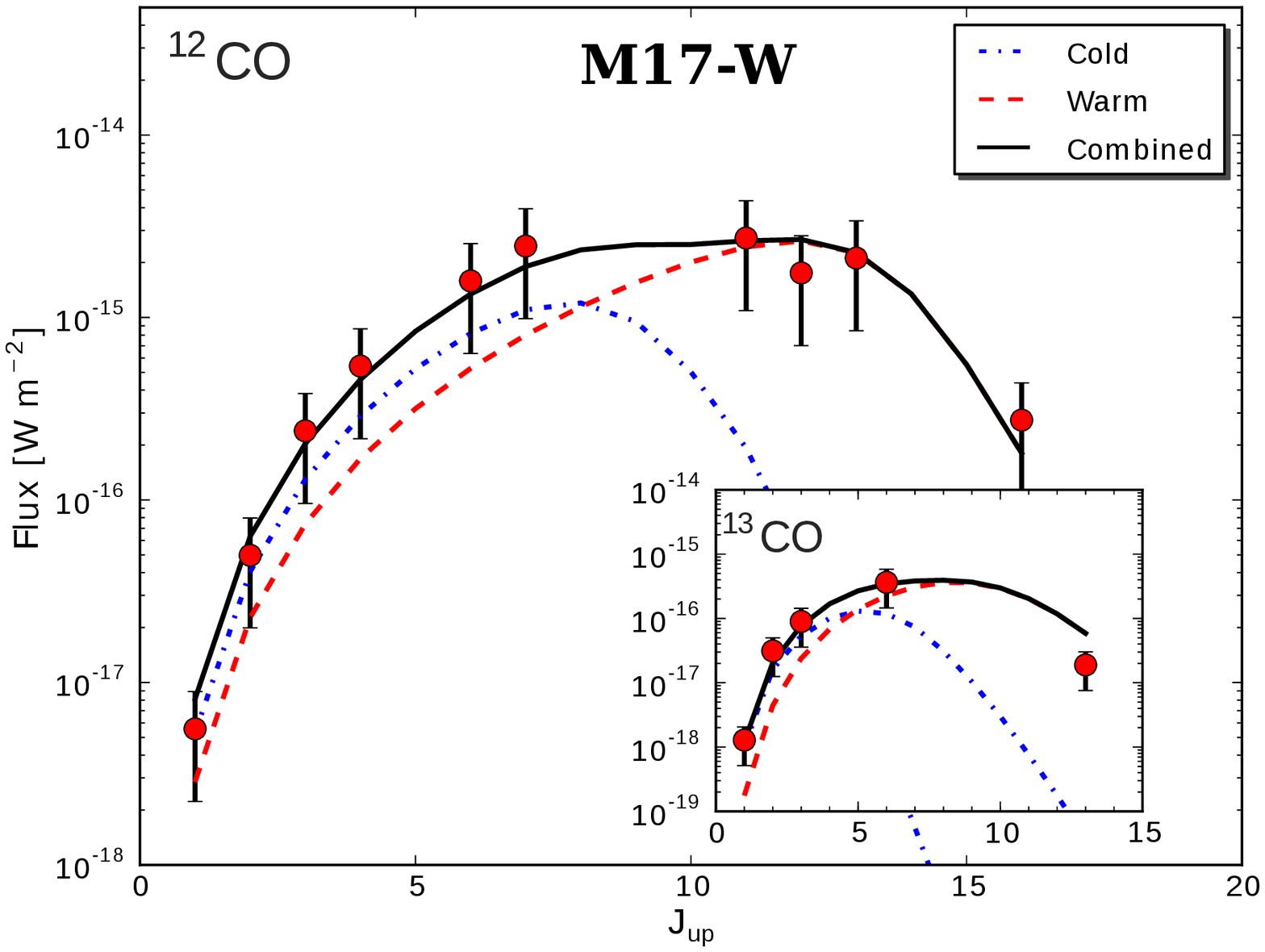,angle=0,width=0.33\linewidth} &
  \hspace{-0.3cm}\epsfig{file=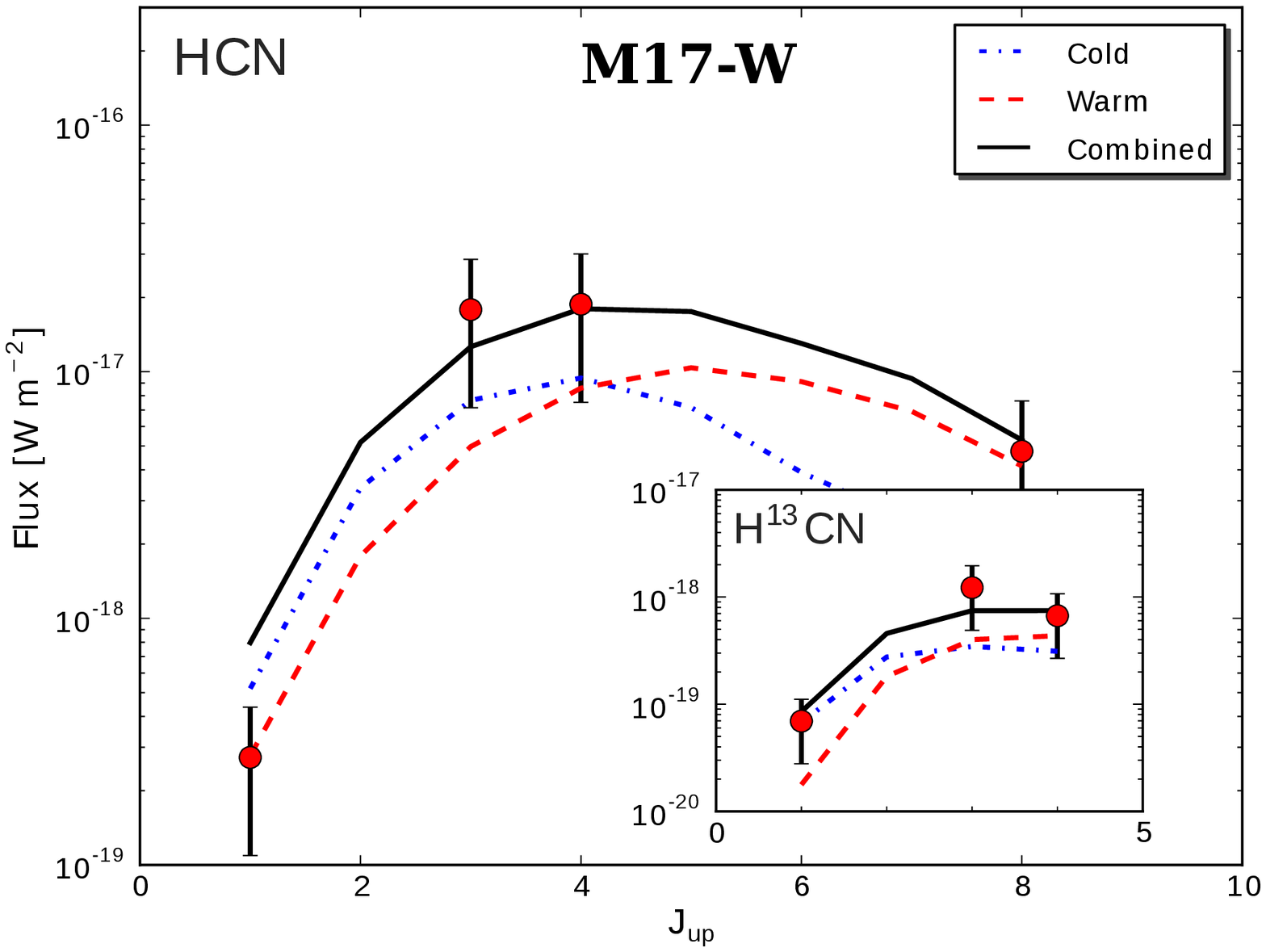,angle=0,width=0.33\linewidth}&
  \hspace{-0.3cm}\epsfig{file=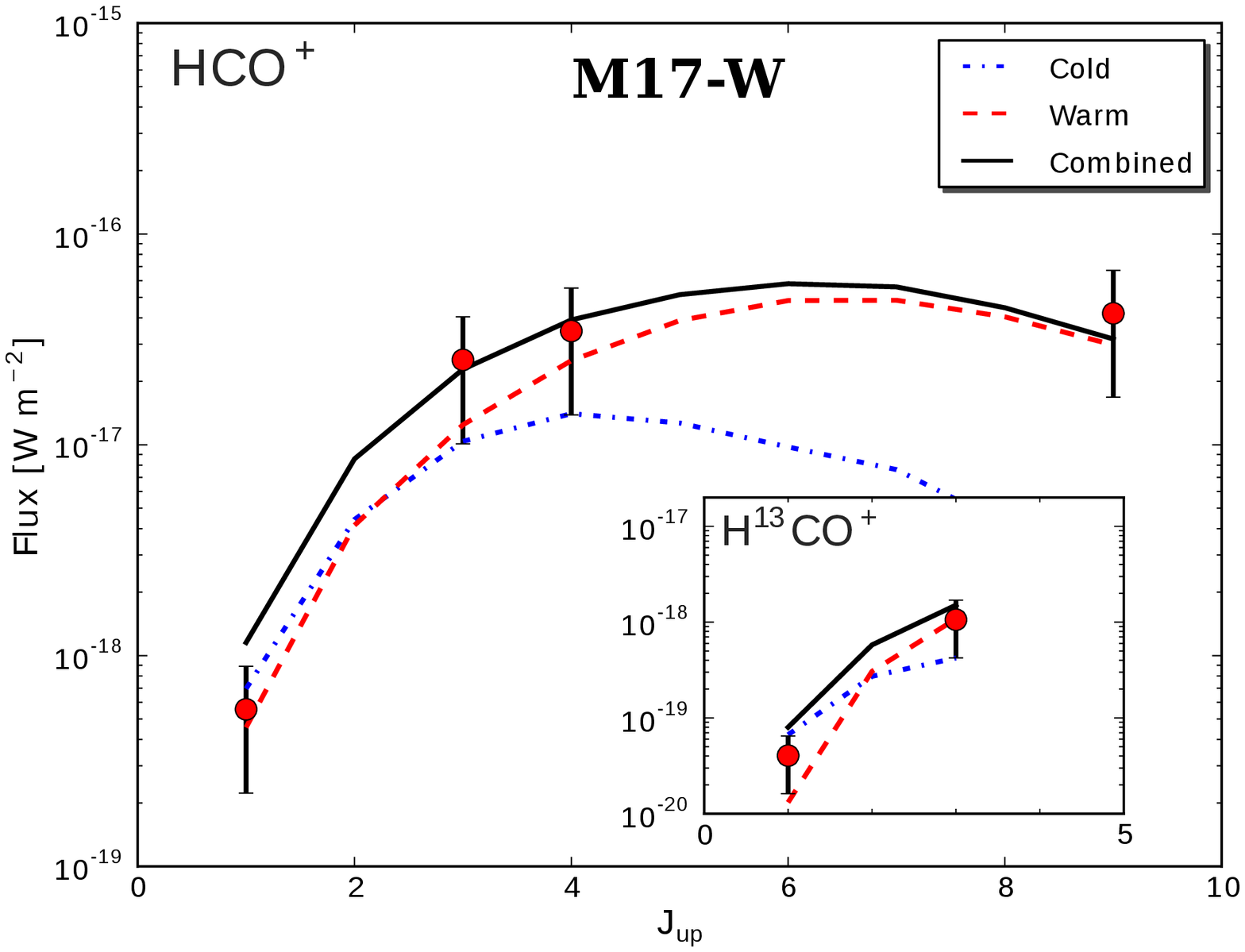,angle=0,width=0.33\linewidth}      
 \end{tabular}

  \caption{\footnotesize{Two component fit of the line spectral energy distribution of the (from left to right) 
  CO, \hcn, and \hcop\ species, for the (from top to bottom) spectra at offset position ($-65''$,$+31''$) toward 
  the peak of the \hcn~$J=8\to7$ intensity, ($-40''$,$+18''$) toward the peak of the \twco~$J=16\to15$ intensity, 
  ($-60''$,$-30''$) toward the PDR, and ($-130''$,$+30''$) toward the molecular gas in M17~SW.
  The fit of the $^{13}$C bearing isotopologues is shown in the insets. 
  The cold and warm components are shown in dotted and dashed lines, respectively. 
  The error bars are as in Fig.~\ref{fig:average-SED-fit}}}

  \label{fig:4pos-SED-fit}
\end{figure*}

The spectra of all the observed lines at these positions are shown in 
Figs.~\ref{fig:CO-spectra-3positions}-\ref{fig:HCOp-spectra-3positions}.
The fluxes obtained from the CO, \hcn\ and \hcop\ spectra at the M17-W position
have the largest uncertainties because the line profiles show the highest
level of sub-structures and self-absorption features. The \hthcn\ lines
at this position have very low S/N, so we consider them as upper limits.
The best fit models for the LSED at these positions are shown in 
Figs.~\ref{fig:4pos-SED-fit}, and the model parameters are summarized in 
Tables~\ref{tab:pos1-LSED-fit}, \ref{tab:pos2-LSED-fit}, 
\ref{tab:pos3-LSED-fit} and \ref{tab:pos4-LSED-fit}. 

The \twco\ LSED at the \hcn-peak shows an increasing trend in flux with $J$, 
with an apparent turn over at the $J=14\to13$ transition (which 
we do not have). The \twco\ LSED of the CO-peak emission seems 
flat at the high-$J$ transitions, and the \twco\ LSED at the PDR position, 
shows a decreasing trend in flux from the $J=11\to10$ transition. 
On the other hand, the LSED of the M17-W position show a flat
distribution of fluxes between the $J=7\to6$ and $J=13\to12$ transitions,
and a sharp (one order of magnitude) decrease at the $J=16\to15$. 
Observations of the $J=9\to8$ transition with APEX would be needed to
confirm the flat distribution of fluxes at the mid- and high-$J$ transitions
of \twco\ at the M17-W position.
These different LSED shapes, specially at the
high-$J$ CO lines obtained with SOFIA/GREAT, are indicative of 
the distinctive excitation conditions dominating mainly the warm 
component of the two-phase model.

%__________________________________________________ One column table
   \begin{table}[!tp]
      \caption[]{LSED fit parameters\tablefootmark{a} for the spectra toward 
      the HCN-peak position ($-65''$,$+31''$).}
         \label{tab:pos1-LSED-fit}
         \centering
         \scriptsize
         %\footnotesize
         \setlength{\tabcolsep}{3.5pt} % Default value: 6pt
         \renewcommand{\arraystretch}{1.0} % Default value: 1
         \begin{tabular}{lcccc}
            \hline\hline
	    \noalign{\smallskip}
            Parameter & CO & HCN  &  HCO$^+$  \\
            \noalign{\smallskip}
            \hline
            \noalign{\smallskip}

$\Phi_{cold}(^{12}{\rm C})$  &   1.00 $\pm$  0.09  &   0.40 $\pm$  0.05  &   0.40 $\pm$  0.05 \\ 
$n_{cold}(\rm H_2)$ [cm$^{-3}$]  &   4.50 $\pm$  0.47  &   4.50 $\pm$  0.47  &   4.50 $\pm$  0.46 \\ 
$T_{cold}$ [K]  &  40.00 $\pm$  3.78  &  40.00 $\pm$  5.37  &  40.00 $\pm$  4.83 \\ 
$N_{cold}$ [cm$^{-2}$]  &  18.90 $\pm$  1.12  &  16.60 $\pm$  1.06  &  16.40 $\pm$  1.08 \\ 
 &  &  &  \\ 
$\Phi_{warm}(^{12}{\rm C})$  &   0.35 $\pm$  0.04  &   0.15 $\pm$  0.01  &   0.15 $\pm$  0.02 \\ 
$n_{warm}(\rm H_2)$ [cm$^{-3}$]  &   6.00 $\pm$  0.73  &   6.00 $\pm$  0.52  &   6.00 $\pm$  0.48 \\ 
$T_{warm}$ [K]  &  130.00 $\pm$ 11.59  &  130.00 $\pm$ 15.33  &  130.00 $\pm$ 14.03 \\ 
$N_{warm}$ [cm$^{-2}$]  &  18.40 $\pm$  0.37  &  15.40 $\pm$  0.54  &  15.20 $\pm$  0.81 \\ 
 &  &  &  \\ 
$\Phi_{cold}(^{13}{\rm C})$  &   1.00 $\pm$  0.11  &   0.40 $\pm$  0.05  &   0.40 $\pm$  0.03 \\ 
$\Phi_{warm}(^{13}{\rm C})$  &   0.35 $\pm$  0.04  &   0.15 $\pm$  0.02  &   0.15 $\pm$  0.01 \\ 
 &  &  &  \\ 
%$^{12}$C/$^{13}$C  &  50.00 $\pm$  5.33  &  50.00 $\pm$  5.50  &  50.00 $\pm$  5.86 \\ 
% &  &  &  \\ 
$\Delta V(^{12}{\rm C})$ [km s$^-1$]  &   4.60   &   7.50   &   6.00  \\ 
$\Delta V(^{13}{\rm C})$ [km s$^-1$]  &   3.50   &   3.90   &   3.70  \\    
		
            \noalign{\smallskip}
            \hline
         \end{tabular}
         
         \tablefoot{
         \tablefoottext{a}{The density and column density values are given in $log_{10}$ scale.}
         }

   \end{table}
%__________________________________________________ One column table

%__________________________________________________ One column table
   \begin{table}[!tp]
      \caption[]{LSED fit parameters\tablefootmark{a} for the spectra toward 
      the CO-peak position ($-40''$,$+18''$).}
         \label{tab:pos2-LSED-fit}
         \centering
         \scriptsize
         %\footnotesize
         \setlength{\tabcolsep}{3.5pt} % Default value: 6pt
         \renewcommand{\arraystretch}{1.0} % Default value: 1
         \begin{tabular}{lcccc}
            \hline\hline
	    \noalign{\smallskip}
            Parameter & CO & HCN  &  HCO$^+$  \\
            \noalign{\smallskip}
            \hline
            \noalign{\smallskip}

$\Phi_{cold}(^{12}{\rm C})$  &   0.50 $\pm$  0.05  &   0.40 $\pm$  0.05  &   0.40 $\pm$  0.04 \\ 
$n_{cold}(\rm H_2)$ [cm$^{-3}$]  &   4.80 $\pm$  0.61  &   4.80 $\pm$  0.46  &   4.80 $\pm$  0.58 \\ 
$T_{cold}$ [K]  &  90.00 $\pm$ 12.59  &  90.00 $\pm$  8.21  &  90.00 $\pm$ 10.18 \\ 
$N_{cold}$ [cm$^{-2}$]  &  18.80 $\pm$  0.35  &  15.30 $\pm$  0.95  &  14.40 $\pm$  1.33 \\ 
 &  &  &  \\ 
$\Phi_{warm}(^{12}{\rm C})$  &   0.10 $\pm$  0.01  &   0.10 $\pm$  0.01  &   0.10 $\pm$  0.01 \\ 
$n_{warm}(\rm H_2)$ [cm$^{-3}$]  &   5.70 $\pm$  0.35  &   5.70 $\pm$  0.62  &   5.70 $\pm$  0.50 \\ 
$T_{warm}$ [K]  &  240.00 $\pm$ 29.77  &  240.00 $\pm$ 26.55  &  240.00 $\pm$ 26.54 \\ 
$N_{warm}$ [cm$^{-2}$]  &  18.50 $\pm$  1.23  &  15.30 $\pm$  1.10  &  15.10 $\pm$  0.87 \\ 
 &  &  &  \\ 
$\Phi_{cold}(^{13}{\rm C})$  &   0.50 $\pm$  0.06  &   0.40 $\pm$  0.04  &   0.40 $\pm$  0.04 \\ 
$\Phi_{warm}(^{13}{\rm C})$  &   0.10 $\pm$  0.01  &   0.10 $\pm$  0.01  &   0.10 $\pm$  0.01 \\ 
 &  &  &  \\ 
%$^{12}$C/$^{13}$C  &  50.00 $\pm$  5.85  &  50.00 $\pm$  5.80  &  50.00 $\pm$  5.75 \\ 
% &  &  &  \\ 
$\Delta V(^{12}{\rm C})$ [km s$^-1$]  &   4.60   &   4.60   &   4.30  \\ 
$\Delta V(^{13}{\rm C})$ [km s$^-1$]  &   3.90   &   3.60   &   3.20  \\ 
		
            \noalign{\smallskip}
            \hline
         \end{tabular}
         
         \tablefoot{
         \tablefoottext{a}{The density and column density values are given in $log_{10}$ scale.}
         }

   \end{table}
%__________________________________________________ One column table

%__________________________________________________ One column table
   \begin{table}[!tp]
      \caption[]{LSED fit parameters\tablefootmark{a} for the spectra toward 
      the southern PDR position ($-60''$,$-30''$).}
         \label{tab:pos3-LSED-fit}
         \centering
         \scriptsize
         %\footnotesize
         \setlength{\tabcolsep}{3.5pt} % Default value: 6pt
         \renewcommand{\arraystretch}{1.0} % Default value: 1
         \begin{tabular}{lcccc}
            \hline\hline
	    \noalign{\smallskip}
            Parameter & CO & HCN  &  HCO$^+$  \\
            \noalign{\smallskip}
            \hline
            \noalign{\smallskip}

$\Phi_{cold}(^{12}{\rm C})$  &   0.90 $\pm$  0.07  &   0.35 $\pm$  0.04  &   0.35 $\pm$  0.04 \\ 
$n_{cold}(\rm H_2)$ [cm$^{-3}$]  &   5.30 $\pm$  0.36  &   5.30 $\pm$  0.49  &   5.30 $\pm$  0.62 \\ 
$T_{cold}$ [K]  &  60.00 $\pm$  5.32  &  60.00 $\pm$  6.82  &  60.00 $\pm$  7.09 \\ 
$N_{cold}$ [cm$^{-2}$]  &  18.95 $\pm$  0.47  &  15.40 $\pm$  0.85  &  14.80 $\pm$  1.13 \\ 
 &  &  &  \\ 
$\Phi_{warm}(^{12}{\rm C})$  &   0.20 $\pm$  0.01  &   0.15 $\pm$  0.02  &   0.15 $\pm$  0.02 \\ 
$n_{warm}(\rm H_2)$ [cm$^{-3}$]  &   5.80 $\pm$  0.37  &   5.80 $\pm$  0.65  &   5.80 $\pm$  0.56 \\ 
$T_{warm}$ [K]  &  110.00 $\pm$ 11.56  &  110.00 $\pm$ 12.33  &  110.00 $\pm$ 13.42 \\ 
$N_{warm}$ [cm$^{-2}$]  &  18.60 $\pm$  0.31  &  15.10 $\pm$  1.10  &  14.70 $\pm$  1.11 \\ 
 &  &  &  \\ 
$\Phi_{cold}(^{13}{\rm C})$  &   0.90 $\pm$  0.09  &   0.35 $\pm$  0.04  &   0.35 $\pm$  0.04 \\ 
$\Phi_{warm}(^{13}{\rm C})$  &   0.20 $\pm$  0.02  &   0.15 $\pm$  0.02  &   0.15 $\pm$  0.02 \\ 
 &  &  &  \\ 
%$^{12}$C/$^{13}$C  &  50.00 $\pm$  6.22  &  50.00 $\pm$  4.69  &  50.00 $\pm$  5.40 \\ 
% &  &  &  \\ 
$\Delta V(^{12}{\rm C})$ [km s$^-1$]  &   4.30   &   5.00   &   4.50  \\ 
$\Delta V(^{13}{\rm C})$ [km s$^-1$]  &   4.30   &   3.80   &   3.90  \\   
		
            \noalign{\smallskip}
            \hline
         \end{tabular}
         
         \tablefoot{
         \tablefoottext{a}{The density and column density values are given in $log_{10}$ scale.}
         }         

   \end{table}
%__________________________________________________ One column table

%__________________________________________________ One column table
   \begin{table}[!tp]
      \caption[]{LSED fit parameters\tablefootmark{a} for the spectra toward 
      the M17-W position ($-130''$,$+30''$).}
         \label{tab:pos4-LSED-fit}
         \centering
         \scriptsize
         %\footnotesize
         \setlength{\tabcolsep}{3.5pt} % Default value: 6pt
         \renewcommand{\arraystretch}{1.0} % Default value: 1
         \begin{tabular}{lcccc}
            \hline\hline
	    \noalign{\smallskip}
            Parameter & CO & HCN  &  HCO$^+$  \\
            \noalign{\smallskip}
            \hline
            \noalign{\smallskip}

$\Phi_{cold}(^{12}{\rm C})$  &   1.00 $\pm$  0.00  &   0.80 $\pm$  0.06  &   0.80 $\pm$  0.09 \\ 
$n_{cold}(\rm H_2)$ [cm$^{-3}$]  &   4.10 $\pm$  0.00  &   4.10 $\pm$  0.35  &   4.10 $\pm$  0.49 \\ 
$T_{cold}$ [K]  &  60.00 $\pm$  0.00  &  60.00 $\pm$  5.17  &  60.00 $\pm$  7.12 \\ 
$N_{cold}$ [cm$^{-2}$]  &  19.00 $\pm$  0.00  &  16.20 $\pm$  0.53  &  15.40 $\pm$  1.01 \\ 
 &  &  &  \\ 
$\Phi_{warm}(^{12}{\rm C})$  &   0.40 $\pm$  0.00  &   0.20 $\pm$  0.02  &   0.20 $\pm$  0.02 \\ 
$n_{warm}(\rm H_2)$ [cm$^{-3}$]  &   5.80 $\pm$  0.00  &   5.80 $\pm$  0.39  &   5.80 $\pm$  0.54 \\ 
$T_{warm}$ [K]  &  80.00 $\pm$  0.00  &  80.00 $\pm$  8.61  &  80.00 $\pm$  8.79 \\ 
$N_{warm}$ [cm$^{-2}$]  &  18.90 $\pm$  0.00  &  15.10 $\pm$  1.21  &  15.00 $\pm$  0.56 \\ 
 &  &  &  \\ 
$\Phi_{cold}(^{13}{\rm C})$  &   1.00 $\pm$  0.00  &   0.80 $\pm$  0.08  &   0.80 $\pm$  0.09 \\ 
$\Phi_{warm}(^{13}{\rm C})$  &   0.40 $\pm$  0.00  &   0.20 $\pm$  0.02  &   0.20 $\pm$  0.02 \\ 
 &  &  &  \\ 
%$^{12}$C/$^{13}$C  &  50.00 $\pm$  0.00  &  50.00 $\pm$  5.17  &  50.00 $\pm$  5.22 \\ 
% &  &  &  \\ 
$\Delta V(^{12}{\rm C})$ [km s$^-1$]  &   6.90   &   6.10   &   7.10  \\ 
$\Delta V(^{13}{\rm C})$ [km s$^-1$]  &   5.50   &   6.10   &   4.40  \\  
		
            \noalign{\smallskip}
            \hline
         \end{tabular}
         
         \tablefoot{
         \tablefoottext{a}{The density and column density values are given in $log_{10}$ scale.}
         }         

   \end{table}
%__________________________________________________ One column table

The highest gas temperature of all selected positions was that of the
warm component of the CO-peak, $\sim$240~K, which is in agreement with 
temperatures previously estimated toward the PDR interface from high- and 
mid-$J$ \twco\ lines \citep[e.g.][]{harris87, stutzki88}. Although the
previous studies used larger beam sizes, probably collecting 
emission from colder gas, their estimates lead to a range of likely 
temperatures between 100~K and 500~K, and densities of few times $10^4~\3cm$.
Instead the warm component of our two-phase models require between one and two 
orders of magnitude higher densities. This discrepancy in densities is 
explained because the previous estimates were based on line ratios of the
\twco~$J=7\to6$ and $J=14\to13$ transitions, and line ratios of only two
transitions are subject to the know dichotomy between density and temperature
of the gas. Our estimates, instead, are constrained in great degree by the
mid- and high-$J$ \thco, \hthcn, and \hthcop\ isotopologue lines, which
require high ($>10^5~\3cm$) densities to reproduce the observations.
Lower densities ($<10^4~\3cm$) and higher temperatures ($>250~K$) produce
low excitation temperatures of the $^{13}$C bearing molecules and, therefore,
low line intensities. This demonstrate the importance of the isotopologue
lines to constrain the excitation conditions of molecular lines.

%---------------------------------------------------------------
%---------------------------------------------------------------

The densities and temperatures derived from our two-phase model for CO also 
fit the excitation conditions for \hcn\ and \hcop, but require lower filling factors 
and molecular column densities. While two components are clearly needed to model the 
excitation towards the selected positions, relaxing the constraint on the 
isotope ratio a single component would also fit the \hcn\ and \hcop\ LSED towards the 
HCN-peak, as well as towards the PDR. These alternative solutions are presented in 
Appendix~\ref{apx:one-comp}. The \hcn\ and \hcop\ LSEDs towards the HCN-peak can
be fit using the ambient conditions of the warm component of the CO LSEDs, and
using the same isotope ratio of 50. However, a lower ratio of 30 and 20 would lead to a
better fit of the \hthcn\ and the \hthcop\ lines, respectively. The LSEDs towards the PDR, 
instead, can be fit using only the ambient conditions of the cold component of the CO, but it
requires a higher isotope ratio of 75 to fit both the \hthcn\ and \hthcop\ lines.

%Fig.~\ref{fig:pos1-SED-fit-WarmComp} and the model parameters are reported in 
%Table~\ref{tab:pos1-LSED-fit-WarmComp}.
 
The need for just a cold component in the \hcn\ and \hcop\ LSEDs 
deep in the PDR may be plausible since the \hcn~$J=8\to7$ and \hcop~$J=9\to8$ are fainter 
at this position, indicating the predominance of gas that is neither warm nor dense enough 
to excite these transitions. Since the gas dominating the excitation of \hcn\ and \hcop\ 
is relatively colder in the PDR, a larger isotope ratio in the \hcn\ and \hcop\ lines could 
be expected \citep[e.g.][]{smith80, langer84}. Likewise, the warm component of the CO LSED
at the M17-W position also needs a larger isotope ratio in order to better fit the upper
limit of the \thco~$J=13\to12$ flux.

%__________________________________________________ One column table
   \begin{table}[!tp]
      \caption[]{Mass and size associated with the cold and warm \twco\ components.}
         \label{tab:clump-mass-size}
         \centering
         %\scriptsize
         \small
         %\footnotesize
         \setlength{\tabcolsep}{3.5pt} % Default value: 6pt
         \renewcommand{\arraystretch}{1.0} % Default value: 1
         \begin{tabular}{lccccc}
            \hline\hline
	    \noalign{\smallskip}
	                 &  & \multicolumn{2}{ c }{Mass} & \multicolumn{2}{ c }{Diameter} \\
            Position\tablefootmark{a} & Offset &  Cold    & Warm    & Cold  & Warm \\
                     &  & [\Msun] & [\Msun] & [pc]  & [pc]  \\
            \noalign{\smallskip}
            \hline
            \noalign{\smallskip}

$200\arcsec$ average  &  $(-100\arcsec,0\arcsec)$  &  2.6$\times$10$^4$  &  102.2  &  2.0  &  5.1$\times$10$^{-3}$ \\ 
\hcn-peak  &  $(-65\arcsec,+31\arcsec)$  &  1.0$\times$10$^2$  &  11.2  &  1.0  &  1.0$\times$10$^{-2}$ \\ 
CO-peak   &  $(-40\arcsec,+18\arcsec)$  &  4.0$\times$10$^1$  &  4.0  &  0.4  &  2.6$\times$10$^{-2}$ \\ 
PDR  &  $(-60\arcsec,-30\arcsec)$  &  1.0$\times$10$^2$  &  10.1  &  0.2  &  2.6$\times$10$^{-2}$ \\ 
M17-W  &  $(-130\arcsec,+30\arcsec)$  &  1.3$\times$10$^2$  &  40.3  &  3.2  &  5.1$\times$10$^{-2}$ \\ 
		
            \noalign{\smallskip}
            \hline
         \end{tabular}
         
         \tablefoot{
         \tablefoottext{a}{Selected positions are in $25''$ beams.}
         }                  

   \end{table}
%__________________________________________________ One column table

\section{Implications}

Based on our accurate estimates of the densities and column densities, 
we can further characterize the physical state of the cloud in terms 
of its parameters and energy balance.

\subsection{Mass and size traced by \twco}

The gas mass of the cloudlets associated with each component of the model can be estimated from the beam averaged 
column density as,
\begin{equation}\label{eq:CO-mass}
M_{i} \approx 1.4 m_{\rm H_2} \frac{ N_{i} }{X_{\rm CO}} A_{beam} \Phi_{i}
\end{equation}
\noindent
where $A_{beam}$ is the area (in cm$^2$) of the emitting region subtended by the source size 
(considered equivalent to the beam size of 25$''$), $\Phi_{i}$ and $N_{i}$ are the beam area filling 
factors and column densities of the two components, and the factor 1.4 multiplying the molecular 
hydrogen mass $m_{\rm H_2}$ accounts for helium and other heavy elements.
We assumed a value $X_{\rm CO}=8\times10^{-5}$ for the [\twco]/[\hh] fractional abundance 
\citep[e.g.][]{frerking82, stutzki90}, although the average value found in warm star-forming 
regions may be larger (e.g., 2.7$\times$10$^{-4}$, as measured in NGC~2024 by \citealt{lacy94}).

The diameter of the cloudlets ($D_{cl}$) associated with 
each component of the models can be estimated based on 
the density and column density as 
$D_{cl} \approx \frac{ N_{i} }{ n({\rm H_2}) X_{\rm CO} }$.
The mass and size estimated for each position, and from 
the average spectra obtained from 
the 200$''$ region, are shown in Table~\ref{tab:clump-mass-size}. 
The molecular mass obtained from the average spectra is 
about 2.6$\times$10$^4$~\Msun\ for the cold component, 
while the mass associated with the warm component is 
only about 100~\Msun. The mass obtained from the cold 
component is about 80\% larger than the mass of 
1.45$\times$10$^4$~\Msun\ obtained by \citet{stutzki90} 
using \ceio\ observations, but is similar (within $\sim$15\%) 
to the mass estimated by \citet{snell84} from CS observations. 
It is also about 6 times larger than the mass we found from 
\cii\ observations \citep{pb15a}. The size of a cloudlet 
obtained from the average spectra is not scaled up in 
the 200$''$ region. It corresponds to the average size 
of a clump associated with the estimated average 
excitation conditions.

The cloudlets associated with the cold components of the \twco\ LSED at the positions of 
the \hcn-peak, CO-peak, and M17-W, have larger sizes than the median 0.11~pc 
and 0.15~pc found by \citet{hobson92} from \hcn\ and \hcop\ $J=3\to2$ observations, respectively.
They are also larger than the extent ($\sim$0.24~pc) corresponding to the $25''$ beam size 
used in our maps. This may only correspond to the depth of the gas and may indicate 
that the cloudlet may not necessarily be symmetric when compared with the size 
projected on the sky plane. 
In other words the cloudlets extend into the plane of the sky, 
not surprising given the edge-on geometry of this PDR. 
Only the cloudlet size associated with the cold component of the PDR at offset ($-60'',-30''$) 
is similar to those found by Hobson et al., and smaller than the spatial scale corresponding to 
the beam size. 
%For comparison, the cloudlets found at the first two 
%positions are between $\sim$20 and 1.8 times smaller than the $\sim$0.5~pc size of the small 
%clouds (SCs) found in some SNRs like IC443, although these SCs are expected to be clumped 
%and have small filling factors in a 45$''$ and 55$''$ FWHM beam \citep[e.g.,][]{lee12}. 
The sizes of the warm components are between two and three orders of magnitude smaller, 
which means they correspond to either a thin layer around the colder and denser cloudlets 
associated with the cold component, or they are actually embedded hot cores.

\subsection{Analysis of the energetics}\label{sec:pressure-balance}

The magnetic field along the line of sight ($B_{los}$) toward M17 was
measured by \citet{brogan01} based on their VLA \h1\ Zeeman and main-line 
OH Zeeman observations. The resolutions of their \h1\ and OH maps 
are $\sim$26$''$ and $\sim$22$''$, respectively, which allow a direct
comparison with our maps. Following their analysis we can now study the
energetics of the cloudlets associated with our selected positions in M17~SW,
using the results obtained from our two-phase models, and assuming that 
the observed magnetic field is conserved throughout the two phases
of the structures associated with our selected positions.

According to \citet{McKee93} the energy balance of a molecular cloud in 
dynamic equilibrium can be described by the virial equation:
\begin{equation}\label{eq:virial}
\mathcal{P}_s + \vert \mathcal{W} \vert = \mathcal{M}_S + \mathcal{M}_w + 2\mathcal{T}
\end{equation}
where $\mathcal{P}_s$ is the external pressure term, $\vert \mathcal{W} \vert$ is the 
gravitational energy, $\mathcal{M}_S$ is the magnetic energy associated with the static 
magnetic field ($B_S$), $\mathcal{M}_w$ is the magnetic wave energy produced 
by the fluctuating magnetic field component ($B_w$), and 2$\mathcal{T}$ is the energy 
contribution from internal motions (or turbulence).
From equating the static magnetic energy to the gravitational energy 
\citet{brogan99} derived the following critical static magnetic field ($B_{S,crit}$) 
as a diagnostic of the importance of $B_S$ in a cloud:
\begin{equation}\label{eq:B-crit}
B_{S,crit} \approx 5\times10^{-21} N_p~~{\rm [\mu G]}
\end{equation}
\noindent
where $N_p$ is the
average proton column density of the cloud, estimated as $2 \times N(\rm H_2)$, where
$N({\rm H_2})\approx N({\rm CO})/X_{{\rm CO}}$ from our two-phase models. 
This is equivalent to the magnetic critical mass ($M_{\phi}$) used by many authors in the
literature \citep[cf.,][]{mouschovias76}.
We show the $B_{S,crit}$ estimated for the four selected positions in 
Table~\ref{tab:pressure-balance}.
If the actual static magnetic field $B_S>B_{S,crit}$ then the cloud can be completely 
supported by $B_S$ (is magnetically subcritical) and further evolution of the 
cloud perpendicular to the field should occur mainly due to ambipolar diffusion. 
If $B_S<B_{S,crit}$, then $B_S$ cannot fully support the cloud (the cloud is magnetically 
supercritical) and internal motions must supply additional support for the cloud to be stable.

%__________________________________________________ One column table
   \begin{table*}[!tp]
      \caption[]{Critical static magnetic field and pressure terms.}
         \label{tab:pressure-balance}
         \centering
         %\scriptsize
         \small
         %\footnotesize
         \setlength{\tabcolsep}{3.5pt} % Default value: 6pt
         \renewcommand{\arraystretch}{1.0} % Default value: 1
         \begin{tabular}{lccccccccccc}
            \hline\hline
	    \noalign{\smallskip}
	                 & \multicolumn{2}{ c }{$B_{S,crit}$} & \multicolumn{1}{ c }{$R_{\rm H}$\tablefootmark{b}} & \multicolumn{1}{ c }{$P_{star}$\tablefootmark{c}} & \multicolumn{1}{ c }{$P_r$} & \multicolumn{2}{ c }{Turbulent pressure\tablefootmark{d}} & \multicolumn{2}{ c }{Thermal pressure} & \multicolumn{2}{ c }{Total internal pressure\tablefootmark{e}} \\
            Position\tablefootmark{a} &  Cold    & Warm &  &  & & Cold  & Warm  & Cold & Warm  & Cold & Warm   \\
                     &  [$\mu$G] & [$\mu$G] & [pc] & [K~cm$^{-3}$] & [K~cm$^{-3}$] & [K~cm$^{-3}$]  & [K~cm$^{-3}$] & [K~cm$^{-3}$]  & [K~cm$^{-3}$] & [K~cm$^{-3}$]  & [K~cm$^{-3}$] \\
            \noalign{\smallskip}
            \hline
            \noalign{\smallskip}

%$200\arcsec$ average  &  3953  &   157  &  0.5  &  4.6$\times$10$^7$  &  1.8$\times$10$^8$  &  6.4$\times$10$^7$  &  1.0$\times$10$^9$  &  2.7$\times$10$^6$  &  1.4$\times$10$^8$  &  2.5$\times$10$^8$  &  1.3$\times$10$^9$ \\ 
HCN-peak  &   993  &   314  &  1.06  &  1.0$\times$10$^7$  &  9.1$\times$10$^6$  &  1.1$\times$10$^7$  &  3.3$\times$10$^8$  &  1.3$\times$10$^6$  &  1.3$\times$10$^8$  &  2.1$\times$10$^7$  &  4.7$\times$10$^8$ \\ 
CO-peak  &   789  &   395  &  0.91  &  1.4$\times$10$^7$  &  9.1$\times$10$^6$  &  2.1$\times$10$^7$  &  1.6$\times$10$^8$  &  5.7$\times$10$^6$  &  1.2$\times$10$^8$  &  3.6$\times$10$^7$  &  2.9$\times$10$^8$ \\ 
PDR  &  1114  &   498  &  1.34  &  6.5$\times$10$^6$  &  9.1$\times$10$^6$  &  5.8$\times$10$^7$  &  1.8$\times$10$^8$  &  1.2$\times$10$^7$  &  6.9$\times$10$^7$  &  7.9$\times$10$^7$  &  2.6$\times$10$^8$ \\ 
M17-W  &  1250  &   993  &  1.58  &  4.6$\times$10$^6$  &  2.5$\times$10$^6$  &  9.4$\times$10$^6$  &  4.7$\times$10$^8$  &  7.6$\times$10$^5$  &  5.0$\times$10$^7$  &  1.3$\times$10$^7$  &  5.2$\times$10$^8$ \\ 
		
            \noalign{\smallskip}
            \hline
         \end{tabular}
         
         \tablefoot{
         \tablefoottext{a}{Selected positions are in $25''$ beams.}
         \tablefoottext{b}{Sky projected distance of the selected positions
         from the cluster of ionizing stars at about offset position $(+35\arcsec,+75\arcsec)$.}
         \tablefoottext{c}{To obtain the pressure terms in units of energy density erg cm$^{-3}$
         you must multiply them by the Boltzmann's constant.}         
         \tablefoottext{d}{The turbulent pressure is calculated from the CO 
         line width corrected for thermal broadening of the CO molecule.}
         \tablefoottext{e}{The total internal pressure is the sum of the internal radiation pressure 
         $P_r$, the turbulent pressure $P_{NT}$, and the thermal pressure $P_{th}$.}         
         %\tablefoottext{b}{At offset position $(-100\arcsec,0\arcsec)$.}         
         %\tablefoottext{b}{At offset position $(-65\arcsec,+31\arcsec)$.}
         %\tablefoottext{c}{At offset position $(-40\arcsec,+18\arcsec)$.}         
         %\tablefoottext{d}{At offset position $(-60\arcsec,-30\arcsec)$.}         
         }                  

   \end{table*}
%__________________________________________________ One column table

The average OH $B_{los}$ detected toward the northern condensation of \hho\ masers, where the
HCN and CO peaks are found, is $-300\pm80$~$\mu$G, while the $B_{los}$ detected in the proximities
of the PDR at offset position ($-60''$,$-30''$) is $-500\pm80$~$\mu$G
\citep[see B-20UC1 and B-20HI in][their Sects.~3.4.6, 3.4.7, and Fig.16]{brogan01}.
There is no quoted measurement of the magnetic field for the M17-W position ($-130''$,$+30''$),
but from Fig.16 by \citet{brogan01} we can infer a lower limit of 
$B_{los}=-300\pm80$~$\mu$G as well, from their HI Zeeman observations.

%Here we use these average values in the analysis of our four selected positions. 
From a statistical view point the measured $B_{los}$ is actually the average over a large 
number of clouds/clumps with magnetic fields randomly oriented with respect to the line of sight.
Hence, it is related to the total magnetic field ($\vert B \vert$) as $B_{los}\approx \vert B \vert /2$ 
\citep[for a more detailed description see][Sect.~5]{crutcher99}. Thus, assuming that the total
magnetic field is the static magnetic field, we would have that $B_S\approx-600\pm160$~$\mu$G
toward the northern condensation (including both the CO- and HCN-peaks) 
and the M17-W position, and about -1000$\pm$160~$\mu$G toward the PDR. 
These values are larger than the $B_{S,crit}$ estimated for the warm component at 
all selected positions, except for the M17-W position. The cloudlets for which 
$B_S>B_{S,crit}$ can be fully supported from collapse against self-gravity (if no other forces are at play). 
The estimated total magnetic fields are comparable to the critical static magnetic field 
estimated for the cold component 
of the CO-peak and the PDR, but the $B_S$ in the northern condensation is about 30\% 
and 25\% smaller than the $B_{S,crit}$ estimated for the HCN-peak and the M17-W position, 
respectively. This means the structures 
associated with the PDR and the CO-peak can be magnetically supported, but the 
magnetic field cannot support the cloudlets at the HCN-peak and M17-W positions. 
In this case the evolution of the cloudlets at these positions should be controlled 
in part by processes that create and dissipate internal motions. Note that
if $\vert B \vert$ is aligned preferentially along the LOS, 
then we would have lower fields by a factor two.

%However, the $B_{los}$ measured toward 
%the northern condensation might be a lower limit, due to the difficulties found to fit 
%the OH absorption line at $\sim$20~\kms\ due to contamination by the OH maser at 
%$\sim$21.4~\kms\ \citep{brogan01}.

\subsubsection{External and internal pressure}\label{sec:pressure}

The interactions of the forces acting on the cloudlets can be compared in terms of the
pressure they exert. 
We first estimate the external pressure produced mainly by the radiation field 
from the most massive O and early B ionizing stars in the NGC~6618 cluster. 
Based on radio observations \citet{felli84} derived a total luminosity by number 
of H-ionizing photons $Q(\rm H)=2.9\times10^{50}$~s$^{-1}$ considering a distance to M17 of 
2.2~kpc. On the other hand, \citet[][their Table~1]{pellegrini07} estimated a 
$Q(\rm H)=1.5\times10^{50}$~s$^{-1}$
for their adopted distance of 1.6~kpc, from the luminosity of ionizing photons per star 
reported by \citet{hanson97}. For the following we have adjusted these values to the 
more recent measurement of 1.98~kpc
for the distance to M17 \citep{xu11} that we consider more accurate. A linear
interpolation between the values reported by \citet{pellegrini07} and \citet{felli84} 
give us the number of ionizing photons emitted by the
star cluster per second $Q(\rm H^0)=2.39\times10^{50}$~s$^{-1}$ at a distance of 1.98~kpc.

Following the analysis by \citet[][their Eq.3]{pellegrini07}, and assuming that 
most of the momentum in the stellar radiation is in the ionizing photons, 
we can estimate the pressure from the stars as 
$P_{star}=Q(H^0)\langle h\nu \rangle / 4\pi R_{\rm H}^2 c k$, where $\langle h\nu \rangle$
is the mean photon energy of an O star, which we assume to be $\sim$15~eV as in Pellegrini et al., 
$R_{\rm H}$ the distance from the star cluster to the position in the cloud where
we want to estimate the radiation pressure, $c$ is the speed of light, and $k$ is 
Boltzmann's constant. For the position of the star cluster we assume
that of the CEN1a and CEN1b stars (see Sect.~\ref{sec:introduction}), 
at an offset of about ($+35''$,$+75''$) (cf., Fig.~\ref{fig:HCN-CO-overlay}). 
The sky projected distances from this stars to the
selected positions in M17~SW, as well as the corresponding stellar 
radiation pressure $P_{star}$,
are shown in Table~\ref{tab:pressure-balance}.

Assuming that the (star-forming) clumps are opaque to their own radiation, 
the internal radiation pressure ($P_r$) due to the dust continuum emission 
(assuming the dust is well mixed in the two gas components) can be estimated as 
$P_r=\tau_{{\rm 100\mu m}} \sigma T_d^4/c k$, where $\tau_{100\mu m}$ the continuum optical depth 
at 100~\mum, $\sigma$ is the Stefan-Boltzmann's constant, and $T_d$ is the dust temperature.
From the maps of dust temperature and optical depth from \citet[][their Fig.~6]{meixner92}, 
we adopt $T_d=50$~K and $\tau_{\rm 100\mu m}=0.106$ for the HCN-peak, CO-peak and PDR positions, 
and $T_d=40$~K and $\tau_{\rm 100\mu m}=0.071$ at the M17-W position. 
From the values presented in Table~\ref{tab:pressure-balance}, we find 
that the internal radiation pressure $P_r$ is lower than the external 
stellar pressure $P_{star}$ at all the selected positions, except at the PDR. 
If these were the only forces acting on the cloudlets, they would collapse 
due to the external radiation pressure and self-gravity, unless the 
magnetic field and other forces support them.

Other sources of internal pressure important to take into account are the internal motions.
We consider two sources of internal motions: turbulence and the kinetic 
temperature. These two sources of motions are responsible for the CO line widths 
in individual cloudlets. However, it is well known from the literature that 
turbulence (that can have several origins) dominates the broadening of the CO lines
in warm molecular clouds. 
In our case, the average line width (FWHM) of the cloudlets we observe at the four selected 
positions is 4.5~\kms, of which the highest kinetic temperature of 240~K, obtained 
for the warm component of the CO LSED observed at the position of the \hcn-peak, 
contributes with only $\sim$0.63~\kms\ (i.e., about 14\%) when 
considering the thermal velocity contributing to the line widths.

The pressure exerted by the thermal and turbulent motions can be estimated as
$P_{th}=n({\rm H_2})T$~K~cm$^{-3}$ and 
$P_{NT}=n({\rm H_2})m_{{\rm H_2}}(\sigma_{NT}^2)/k$~K~cm$^{-3}$, where $T$ is the 
kinetic temperature, $k$ is the Boltzmann's constant, and $\sigma_{NT}$ is the 
non-thermal component of the one dimensional velocity dispersion of an individual cloudlet.
Other authors use the total velocity dispersion (assumed equal to the line width) 
to estimate the total internal pressure \citep[cf.,][]{blitz91}. However, 
we prefer to separate these terms, since many other authors use the thermal 
pressure (alone) in their analysis, and to compare their relative 
contributions in the two components of our models. 
The estimates of these terms are presented in Table~\ref{tab:pressure-balance}.
For the selected positions, $P_{NT}/P_{th}$ ranges between 4 and 12 for the cold
component, and between 1.4 and 9 in the warm component. Whether thermal or turbulent,
is clear that the internal pressure in the warm component is higher than in the cold 
component. This is also reflected in the total internal pressure estimated for the
two phases (cf., Table~\ref{tab:pressure-balance}). Note as well that the turbulent
pressure is comparable (within a factor two) to the external stellar 
radiation pressure $P_{star}$ toward all selected positions, except at the southern PDR.

At first look, the total internal pressure of these cloudlets seem quite high. 
In fact, it is about four orders of magnitude higher than the median pressure 
$\sim$3000~Kcm$^{-3}$ estimated toward a number of molecular clouds with 
densities between a few hundreds and 10$^3~\3cm$, and temperatures of $\sim$50~K
\citep[e.g.,][]{crutcher99, jenkins01, wolfire03, heiles04, heiles05, troland08}.
Most of these studies were done based on observations of \h1\ in absorption and
other tracers of the diffuse gas. However, higher gas densities of the order of 
10$^5~\3cm$ and 10$^6~\3cm$, but with lower temperatures ($T_k<20$~K) have also 
been inferred from CN Zeeman observations \citep[][]{crutcher96, crutcher99L, falgarone08}.
These higher densities and temperatures lead to an upper limit for the thermal 
pressure of $P_{th}<2\times10^7$~K~cm$^{-3}$, which is similar to the values
we found in M17~SW.
It is important then to explore whether the high internal pressure estimated 
toward our selected positions can be supported by the ambient magnetic field 
and by self-gravity.

The magnetic pressure can be estimated from the magnetic energy density as 
$P_M/k=\vert B \vert^2/(8\pi k)$~K~cm$^{-3}$, where $\vert B \vert^2\approx3 B_{los}^2$
\citep[see][Sect.~5, for a detailed description of this approximation]{crutcher99}.
Using the same average values of $B_{los}$ as above, we find that the magnetic pressure 
is $\sim$8$\times$10$^7$~K~cm$^{-3}$ in the northern condensation of \hho\ masers, and
$\sim$2$\times$10$^8$~K~cm$^{-3}$ in the proximities of the PDR position. 
Thus, the magnetic pressure is larger than the total internal pressure of the cold
component in all the selected positions, but it would be lower than the internal pressure
of the warm components by factors $\sim$6, $\sim$4, $\sim$1.3 and $\sim$2.6 toward the 
HCN-peak, the CO-peak, the PDR, and the M17-W position, respectively. This mean that, from the 
pressure balance point of view, the cold components are magnetically supported 
in all the positions, but the warm components are not, if the magnetic
field remains the same throughout all the gas phases. 

This seems to contradict the results we found above based on the critical static magnetic 
field criteria, where two of the warm components were classified as subcritical and fully 
supported by the magnetic field. This likely means that the assumption of equipartition 
between the static magnetic energy and the gravitational energy used to estimate 
$B_{S,crit}$ by \citet{brogan99} does not hold true in our case. In fact, the 
equipartition assumption $\vert \mathcal{W} \vert = \mathcal{M}_S$ does not take 
into account the effects of the thermal and turbulent pressure, which indeed 
play a significant role in the energy balance of dense and warm gas, as we have 
just shown. Since the turbulent pressure dominates the total internal pressure at 
all the selected positions, it is relevant to investigate what causes non-thermal 
turbulence in the cloudlets.

\subsubsection{Origin of the turbulence}\label{sec:mach-numbers}

%__________________________________________________ One column table
   \begin{table}[!tp]
      \caption[]{Sonic and Alf\'enic Mach numbers.}
         \label{tab:mach-numbers}
         \centering
         %\scriptsize
         \small
         %\footnotesize
         \setlength{\tabcolsep}{3.5pt} % Default value: 6pt
         \renewcommand{\arraystretch}{1.0} % Default value: 1
         \begin{tabular}{lcccccc}
            \hline\hline
	    \noalign{\smallskip}
	                 & \multicolumn{2}{ c }{$m_s$} & \multicolumn{2}{ c }{$m_A$} & \multicolumn{2}{ c }{$\beta_p$} \\
            Position &  Cold & Warm   &  Cold & Warm   &  Cold & Warm   \\
            \noalign{\smallskip}
            \hline
            \noalign{\smallskip}

HCN-peak  &  8.991  &  4.987  &  0.770  &  4.327  &  0.015  &  1.506 \\ 
CO-peak  &  5.994  &  3.671  &  0.652  &  1.838  &  0.024  &  0.502 \\ 
PDR  &  6.862  &  5.068  &  1.807  &  3.213  &  0.139  &  0.804 \\ 
M17-W  &  11.012  &  9.536  &  0.728  &  5.156  &  0.009  &  0.585 \\ 
		
            \noalign{\smallskip}
            \hline
         \end{tabular}

   \end{table}
%__________________________________________________ One column table

We explore two possible mechanisms that can cause non-thermal motions. 
%First, virialization. If the mass of the cloudlets is high enough to produce 
%collapse by gravity, it should show in the one-dimensional velocity dispersion.
First, the presence of outflows. 
The \hcn\ lines (and in a lower degree the \hcop), show 
asymmetries in the line profiles that are not present in the \hthcn\ and 
\hthcop\ isotope lines, at least at the S/N level of the isotope lines. 
In particular the $J_{up}\geq3$ lines at the \hcn-peak 
(top panel in Figs.~\ref{fig:HCN-spectra-3positions} and 
\ref{fig:HCOp-spectra-3positions}) show a top-flat shape, while most of the 
\hcn\ and \hcop\ lines in the other positions show an asymmetric line 
profile with a red-shifted peak stronger than the blue-shifted emission.
At first look this line profiles are characteristic of outflowing motions. 
However, we know the \hcn\ lines are optically thick and can be affected by 
self-absorption. Besides, due to the highly clumpy structure of M17~SW the 
asymmetric line profiles could be due to different cloudlets overlapping
in velocity along the line of sight. However, the current S/N of the \hthcn\ 
and \hthcop\ lines is not high enough to confirm this. Outflowing motions,
on the other hand, need to be confirmed by detection of additional molecular 
tracers of this process (e.g., SiO, SO, CH$_3$CN, HNCO) at the selected positions.

Due to the relatively strong magnetic field measured toward M17~SW, a second 
possible source of turbulence is Alfv\'enic waves or MHD waves generated in a 
turbulent velocity field.
%Although it has been stated that the time dependent magnetic field can be appreciated
%at large scales in molecular clouds but it has no effect on "cores" \citep{bertoldi96}, 
%the significantly strong magnetic field observed toward M17~SW and the dominance
%of turbulent pressure, rise the question about what fraction of the observed magnetic 
%field is actually due to the magnetic wave $B_w$.
To explore this alternative we estimate and compare the sonic and Alfv\'enic numbers.
The sonic Mach number is $m_s=\sqrt{3} \sigma /c_s$, where $\sigma$ is the observed 
one-dimensional velocity dispersion ($\sigma=\Delta V/ \sqrt{8 ln 2}$, and 
$\Delta V$ the FWHM line width), and $c_s=(k T_k/\mu)^{1/2}$ is the isothermal 
sound speed (with $\mu=2.33 \rm m_H$ the mean particle mass for a region of 
molecular hydrogen including 10\% helium).
The Alfv\'enic Mach number is $m_A=\sqrt{3} \sigma / V_A$, where 
$V_A=\vert B \vert / (4 \pi \rho)^{1/2}$ is the Alfv\'en speed, 
$\vert B \vert \approx 2 B_{los}$ the strength of the magnetic field, 
and $\rho=1.4 m_{\rm H_2} n({\rm H_2})$ is the gas density 
accounting for helium and other heavy elements.
From these Mach numbers we can also estimate the ratio of thermal 
to magnetic pressures, $\beta_p=2(m_A/m_s)^2=2(c_s/V_A)^2$, which
is a crucial parameter in theory or supercomputer simulations 
of the structure and evolution of magnetic clouds. The estimates of these
parameters for the cold and warm components are shown in Table~\ref{tab:mach-numbers}.

%__________________________________________________ One column table
   \begin{table*}[!tp]
      \caption[]{Energy terms of the Virial equation.}
         \label{tab:energy-balance}
         \centering
         %\scriptsize
         \small
         %\footnotesize
         \setlength{\tabcolsep}{3.5pt} % Default value: 6pt
         \renewcommand{\arraystretch}{1.0} % Default value: 1
         \begin{tabular}{lcccccccccc}
            \hline\hline
	    \noalign{\smallskip}
   
	                 & \multicolumn{2}{ c }{$P_s$} & \multicolumn{2}{ c }{$\mathcal{M}_S$} & \multicolumn{2}{ c }{$\mathcal{M}_w$} & \multicolumn{2}{ c }{$2\mathcal{T}$} & \multicolumn{2}{ c }{$\vert W \vert$} \\
            Position\tablefootmark{a} &  Cold    & Warm    & Cold  & Warm  & Cold & Warm  & Cold & Warm & Cold & Warm  \\
                     &  [ergs] & [ergs] & [ergs]  & [ergs] & [ergs]  & [ergs] & [ergs]  & [ergs] & [ergs]  & [ergs] \\
            \noalign{\smallskip}
            \hline
            \noalign{\smallskip}

HCN-peak  &  1.3$\times$10$^{47}$  &  2.3$\times$10$^{45}$  &  1.0$\times$10$^{47}$  &  1.0$\times$10$^{41}$  &  1.1$\times$10$^{46}$  &  1.3$\times$10$^{45}$  &  2.3$\times$10$^{46}$  &  2.5$\times$10$^{45}$  &  1.2$\times$10$^{45}$  &  1.5$\times$10$^{45}$ \\ 
CO-peak  &  2.0$\times$10$^{46}$  &  1.3$\times$10$^{45}$  &  6.6$\times$10$^{45}$  &  1.7$\times$10$^{42}$  &  4.6$\times$10$^{45}$  &  4.6$\times$10$^{44}$  &  9.1$\times$10$^{45}$  &  9.1$\times$10$^{44}$  &  4.9$\times$10$^{44}$  &  7.8$\times$10$^{43}$ \\ 
PDR  &  2.4$\times$10$^{46}$  &  2.5$\times$10$^{45}$  &  1.6$\times$10$^{45}$  &  4.6$\times$10$^{42}$  &  1.0$\times$10$^{46}$  &  1.0$\times$10$^{45}$  &  2.0$\times$10$^{46}$  &  2.0$\times$10$^{45}$  &  7.1$\times$10$^{45}$  &  4.9$\times$10$^{44}$ \\ 
M17-W  &  3.4$\times$10$^{48}$  &  2.7$\times$10$^{46}$  &  3.3$\times$10$^{48}$  &  1.3$\times$10$^{43}$  &  3.2$\times$10$^{46}$  &  1.0$\times$10$^{46}$  &  6.5$\times$10$^{46}$  &  2.1$\times$10$^{46}$  &  6.2$\times$10$^{44}$  &  3.9$\times$10$^{45}$ \\ 
		
            \noalign{\smallskip}
            \hline
         \end{tabular}
         
         \tablefoot{
         \tablefoottext{a}{Selected positions are in $25''$ beams.}
         %\tablefoottext{b}{At offset position $(-100\arcsec,0\arcsec)$.}         
         %\tablefoottext{b}{At offset position $(-65\arcsec,+31\arcsec)$.}
         %\tablefoottext{c}{At offset position $(-40\arcsec,+18\arcsec)$.}         
         %\tablefoottext{d}{At offset position $(-60\arcsec,-30\arcsec)$.}         
         }                  

   \end{table*}
%__________________________________________________ One column table

The internal motions at all the selected positions are supersonic ($m_s>1$)
by factors between 4 and 10 in the warm components, and by factors
between 6 and 11 in the cold components. Instead the magnetic waves 
are sub-Alfv\'enic ($m_A<1$) in the cold components toward all the positions,
except at the PDR where it is super-Alfv\'enic by about factor two. 
The warm components are all super-Alfv\'enic by factors between 2 and 5.
The Mach numbers obtained for the cold components are similar to those
obtained by \citet{crutcher99} toward M17~SW and other molecular clouds.
Supersonic motions would be attributed to hydrodynamic turbulence, 
but Alfv\'enic Mach numbers $m_A\lesssim 1$ in the cold components 
of three positions, indicates that motions are Alfv\'en waves 
(or MHD turbulence). In fact, the plasma parameter $\beta_p<1$ in both cold
and warm components toward most of the positions indicates that magnetic 
pressure dominates thermal pressure. Only toward the HCN-peak seems the 
warm component dominated by thermal pressure.
These results agree with the previous estimates of the magnetic pressure 
$P_M/k$ obtained from the average magnetic field measured toward the selected 
positions compared to the explicitly estimated thermal pressure $P_{th}$
(see Sect.~\ref{sec:pressure} and Table~\ref{tab:pressure-balance}).

What remains to be determined then is whether the net contributions of 
$\mathcal{M}_S$, $\mathcal{M}_w$ and $\mathcal{T}$ can support the cloudlets at 
the selected positions against gravity $\mathcal{W}$.

\subsubsection{Energy balance in the cloudlets}\label{sec:energy-balance}

The virial terms in Eq.(\ref{eq:virial}) are 
$\mathcal{T}=(3/2)M\sigma^2$, $\mathcal{M}_S=(1/3)b \vert B \vert^2 R^3$,
and $\mathcal{W}=(3/5)aG M^2/R$, for the kinetic, magnetic, and gravitational
energy, respectively. In this case $\sigma$ is the total (i.e., thermal and non-thermal) 
one-dimensional velocity dispersion ($\sigma=\Delta v_{{\rm FWHM}}/\sqrt{8 ln(2) }$). Following \citet{crutcher99} we use the values
$a=1.2$ and $b=0.3$ to correct the virial terms for the geometry, central density 
concentration, and magnetic gradient effects (see also Tomisaka et al. 1998 for a detailed
estimate of these parameters). Because an equipartition between internal motions 
and $B_W$ is likely to be at play in M17~SW we also assume that the magnetic 
wave energy is half the kinetic energy, $\mathcal{M}_w=\mathcal{T}$  \citep{brogan99}.
Having all these terms of the virial equation allow us to estimate the term of
the external pressure from Eq.(\ref{eq:virial}), assuming dynamical equilibrium.
The values estimated for all these terms are shown in Table~\ref{tab:energy-balance}.

The static magnetic energy dominates all the energy 
terms of the cold components in most of the selected positions, while the 
gravitational energy is between one and two orders of magnitude lower than $\mathcal{M}_S$.
The static magnetic energy of the warm component is one or two orders of magnitude
lower than $\mathcal{W}$, meaning that the wave magnetic energy and internal motions
support the cloudlets associated with the warm components.
Only at the PDR position the gravity in the cold component dominates (by factor $\sim$7)
the static magnetic energy.
The kinetic energy of the cold component at the PDR is $\sim$40\% larger than the 
static magnetic energy, indicative of the internal motions playing the mayor role 
supporting the gas against gravity at this position, despite the $B_{los}$ magnetic 
field being the strongest close to the PDR.

Since we are using an average magnetic field measured towards the northern condensation
of \hho\ masers and close to the PDR, $B_S$ may actually be stronger (or weaker) toward 
some of the selected positions. 
%Note that the wave
%magnetic energy $\mathcal{M}_w$ is one order of magnitude larger than the static magnetic
%energy in the cold component of the HCN peak, opposite to what we obtained for the other 
%selected positions. In order to support the effect of gravity $B_S$ can be estimated
%from $\mathcal{M}_S\approx \mathcal{W} - 2 \mathcal{T}$ (the two dominant virial terms
%for the HCN peak), which gives a static magnetic 
%field of $\sim$2700~$\mu$G, or a $B_{los}\approx1600$~$\mu$G, that is, about a factor three
%times larger than the average $B_{los}$ used in our estimates. 
We explore scenarios where the magnetic field can be weaker or stronger than considered here.

\subsubsection{Alternative scenarios}\label{sec:alternative-scenarios}

One possibility that the actual magnetic
field is weaker than we have estimated, is if the field lines are mostly
aligned along the line of sight and just a relatively weak field is parallel to
the plane of the sky. If this holds true the total magnetic field would (approximately)
be the observed $B_{los}$, and the magnetic pressure and the static magnetic energy 
would be a factor 3 lower than estimated. In this case the magnetic energy would 
still be sufficient to support the gas in the cold components against collapse for 
most of the selected positions, except for that in the southern PDR.

On the other hand, there are two arguments favoring stronger magnetic fields 
than we estimated.
First, there is evidence for increasing strength
(between factor 2 and 4) in the magnetic field in small scale structures, 
as observed with higher resolution ($22''$) with respect to earlier Zeeman 
observations done with coarser resolutions ($60''$) \citep{brogan01}. 
Second, and related with the previous argument, the maximum magnetic field 
strength in a cloud is expected to scale 
with density as $B \approx B_0(n/n_0)^{2/3}$ for densities $n>300~\3cm$, 
where $n_0=300~\3cm$ for diffuse clouds and $B_0$ is the magnetic field
at those densities \citep{crutcher10}. 
In our case we have that the OH observations toward the northern \hho\ 
condensation arise from gas with densities $\sim4\times10^4~\3cm$ \citep{brogan01}.
Hence, assuming $B_0=B_{los}=-300$~$\mu$G and $n_0$ the density for the OH bearing gas, 
the magnetic field strength in the cold component of the, e.g., CO-peak,
with density $n\sim6.3\times10^4~\3cm$ (cf., Table~\ref{tab:pos1-LSED-fit}), 
would be $B\approx-400~\mu$G. The total magnetic field at this position 
would then have a strength of about $-800~\mu$G (assuming $B\approx2B_{los}$ as above).
%If this was the case, a density $n\sim3\times10^4~\3cm$ 
%for the cold component at the HCN peak (cf., Table~\ref{tab:pos1-LSED-fit}) 
%would imply $B\approx6500$~$\mu$G,
%assuming that the average measured $B_{los}=300$~$\mu$G at the northern
%condensation of \hho\ masers correspond to the total magnetic field $B_0$ 
%associated with the diffuse \h1\ halo gas reported by \citep{meixner92}.
Such a strong magnetic field would support the cloud structures in the cold 
component of the CO-peak against self-gravity (comparing with the critical
static magnetic field $B_{S,crit}$), but not at the HCN-peak nor at the M17-W 
positions. The magnetic field would also support the cloudlets of the cold 
component at the PDR position, as well as the warm components at all selected 
positions, since their densities are larger.
The pressure exerted by the scaled up magnetic fields would match even the
high total internal pressure of the warm component 
(cf., Table~\ref{tab:pressure-balance}) at all selected positions.

Higher resolution observations, as well as high sensitivities to detect
low Stokes I and V intensities in molecular line Zeeman observations are
needed to test the hypothetically strong magnetic fields in dense and small
structures within M17~SW.

%Due to the dichotomy between temperature and density, the LSED fit we found 
%may not necessarily be unique solutions. Hence, there is the possibility of 
%finding alternative solutions with higher filling factors and lower densities
%and column densities at some positions. Thus, the observed magnetic field 
%may be strong enough to support the cloudlets against gravity in all selected 
%positions, if ambient conditions with lower densities and column densities 
%would be found.

\section{Conclusions}

We presented new velocity-resolved maps (2.0~pc~$\times$~3.1~pc) of the $J=11\rightarrow10$, $J=12\rightarrow11$ 
and $J=16\rightarrow15$ transitions of \twco, as well as the \thco~$J=13\to12$ isotopologue line, observed toward 
M17~SW using the dual-color Terahertz receiver GREAT on board of SOFIA. We also completed 
extended maps of the high density tracers \hcn\ and \hcop\ using the EMIR receivers on the IRAM 30m telescope, 
and the FLASH$^+$ and CHAMP$^+$ receivers on APEX.

Combining this new data set we obtained the most complete 
large-scale \twco, \hcn\ and \hcop\ line spectral energy 
distributions (LSEDs) reported so far on M17~SW and on any other 
star-forming region, in terms of transitions covered.

\subsection{Line intensity ratios}

We found that the \hcop/\hcn\ $J=1\to0$ line ratio is lower than 
unity in all the region mapped, while the ratios between the 
higher-$J$ lines is larger than unity. The ratios lower than 
unity in the $J=1\to0$ cold be explained with infrared pumping 
of the low-$J$ transition of \hcn, but IR observations are 
needed to check for absorption at 14.0~\mum\ wavelength of the 
first bending state of \hcn. 
The ratios larger than unity in the higher-$J$ lines cannot be 
explained with a standard PDR model. We conclude that the most 
likely mechanism that can produce the bright \hcop\ emission 
is a lower excitation temperature of \hcn, which requires larger 
critical densities than \hcop.

\subsection{Excitation conditions from LSEDs}

The excitation conditions of four selected positions in the maps were  
determined by fitting a two-phase non-LTE radiative transfer model 
to the CO, \hcn\ and \hcop\ LSEDs. 
The different LSED shapes associated with the selected 
positions are indicative of the distinct excitation conditions that 
dominate the warm component of the models needed to fit the high-$J$ CO 
lines observed with SOFIA/GREAT. The gas temperature exciting the gas in 
the cloudlet associated with the peak emission of the \twco~$J=16\to15$ line 
is $\sim$240~K, in agreement with previous estimates from the literature. 
This high temperature is also in line with the proximity of the M17 ultra-
compact \hii\ region, an \hho\ maser, and at least three deeply embedded 
X-ray sources.

The \hcn\ and \hcop\ LSEDs can be fit using the same ambient conditions 
found for the CO lines, but with lower beam filling factors and column 
densities. All LSEDs could be fit using the same isotope ratio of 50 for
the three species, in line with what would be expected for equilibrium 
chemistry of evolved gas. However, lower isotope 
ratios of 20 and 30 would lead to a better fit of the \hthcn\ and 
\hthcop\ lines. Inhomogeneities and non-equilibrium chemistry (in the 
proximity of YSOs, hot cores and UC-\hii\ regions) may explain an enhanced 
abundance of the rare isotopologues. An alternative solution using only a 
cold component, and a larger isotope ratio of 75, is also possible for the 
\hcn\ and \hcop\ lines at the southern PDR position.

The diameter obtained for the warm components indicates they may correspond 
to a thin layer around the cold cloudlets, or to an embedded hot core or 
UC-\hii\ region.

We also constrained the excitation conditions of the average LSEDs obtained over 
an area of 200~arcsec$^2$. The density and temperature found for the cold and warm
components of the average spectra are similar to those found for the \hcn-peak, but
with a larger CO column density for the cold component. The shape of the CO LSEDs, 
however, resembles that observed at the offset position of the southern PDR. Therefore,
the LSED shape and ambient conditions derived for an unresolved source (as in the case
of extragalactic sources with ambient conditions similar to M17~SW) would be dominated 
by a dense PDR like component, while the distinctive features of the high-$J$ CO lines 
obtained with SOFIA/GREAT for dense cores like the \hcn-peak and dense and warm gas 
like that found in the proximities of the UC-\hii\ region, would be smeared out.

\subsection{Energy balance and stability of the cloudlets}

We estimated the stability and energy balance of the cloudlets associated
with both components in our models, considering measurements of the magnetic 
field based on OH and HI Zeeman observations toward M17~SW, as well as the 
external radiation pressure exerted by the cluster of ionizing stars.

We found that the cloudlets associated with the cold component of the models are 
magnetically subcritical ($\mathcal{M}_S > \vert \mathcal{W} \vert$) and 
supervirial ($2 \mathcal{T}/\vert \mathcal{W} \vert > 1$) at all the
selected positions, except for the southern PDR. 
The warm cloudlets are all supercritical ($\mathcal{M}_S < \vert \mathcal{W} \vert$)
and also supervirial.

Considering only gravity, the critical magnetic field, and the static magnetic field, 
we found that the cold cloudlets associated with the southern PDR and CO-peak 
positions can be magnetically supported, but the magnetic field cannot support 
the cloudlets at the HCN-peak and M17-W positions. Thus, the evolution of these 
cloudlets is expected to be partially controlled by processes that create and
dissipate internal motions.

Turbulent pressure dominates over thermal pressure in both cold and warm components 
at all selected positions. The total internal pressure of the cold components 
is between a factor of two and three larger than the external radiation pressure 
at four selected positions, and is one order of magnitude larger 
at the southern PDR position.

The gas velocities of both components are supersonic at all selected positions 
by factors between 4 and 11. The velocities of the cold component are 
sub-Alfv\'enic at all the positions, except at the southern PDR. The velocities of the 
warm components are super-Alfv\'enic by factors between 2 and 5 at all selected 
positions. This suggests that internal motions of at least the cold components 
are due to MHD waves.
The ratio of thermal to magnetic pressures $\beta_p<1$ in both 
gas components indicates that magnetic effects dominate thermal effects at 
most of the selected positions.

Assuming that the static magnetic field remains 
constant throughout all the gas phases, the magnetic pressure can support 
the total internal pressure of the cold components in all the positions, 
but it cannot support the internal pressure of the warm components. 

If the magnetic field scales with density as $B \propto n^{2/3}$, then the actual 
magnetic field associated with the cold components of our two phase model would be 
strong enough to support the cloudlets against gravity at two of the selected positions.
The same holds true for the warm components at all selected positions.
This mean that ambipolar diffusion would dominate the evolution of these cloudlets.

\begin{acknowledgements}
We are grateful to the teams of SOFIA/GREAT, MPIfR, IRAM 30m and APEX staff members for their help and support 
during and after the observations. We thank Friedrich Wyrowski and Silvia Leurini for helpful discussions.
We thank the referee for the careful reading of the manuscript and constructive comments.
Molecular Databases that have been helpful include the NASA/JPL spectroscopy line catalog and the University of 
Leiden's LAMDA databases.
\end{acknowledgements}

\bibliographystyle{aa}
\setlength{\bibsep}{-2.1pt}
\bibliography{m17sw}

%\addcontentsline{toc}{chapter}{\textmd{APPENDIX}}
%\begin{appendix}
%\input{aa}
%\end{appendix}

\begin{appendix}

%\newpage

\section{Gaussian fit of spectral lines}

The line profiles of the observed \twco\ and \thco\ lines are the result of 
several components overlapping along the line of sight and in the velocity 
space, which are affected by self-absorption, mainly in the lower-$J$ transitions. 
The lower- and mid-$J$ \twco\ lines can be fit with several Gaussian 
components, but the higher-$J$ ($J_{up}\geq11$) can be fit with just one 
or two components. We selected the Gaussian component of the lower- and mid-$J$ 
lines based on the central velocity and line width obtained for the single 
Gaussian used to fit the \twco\ $J=16\to15$ line.

With the exception of the $J=1\to0$ transitions, the higher-$J$ \hthcn\ and 
\hthcop\ lines seem to consist of a single component that can be fit with one 
Gaussian. Because the main \hcn\ and \hcop\ line profiles show more structure 
than that of their rare isotopologues, the single component of $^{13}$C bearing lines 
is indicative of self absorption in the main lines, at least at the rms level 
of our \hthcn\ and \hthcop\ spectra. Therefore, we fit a single Gaussian component
to the main \hcn\ and \hcop\ lines masking the line centers and using the less
self-absorbed line wings, in order to reproduce the missing flux. 

The Gaussian parameters of the fit is summarized in Tables~\ref{tab:gaussian-fit-pos1}, 
\ref{tab:gaussian-fit-pos2} and \ref{tab:gaussian-fit-pos3}, for all the observed 
transitions at the four selected positions.

\newpage

%__________________________________________________ One column table
   \begin{table}[!htbp]
      \caption[]{Gaussian fit parameters for the spectra toward the HCN-peak position ($-65''$,$+31''$).}
         \label{tab:gaussian-fit-pos1}
         \centering
         \scriptsize
         %\footnotesize
         \setlength{\tabcolsep}{3.5pt} % Default value: 6pt
         \renewcommand{\arraystretch}{1.0} % Default value: 1
         \begin{tabular}{lcccc}
            \hline\hline
	    \noalign{\smallskip}
            Transition & $\int T_{mb} dV$ & $T_{mb}^{peak}$  &  $V_0$  & $\Delta V$ (FWHM) \\
                       &   [\Kkms]        &    [K]           &  [\kms]  &    [\kms]  \\
            \noalign{\smallskip}
            \hline
            \noalign{\smallskip}
            \multicolumn{5}{c} {\twco} \\
            \noalign{\smallskip}
            \hline
            \noalign{\smallskip}

$J=1\to0$ & 140.4$\pm$135.2 & 33.77$\pm$3.37 & 20.14$\pm$ 0.50 &  3.91$\pm$ 1.27\\ 
$J=2\to1$ & 175.7$\pm$ 10.6 & 42.02$\pm$2.88 & 19.65$\pm$ 0.03 &  3.93$\pm$ 0.13\\ 
$J=3\to2$ & 289.3$\pm$  1.2 & 68.02$\pm$0.37 & 19.95$\pm$ 0.01 &  4.00$\pm$ 0.01\\ 
$J=4\to3$ & 275.9$\pm$  5.0 & 64.09$\pm$1.38 & 19.57$\pm$ 0.02 &  4.05$\pm$ 0.05\\ 
$J=6\to5$ & 256.8$\pm$  0.0 & 48.46$\pm$0.00 & 20.67$\pm$ 0.00 &  4.98$\pm$ 0.00\\ 
$J=7\to6$ & 329.1$\pm$ 13.9 & 60.46$\pm$2.83 & 20.67$\pm$ 0.03 &  5.11$\pm$ 0.10\\ 
$J=11\to10$ & 158.8$\pm$ 31.5 & 31.94$\pm$7.13 & 20.79$\pm$ 0.11 &  4.67$\pm$ 0.47\\ 
$J=12\to11$ & 148.6$\pm$  8.4 & 25.99$\pm$2.11 & 19.98$\pm$ 0.15 &  5.37$\pm$ 0.31\\ 
$J=13\to12$ & 139.6$\pm$  7.7 & 25.72$\pm$2.15 & 20.68$\pm$ 0.13 &  5.10$\pm$ 0.32\\ 
$J=16\to15$ &  36.4$\pm$  1.4 &  8.20$\pm$0.52 & 20.42$\pm$ 0.08 &  4.17$\pm$ 0.21\\

            \noalign{\smallskip}
            \hline
            \noalign{\smallskip}
            \multicolumn{5}{c} {\thco} \\
            \noalign{\smallskip}
            \hline
            \noalign{\smallskip}

$J=1\to0$ &  64.6$\pm$  0.7 & 15.51$\pm$0.29 & 19.37$\pm$ 0.02 &  3.92$\pm$ 0.06\\ 
$J=2\to1$ & 129.3$\pm$ 29.5 & 31.42$\pm$7.58 & 19.60$\pm$ 0.05 &  3.87$\pm$ 0.30\\ 
$J=3\to2$ & 104.9$\pm$  0.1 & 26.26$\pm$0.04 & 19.61$\pm$ 0.00 &  3.75$\pm$ 0.00\\ 
$J=6\to5$ &  99.9$\pm$  0.4 & 18.72$\pm$0.12 & 18.88$\pm$ 0.01 &  5.01$\pm$ 0.02\\ 
$J=13\to12$ &   8.3$\pm$  1.0 &  2.55$\pm$0.44 & 21.22$\pm$ 0.19 &  3.09$\pm$ 0.39\\

            \noalign{\smallskip}
            \hline
            \noalign{\smallskip}
            \multicolumn{5}{c} {\hcn} \\
            \noalign{\smallskip}
            \hline
            \noalign{\smallskip}

$J=1\to0$ & 104.4$\pm$  2.4 & 23.71$\pm$0.71 & 18.91$\pm$ 0.05 &  4.14$\pm$ 0.08\\ 
$J=3\to2$ & 130.4$\pm$  1.3 & 15.39$\pm$0.24 & 19.24$\pm$ 0.03 &  7.96$\pm$ 0.09\\ 
$J=4\to3$ &  86.0$\pm$  0.3 & 11.73$\pm$0.07 & 19.39$\pm$ 0.01 &  6.90$\pm$ 0.03\\ 
$J=8\to7$ &  10.7$\pm$  0.1 &  2.09$\pm$0.05 & 19.65$\pm$ 0.04 &  4.80$\pm$ 0.10\\ 

            \noalign{\smallskip}
            \hline
            \noalign{\smallskip}
            \multicolumn{5}{c} {\hthcn} \\
            \noalign{\smallskip}
            \hline
            \noalign{\smallskip}
            
$J=1\to0$ &   8.8$\pm$  0.5 &  1.98$\pm$0.20 & 18.14$\pm$ 0.11 &  4.20$\pm$ 0.34\\ 
$J=3\to2$ &  11.0$\pm$  0.6 &  2.54$\pm$0.25 & 19.38$\pm$ 0.12 &  4.10$\pm$ 0.31\\ 
$J=4\to3$ &   6.3$\pm$  0.1 &  1.64$\pm$0.07 & 19.43$\pm$ 0.04 &  3.63$\pm$ 0.11\\ 

            \noalign{\smallskip}
            \hline
            \noalign{\smallskip}
            \multicolumn{5}{c} {\hcop} \\
            \noalign{\smallskip}
            \hline
            \noalign{\smallskip}
            
$J=1\to0$ & 115.5$\pm$  0.7 & 28.44$\pm$0.28 & 18.65$\pm$ 0.01 &  3.82$\pm$ 0.03\\ 
$J=3\to2$ & 157.7$\pm$  0.6 & 22.45$\pm$0.14 & 20.38$\pm$ 0.01 &  6.60$\pm$ 0.03\\ 
$J=4\to3$ & 147.0$\pm$  0.1 & 21.48$\pm$0.03 & 20.07$\pm$ 0.01 &  6.43$\pm$ 0.01\\ 
$J=9\to8$ &  27.3$\pm$  0.4 &  5.30$\pm$0.12 & 19.60$\pm$ 0.03 &  4.84$\pm$ 0.08\\ 

            \noalign{\smallskip}
            \hline
            \noalign{\smallskip}
            \multicolumn{5}{c} {\hthcop} \\
            \noalign{\smallskip}
            \hline
            \noalign{\smallskip}
            
$J=1\to0$ &  15.8$\pm$  0.4 &  4.86$\pm$0.22 & 19.06$\pm$ 0.04 &  3.07$\pm$ 0.11\\ 
$J=3\to2$ &  26.1$\pm$  0.7 &  6.57$\pm$0.30 & 19.87$\pm$ 0.05 &  3.74$\pm$ 0.13\\ 

            \noalign{\smallskip}
            \hline
         \end{tabular}

   \end{table}
%__________________________________________________ One column table

%__________________________________________________ One column table
   \begin{table}[!htbp]
      \caption[]{Gaussian fit parameters for the spectra toward the CO-peak position ($-40''$,$+18''$).}
         \label{tab:gaussian-fit-pos2}
         \centering
         \scriptsize
         %\footnotesize
         \setlength{\tabcolsep}{3.5pt} % Default value: 6pt
         \renewcommand{\arraystretch}{1.0} % Default value: 1
         \begin{tabular}{lcccc}
            \hline\hline
	    \noalign{\smallskip}
            Transition & $\int T_{mb} dV$ & $T_{mb}^{peak}$  &  $V_0$  & $\Delta V$ (FWHM) \\
                       &   [\Kkms]        &    [K]           &  [\kms]  &    [\kms]  \\
            \noalign{\smallskip}
            \hline
            \noalign{\smallskip}
            \multicolumn{5}{c} {\twco} \\
            \noalign{\smallskip}
            \hline
            \noalign{\smallskip}

$J=1\to0$ & 128.3$\pm$  4.3 & 25.99$\pm$1.29 & 20.06$\pm$ 0.06 &  4.64$\pm$ 0.17\\ 
$J=2\to1$ & 126.8$\pm$ 29.6 & 30.99$\pm$7.93 & 19.45$\pm$ 0.06 &  3.84$\pm$ 0.40\\ 
$J=3\to2$ & 215.1$\pm$  0.3 & 40.40$\pm$0.07 & 20.09$\pm$ 0.00 &  5.00$\pm$ 0.00\\ 
$J=4\to3$ & 214.9$\pm$  2.9 & 38.98$\pm$0.86 & 19.99$\pm$ 0.03 &  5.18$\pm$ 0.09\\ 
$J=6\to5$ & 327.1$\pm$  0.5 & 65.24$\pm$0.10 & 20.60$\pm$ 0.00 &  4.71$\pm$ 0.00\\ 
$J=7\to6$ & 291.2$\pm$  0.0 & 63.87$\pm$0.03 & 20.10$\pm$ 0.00 &  4.28$\pm$ 0.00\\ 
$J=11\to10$ & 202.9$\pm$  6.2 & 39.29$\pm$1.81 & 20.34$\pm$ 0.07 &  4.85$\pm$ 0.17\\ 
$J=12\to11$ & 123.2$\pm$  9.6 & 25.09$\pm$2.52 & 19.77$\pm$ 0.17 &  4.61$\pm$ 0.29\\ 
$J=13\to12$ & 104.5$\pm$  2.6 & 20.14$\pm$0.80 & 20.46$\pm$ 0.06 &  4.88$\pm$ 0.15\\ 
$J=16\to15$ &  53.8$\pm$  3.0 & 14.50$\pm$1.06 & 19.71$\pm$ 0.05 &  3.49$\pm$ 0.16\\

            \noalign{\smallskip}
            \hline
            \noalign{\smallskip}
            \multicolumn{5}{c} {\thco} \\
            \noalign{\smallskip}
            \hline
            \noalign{\smallskip}

$J=1\to0$ &  24.7$\pm$  0.8 &  5.82$\pm$0.32 & 19.64$\pm$ 0.06 &  3.99$\pm$ 0.17\\ 
$J=2\to1$ & 109.1$\pm$  5.2 & 29.14$\pm$1.76 & 19.90$\pm$ 0.00 &  3.52$\pm$ 0.13\\ 
$J=3\to2$ &  72.0$\pm$  2.0 & 18.27$\pm$0.67 & 20.02$\pm$ 0.04 &  3.70$\pm$ 0.09\\ 
$J=6\to5$ & 134.5$\pm$  2.2 & 33.30$\pm$0.61 & 20.40$\pm$ 0.00 &  3.79$\pm$ 0.02\\ 
$J=13\to12$ &  10.6$\pm$  1.2 &  2.42$\pm$0.45 & 20.48$\pm$ 0.25 &  4.14$\pm$ 0.58\\

            \noalign{\smallskip}
            \hline
            \noalign{\smallskip}
            \multicolumn{5}{c} {\hcn} \\
            \noalign{\smallskip}
            \hline
            \noalign{\smallskip}

$J=1\to0$ &  43.8$\pm$  0.7 & 10.48$\pm$0.25 & 18.83$\pm$ 0.04 &  3.93$\pm$ 0.06\\ 
$J=3\to2$ &  45.7$\pm$  0.6 &  7.80$\pm$0.18 & 19.14$\pm$ 0.04 &  5.51$\pm$ 0.10\\ 
$J=4\to3$ &  33.0$\pm$  0.4 &  6.30$\pm$0.11 & 19.26$\pm$ 0.02 &  4.93$\pm$ 0.06\\ 
$J=8\to7$ &   5.2$\pm$  0.1 &  1.48$\pm$0.06 & 19.11$\pm$ 0.05 &  3.34$\pm$ 0.11\\ 

            \noalign{\smallskip}
            \hline
            \noalign{\smallskip}
            \multicolumn{5}{c} {\hthcn} \\
            \noalign{\smallskip}
            \hline
            \noalign{\smallskip}
            
$J=1\to0$ &   3.0$\pm$  0.5 &  0.86$\pm$0.26 & 18.01$\pm$ 0.32 &  3.27$\pm$ 0.76\\ 
$J=3\to2$ &   5.1$\pm$  0.7 &  1.25$\pm$0.29 & 18.96$\pm$ 0.27 &  3.88$\pm$ 0.69\\ 
$J=4\to3$ &   2.5$\pm$  0.1 &  0.73$\pm$0.07 & 18.95$\pm$ 0.09 &  3.31$\pm$ 0.23\\ 

            \noalign{\smallskip}
            \hline
            \noalign{\smallskip}
            \multicolumn{5}{c} {\hcop} \\
            \noalign{\smallskip}
            \hline
            \noalign{\smallskip}

$J=1\to0$ &  41.3$\pm$  1.1 & 10.72$\pm$0.44 & 19.00$\pm$ 0.05 &  3.62$\pm$ 0.11\\ 
$J=3\to2$ &  57.7$\pm$  0.7 & 12.31$\pm$0.27 & 20.08$\pm$ 0.03 &  4.40$\pm$ 0.08\\ 
$J=4\to3$ &  53.0$\pm$  0.4 & 11.17$\pm$0.16 & 19.99$\pm$ 0.02 &  4.46$\pm$ 0.05\\ 
$J=9\to8$ &  14.1$\pm$  0.6 &  3.28$\pm$0.23 & 19.35$\pm$ 0.06 &  4.04$\pm$ 0.21\\ 

            \noalign{\smallskip}
            \hline
            \noalign{\smallskip}
            \multicolumn{5}{c} {\hthcop} \\
            \noalign{\smallskip}
            \hline
            \noalign{\smallskip}
            
$J=1\to0$ &   4.4$\pm$  0.2 &  1.79$\pm$0.16 & 18.92$\pm$ 0.06 &  2.35$\pm$ 0.17\\ 
$J=3\to2$ &   7.2$\pm$  1.5 &  2.16$\pm$0.69 & 19.48$\pm$ 0.35 &  3.16$\pm$ 0.74\\ 

            \noalign{\smallskip}
            \hline
         \end{tabular}

   \end{table}
%__________________________________________________ One column table

%\newpage

%__________________________________________________ One column table
   \begin{table}[!htbp]
      \caption[]{Gaussian fit parameters for the spectra toward 
      the southern PDR position ($-60''$,$-30''$).}
         \label{tab:gaussian-fit-pos3}
         \centering
         \scriptsize
         %\footnotesize
         \setlength{\tabcolsep}{3.5pt} % Default value: 6pt
         \renewcommand{\arraystretch}{1.0} % Default value: 1
         \begin{tabular}{lcccc}
            \hline\hline
	    \noalign{\smallskip}
            Transition & $\int T_{mb} dV$ & $T_{mb}^{peak}$  &  $V_0$  & $\Delta V$ (FWHM) \\
                       &   [\Kkms]        &    [K]           &  [\kms]  &    [\kms]  \\
            \noalign{\smallskip}
            \hline
            \noalign{\smallskip}
            \multicolumn{5}{c} {\twco} \\
            \noalign{\smallskip}
            \hline
            \noalign{\smallskip}

$J=1\to0$ & 131.2$\pm$ 16.7 & 34.84$\pm$4.67 & 21.42$\pm$ 0.08 &  3.54$\pm$ 0.15\\ 
$J=2\to1$ & 195.0$\pm$ 26.7 & 45.90$\pm$6.55 & 21.28$\pm$ 0.10 &  3.99$\pm$ 0.16\\ 
$J=3\to2$ & 262.3$\pm$  0.0 & 68.82$\pm$0.01 & 21.29$\pm$ 0.00 &  3.58$\pm$ 0.00\\ 
$J=4\to3$ & 273.9$\pm$  8.1 & 65.37$\pm$2.29 & 20.79$\pm$ 0.02 &  3.94$\pm$ 0.07\\ 
$J=6\to5$ & 316.3$\pm$  3.6 & 75.79$\pm$0.96 & 21.82$\pm$ 0.01 &  3.92$\pm$ 0.02\\ 
$J=7\to6$ & 352.5$\pm$  9.4 & 80.00$\pm$2.56 & 21.45$\pm$ 0.03 &  4.14$\pm$ 0.07\\ 
$J=11\to10$ & 163.9$\pm$  2.1 & 31.37$\pm$0.61 & 20.83$\pm$ 0.04 &  4.91$\pm$ 0.07\\ 
$J=12\to11$ & 102.4$\pm$  3.6 & 18.00$\pm$0.97 & 19.68$\pm$ 0.10 &  5.35$\pm$ 0.21\\ 
$J=13\to12$ &  76.2$\pm$  9.1 & 14.60$\pm$2.47 & 20.46$\pm$ 0.28 &  4.91$\pm$ 0.58\\ 
$J=16\to15$ &  11.3$\pm$  1.5 &  2.51$\pm$0.48 & 20.15$\pm$ 0.32 &  4.27$\pm$ 0.58\\

            \noalign{\smallskip}
            \hline
            \noalign{\smallskip}
            \multicolumn{5}{c} {\thco} \\
            \noalign{\smallskip}
            \hline
            \noalign{\smallskip}

$J=1\to0$ &  65.2$\pm$  1.8 & 12.77$\pm$0.40 & 19.60$\pm$ 0.06 &  4.80$\pm$ 0.07\\ 
$J=2\to1$ & 135.8$\pm$  3.0 & 40.13$\pm$1.17 & 21.20$\pm$ 0.00 &  3.18$\pm$ 0.06\\ 
$J=3\to2$ & 151.9$\pm$  1.2 & 45.56$\pm$0.47 & 21.20$\pm$ 0.00 &  3.13$\pm$ 0.02\\ 
$J=6\to5$ & 169.8$\pm$  0.2 & 45.32$\pm$0.11 & 21.24$\pm$ 0.00 &  3.52$\pm$ 0.01\\ 
$J=13\to12$ &   3.7$\pm$  1.3 &  1.75$\pm$0.64 & 18.95$\pm$ 0.37 &  2.00$\pm$ 0.00\\

            \noalign{\smallskip}
            \hline
            \noalign{\smallskip}
            \multicolumn{5}{c} {\hcn} \\
            \noalign{\smallskip}
            \hline
            \noalign{\smallskip}

$J=1\to0$ &  70.5$\pm$  0.8 & 14.84$\pm$0.23 & 19.33$\pm$ 0.03 &  4.47$\pm$ 0.05\\ 
$J=3\to2$ &  74.0$\pm$  0.9 & 11.63$\pm$0.19 & 19.18$\pm$ 0.03 &  5.98$\pm$ 0.06\\ 
$J=4\to3$ &  43.9$\pm$  0.3 &  8.04$\pm$0.09 & 19.81$\pm$ 0.01 &  5.13$\pm$ 0.04\\ 
$J=8\to7$ &   3.0$\pm$  0.3 &  0.69$\pm$0.11 & 19.23$\pm$ 0.22 &  4.16$\pm$ 0.47\\ 

            \noalign{\smallskip}
            \hline
            \noalign{\smallskip}
            \multicolumn{5}{c} {\hthcn} \\
            \noalign{\smallskip}
            \hline
            \noalign{\smallskip}
            
$J=1\to0$ &   7.3$\pm$  0.6 &  0.99$\pm$0.13 & 19.17$\pm$ 0.27 &  6.94$\pm$ 0.72\\ 
$J=3\to2$ &   4.2$\pm$  0.7 &  1.13$\pm$0.25 & 19.46$\pm$ 0.32 &  3.56$\pm$ 0.55\\ 
$J=4\to3$ &   1.6$\pm$  0.1 &  0.38$\pm$0.06 & 19.87$\pm$ 0.20 &  3.95$\pm$ 0.44\\ 

            \noalign{\smallskip}
            \hline
            \noalign{\smallskip}
            \multicolumn{5}{c} {\hcop} \\
            \noalign{\smallskip}
            \hline
            \noalign{\smallskip}
            
$J=1\to0$ &  70.2$\pm$  0.5 & 15.11$\pm$0.17 & 19.76$\pm$ 0.02 &  4.37$\pm$ 0.03\\ 
$J=3\to2$ &  79.2$\pm$  0.5 & 15.96$\pm$0.13 & 20.44$\pm$ 0.01 &  4.66$\pm$ 0.02\\ 
$J=4\to3$ &  62.7$\pm$  0.2 & 12.38$\pm$0.06 & 20.50$\pm$ 0.01 &  4.76$\pm$ 0.01\\ 
$J=9\to8$ &   6.3$\pm$  1.2 &  1.47$\pm$0.46 & 19.82$\pm$ 0.35 &  4.05$\pm$ 0.97\\ 

            \noalign{\smallskip}
            \hline
            \noalign{\smallskip}
            \multicolumn{5}{c} {\hthcop} \\
            \noalign{\smallskip}
            \hline
            \noalign{\smallskip}
            
$J=1\to0$ &   6.3$\pm$  0.3 &  1.51$\pm$0.12 & 19.69$\pm$ 0.11 &  3.96$\pm$ 0.23\\ 
$J=3\to2$ &   8.4$\pm$  0.9 &  2.04$\pm$0.31 & 19.87$\pm$ 0.22 &  3.91$\pm$ 0.42\\ 

            \noalign{\smallskip}
            \hline
         \end{tabular}

   \end{table}
%__________________________________________________ One column table

%__________________________________________________ One column table
   \begin{table}[!htbp]
      \caption[]{Gaussian fit parameters for the spectra toward 
      the M17-W position ($-130''$,$+30''$).}
         \label{tab:gaussian-fit-pos4}
         \centering
         \scriptsize
         %\footnotesize
         \setlength{\tabcolsep}{3.5pt} % Default value: 6pt
         \renewcommand{\arraystretch}{1.0} % Default value: 1
         \begin{tabular}{lcccc}
            \hline\hline
	    \noalign{\smallskip}
            Transition & $\int T_{mb} dV$ & $T_{mb}^{peak}$  &  $V_0$  & $\Delta V$ (FWHM) \\
                       &   [\Kkms]        &    [K]           &  [\kms]  &    [\kms]  \\
            \noalign{\smallskip}
            \hline
            \noalign{\smallskip}
            \multicolumn{5}{c} {\twco} \\
            \noalign{\smallskip}
            \hline
            \noalign{\smallskip}

$J=1\to0$ & 426.2$\pm$  5.8 & 44.94$\pm$1.28 & 20.81$\pm$ 0.09 &  8.91$\pm$ 0.22\\ 
$J=2\to1$ & 476.0$\pm$ 26.1 & 61.97$\pm$7.58 & 21.57$\pm$ 0.29 &  7.22$\pm$ 0.79\\ 
$J=3\to2$ & 678.3$\pm$  2.1 & 64.33$\pm$0.45 & 21.59$\pm$ 0.03 &  9.91$\pm$ 0.06\\ 
$J=4\to3$ & 648.4$\pm$  3.6 & 65.30$\pm$0.84 & 21.21$\pm$ 0.04 &  9.33$\pm$ 0.11\\ 
$J=6\to5$ & 562.7$\pm$  1.6 & 66.06$\pm$0.42 & 21.87$\pm$ 0.02 &  8.00$\pm$ 0.05\\ 
$J=7\to6$ & 550.4$\pm$  2.7 & 64.37$\pm$0.49 & 21.60$\pm$ 0.02 &  8.03$\pm$ 0.05\\ 
$J=11\to10$ & 157.1$\pm$  3.1 & 25.05$\pm$0.80 & 21.24$\pm$ 0.06 &  5.89$\pm$ 0.15\\ 
$J=12\to11$ &  77.9$\pm$  2.4 & 13.81$\pm$0.67 & 20.11$\pm$ 0.08 &  5.30$\pm$ 0.19\\ 
$J=13\to12$ &  73.9$\pm$  6.0 & 11.52$\pm$1.59 & 20.97$\pm$ 0.22 &  6.03$\pm$ 0.68\\ 
$J=16\to15$ &   5.1$\pm$  0.8 &  1.93$\pm$0.47 & 21.39$\pm$ 0.23 &  2.50$\pm$ 0.43\\

            \noalign{\smallskip}
            \hline
            \noalign{\smallskip}
            \multicolumn{5}{c} {\thco} \\
            \noalign{\smallskip}
            \hline
            \noalign{\smallskip}

$J=1\to0$ & 112.4$\pm$  0.6 & 17.20$\pm$0.16 & 20.47$\pm$ 0.02 &  6.14$\pm$ 0.04\\ 
$J=2\to1$ & 342.0$\pm$  3.2 & 46.82$\pm$0.68 & 20.61$\pm$ 0.03 &  6.86$\pm$ 0.07\\ 
$J=3\to2$ & 290.9$\pm$  9.3 & 39.04$\pm$1.49 & 20.76$\pm$ 0.09 &  7.00$\pm$ 0.14\\ 
$J=6\to5$ & 147.8$\pm$  0.1 & 33.95$\pm$0.02 & 20.29$\pm$ 0.00 &  4.09$\pm$ 0.00\\ 
$J=13\to12$ &   1.1$\pm$  0.3 &  0.28$\pm$0.06 & 21.00$\pm$ 6.30 &  4.0$\pm$ 1.20\\

            \noalign{\smallskip}
            \hline
            \noalign{\smallskip}
            \multicolumn{5}{c} {\hcn} \\
            \noalign{\smallskip}
            \hline
            \noalign{\smallskip}

$J=1\to0$ &  45.9$\pm$  1.9 & 12.30$\pm$0.62 & 19.22$\pm$ 0.04 &  3.51$\pm$ 0.10\\ 
$J=3\to2$ & 111.3$\pm$  6.8 & 14.49$\pm$0.97 & 20.76$\pm$ 0.06 &  7.22$\pm$ 0.20\\ 
$J=4\to3$ &  49.4$\pm$  0.7 &  7.36$\pm$0.14 & 21.15$\pm$ 0.02 &  6.31$\pm$ 0.07\\ 
$J=8\to7$ &   1.5$\pm$  0.3 &  0.31$\pm$0.10 & 20.25$\pm$ 0.58 &  4.82$\pm$ 1.10\\

            \noalign{\smallskip}
            \hline
            \noalign{\smallskip}
            \multicolumn{5}{c} {\hthcn} \\
            \noalign{\smallskip}
            \hline
            \noalign{\smallskip}
            
$J=1\to0$ &  12.6$\pm$  1.0 &  0.95$\pm$0.12 & 19.48$\pm$ 0.50 & 12.52$\pm$ 1.21\\ 
$J=3\to2$ &   8.2$\pm$  1.5 &  0.80$\pm$0.23 & 20.85$\pm$ 0.85 &  9.70$\pm$ 2.13\\ 
$J=4\to3$ &   1.9$\pm$  0.2 &  0.25$\pm$0.04 & 21.33$\pm$ 0.37 &  7.14$\pm$ 0.79\\

            \noalign{\smallskip}
            \hline
            \noalign{\smallskip}
            \multicolumn{5}{c} {\hcop} \\
            \noalign{\smallskip}
            \hline
            \noalign{\smallskip}
            
$J=1\to0$ &  91.9$\pm$  2.3 & 11.65$\pm$0.36 & 20.50$\pm$ 0.04 &  7.41$\pm$ 0.13\\ 
$J=3\to2$ & 154.7$\pm$  9.1 & 25.19$\pm$1.61 & 21.54$\pm$ 0.04 &  5.77$\pm$ 0.14\\ 
$J=4\to3$ &  89.4$\pm$  0.8 & 14.43$\pm$0.17 & 21.60$\pm$ 0.02 &  5.82$\pm$ 0.04\\ 
$J=9\to8$ &   9.5$\pm$  1.8 &  0.91$\pm$0.24 & 21.11$\pm$ 0.75 &  9.90$\pm$ 1.83\\ 

            \noalign{\smallskip}
            \hline
            \noalign{\smallskip}
            \multicolumn{5}{c} {\hthcop} \\
            \noalign{\smallskip}
            \hline
            \noalign{\smallskip}
            
$J=1\to0$ &   7.2$\pm$  0.2 &  1.17$\pm$0.07 & 21.11$\pm$ 0.11 &  5.83$\pm$ 0.25\\ 
$J=3\to2$ &   7.0$\pm$  1.0 &  1.51$\pm$0.33 & 22.15$\pm$ 0.35 &  4.39$\pm$ 0.69\\

            \noalign{\smallskip}
            \hline
         \end{tabular}

   \end{table}
%__________________________________________________ One column table

\newpage

\section{Alternative solution for the LSED fit}\label{apx:one-comp}

\begin{figure*}[!ht]
 \centering
 \begin{tabular}{ccc}
  \hspace{-0.3cm}\epsfig{file=figs/M17SW_CO_SED_2comps_final_fit_wBg_-65_+31.eps,angle=0,width=0.33\linewidth} &
  
  \hspace{-0.3cm}\epsfig{file=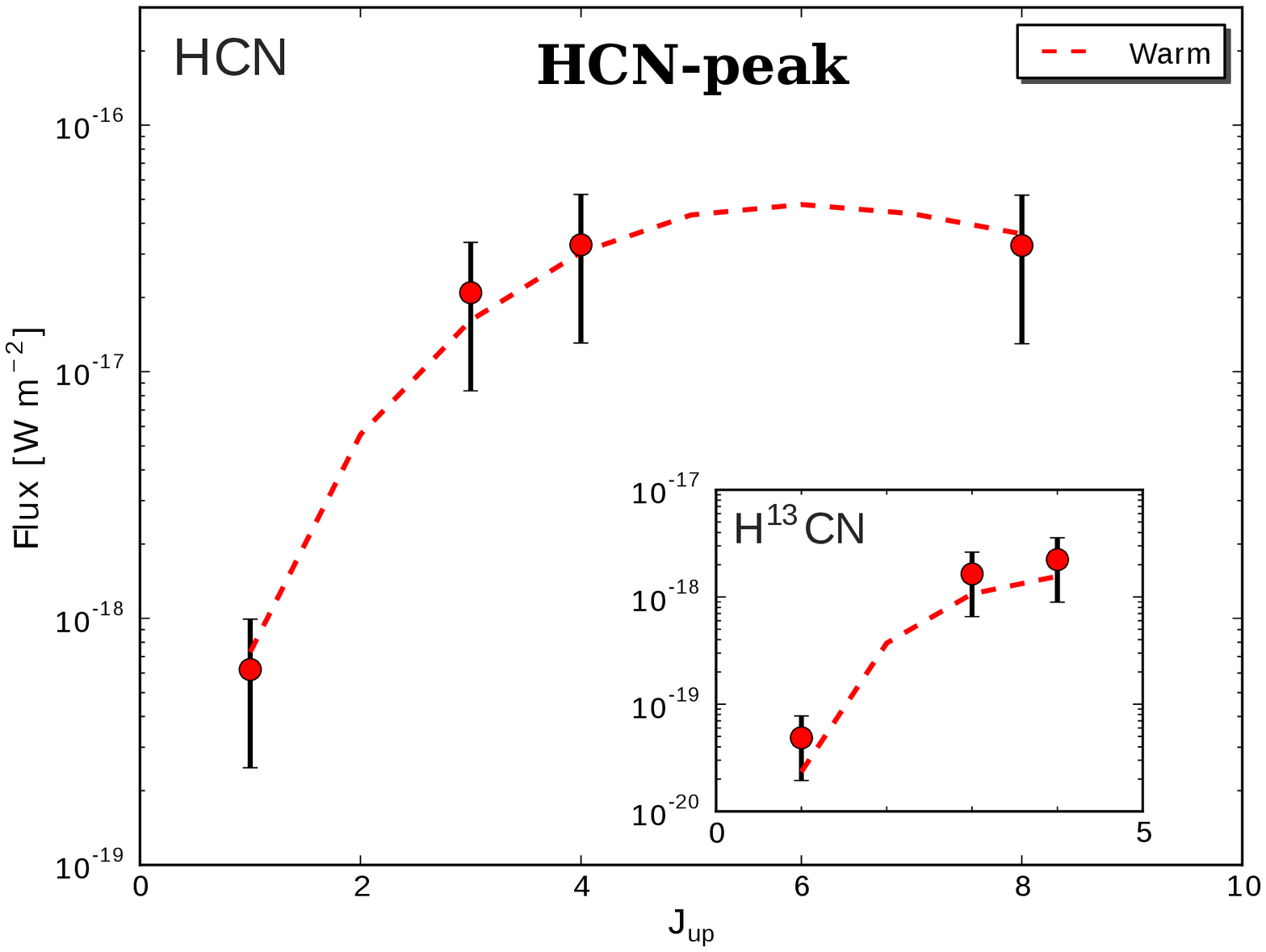,angle=0,width=0.33\linewidth} &
  
  \hspace{-0.3cm}\epsfig{file=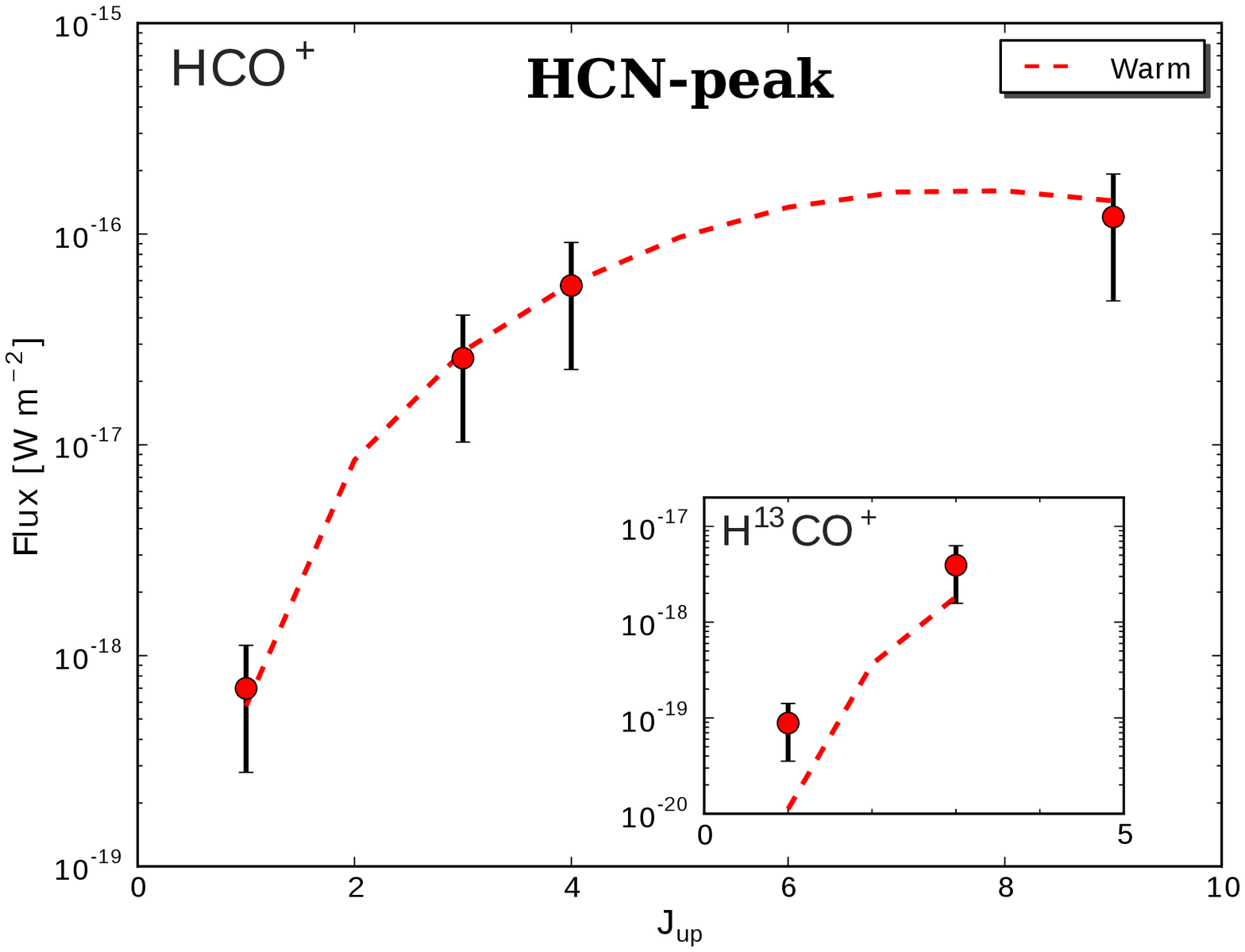,angle=0,width=0.33\linewidth} 
 \end{tabular}

  \caption{\footnotesize{Two component fit of the line spectral energy distribution of the CO, \hcn, and \hcop\ 
  species, for the spectra at position ($-65''$,$+31''$) toward the HCN-peak. 
  The fit of the $^{13}$C bearing isotopologues is shown in the 
  insets. The cold and warm components are shown in dashed and dotted lines, respectively. The error bars are as 
  in Fig.~\ref{fig:average-SED-fit}.}}

  \label{fig:pos1-SED-fit-WarmComp}
\end{figure*}

The \hcn\ and \hcop\ LSEDs at the position of the peak 
\hcn~$J=8\to7$ emission, at offset position ($-65''$,$+31''$), can be fit 
using only the ambient conditions found for the warm component of the \twco\ 
LSED. The parameters of the model
are presented in Table~\ref{tab:pos1-LSED-fit-WarmComp} and the LSED fit
is shown in Fig.~\ref{fig:pos1-SED-fit-WarmComp}. 
Here we used the same isotope ratio of 50, but lower ratios of 30 and 20 would 
lead to a better fit of the \hthcn\ and \hthcop\ lines, respectively.

\begin{figure*}[!ht]
 \centering
 \begin{tabular}{ccc}
  \hspace{-0.3cm}\epsfig{file=figs/M17SW_CO_SED_2comps_final_fit_wBg_-60_-30.eps,angle=0,width=0.33\linewidth} &
  
  \hspace{-0.3cm}\epsfig{file=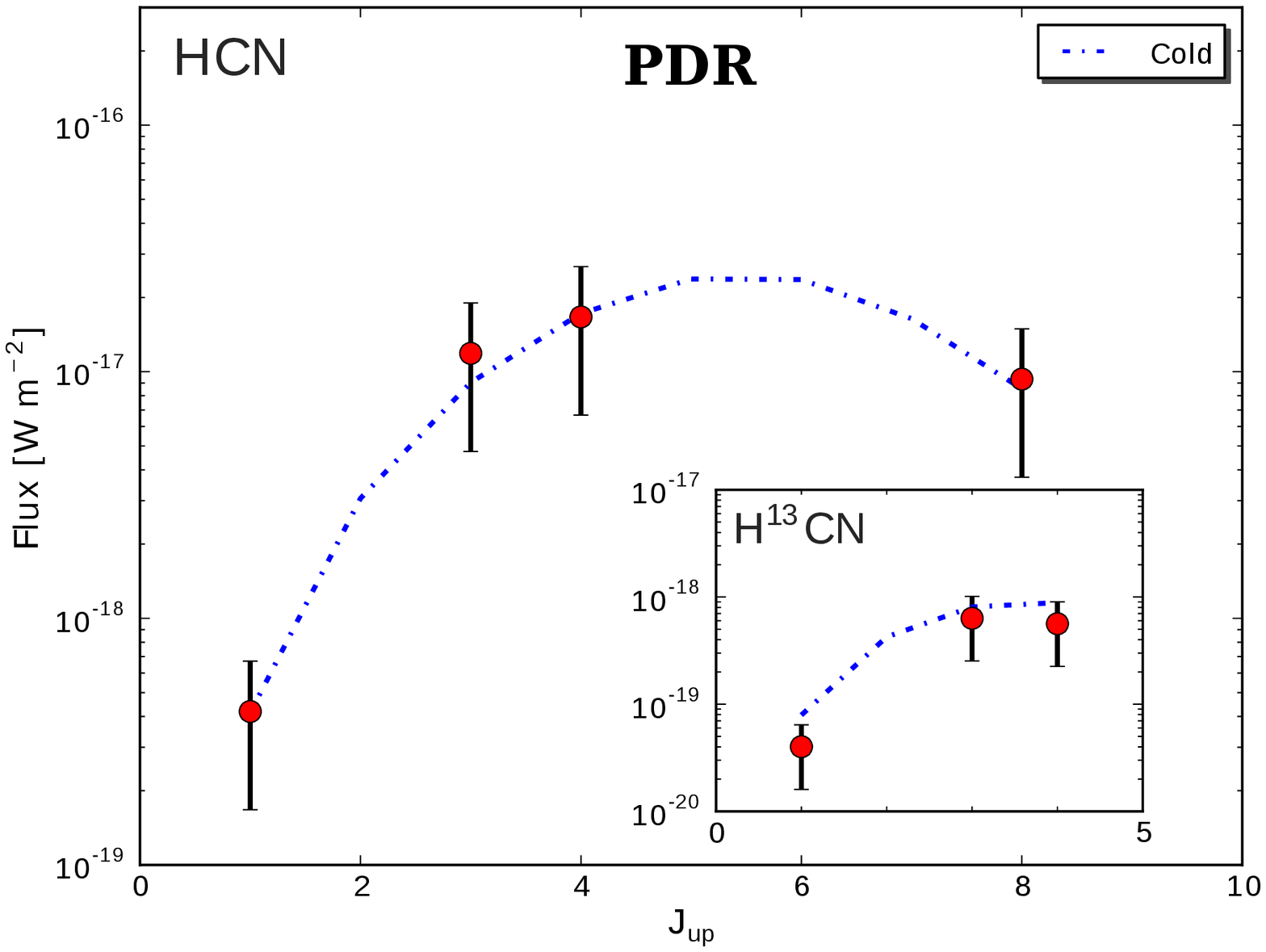,angle=0,width=0.33\linewidth} &
  
  \hspace{-0.3cm}\epsfig{file=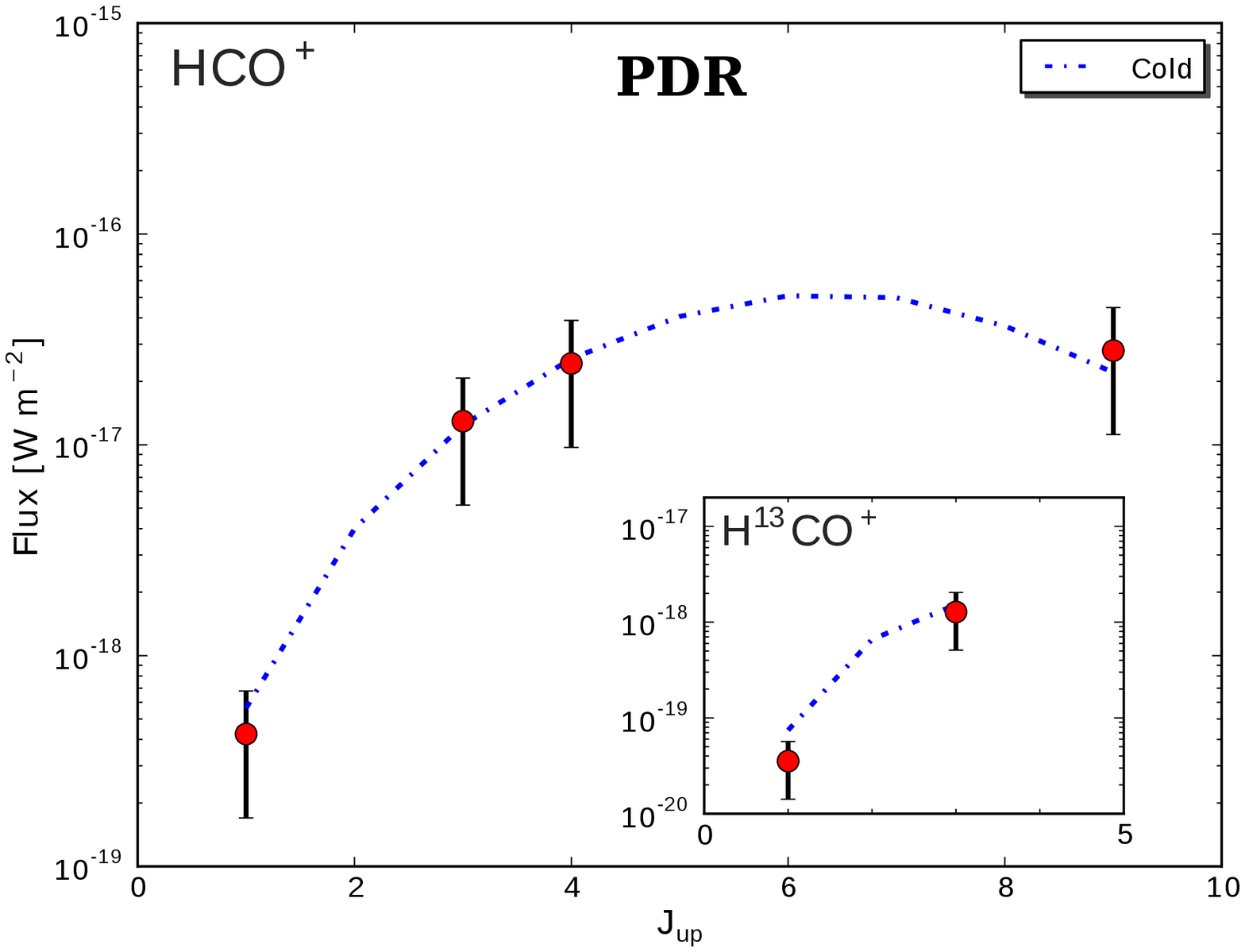,angle=0,width=0.33\linewidth} 
 \end{tabular}

  \caption{\footnotesize{Two component fit of the line spectral energy distribution of the CO, \hcn, and \hcop\ 
  species, for the spectra at position ($-60''$,$-30''$) toward the southern PDR. 
  The fit of the $^{13}$C bearing isotopologues is shown in the 
  insets. The cold and warm components are shown in dashed and dotted lines, respectively. The error bars are as 
  in Fig.~\ref{fig:average-SED-fit}.}}

  \label{fig:pos3-SED-fit-ColdComp}
\end{figure*}

The LSED of the \hcn\ and \hcop\ lines observed deeper into the PDR at offset 
position ($-60''$,$-30''$) can also be fit using just the temperature and density
of the cold component found for the \twco\ LSED, but it requires a larger isotope
ratio of 75 in order to fit the \hthcn\ and \hthcop\ lines. The parameters of the model
are presented in Table~\ref{tab:pos3-LSED-fit-ColdComp} and the LSED fit
is shown in Fig.~\ref{fig:pos3-SED-fit-ColdComp}. 

\newpage
			
%-------------------------------------------------------------
%---------------------------------------------------------------

%-------------------------------------------------------------
%---------------------------------------------------------------

%\newpage

%__________________________________________________ One column table
   \begin{table}[h]
      \caption[]{LSED fit parameters\tablefootmark{a} for the spectra toward 
      the HCN-peak position ($-65''$,$+31''$).}
         \label{tab:pos1-LSED-fit-WarmComp}
         \centering
         \scriptsize
         %\footnotesize
         \setlength{\tabcolsep}{3.5pt} % Default value: 6pt
         \renewcommand{\arraystretch}{1.0} % Default value: 1
         \begin{tabular}{lcccc}
            \hline\hline
	    \noalign{\smallskip}
            Parameter & CO & HCN  &  HCO$^+$  \\
            \noalign{\smallskip}
            \hline
            \noalign{\smallskip}

$\Phi_{cold}(^{12}{\rm C})$  &   0.90 $\pm$  0.09  &     &    \\ 
$n_{cold}(\rm H_2)$ [cm$^{-3}$]  &   4.50 $\pm$  0.43  &     &    \\ 
$T_{cold}$ [K]  &  40.00 $\pm$  4.52  &    &   \\ 
$N_{cold}$ [cm$^{-2}$]  &  19.40 $\pm$  1.10  &    &   \\ 
 &  &  &  \\ 
$\Phi_{warm}(^{12}{\rm C})$  &   0.35 $\pm$  0.04  &   0.35 $\pm$  0.04  &   0.35 $\pm$  0.02 \\ 
$n_{warm}(\rm H_2)$ [cm$^{-3}$]  &   6.00 $\pm$  0.63  &   6.00 $\pm$  0.42  &   6.00 $\pm$  0.32 \\ 
$T_{warm}$ [K]  &  130.00 $\pm$  9.62  &  130.00 $\pm$ 14.11  &  130.00 $\pm$ 15.78 \\ 
$N_{warm}$ [cm$^{-2}$]  &  18.40 $\pm$  0.33  &  15.10 $\pm$  0.39  &  14.90 $\pm$  0.34 \\ 
 &  &  &  \\ 
$\Phi_{cold}(^{13}{\rm C})$  &   0.90 $\pm$  0.10  &   &  \\ 
$\Phi_{warm}(^{13}{\rm C})$  &   0.35 $\pm$  0.03  &   0.35 $\pm$  0.04  &   0.35 $\pm$  0.03 \\ 
 &  &  &  \\ 
%$^{12}$C/$^{13}$C  &  50.00 $\pm$  3.76  &  50.00 $\pm$  6.85  &  50.00 $\pm$  5.86 \\ 
% &  &  &  \\ 
$\Delta V(^{12}{\rm C})$ [km s$^-1$]  &   4.60   &   7.50   &   6.00  \\ 
$\Delta V(^{13}{\rm C})$ [km s$^-1$]  &   3.50   &   3.90   &   3.70  \\  
		
            \noalign{\smallskip}
            \hline
         \end{tabular}
         
         \tablefoot{
         \tablefoottext{a}{The density and column density values are given in $log_{10}$ scale.}
         }         

   \end{table}
%__________________________________________________ One column table

%__________________________________________________ One column table
   \begin{table}[h]
      \caption[]{LSED fit parameters\tablefootmark{a} for the spectra toward 
      the southern PDR position ($-60''$,$-30''$).}
         \label{tab:pos3-LSED-fit-ColdComp}
         \centering
         \scriptsize
         %\footnotesize
         \setlength{\tabcolsep}{3.5pt} % Default value: 6pt
         \renewcommand{\arraystretch}{1.0} % Default value: 1
         \begin{tabular}{lcccc}
            \hline\hline
	    \noalign{\smallskip}
            Parameter & CO & HCN  &  HCO$^+$  \\
            \noalign{\smallskip}
            \hline
            \noalign{\smallskip}

$\Phi_{cold}(^{12}{\rm C})$  &   0.90 $\pm$  0.07  &   0.35 $\pm$  0.04  &   0.35 $\pm$  0.04 \\ 
$n_{cold}(\rm H_2)$ [cm$^{-3}$]  &   5.50 $\pm$  0.60  &   5.30 $\pm$  0.51  &   5.30 $\pm$  0.55 \\ 
$T_{cold}$ [K]  &  60.00 $\pm$  8.21  &  60.00 $\pm$  7.01  &  60.00 $\pm$  6.16 \\ 
$N_{cold}$ [cm$^{-2}$]  &  18.80 $\pm$  0.67  &  16.00 $\pm$  0.56  &  15.50 $\pm$  0.53 \\ 
 &  &  &  \\ 
$\Phi_{warm}(^{12}{\rm C})$  &   0.20 $\pm$  0.03  &     &    \\ 
$n_{warm}(\rm H_2)$ [cm$^{-3}$]  &   5.80 $\pm$  0.52  &     &    \\ 
$T_{warm}$ [K]  &  110.00 $\pm$ 10.18  &    &   \\ 
$N_{warm}$ [cm$^{-2}$]  &  18.60 $\pm$  1.37  &    &   \\ 
 &  &  &  \\ 
$\Phi_{cold}(^{13}{\rm C})$  &   0.90 $\pm$  0.12  &   0.35 $\pm$  0.04  &   0.35 $\pm$  0.04 \\ 
$\Phi_{warm}(^{13}{\rm C})$  &   0.20 $\pm$  0.02  &     &    \\ 
 &  &  &  \\ 
%$^{12}$C/$^{13}$C  &  50.00 $\pm$  5.01  &  75.00 $\pm$  8.29  &  75.00 $\pm$  8.88 \\ 
% &  &  &  \\ 
$\Delta V(^{12}{\rm C})$ [km s$^-1$]  &   4.30   &   5.00   &   4.50  \\ 
$\Delta V(^{13}{\rm C})$ [km s$^-1$]  &   4.30   &   3.80   &   3.90  \\     
		
            \noalign{\smallskip}
            \hline
         \end{tabular}
         
         \tablefoot{
         \tablefoottext{a}{The density and column density values are given in $log_{10}$ scale.}
         }         

   \end{table}
%__________________________________________________ One column table

\end{appendix}

\end{document}